\documentclass[]{aa}  
\pdfoutput=1
\usepackage{graphicx}
\usepackage{caption}
\usepackage{subcaption}
\usepackage{epsf}
\usepackage{longtable,lscape}
\usepackage{dcolumn}
\usepackage{booktabs} 
\usepackage{url}
\usepackage{float}
\usepackage{microtype}
\usepackage[varg]{txfonts}
\usepackage{natbib,twoopt} 
\usepackage{array,multirow}
\usepackage{xcolor}
\def\apjl{ApJ}%
\def\apjs{ApJS}%
\def\ao{Appl.~Opt.}%
\def\aap{A\&A}%
\bibpunct{(}{)}{;}{a}{}{,} % citation like A&A style

\let\oldhat\hat

\renewcommand{\hat}[1]{\oldhat{\mathbf{#1}}}
\newcommand{\kms}{km\,${\rm s}^{-1}$}
\newcommand{\smy}{$[M_\odot\,{\rm yr}^{-1}]$\,}

\newcommand{\myr}{\mbox{$M_\odot\,{\rm yr}^{-1}$}}

\newcommand{\lum}{erg\,s$^{-1}$}

\newcommand\Tstrut{\rule{0pt}{2.6ex}}         % = `top' strut
\begin{document}

   \title{The Wolf-Rayet binaries of the nitrogen sequence in the Large Magellanic Cloud}

   \subtitle{Spectroscopy, orbital analysis, formation, and evolution}

   \author{T. Shenar\inst{1} 
          \and D.\ P.\ Sablowski\inst{2}   
          \and R.\ Hainich\inst{3}
          \and H.\ Todt\inst{3}                            
          \and A.\ F.\ J.\ Moffat\inst{4}   
          \and L.\ M.\ Oskinova\inst{3}                 
          \and V.\ Ramachandran\inst{3} 
          \and H.\ Sana\inst{1}                     
          \and A.A.C Sander\inst{5}       
          \and O.\ Schnurr\inst{6}    
          \and N.\ St-Louis\inst{4}             
          \and D.\ Vanbeveren\inst{7}   
          \and Y.\ G\"otberg\inst{8}            
          \and W.-R.\ Hamann\inst{3}               
          }

   \institute{\inst{1}{Institute of Astrophysics, KU Leuven, Celestijnlaan 200D, 3001 Leuven, Belgium}\\     
              \inst{2}{Leibniz-Institut f\"ur Astrophysik Potsdam, An der Sternwarte 16, 14482 Potsdam, Germany}\\      
              \inst{3}{Institut f\"ur Physik und Astronomie, Universit\"at Potsdam,
                       Karl-Liebknecht-Str. 24/25, D-14476 Potsdam, Germany}\\                        
              \inst{4}{D\'epartement de physique and Centre de Recherche en Astrophysique 
                du Qu\'ebec (CRAQ), Universit\'e de Montr\'eal, C.P. 6128, Succ.~Centre-Ville, Montr\'eal, Qu\'ebec, H3C 3J7, Canada}\\                  
              \inst{5}{Armagh Observatory, College Hill, BT61 9DG, Armagh, Northern Ireland}        \\
              \inst{6}{Cherenkov Telescope Array Observatory gGmbH, Via Piero Gobetti 93/3, I-40126 Bologna, Italy             }        \\                      
              \inst{7}{Astronomy and Astrophysics Research Group, Vrije Universiteit Brussel, Pleinlaan 2, 1050, Brussels, Belgium}\\
              \inst{8}{The observatories of the Carnegie institution for science, 813 Santa Barbara St., Pasadena, CA 91101, USA}\\
              \email{tomer.shenar@kuleuven.be}   
              }
   \date{Received ? / Accepted ?}

%-------------------  Abstract --------------------

\abstract
% context heading (optional) 
{ 
Massive Wolf-Rayet (WR) stars dominate the radiative and mechanical  
energy budget of galaxies and  probe a critical phase in the evolution  
of massive stars prior to  core-collapse. It is not known  
whether core He-burning WR  stars (classical WR, cWR) form predominantly through 
wind-stripping (\mbox{w-WR}) or binary stripping (\mbox{b-WR}). With  
spectroscopy of WR binaries 
so-far largely avoided due to its complexity,  our study focuses on the 44 WR binaries / binary candidates 
of the Large Magellanic Cloud (LMC, metallicity $Z \approx 0.5\,Z_\odot$), identified on the basis 
of radial velocity variations, composite spectra, or high X-ray luminosities.
}
% aims heading (mandatory)
{
Relying on a diverse spectroscopic database, we aim to derive the physical and orbital parameters of our targets, confronting
evolution models of evolved massive stars at sub-solar metallicity, 
and constraining the impact of binary interaction in forming them.
}
% methods heading (mandatory)
{
Spectroscopy is performed using the Potsdam Wolf-Rayet (PoWR) code and cross-correlation techniques. 
Disentanglement is performed using the code \texttt{Spectangular} or the shift-and-add algorithm.
Evolutionary status is interpreted using the Binary Population 
and Spectral Synthesis (BPASS) code, exploring binary interaction and chemically-homogeneous evolution.
}
% results heading (mandatory)
{
 Among our sample, 28/44 objects show composite spectra and are analysed as such.  Additional five targets show 
periodically-moving WR primaries but no detected companions (SB1), with two (BAT99~99 and 112) being potential WR + compact-object candidates 
due to their high X-ray luminosities. 
We cannot confirm the binary nature of the remaining  11 candidates. 
About 2/3 of the WN components in binaries  
are identified as cWR, and 1/3 as hydrogen-burning WR stars.
metallicity-dependent mass-loss recipes are established and broadly agree with those recently derived for single WN stars, with 
so-called WN3/O3 stars being clear outliers. We estimate that $45{\pm}30\%$ of the cWR stars in our sample have interacted with a companion 
via mass-transfer. However, only
${\approx}12{\pm}7\%$ of the cWR stars in our sample 
naively appear to have formed purely due to stripping via a companion (12\% b-WR). Assuming that 
apparently-single WR stars truly formed as single stars, 
this makes $\approx 4\%$ of the whole LMC WN population, about ten times less than expected.
No obvious differences in the properties of single and binary WN stars 
are apparent, whose luminosities 
extend down to $\log L {\approx}5.2\,[L_\odot]$.
With the exception of a few systems (BAT99\,19, 49, and 103), the equatorial rotational velocities of the OB-type companions 
are moderate ($v_\text{eq} \lesssim 250\,$\kms) and challenge standard formalisms of angular-momentum accretion.
For most objects, chemically-homogeneous evolution can be rejected for the secondary, but not for the 
WR progenitor.
}
{
No obvious dichotomy in the locations of apparently-single and binary WN stars on the 
Hertzsprung-Russell diagram is apparent. According to commonly used 
stellar evolution models (BPASS, Geneva), most apparently-single WN stars could not have formed as single 
stars, implying that they were stripped by an undetected companion. Otherwise, it must follow that pre-WR mass-loss/mixing 
(e.g., during the red supergiant phase) are strongly 
underestimated in standard stellar evolution models.
}
% }
\keywords{stars: massive -- stars: Wolf-Rayet -- Magellanic Clouds -- Binaries: close -- Binaries: spectroscopic -- Stars: evolution}

\maketitle

\section{Introduction}
\label{sec:introduction}
Through their stellar winds, intense radiation, and supernova (SN) explosions, 
massive stars ($M_\text{i} \gtrsim 8\,M_\odot$) dominate the energy budget of their host galaxies. 
Among them, massive Wolf-Rayet (WR) stars define a spectral class 
of stars with emission-dominated spectra that 
are physically characterized by strong, radiatively 
driven winds \citep[see][for a review]{Crowther2007}. They are subdivided in three main flavours: the nitrogen-sequence (WN),  
the carbon-sequence (WC), or the very rare oxygen-sequence (WO), 
depending on whether their atmospheres are N-rich (CNO cycle products),  
or C / O-rich (He-burning products). Most known WR stars are classical WR stars\footnote{For example, $\approx 90\%$ of the WR 
stars in the Galaxy are cWR stars, as can be estimated from the number of WC stars and hydrogen free/depleted WN stars - see 
\citealt{Crowther2007, Hamann2019}.} (cWR), defined as evolved, core He-burning (or rarely C-burning) 
WR stars.  However, very massive stars can already appear as WR stars on the main sequence \citep{deKoter1997}.
As immediate progenitors of black holes (BHs) and neutron stars (NS), the attributes of WR stars largely determine 
observed properties of SN explosions and gravitational-wave (GW) detections arising from the merging of compact 
objects. Studying WR stars is hence essential for understanding the evolution of massive 
stars \citep[e.g.,][]{Hamann2006, Tramper2015, Shenar2016, Sander2019}
, the energy budget of galaxies \citep[e.g.,][]{Doran2013, Ramachandran2018}, 
the upper-mass limit of stars \citep[e.g.,][]{Bestenlehner2011, Shenar2017, Tehrani2019}, and 
the properties of compact objects and SNe \citep[e.g.,][]{Woosley2002, Langer2012, DeMink2014, Marchant2016, Hainich2018b}.  
Despite this, their formation, especially in the context of binary interaction, is still considered poorly understood.

\begin{figure}[!htb]
\centering
  \includegraphics[width=0.5\textwidth]{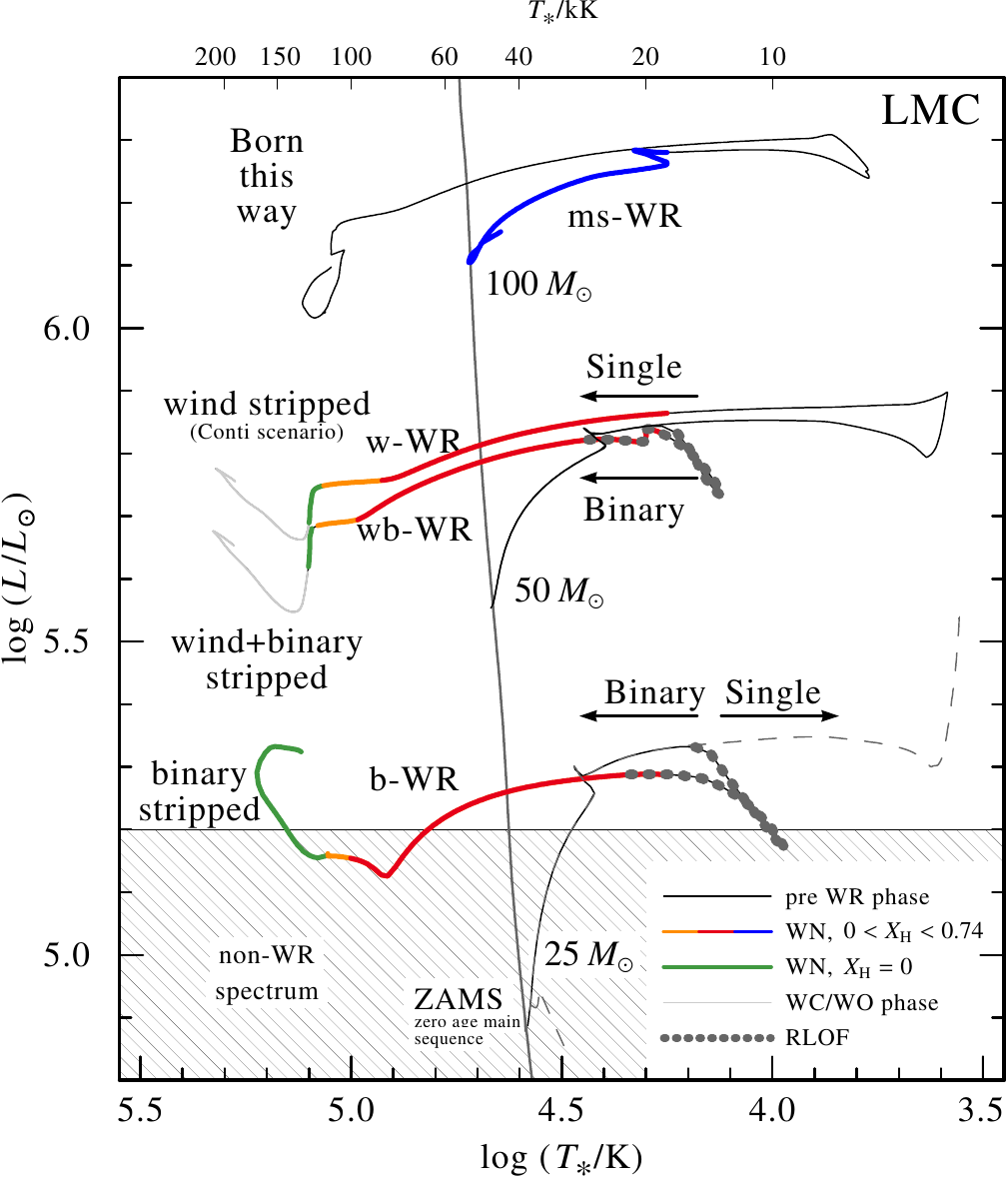}
  \caption{Illustration of the ms-WR, w-WR, wb-WR, and w-WR formation channels of WR stars. 
  Shown are evolution tracks calculated at a metallicity $Z=0.008$ (${\approx}Z_\text{LMC}$) 
  with the BPASS code for single stars with initial masses $M_\text{i} = 100, 50,$ and 
  $25\,M_\odot$ (upper, middle, and dashed lower tracks, respectively). Also plotted are BPASS binary evolution 
  tracks with $M_\text{i} = 50\,M_\odot, P_\text{i} = 25\,d$, $q_\text{i} = 0.7$ (upper) and 
  $M_\text{i} = 25\,M_\odot, P_\text{i} = 25\,d$, $q_\text{i} = 0.7$ (lower). The colours correspond 
  to surface hydrogen mass-fractions (blue: $0.4 < X_\text{H} < 0.74$, red: $0.2 < X_\text{H} < 0.4$, 
  orange: $0.05< X_\text{H} < 0.2$, green: $X_\text{H} < 0.05$).
  Empirically, stars found in the dashed region are 
  not expected to show a WR spectrum (see Sect.\,\ref{subsubsec:strippedstars}). 
  }
\label{fig:WRformation}
\end{figure} 

We discern among four distinct formation channels for WR stars.
These formation channels are illustrated in a Hertzsprung-Russell diagram (HRD) in 
Fig.\,\ref{fig:WRformation} using evolution tracks calculated with the 
BPASS\footnote{bpass.auckland.ac.nz} (Binary Population and Spectral Synthesis) code  V2.0  
\citep{Eldridge2008, Eldridge2016}, and are defined as follows:

\begin{enumerate}
 
 \item {\bf Main sequence WR stars} (ms-WR, ``born this way'') 
 are core H-burning WR stars.
 They typically exhibit WN spectra already on the main sequence by virtue of their very large 
 masses ($\gtrsim 60\,M_\odot$ at solar metallicity) and correspondingly strong winds 
 \citep{deKoter1997, Crowther2011}. 
 Spectroscopically, they are associated with    
 weaker-wind ``slash WR stars'' (/WN), hydrogen-rich WN stars (WNh), and 
 luminous blue variables (LBVs). Examples include \object{WR\,24} in the Galaxy (WN6h), 
 the two components of BAT99\,119 (WN6h + O3.5~If/WN7) in the Large Magellanic Cloud (LMC), and 
 probably the two components of  \object{HD\,5980} (WN6h + WN6-7h) in the Small Magellanic Cloud (SMC).
 \item  {\bf Wind-stripped WR stars} (\mbox{w-WR}) are  cWR stars that 
 formed through \emph{intrinsic} mass-loss, 
 i.e., stellar winds or eruptions \citep{Conti1976, Smith2014}.
 Only stars that are sufficiently massive can become \mbox{w-WR} stars. 
 The minimum initial mass $M_\text{i,w-WR}$ is a strong function 
 of the metallicity $Z$. It is estimated to be ${\approx}20-30\,M_\odot$ at solar 
 metallicity, ${\approx}30{-}60\,M_\odot$ at LMC metallicity (${\approx}1/3\,Z_\odot$), 
 and $45{-}100\,M_\odot$ at SMC metallicity (${\approx}1/5\,Z_\odot$) \citep[][]{Crowther2006, Maeder2002, Hainich2015}, keeping 
 in mind that these values are strongly dependent on the mass-loss and mixing prescriptions in evolution models.  
 Examples include 
 \object{WR\,6} in the Galaxy (WN4b) or \object{BAT99\,7} in the LMC (WN4b).

 \item {\bf Wind+binary-stripped WR} (\mbox{wb-WR}) stars are cWR stars that were originally massive 
 enough to become WR stars as single-stars (i.e., $M_\text{i} {\ge} M_\text{i,w-WR}$), but were further 
 stripped by a companion, either via Roche lobe overflow (RLOF) or via common-envelope evolution 
 \citep[CEE,][]{Paczynski1973, Vanbeveren1998b}. The WR primary of the Galactic 
 binary \object{WR\,139} (V444\,Cyg, WN5+O6) likely started its life with $M_{\rm i} \gtrsim 30\,M_\odot$ and was partly stripped 
 by the secondary star  \citep{Vanbeveren1998}, making it a good candidate for a wb-WR star. Other examples include 
 most confirmed WR binaries in the SMC \citep{Shenar2016}.

 \item {\bf Binary-stripped WR stars} (\mbox{b-WR}) are cWR stars that could 
 \emph{only} form as a result of binary interaction. That is, a b-WR star would not become 
 a WR star without a companion. 
 The \mbox{b-WR} channel extends the minimum initial mass of WR stars to lower values, bounded from below by 
 the initial mass $M_\text{i,WR}$
 at which the stripped product no longer exhibits a WR spectrum
 (see Fig.\,\ref{fig:WRformation} and Sect.\,\ref{subsubsec:strippedstars}). Only few candidates for b-WR exist. 
  The peculiar primary of the system \object{HD\,45166} (WN7 + B7~V or qWR + B7~V), which was reported as a short-period ($\approx 1.6\,$d) 
  $4.2\,M_\odot + 4.8\,M_\odot$ WR binary seen at a very low inclination of $0.77^\circ$, is probably 
  the best-known candidate for a b-WR star \citep{Steiner2005, Groh2008}.

\end{enumerate}

  By construction, each WN star belongs uniquely to one of these categories.
Note that w-WR, wb-WR, and b-WR stars are all cWR stars.
Spectroscopically, classical WN stars would tend to early types (WNE; WN2-5), while 
ms-WR stars to late-types (WNL, WN6-11), but this does not hold strictly 
(e.g., in the Galaxy: WR\,123 - a hydrogen-free WN8 star; WR\,3 - a hydrogen-rich WN3 star,
\citealt{Hamann2006}). WC/WO stars, which will be the subject of future studies, 
are always H-free and are therefore always cWR stars.
While the spectroscopic classification of WR stars is fairly unambiguous,
it is not straight forward to identify their evolutionary channel.

One of the central problems in this context is to correctly estimate the frequency of  binary stripped (\mbox{b-WR})
stars among a population of WR stars  
as a function of $Z$.    
It is now widely accepted that the majority of massive stars will interact with a companion star during their lifetime 
\citep{Sana2012}. Among the Galactic WR stars, 
about $40\,\%$ are observed to be binaries \citep{Vanderhucht2001}, {comparable to binary fraction recently reported for the M31 and M33 galaxies
\citep{Neugent2014}.} Considering the rapid power-law increase of the initial mass function (IMF) towards lower initial masses,  the 
longer lifetimes of lower-mass stars, and the high frequency of 
interacting binaries, \mbox{b-WR} stars should be abundant in 
our Universe, which may significantly affect the energy budget of galaxies \citep{Goetberg2017}. 
 However, to date, only a few WR stars are considered good candidates for b-WR stars 
\citep{Groh2008, Richardson2011}.

It is by now empirically \citep{Nugis2007, Mokiem2007, Hainich2015} as well as theoretically \citep{Kudritzki1987, Vink2001}  
established that the intrinsic mass-loss rates of massive stars decreases with decreasing surface metallicity, $\dot{M} {\propto} Z^\alpha$, with 
$0.5{\lesssim}\alpha{\lesssim}1$. This immediately implies that it is harder for stars at 
lower metallicity to intrinsically peel off their outer 
layers and become \mbox{w-WR} stars. In other words, 
the intrinsic formation channel becomes increasingly inefficient with decreasing metallicity. 
In contrast, no evidence exists that the efficiency of binary-stripping strongly depends on 
metallicity \cite[e.g.][]{Sana2013, Neugent2014}\footnote{At low metallicity, the radiation force exerted on the 
stellar layers is smaller, and thus so are the stellar radii 
for a given initial mass and age, which in turn 
reduces the likelihood of binary interaction. 
However, this effect is negligible compared to the sensitivity of $\dot{M}$ to $Z$.}. 
One may therefore expect that the fraction of \mbox{b-WR} stars in a population of WR stars 
should grow with decreasing metallicity.

Motivated by such predictions, \citet{Bartzakos2001}, \citet{Foellmi2003SMC}, \citet*[][FMG03 hereafter]{Foellmi2003LMC},
and \citet[][S08 hereafter]{Schnurr2008}, conducted 
a large spectroscopic survey in the SMC and LMC
with the goal of measuring the binary fraction in their WR populations and deriving the binary orbits (sensitive 
to periods up to ${\approx}200\,$d). The LMC and SMC are 
both known to have a sub-solar metallicity of a factor $\sim 1/3$ and 
$\sim 1/5$ solar, respectively \citep{Dufour1982, Larsen2000}. 
Following the reasoning of the previous paragraph, 
it is expected that the fraction of WR stars formed via the binary channel 
will be relatively large in the LMC, and even larger in the SMC.  
It was therefore surprising that 
FMG03 and S08 measured a WN binary fraction of ${\approx}40\%$ in the SMC and ${\approx}30\%$ in the LMC,
comparable to the Galactic fraction.

In \citet{Shenar2016, Shenar2018}, we performed spectroscopic analyses of the five confirmed WR binaries in the SMC, with the seven remaining  
apparently single WR stars analysed by \citet{Hainich2015}.
The results only worsened the problem. Although indications for past mass-transfer were found in the binaries,
\emph{all} WR stars were found to have very large initial masses of $M_\text{i} \gtrsim 60\,M_\odot$. While their evolution 
depends on the detailed treatment of rotation and the mass-loss prescription, within the uncertainties, such stars may already reach the 
WR phase intrinsically even at SMC metallicity \citep[see details in][]{Shenar2016}. Using the terminology introduced 
above, all WR binary components in the SMC were found to be wb-WR stars, while  
no b-WR stars could be identified. Thus, binary interaction does not seem to be 
responsible for the number of observed WR stars in the SMC.

In this study, we extend our analysis to the LMC, for which we adopt a 
distance of $d = 49.97\,$kpc \citep{Pietrzynski2013}.
Owing to both its higher metallicity and larger size, the sample of binary candidates 
amounts to 44, virtually ten-folding the SMC sample.
The goal of this study is threefold. First, through quantitative spectroscopy, we wish to supply physical
parameters for the LMC WN binaries. Second, we wish to exploit existing orbital solutions and, if possible, 
derive improved ones through radial-velocity (RV) measurements. 
This method is indispensable for weighing WR stars, 
whose surfaces are concealed by their thick winds, rendering a spectroscopic measurement 
of their mass via measurement of their surface gravity impossible.  Thirdly, being no longer limited by low-number statistics, 
we aim to estimate the impact of binary interaction 
in forming WR stars at sub-solar metallicity. 

The paper is organized as follows: In Sects.\,\ref{sec:sample} and \ref{sec:obs}, we briefly describe the selection of our targets 
and the observational data used in our study. In Sect.\,\ref{sec:analysis}, we present the methods and assumptions for our spectroscopic and orbital 
analyses, while Sect.\,\ref{sec:results} contains our results. We discuss our results in Sect.\,\ref{sec:disc} and summarize our findings in 
Sect.\,\ref{sec:sum}. In Appendix\,\ref{sec:comments}, we provide a detailed discussion for each of our targets, 
in Appendix\,\ref{sec:specfits}  we present the spectral fits for the whole sample, and 
in Appendix\,\ref{sec:obslog} we give a log of the observational data used and measured RVs.

\section{The sample}
\label{sec:sample}

\begin{figure*}[!htb]
\centering
  \includegraphics[width=\textwidth]{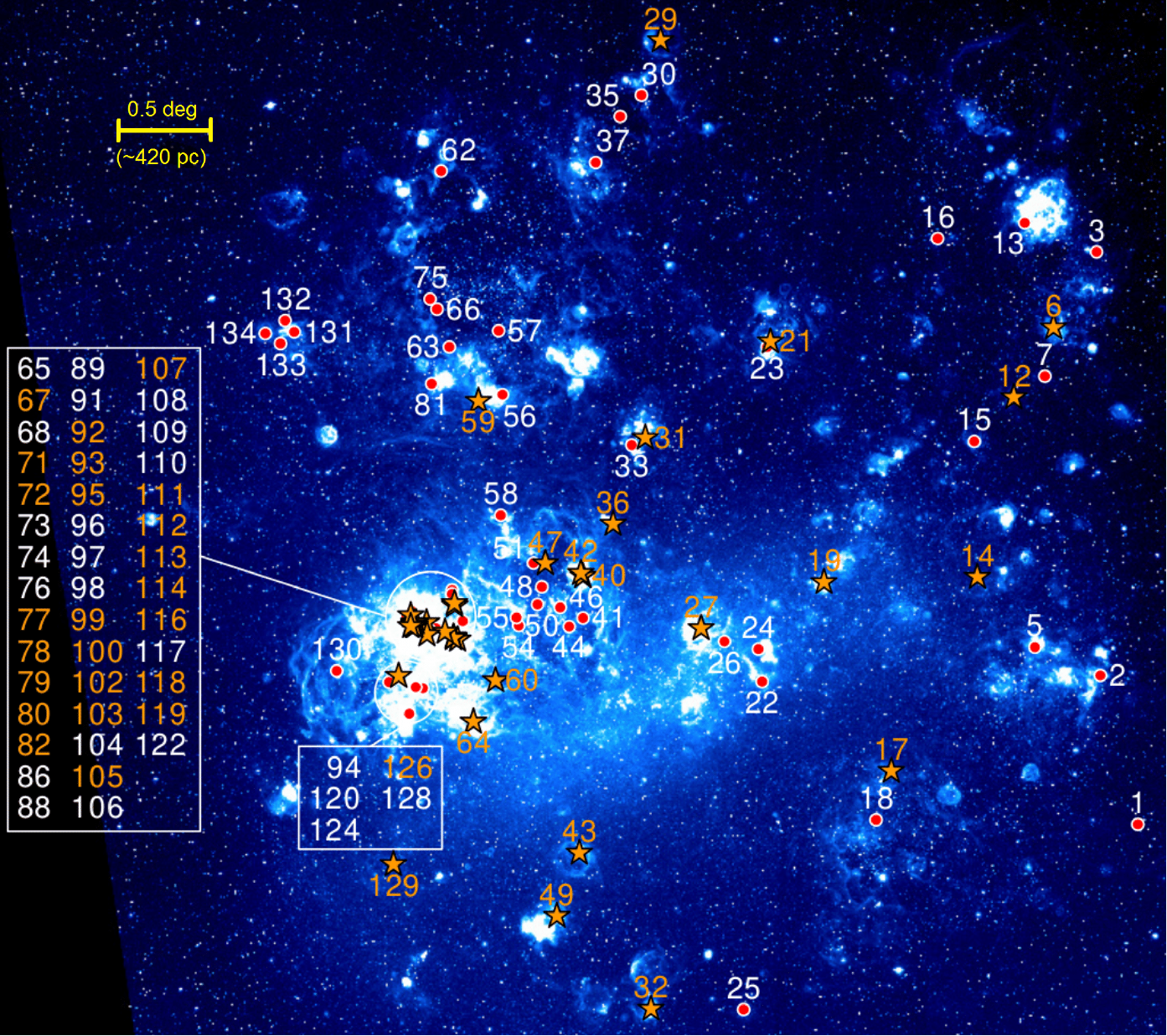}
  \caption{$H\alpha$ image of the LMC \citep{Smith2005}. Marked are the locations of all  putatively single (red circles) 
  and binary-candidate (yellow stars / labels) WN stars in the LMC BAT99 catalogue. The WN binaries constitute our sample.}
\label{fig:LMCpic}
\end{figure*} 
% \pagebreak

The LMC hosts 154 known WR stars \citep{Massey2014,Neugent2018}, $82\%$ of which belonging to the nitrogen sequence (WN). 
The WC/WO stars in the LMC, which comprise about 18\% of the total WR content with only a few confirmed binaries, 
will be the subject of future studies.
The 109 WN stars listed in the fourth catalogue of LMC WN stars \citep[][BAT99 hereafter]{Breysacher1999}
were previously analysed as single stars by \citet[][H14 thereafter]{Hainich2014}. Among them, 
H14 marked 43 that are either known binary/multiple systems or binary candidates. These objects constitute our sample. 
Two additional objects, BAT99\,17 and 60, are included here based on the presence of 
absorption features in their spectra, which are indicative of the presence of a companion.
Similarly to BAT99\,116 and 119, which were analysed in separate studies \citep{Shenar2017, Tehrani2019}, 
the very luminous ms-WR + ms-WR system BAT99\,118 requires a more in-depth analysis and is therefore delayed to future studies. Two 
additional WN binaries discovered by \citet{Neugent2017}, \object{LMC 143-1} and \object{LMC 173-1}, are not included here 
due to lack of data. 
Altogether, our final sample comprises 44 WN binaries / binary candidates.

The binary nature of 22 objects in our sample was established in previous studies
via periodic RV variation, and seven additional targets are considered binary 
candidates on the basis of RV variations ($\sigma_\text{RV} > \sigma_\text{err}$) 
for which no period could be found (FMG03, S08). A few additional candidates are included 
due to the presence of spectral features that are strongly indicative of a companion (e.g., BAT99\,17 and 60).
H14 identified further binary
candidates based on their X-ray properties. 
Single WN stars are generally known to exhibit faint X-ray luminosities 
\citep{Ignace2000, Skinner2012, Oskinova2012} not exceeding ${\approx}10^{32}\,$\lum. 
In contrast, wind-wind collisions (WWC) in WR binaries can yield X-ray luminosities that are a few orders 
of magnitude larger, ${\approx}10^{33}-10^{34}$\,\lum \citep{Moffat1998}. Even more X-ray luminous are the rarely-observed
WR binaries with accreting compact objects \citep[e.g., \object{Cyg~X-3},][]{Lommen2005}.
Every WN star that was detected in X-rays in surveys by \citet{Guerrero2008, Guerrero2008b} and correspondingly 
exhibits X-ray luminosities of at least ${\approx}10^{33}\,$\lum\,is considered here to be a binary candidate. We note, however,
that the presence of X-rays does not necessarily imply binarity, and vice versa. For example, long-period 
binaries may only emit faint WWC X-ray emission, while single ms-WR stars may emit significant
X-rays intrinsically \citep[e.g.][]{Pollock1995, Huenemoerder2015}.

\begin{table*}[!htb]
% \tiny
% \scriptsize
% \small 
% \footnotesize
\setlength\tabcolsep{3pt}
\caption{Overview on our sample: the LMC WN binaries / binary candidates} 
\label{tab:catalog}
\begin{center}
% \resizebox{0.9\textwidth}{!}{
{\tiny 
\begin{tabular}{lclccccccccc}
\hline 
\hline
\multirow{3}{*}       &                &                                                                          &                                                                                                        &       &  \multicolumn{5}{c}{binary-candidate criteria}                 \\\cline{6-10} \Tstrut     %             & confirmed here \\ 
BAT99                 & Reference      &  Aliases                                                                 & Spectral type\tablefootmark{t}                                                                         &   $v$\,[mag] & $L_\text{X}$\tablefootmark{a}[\lum]   & composite     & eclipsing                   & $\sigma_\text{RV}$       & $P$\,[d] & configuration                            \\ %       & vis.\ bin.\                 \\ 
%                       &                &                                                                          &                                                                                                        & [mag] &    [\lum]                              &                          &                         &    [d]            \\ %        &  &      &                                        \Tstrut         \\    %             &                      \\ 
\hline                                                                                                                                                                                                                                                                                                                                                \\ %            &                                               %
{\bf 6} &     b,c        & \object{Br~5}, \object{SK\,-67~18}                                                     & \begin{tabular}{c}O3~If*/WN7\tablefootmark{c} + OB \\ + (O7\tablefootmark{c} + ?)\end{tabular}        &  12.2 & $3{\cdot}10^{33}$\,\tablefootmark{c} & yes                  & yes                         & yes                      & 2.0      & SB1 + SB1:                            \\ %        & no                    \\    %               & yes \\                & 
% \hline                             &                                                                                                                                                                                                                                     & yes                                                                           \\ %       &                                                                \\    %      
12      &     d          & \object{Br~10a}, \object{SK\,-67~22}                                                   & O2~If*/WN5                                                                                              &  13.7 & $<6{\cdot}10^{33}$                   & no                   & no                          & yes                      & 3.2      & SB1: or single:                      \\ %        & no                    \\    %               & no \\  
%                        &                                                                                                                                                                                                                                                                                                                           \\ %             &                                              \\    %             
{\bf 14}&     e,c        & \object{Br~11}, \object{SK\,-68~19}                                                    & WN4 + O9~V\tablefootmark{c}                                                                        &  13.7 & $\lesssim10^{33}$                    & yes                  & no                          & marginal                 & -        & ?                                   \\ %        & no                      \\    %             & yes \\  
%              ,c        &                                                                                                                                                                                                                                                                                                                                            &                                     \\ %                               \\    %      
{\bf 17}&     e,c        & \object{Br\,14}                                                                        & WN4 + B0~V\tablefootmark{c}                                                                            &  14.4 & -                                    & yes           & -                           & no                       & -        & ?                                   \\ %       & no                           \\    %               & yes \\ 
{\bf 19}&     e,c        & \object{Br\,16}                                                                        & WN3 + O6~V\tablefootmark{c}                                                                            &  13.8 & $3.8{\cdot}10^{34}$                  & yes                  & yes                         & yes                      & 18       & SB2                                 \\ %       & no                           \\    %               & yes \\ 
%              ,c        &                                                                                                                                                                                                                                                                                                                                            &                                     \\ %                               \\    %      
{\bf 21}&     e,c        & \object{Br\,17}, \object{SK\,-67~63}                                                    & WN4 + O9~III\tablefootmark{c}                                                                         &  13.1 & $<8.2{\cdot}10^{33}$                 & yes                  & -                           & marginal                 & -        & ?                                   \\ %       & yes                \\    %                & yes \\ 
% %              ,c        &                                                                                                                                                                                                                                                                                                                                          &                                     \\ %                              \\    %          
{\bf 27}&     e          & \object{Br\,21}, \object{SK\,-69~95}                                                    & WN4 + B1~Ia                                                                                           &  11.3 & $<2.4{\cdot}10^{33}$                 & yes           & no                          & no                       & -        & ?                                   \\ %       & yes                \\    %               & yes \\ 
{\bf 29}&     e,c        & \object{Br\,23}, \object{SK\,-65~45}                                                    & WN3 + B1.5~V\tablefootmark{c}                                                                         &  14.6 & $\lesssim10^{33}$                    & yes                  & no                          & yes                      & 2.2      & SB2:\tablefootmark{s}               \\ %        & no                 \\    %             & yes \\ 
31      &     e          & \object{Br\,25}                                                                        & WN3                                                                                                    &  15.5 & diffuse                              & no                   & no                          & marginal\tablefootmark{f}& -        & single:                             \\ %        & no                         \\    %             & no \\ 
{\bf 32}&     c,c        & \object{Br\,26}, \object{SK\,-71~21}                                                    & WN5(h)\tablefootmark{c}   + WN6(h):\tablefootmark{c} (+abs)                                               &  12.7 & $<3{\cdot}10^{33}$                & yes                  & -                           & yes                       & 1.9     & SB2 + (single:)                      \\ %        & no          \\    %                 & yes \\ 
36      &     e,c        & \object{Br\,29}, \object{SK\,-68~77}                                                    & WN3/WCE(+OB?)                                                                                          &  14.8 & $<8.9{\cdot}10^{33}$                 & marginal & no                          & no                       & -        & single:                              \\ %        & no                          \\    %              & no \\ 
40      &     e          & \object{Br\,33}                                                                        & WN4                                                                                                 &  15.0 & $5{\cdot}10^{33}$\,\tablefootmark{g} & no                   & no                          & no                       & -        & single:                             \\ %        & no                     \\    %               & no \\ 
{\bf42 }&     e,c        & \object{Br\,34}, \object{SK\,-68~82}                                                    & WN5 + B3~I\tablefootmark{c} + ?                                                                    &  9.9  & $1{\cdot}10^{34}$                    & yes                   & no                          & yes                      & -       & SB1: + single:                       \\ %      & yes                          \\    %               & yes \\ 
{\bf43 }&     e,c        & \object{Br\,33}, \object{SK\,-70~92}                                                    & WN3 + O9~V\tablefootmark{c}                                                                           &  14.2 & $<1.1{\cdot}10^{34}$                 & yes                  & marginal                    & yes                      & 2.8      & SB2:\tablefootmark{s}               \\ %        & no            \\    %               & yes \\ 
47      &     e          & \object{Br\,39}, \object{SK\,-68~98}                                                    & WN3                                                                                                   &  14.1 & $5{\cdot}10^{33}$                    & no                   & no                          & no                       & -        & single:                             \\ %       & no                                \\    %                 & no \\ 
{\bf49 }&     e,c        & \object{Br\,40a}, \object{SK\,-71~34}                                                   & WN3 + O8~V\tablefootmark{c}                                                                          &  13.6 & $<9.1{\cdot}10^{33}$                 & yes                   & -                           & yes                      & 32      & SB2                                  \\ %      & no                   \\    %               & yes \\ 
{\bf59 }&     e,c        & \object{Br\,48}, \object{SK\,-67~184}                                                   & WN3 + O6~III\tablefootmark{c}                                                                       &  13.3 & -                                    & yes                  & -                           & yes                      & 4.7      & SB2:\tablefootmark{s}               \\ %        & no                               \\    %               & yes \\ 
{\bf60 }&     c          & \object{Br\,49}                                                                        & WN3\tablefootmark{c} + O9~V\tablefootmark{c}                                                         &  14.6 & $<4.2{\cdot}10^{33}$                 & yes             & no                          & no                       & -        & ?                                   \\ %       & no                         \\    %             & yes \\ 
{\bf64 }&     e,c        & \object{Br\,53}, \object{SK\,-69~198}                                                   & WN3 + O9~V\tablefootmark{c}                                                                           &  14.4 & -                                    & yes                  & yes                         & yes                      & 38       & SB2:\tablefootmark{s}               \\ %       & no                         \\    %             & yes \\ 
67      &     e          & \object{Br\,56}                                                                        & WN5                                                                                                 &  13.9 & $2{\cdot}10^{33}$                    & no                   & no                          & marginal                 & -        & single:                             \\ %        & no                           \\    %               & no \\ 
{\bf71 }&     e,c        & \object{Br\,60}                                                                        & WN3 + O6.5~V\tablefootmark{c}                                                                           &  15.1 & $<6.5{\cdot}10^{33}$                 & yes                  & marginal\tablefootmark{h}   & yes                      & 5.2      & SB2:\tablefootmark{s}               \\ %        & no              \\    %                & yes \\ 
   72   &     e,c        & \object{Br\,61}                                                                        & WN4 + O3.5~V\tablefootmark{c,i}                                                                         &  15.8 & $<5.6{\cdot}10^{33}$                & marginal             & no                          & marginal                 & -        & SB2:                                \\ %       & no         \\    %                     & marginal \\ 
{\bf77 }&     d,c        & -                                                                                      & WN7 + O7.5~III\tablefootmark{c}                                                                       &  13.3 & $1.4{\cdot}10^{33}$                  & yes                  & -                           & yes                      & 3.0      & SB2                                 \\ %        & no                     \\    %           & yes \\ 
78      &     d          & \object{Br\,65b}                                                                       & WN4                                                                                                     &  13.1 & $7{\cdot}10^{32}$                    & no                   & -                           & no                       & -        & single:                             \\ %       & no                              \\    %             & no \\                                   
{\bf79 }&     d,c        & \object{Br\,57}                                                                        & WN7 + O9~I\tablefootmark{c}                                                                            &  13.6 & $7{\cdot}10^{32}$                    & yes                  & -                           & no                       & -        & ?                                   \\ %       & no                           \\    %             & yes \\ 
{\bf80 }&     d,c        & \object{Br\,65c}                                                                       & WN5 + O9.5~III\tablefootmark{c}                                                                      &  13.2 & $1{\cdot}10^{33}$                    & yes                  & no                          & no                       & -        & ?                                   \\ %       & no                     \\    %           & yes \\ 
82      &     e          & \object{Br\,66}                                                                        & WN3                                                                                                    &  16.1 & $2{\cdot}10^{33}$                    & no                   & no                          & no                       & -        & single:                             \\ %       & no                           \\    %             & no \\ 
% 86      &     c          & \object{Br\,69}                                                                        & WN3 + OB?\tablefootmark{c}                                                                              &  16.4 & $<6.3{\cdot}10^{33}$                 & marginal             & no                          & marginal                 & -      &                                       \\ %       & no                     \\    %           & yes \\ 
{\bf92 }&     d,c        & \object{Br\,72}, \object{SK\,-69~235}, \object{R\,130}                                 & \begin{tabular}{c} WC4\tablefootmark{c, j} + B1~Ia\end{tabular}                                            &  11.5 & $4{\cdot}10^{33}$                 & yes                  & -                           & yes                      & 4.3      & SB2 + single:                       \\ %        & no                                           \\    %             & yes \\ 
93      &     d          & \object{Br\,74a}, \object{VFTS~180}                                                    &       O3~If*                                                                                                 &  13.8 & $8{\cdot}10^{32}$                    & no                   & -                           & no                       & -        & single:                             \\ %       & no                         \\    %               & no \\ 
{\bf95} &     d,c        & \object{Br\,80}, \object{R\,135}, \object{VFTS\,402}                                    & WN5:\tablefootmark{c}  + WN7\tablefootmark{c}                                                     &  13.2 & $<6.4{\cdot}10^{32}$                 & yes                  & -                           & yes                      & 2.1      & SB2:                                \\ %        & no                  \\    %                & yes \\ 
99      &     d          & \object{Br\,78}, \object{Mk\,39}                                                        & O2.5~If*/WN6                                                                                            &  13.0 & $2{\cdot}10^{34}$                    & no                   & -                           & yes                      & 93       & SB1:                                \\ %       & no                          \\    %           & marginal \\ 
100     &     d          & \object{Br\,75} , \object{R\,134}, \object{VFTS\,1001}                                & WN6h                                                                                                      &  12.8 & $2{\cdot}10^{33}$                    & no                   & -                           & no                       & -        & single:                             \\ %      & no                            \\    %          & no \\ 
102     &     d          & \object{R\,140a}, \object{VFTS\,507}                                                 & WN6                                                                                                       &  13.0 & $2{\cdot}10^{35}$                    & no                   & -                           & no                       & -        & single:                             \\ %       & no                             \\    %               & no \\ 
{\bf103}&     d,c        & \object{R\,140b}, \object{VFTS\,509}                                                 & WN5(h) + O4~V\tablefootmark{c}                                                                          &  13.0 & $1{\cdot}10^{33}$                      & yes                  & -                           & yes                      & 2.8      & SB2                                 \\ %        & no                        \\    %               & yes \\ 
105     &     d,k        & \object{Br\,77},  \object{Mk\,42}                                                    & O2~If*                                                                                               &  12.8 & $4{\cdot}10^{33}$                    & no                   & -                           & yes                      & -        & SB1:                                \\ %      & no                           \\    %               & no \\ 
{\bf107}&     d,l        & \object{Br\,86}, \object{R\,139}, \object{VFTS\,527}                              & O6.5Iafc + O6Iaf                                                                                             &  12.1 & $3{\cdot}10^{33}$                    & yes                  & no                          & yes                      & 154      & SB2                                 \\ %        & no                             \\    %             & yes \\ 
111     &     m,n        & \object{R\,136b}                                                                      &  O4~If/WN8\tablefootmark{m}                                                                                                &  13.4 & $\lesssim10^{33}$\tablefootmark{p}   & no                   & -                           & no                       & -        & single:                             \\ %       & no              \\    %              & marginal \\ 
112     &     m          & \object{R\,136c}                                                                      & WN4.5h                                                                                                     &  13.6 & $6{\cdot}10^{34}$                    & no                   & -                           & yes                      & 8.2      & SB1                                 \\ %        & no                            \\    %               & marginal \\ 
{\bf113}&     m,k,c      & \object{VFTS\,542}, \object{Mk\,30}                                                   & O2~If*/WN5+ B0~V\tablefootmark{c}                                                                        &  13.6 & $<1.3{\cdot}10^{33}$                 & yes                  & -                           & yes                      & 4.7      & SB2                                 \\ %        & no                           \\    %               & yes \\ 
114     &     d,k        & \object{VFTS\,545}, \object{Mk\,35}                                                   & O2~If*/WN5                                                                                               &  13.6 & $1{\cdot}10^{33}$                    & no                   & -                           & marginal                 & -        & SB1:                                \\ % & no                         \\    %            & no \\ 
{\bf116}&     o          & \object{Br\,84}, \object{Mk\,34}                                                        & WN5h\tablefootmark{o} +  WN5h\tablefootmark{o}                                                                                           &  13.6 & $2{\cdot}10^{35}$                    & yes                  & -                           & yes                      & 151      & SB2                                 \\ %        & no                          \\    %            & yes \\ 
% 1      \begin{tabular}{c}\object{Br\,89}, \object{Sk\,-69~246}\\ \object{R\,144}\end{tabular} & WN                                                                                                   &  11.1 & $2{\cdot}10^{33}$                 & yes  & -                                          & yes                       & no           &           & no    &                                   & \\ %          \\    %            & yes \\ 
{\bf119}&     q          & \object{Br\,90}, \object{VFTS\,695}, \object{R\,145}                               & WN6h  + O3.5~If*/WN7                                                                                        &  12.2   & $2{\cdot}10^{33}$                  & yes                  & no                          & yes                      & 159      & SB2                                 \\ %        & no                          \\    %             & yes \\ 
{\bf126}&     e,c        & \object{Br\,95}                                                                    & WN3  + (O7~V + O)\tablefootmark{c}                                                                                &  13.3 & $1{\cdot}10^{33}$               & yes                  & -                           & yes                      & 25       & SB1 + SB2:                          \\ %       & no                                    \\    %                  & yes \\ 
{\bf129}&     r          & \object{Br\,97}                                                                       & WN3  + O5~V                                                                                           &  14.9  & -                                   & yes                  & yes                         & yes                      & 2.8      & SB2                                 \\ %        & no                                                 \\    %                  & yes \\ 
\hline
\end{tabular}}
\tablefoot{
The table gives all binary candidates among the WN stars in the BAT99 catalogue (excluding BAT99\,118, see text for details). The columns are, 
from left to right: BAT99 catalogue number, reference, aliases, spectral types, and Smith visual magnitudes. 
In columns 6-10, we list X-ray luminosities,  whether the systems show the presence of two or more components in their spectra (composite), 
and whether they are eclipsing (wind/photospheric),  
significantly RV-variable (cf. FMG2003, S2008), and periodically RV-variable. The final column shows the suggested configuration based on this study 
and previous studies.
Bolded entries correspond to objects that are confirmed here as binary or multiple. Entries taken from references in column 2, with the exception unless 
otherwise noted with a footmark. Colons stand for uncertain entries.
\tablefoottext{a}{Taken from \citet{Guerrero2008, Guerrero2008b}, except for BAT99\,6, which is derived here.}
\tablefoottext{b}{\citet{Niemela2001}, \citet{Koenigsberger2003}}
\tablefoottext{c}{This study}
\tablefoottext{d}{S2008}
\tablefoottext{e}{FMG03}
\tablefoottext{f}{Not confirmed by our study}
\tablefoottext{g}{Reported by FMG03, but reported as undetected by \citet{Guerrero2008}}
\tablefoottext{h}{A single faint eclipse is obtained with the period $P = 5.2\,d$, unequal to the spectroscopically derived period ($2.3\,$d)}
\tablefoottext{i}{Uncertain, possibly single}
\tablefoottext{j}{Revised from WN4b to WC4 in our study}
\tablefoottext{k}{\citet{Crowther2011}}
\tablefoottext{l}{\citet{Taylor2011}}
\tablefoottext{m}{\citet{Schnurr2009}}
\tablefoottext{n}{\citet{Crowther2016}}
\tablefoottext{o}{\citet{Pollock2018, Tehrani2019}}
\tablefoottext{p}{A potential detection was reported by \citet{Townsley2006} but was not confirmed by \citet{Guerrero2008}}
\tablefoottext{q}{\citet{Shenar2017}}
\tablefoottext{r}{\citet{Foellmi2006}}
\tablefoottext{s}{Periodic RV variability for WR star; better data needed to confirm periodicity of OB-type companion.}
\tablefoottext{t}{Spectral types adopted from \citet{Neugent2018}, unless otherwise stated}
% \tablefoottext{s}{Due to composite spectrum - line-of-sight contamination / crowding possible. RV variability should be monitored}
}
\end{center}
\end{table*}

Fig.\,\ref{fig:LMCpic} shows the positions of the 109 known WN stars of the BAT99 catalogue 
on an image of the LMC, marking also the binary candidates.
In Table\,\ref{tab:catalog}, we list all LMC WN binary candidates. 
We also give their aliases, spectral types, Smith v-band magnitudes, where 
the classification procedure is described in Sect.\,\ref{subsubsec:specclass}. 
The status of various 
binary-candidate criteria is given for each of the targets: X-ray luminosities, composite spectra, eclipses, and RV variations.
Finally, we give the suggested configuration for each system based on our study and previous studies. Here, SB2 refers to systems in which 
two components are seen in the spectrum and move periodically in anti-phase, while SB1 refers to systems in which only one component
(here always the WR star) is moving periodically. 
Per definition, each target in our sample is positive on at least one of these criteria. 
Bold entries correspond to objects that are confirmed as binaries in our study. 

It is important to stress that the non-confirmed candidates, as well as other apparently-single WN stars, 
may still be binaries that were not observed as such due to, e.g., low-mass companions, long periods, or inclination effects.
Detection biases are discussed in length in FMG03 and S2008. Loosely speaking, the binary sample is estimated to be ${\approx}70\%$ complete
to binaries with periods $P \lesssim 200\,$d and  secondaries with $M_2 \gtrsim 8\,M_\odot$. 
The true binary fraction for WR binaries with larger periods / 
lower-mass secondaries remains unconstrained.

\section{Observations}
\label{sec:obs}

The spectral analysis of the objects in our sample relies on various observational datasets, as described below.
In Sect.\,\ref{sec:obslog} in the appendix, 
we compile all spectra used for the spectral/orbital analysis.

Previously unpublished data were collected  by a member of our team (O.\ Schnurr) for all short-period ($P \leq 5$\,d) WNL binaries in the LMC located outside the too crowded 
\object{R\,136} cluster at the center of the giant H{\sc ii} region 30 Doradus: 
BAT99\,12, 32, 77, 92 (in fact a WC binary erroneously classified as WN previously, see Sect.\,\ref{sec:comments}), 95, 103, and 113 \citep[][S08]{Moffat1989}.
These data were obtained during a 5-night
observing run at Cerro Tololo Inter-American Observatory (CTIO), Chile, from December 14 to 18 2005, 
using the Ritchey-Chr\'etien (R-C) Spectrograph 
attached to the CTIO-4m telescope. The R-C spectrograph set-up used the 
blue Schmidt camera and the G450 grating (450 l/mm) set up in second 
order, to cover a spectral wavelength range from 3700 to 5200$\AA$. 
With a slit width of 150 $\mu$m 
(corresponding to 1'' on-sky, to match the ambient seeing), a linear 
dispersion of $0.95\AA$ per pixel was reached; the three-pixel 
spectral (velocity) resolving power was thus $R \sim 2,400$.
For each object, the goal was to obtain at 
least one high-quality spectrum per night with S/N$\sim200$ in the continuum. To achieve this, 
exposure times ranged from 2\,250 to 4\,500 sec. 
For better cosmic-ray rejection, exposures were split into three sub-exposures.
At the beginning of each night, bias frames and high-SNR internal 
(quartz lamp) flat-field frames were also taken, and averaged for better 
statistics. No dark-frames were taken. 
The data were reduced by standard procedure within \texttt{MIDAS} in context ``long''. 
These observations are summarized in Tables\,\ref{tab12} - \ref{tab113}.

For all objects, visual spectra obtained during the campaigns 
by FMG03 and S08 were used, taken with different telescopes and instruments (see FMG03 and S08). They 
typically cover ${\approx}4000-6800\,\AA$ and have an average resolving power of $R\approx1000$ (for 
more details, see FMG03 and S08). 
As in \citet{Shenar2016}, we 
only use spectra that were co-added in the frame-of-reference of the WR star to enhance the signal to noise to $S/N=100{-}150$, 
since the original data can no longer be retrieved.  
Co-adding the spectra in the frame of
the WR star may cause the companion’s spectral features to smear when significant RV variations are present. 
The companion's spectrum is therefore subject 
to additional broadening which scales with the sum of the RV amplitudes 
$K_1 + K_2$. To roughly account for this broadening, we convolve the companion's model with box profiles with a width of $K_1 + K_2$ 
\cite[see][]{Shenar2016}. In cases where $K_1 + K_2$ is large, the secondary's projected rotational velocity and surface gravity 
could be poorly constrained - see Appendix\,\ref{sec:comments}.

For a large fraction of our sample, UV spectra were retrieved from the Mikulski Archive for Space Telescopes (MAST). In almost all cases, 
the spectra were obtained using the International Ultraviolet Explorer (IUE), 
covering the spectral range $1200 - 2000\,\AA$ from the MAST archive. 
When available, high resolution spectra are preferred, binned at intervals of $0.05\,\AA$ to achieve an S/N$\approx 20$. Otherwise, 
low resolution spectra are used ($\text{FWHM} \approx 6\,\AA$, $\text{S/N} \approx 20$). 
Low resolution, flux calibrated IUE spectra in the range $2000 - 3000\,\AA$ are not used 
for detailed spectroscopy because of their low S/N ($\approx 5-10$), but rather to cover the spectral energy distribution (SED) of the targets.
Optical low resolution spectra taken by \citet{Torres1988} are also used for the SEDs of our targets.
When available, flux calibrated, high resolution Far Ultraviolet Spectroscopic Explorer (FUSE) spectra covering the spectral range $960-1190\,\AA$
are also retrieved from the MAST archive and binned at $0.05\,\AA$ to achieve an $\text{S/N}\approx 30$.  The IUE and FUSE spectra are normalized with the reddened model continuum.

For a significant number of objects, we retrieved additional spectra taken with the Fibre Large Array Multi Element Spectrograph (FLAMES) 
mounted on the  Very Large Telescope's (VLT) UT2. 
The FLAMES spectra
(072.C-0348, Rubio; 182.D-0222, Evans; 090.D-0323, Sana; 092.D-0136, Sana)
were secured between 2004 and 2014 with the FLAMES instrument mounted on the Very Large Telescope (VLT), Chile, partly in the 
course of two programs: the VLT FLAMES Tarantula Survey \citep{Evans2011} and the Tarantula Massive Binary Monitoring (TMBM) project. 
They cover the spectral range $3960 - 4560\,\AA$, and typically have S/N$\gtrsim$100 and $R \approx 7000$.
The spectra are rectified 
using an automated routine that fits a piecewise first-order polynomial to the apparent continuum and 
cleaned from cosmic events using a self-written Python routine.

Two archival spectra of BAT99\,6, taken with the FEROS spectrograph mounted on the 2.2m telescope in La Silla on 3 Oct.\ 2005 and 27 Jun.\ 2006, were 
retrieved from the ESO archives. These spectra have a resolving power of $R=48000$, a S/N${\approx}50$, and a spectral coverage of 
$3900-7000\,\AA$. 
A single spectrum of BAT99\,12, taken with the UVES spectrograph mounted on the VLT on 29 Nov.\ 2004, 
was retrieved from the ESO archives. The spectrum has $R{\approx}60000$ and S/N${\approx}80$.

Photometry for all our objects was extracted from the literature using the Vizier tool\footnote{vizier.u-strasbg.fr/viz-bin/VizieR}. 
$UBV$, $JHK$, and IRAC photometry was obtained from compilations by \citet{Bonanos2009}, \citet{Zacharias2005}, 
\citet{Zaritsky2004}, \citet{Ulaczyk2012}, \citet{Delmotte2002}, \citet{Kato2007}, \citet{Popescu2012}, \citet{Massey2000}, 
\citet{Nascimbeni2016}, \citet{Parker1992}, \citet{Evans2011}, and \citet{Roeser2008}. 
WISE photometry 
was obtained from \citet{Cutri2012,Cutri2013}. UBV Photometry compiled in \citet{Tehrani2019} was taken for BAT99\,116.

\section{Analysis}
\label{sec:analysis}

\subsection{Spectral analysis}
\label{subsec:specan}

\subsubsection{The PoWR code}
\label{subsubsec:PoWR}

We performed a spectral analysis of the available spectra for each of the targets listed in Table\,\ref{tab:catalog}. Unless otherwise stated, the analysis 
accounts for all known components in the system. The analysis is performed with the 
Potsdam Wolf-Rayet (PoWR) model atmosphere
code, especially suitable for hot stars with expanding atmospheres\footnote{\label{footnote:PoWR}PoWR models of Wolf-Rayet stars can be downloaded at
www.astro.physik.uni-potsdam.de/PoWR}. PoWR iteratively solves the co-moving frame
radiative transfer and the statistical balance equations in spherical symmetry under the constraint
of energy conservation without assuming local thermodynamic equilibrium (non-LTE). 
A more detailed description of the assumptions and methods used
in the code is given by \citet{Graefener2002} and \citet{Hamann2004}. By
comparing synthetic spectra generated by PoWR to observations, the stellar
parameters can be derived.

\begin{figure*}[!htb]
\centering
  \includegraphics[width=0.9\textwidth]{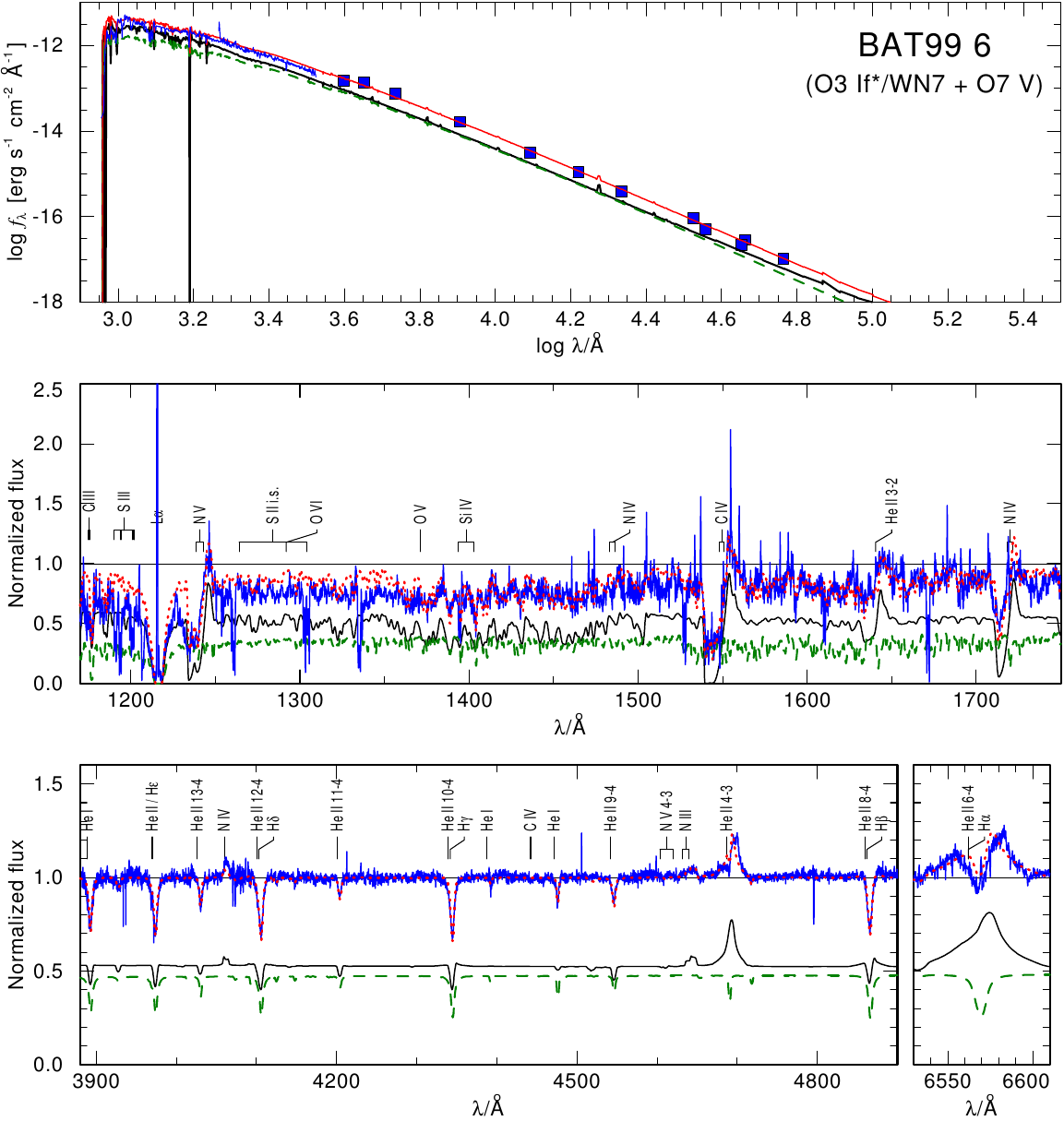}
  \caption{A spectral analysis of the system BAT99 6. The observed photometry and spectra (archival IUE, FEROS) 
 of BAT99 6 are shown in blue. The composite synthetic spectrum (red dotted line) is the sum of the WR (black solid line) 
 and O (green dashed line) models. The relative offsets of the model continua correspond to the 
 light ratio between the two stars. }
\label{fig:ideal006}
\end{figure*}

The inner boundary of the model, 
referred to as the stellar radius $R_*$,
is defined at the Rosseland continuum optical depth $\tau_\mathrm{Ross}$=20, where LTE can be safely assumed. $R_*$ 
is supposed to represent the radius at the hydrostatic layers of the star.
In the subsonic region, the velocity
field is defined so that a hydrostatic density stratification is
approached \citep{Sander2015}. In the supersonic wind region, the
pre-specified wind velocity field $v(r)$ generally takes the form of a $\beta$-law \citep{CAK1975}

\begin{equation}
 v(r) = v_\infty \left(1 - \frac{R_*}{r+r_0}\right)^{\beta}.
\label{eq:2betalaw}
\end{equation}
Here, $v_\infty$ is the terminal velocity, 
and  $r_0{\ll}R_\ast$ is a constant determined so as
to achieve a smooth transition between 
the subsonic and supersonic regions. 
For OB-type stars, we adopt the usual value of $\beta=0.8$ 
\citep[e.g.,][]{Kudritzki1989}. The value of $\beta$ for WR stars is heavily debated. 
Values of the order of unity are reported for some \citep[e.g.,][]{Chene2008, Graefener2008}, while values 
in the excess of four are reported for others \citep[e.g.,][]{Lepine1999, Dessart2005}. In fact, 
hydrodynamically-consistent models suggest that the $\beta$-law may be too simplistic in the 
case of WR stars \citep{Graefener2005, Sander2017}. To avoid an excess of free parameters, we 
follow the convention of $\beta=1$ (see Appendix\,\ref{sec:comments}). 
This has the advantage of direct comparability with the majority 
of other studies \citep[e.g.,][H14]{Crowther1997}. An underestimation of $\beta$, especially for winds that 
are very optically thick, generally results in an underestimation of $T_*$ and $v_\infty$, both of which are not 
expected to alter our main conclusions.

Beside the velocity law and chemical composition, 
four fundamental input parameters are needed to define a model atmosphere: 
the effective temperature $T_*$ of the hydrostatic star,
its surface gravity $g_*$, the mass-loss rate $\dot{M}$, and the stellar luminosity $L$.
The effective temperature relates to $R_*$ and $L$ via 
the Stefan-Boltzmann law: $L = 4\,\pi\,\sigma\,R_*^2\,T_*^4$.  We stress that, for WR stars,  $T_*$ may significantly 
differ from the photospheric effective temperature $T_\text{2/3}$, defined relative to $R_\text{2/3}$  at 
$\tau_\text{Ross} = 2/3$. When comparing to evolution tracks, which generally do not account for stellar winds, 
it is $T_*, g_*,$ and $R_*$ - and not $T_\text{2/3}, g_\text{2/3}$ and $R_{2/3}$ - which should be 
used \citep[see e.g.,][]{Groh2014}.
The gravity $g_*$ relates to the radius $R_*$ and mass $M_*$  via the usual definition: $g_* = g(R_*) = G\,M_* R_*^{-2}$. 
For the vast majority of WR models, the value of $g_*$ bears no significant effects on the synthetic spectrum, 
which originates primarily in the wind, and is therefore not included in the fitting procedure. 
The outer boundary is taken to be 
$R_\text{max} = 100\,R_*$  for O models and $1000\,R_*$ for WR models, 
which were tested to be sufficiently large.

During the iterative solution, the line opacity and emissivity profiles at each radial layer are 
Gaussians with a constant Doppler width $v_\text{Dop}$. This parameter
is set to $30$ and $100\,$\kms\, for O and WR models, respectively.
In the formal integration, the
Doppler velocity is decomposed to depth-dependent thermal motion and microturbulence $\xi(r)$. 
We assume $\xi(r)$ grows with the wind velocity up to
\mbox{$\xi(R_\text{max}) = 0.1\,v_\infty$}, and set $\xi(R_*) = 20$ and 100\,\kms\, for OB  
and WR models, respectively \citep[e.g.,][]{Hamann2006, Shenar2015}.
We assume a macroturbulent velocity 
of $30\,$\kms\, for all O components \citep[e.g.,][]{Markova2008, SimonDiaz2010, Bouret2012}, accounted for by 
convolving the profiles with radial-tangential profiles \citep[e.g.,][]{Gray1975}. 
Rotational broadening  is typically accounted for 
via convolution with rotation profiles (see Sect.\,\ref{subsubsec:sinan}). The synthetic spectra are 
further convolved with Gaussians that mimic the instrumental profiles.

It is a consensus that winds of hot massive stars are not smooth, but 
rather clumped
\citep{Moffat1988, Lepine1999, Prinja2010, Surlan2013}.
An approximate treatment of optically thin clumps using the so-called microclumping approach was introduced by 
\cite{Hillier1984} and systematically implemented by \cite{Hamann1998}, 
where the population numbers of the atomic levels are calculated in clumps which are a factor of $D$ denser 
than the equivalent smooth wind ($D = 1 / f$, where $f$ is the filling factor). Unless otherwise 
stated in the individual comments (Sect.\,\ref{sec:comments}), 
we fix $D$ to $10$ for both WR 
and O-type stars: this value generally agrees well with the observations and is consistent with previous studies 
\citep{Shenar2015, Hainich2015}. To first order, the mass-loss rates can be scaled as $\dot{M}{\propto}D^{-1/2}$ if other 
clumping parameters are found to be more adequate in the future. To avoid further free parameters 
that cannot be constrained with our dataset, optically thick clumps, or macroclumping 
\citep{Oskinova2007, Sundqvist2011, Surlan2013}, is not accounted for here, and may result in an 
underestimation of mass-loss rates by up to a factor of ${\approx}2$.

Because optical WR spectra  are dominated by recombination lines, 
it is customary 
to parametrize their atmospheric models using the so-called transformed radius \citep{Schmutz1989},

\begin{equation}
 R_\text{t} = R_* \left[ \frac{v_\infty}{2500\,{\rm km}\,{\rm s}^{-1}\,}  \middle/  
 \frac{\dot{M} \sqrt{D}}{10^{-4}\,M_\odot\,{\rm yr}^{-1}}  \right]^{2/3},
\label{eq:Rt}
\end{equation}
defined such that equivalent widths of recombination lines of models  
with given $R_\text{t}$ and $T_*$ are 
approximately preserved, independently 
of $L$, $\dot{M}$, $D$, and $v_\infty$. $R_\text{t}$ is thus a proxy for the mass-loss rate, 
normalized to the surface area of the star.

X-rays can alter the ionization structure in the wind via Auger ionization. We account for X-rays
in a few targets in which we found evidence for this effect (e.g., the presence of the N\,{\sc v} 
resonance line in the UV for late-type OB stars). X-rays are accounted for by assuming 
a spherical distribution of an optically-thin plasma \citep{Baum1992}. The onset radius was always 
fixed to 1.1\,$R_*$, and the X-ray temperature and filling factor are chosen so that 
a luminosity comparable to that observed is obtained. 

Our models include complex model atoms for H, He, C, N, O, Mg, Si, P, S, and the iron group elements 
(dominated by Fe). In this study, we fit the hydrogen and nitrogen mass fractions 
$X_\text{H}$, $X_\text{N}$ (from which $X_\text{He}$ follows) for WN stars; 
the remaining abundances are kept fixed (with a few exceptions, see Sect.\,\ref{sec:comments}). For OB-type stars, 
all abundances are kept fixed to base LMC values unless stated otherwise in Sect.\,\ref{sec:comments}.
Following H14, the base values for C, N, O, Mg, Si, and Fe and are adopted from studies  
for by \citet{Korn2005}, \citet{Hunter2007}, and \citet{Trundle2007}, and the remainder 
are fixed by a scaling of 1/2 solar, resulting in: $X_\text{H} = 0.74$, $X_\text{C} = 4.75\cdot10^{-4}$, $X_\text{N} = 7.83\cdot10^{-5}$, 
$X_\text{O} = 2.64\cdot10^{-3}$, $X_\text{Mg} = 2.06\cdot10^{-4}$, $X_\text{Si} = 3.21\cdot10^{-4}$, 
$X_\text{P} = 2.91\cdot10^{-6}$, $X_\text{S} = 1.55\cdot10^{-4}$, and $X_\text{Fe} = 7.02\cdot10^{-4}$. 
For WN-type stars, the CNO abundances are fixed by assuming a CNO-cycle equilibrium, in which most of the 
carbon and oxygen were converted to nitrogen: $X_\text{N} = 4\cdot10^{-3}$, $X_\text{C} = 7\cdot10^{-5}$ (see detailed discussion in H14). 
Oxygen is usually not included in the calculation of WN 
models because no corresponding lines are observed in spectra of WN stars.

\subsubsection{Spectroscopy of single stars}
\label{subsubsec:sinan}

For single stars, $T_*$ is derived from the ionization balance (primarily He lines for OB-type stars and 
N lines for WN stars). The surface gravity $\log g_*$, which usually cannot be derived for WR stars, 
is inferred from the strength and shape of pressure broadened lines, primarily 
belonging to the hydrogen Balmer series. 
The wind parameters $\dot{M}$ and $v_\infty$ are derived from the strength and widths (respectively) of 
resonance and recombination wind lines in the spectra.
Abundances are derived from the overall strength of the corresponding spectral lines.
For OB-type stars, $v \sin i$ is derived by convolving the synthetic spectra 
with rotational profiles and fitting them to the observations. 
In cases where the WR star exhibits lines that form relatively close to the hydrostatic core, 
its $v \sin i$ can also be constrained. For this, we
utilize a 3D integration routine in the formal integration that assumes co-rotation up to $\tau_\text{Ross} = 2/3$ and angular momentum 
conservation beyond \citep[see][]{Shenar2014}. 

The luminosity $L$ and reddening $E_{B - V}$ are derived by fitting 
the spectral energy distribution (SED) of the model spectra to observed photometry or flux-calibrated spectra. 
For the reddening, we assume two contributions. The first follows a Seaton reddening law \citep{Seaton1979} 
with $R_\text{V} = 3.1$ and a constant $E_{B-V}^{\rm Gal} = 0.03\,$mag, mimicking the Galactic absorption in the direction 
of the LMC. The second contribution follows reddening laws published for the LMC by 
\citet{Howarth1983} with a fixed $R_{\rm V} = 3.1$ , where $E_{B-V}^{\rm LMC}$ is fit individually for each target. In the results, we 
give the total extinction $E_{B-V} = E_{B-V}^{\rm Gal} + E_{B-V}^{\rm LMC}$.

\subsubsection{Binary spectral analysis}
\label{subsubsec:binan}

The PoWR code is a tool designed for the analysis of single stars. 
However, in this work, we need to cope with the analysis 
of composite spectra originating in binary or multiple systems. 
In some cases, given sufficient data, 
the spectroscopy of binaries can be reduced to the analysis of single stars.
The analysis procedure of binaries using PoWR was thoroughly 
described in \citet{Shenar2016}. Here, we only repeat the essentials. 

The first challenge of binary analysis is that the number of free parameters is essentially multiplied by the number 
of components. This problem can be overcome 
if the components can be unambiguously identified in the spectrum. For some 
of our targets, we could 
not identify any signature from a binary companion (see  ``composite'' column in Table\,\ref{tab:catalog}). This can have various reasons: 
the companion cannot be seen at the S/N level of the data, the companion is a compact object, 
or a companion is not present at all. For such targets, the parameters derived are expected to be 
similar to those obtained by H14 in their single-star analyses, and they were therefore not re-analysed  here

Ideally, one possesses a time series of the spectra with good phase coverage. 
In this case, it is possible to disentangle the composite spectrum into its constituent spectra (see Sect.\,\ref{subsec:disen}), 
significantly simplifying the analysis procedure. 
This was only possible for five systems - BAT99\,19, 32, 77, 103, and 113 (Sect.\,\ref{subsec:disen}). 
In cases where only co-added spectra were available, or when phase-resolved spectra did not yield 
plausible disentangled spectra (e.g., BAT99 95), the spectra were 
analysed by adding up model spectra that represent the system's components. 

Another challenge is introduced by the unknown light ratio of the stars, e.g., in the visual band. 
A dedicated photometric analysis is only possible for a few targets 
and will be the focus of future studies. 
However, the light ratio can also be estimated spectroscopically.
Specific spectral features that do not change significantly in the relevant parameter domain can help to assess the true 
light ratio of the system (see section 4.3 in \citealt{Shenar2016}).
Another method to constrain the light ratios is by comparing the observed equivalent widths of specific spectral features
with those expected for the star's spectral type. 
The applicability of these methods for each target is discussed in Appendix\,\ref{sec:comments}.

Once the two components are unambiguously identified in the spectrum and their relative light contribution is constrained, 
the analysis of the multiple system in principle reduces to the analysis of single stars. The individual model spectra calculated for 
the system's components are added together to reproduce the observed SED and normalized spectra. 

For an efficient analysis procedure and a reliable error estimate, we utilized dense grids calculated for WN stars \citep{Todt2015}
and OB-type \citep{Hainich2018} stars at LMC metallicity, which are available 
on the PoWR homepage$^{\rm \ref{footnote:PoWR}}$. The 
grids are 2D and span the $T_*-R_{\rm t}$ plane for WR stars and the $T_*-\log g$ plane for OB-type stars.
This enables us to obtain a first good guess for the system's 
parameters, as well as an impression of the errors (see error discussion in Sect.\,\ref{subsubsec:errors}). 
However, in all cases, tailored models were calculated 
to improve the quality of the fit and to better constrain the errors. An example for a spectral fit 
of the binary BAT99\,6 is shown in Fig.\,\ref{fig:ideal006}.

PoWR models are limited to spherical symmetry, which may 
break in the case of binaries. 
Firstly, the stellar surface of components of tight binaries may deviate from spherical-symmetry due to tidal forces or rapid rotation. 
Such deformations may be important especially for OB-type companions in binaries with periods of the order of a few days, and may result 
in pole-to-equator temperature and gravity differences of the order of up to a few kK and 0.2\,dex, respectively, amounting to an overall 
error of $\approx 1\,$kK in $T_*$ and $\approx 0.1\,$dex in $\log g$ \citep{VonZeipel1924}. 
Reflection effects may be present in the case of binary components with large temperature differences, but usually amount to errors of the order 
$\Delta T_* \approx 100\,$K, which are much smaller than our reported errors \citep{Wilson1990, Palate2013PhD}.
If both components possess significant stellar winds, WWCs may occur and 
result in excess emission \citep[e.g.,][]{Luehrs1997, Moffat1998}.
While such phenomena may be
significant or even dominant in the case of specific lines \citep[e.g.,][]{Bartzakos2001}, 
they typically amount to flux variations
of the order a few percent \citep{Hill2000}. Given the number of analysed objects and the conservative 
errors we report, we expect that neglecting these effects  would not impact our main results and conclusions.

\subsection{Orbital analysis}
\label{subsec:orban}

With the hydrostatic layers of WR stars typically hidden behind thick winds, masses of WR stars are 
notoriously difficult to measure via spectroscopy. 
One of the most important advantages of WR binaries is that they enable a derivation of the mass via orbital analyses.
If the orbital inclination $i$ and both RV curves can be obtained, the companions' 
masses can be calculated from Newtonian dynamics. This method is indispensable in the case 
of WR stars. Knowledge of these masses provides a critical test not only of stellar evolution models, 
but also of mass-luminosity relations which exist for WR stars \citep{Langer1989, Graefener2011}. 

The orbital analysis follows a similar pattern to the one outlined by \citet{Shenar2017}. The first step in determining an orbit is the measurement of RVs. 
For single-lined targets, which always have WR-like spectra in our case, the RVs were measured by cross-correlating
specific spectral lines or whole regions of lines (see below) with a template, and fitting a parabola to the maximum region of the cross-correlation function \citep[e.g.,][]{Zucker2003}. 
The template in this case was always chosen initially to be one of the observations, from which preliminary RVs were determined.
A big advantage of using an observation as a template is that it is not affected by the fact that different spectral lines of WR stars may imply different RVs due to their varying formation regions 
and asymmetric profiles. The 
spectra are then co-added in the frame-of-reference of the WR star using these RVs to create a high S/N template, which is then used 
to iterate on the RV measurement, thereby reducing the statistical measuring errors. The absolute values of the RVs are obtained by cross-correlating the template with a suitable PoWR model. The absolute values 
are therefore less certain ($\sigma \approx 30\,$\kms) than the relative RVs, but this has no bearing on the orbit determination and binary identification.

For SB2 binaries, two different approaches are used. If unique spectral lines can be identified that originate only in one component, 
these lines are used to measure the RVs of the individual components. For the WR component, we repeat exactly the same procedure as done for single-lined binaries. 
For an OB-type component, we use suitable PoWR models as templates. 
If all spectral lines of high enough S/N show contributions of both components, 
we implement a 2D cross-correlation technique following \citet{Zucker1994}.
In this case, the template is constructed from two templates, one for each component, 
each shifted across the velocity space. Since an observation cannot be used as a template for the WR star (because it is entangled with the companion), 
suitable PoWR models to derive the preliminary RVs instead. If the spectra could be disentangled, 
the RVs were derived again using the disentangled spectra as templates (see Sect.\,\ref{subsec:disen}). In all cases, errors are calculated as in \citet{Zucker2003}.

The choice of lines/region to cross-correlate with depend on the target. 
Generally, the He\,{\sc ii}\,$\lambda 4686$ line, despite typically being the strongest 
spectral lines for WR stars, should be avoided for RV measurements if possible, because it is very susceptible to wind variability and WWCs
and is generally not a good 
tracer for the RVs of the star \citep[see, e.g., Figure 5 in][]{Shenar2018}. For SB1 or apparently single WR stars (BAT99\, 12, 31, 102), we use a large spectral region covering $\approx4000-4600\,\AA$, which includes 
Balmer lines, the He\,{\sc i}\,$\lambda 4388$ and $\lambda 4471$, He\,{\sc ii}\,$\lambda 4200$ and $\lambda 4542$, N\,{\sc iv}\,$\lambda 4060$ lines, 
and the Si\,{\sc iv} $\lambda \lambda 4089, 4116$ doublet (depending on the target). This enabled us 
to boost the measurement accuracy. No significant differences were obtained by exploring specific lines instead. 

For the SB2 systems BAT99\,19, 103, and 107, the RVs of both components are derived using a 
2D cross-correlation technique on the whole available spectral region ($\approx 4000-4600\,\AA$). The initial templates were chosen to be 
suitable PoWR models, and after disentanglement (Sect.\,\ref{subsec:disen}), they are replaced with the disentangled spectra. Similarly, the RVs of the companions in BAT99\,95 were measured 
through a 2D cross-correlation, but this time using the sharp N\,{\sc iv}\,$\lambda 4060$ line alone, since no reliable template spectra could be established for other lines. 

Due to the limited quality and small number of spectra available, the RVs of the SB2 systems BAT99\,32 and 77 could only be established for the primary WR star using the N\,{\sc iv}\,$\lambda 4060$ line. 
The RVs of the WR primary in the SB2 system BAT99\,113 is measured from standard cross-correlation of the whole spectrum, since the secondary contributes only $\approx 10\%$ to the total light. The 
RVs of the secondary are measured by performing standard cross-correlation with the He\,{\sc i}\,$\lambda 4471$ line. The same technique is used for the potential SB2 system BAT99\,92, although this time, 
the He\,{\sc ii}\,$\lambda 4686$ is used for the WR primary, while the region $4000-4600\,\AA$ is used for the secondary. The He\,{\sc ii}\,$\lambda 4686$ line is used here because the WR primary 
is strongly diluted by the secondary, and is the only line clearly visible for the WR star.
A compilation of the final measured RVs for each spectrum, as well as the lines/regions used, are given in Tables \ref{tab:ObsLogFUV} - 
\ref{tab:ObsLogOpt4}.

Once the RVs have been established, 
an SB2 orbit is then fit 
to the derived RVs of both components simultaneously, constraining the orbital period $P$, eccentricity $e$, 
RV amplitudes $K_1 \equiv K_\text{WR}$ and $K_2$, periastron time $T_0$, and argument of periapsis $\omega$ (for 
non-circular orbits). 
The fitting is done using a self-written Python tool that relies on the minimization package lmfit\footnote{https://lmfit.github.io/lmfit-py/}.  The tool finds the best-fitting 
RV curves for both sets of RVs simultaneously through the Levenberg-Marquardt algorithm, which is a damped least-squares minimization technique.
For the RVs of the WR component, we allow for a constant velocity shift, which is fixed by the relative offset of the 
O-star RVs, since the O-star is much more reliable for absolute RV measurements.

\subsection{Spectral disentanglement}
\label{subsec:disen}

Spectral disentanglement is a powerful mathematical tool that separates composite spectra to 
their constituent spectra\footnote{We use the terms disentangling and spectral separation interchangeably, regardless 
of whether prior knowledge of RV measurements is required.} \cite[e.g.,][]{Bagnuolo1991, Hadrava1995, Marchenko1998}.
For this to work, spectra with a sufficient phase coverage (typically ${\approx}$5-10 spread roughly homogeneously in RV) are necessary. 
This condition is met for eight of our targets: BAT99\,12, 19, 32, 77, 92, 95, 103, and 113. 
However, disentanglement attempts of BAT99\,12, 92, and 95 did not yield plausible results. BAT99\,12 and 92 are not 
found to show significant RV variation in the few CTIO spectra at hand. 
BAT99\,95 is clearly a binary, but the FLAMES data at hand imply that it potentially consists of 
two WR stars, and requires a better phase coverage to disentangle.

BAT99\,19, 103, and 113 were disentangled using a 
self-written Python tool that applies the ``shift-and-add'' algorithm described in detail in \citet{Marchenko1998}.
This method already assumes knowledge of the RV orbits (Sect.\,\ref{subsec:orban}). 
The shift-and-add technique relies on an iterative co-adding of all composite spectra 
in the frame of reference of star A, subtracting this (Doppler-shifted) template of star A from all composite spectra, 
and then co-adding the residual spectra in the frame of reference of star B. 
This results in two templates,  one for star A and one for B. 
This iteration is performed until no further difference can be seen in the solution, which typically takes 3-4 iterations.

Since the RVs of BAT99\,32 and 77 could not be derived, we attempted their disentanglement 
using the code \texttt{Spectangular}, which is based on singular value decomposition (SVD) in the wavelength domain 
\citep{Sablowski2017}, first applied to WR stars in \citet{Shenar2017}. 
This procedure needs spectra spread over the orbital period and optimizes the orbital parameters or the RVs to minimize the 
residuals between the disentangled spectra and the observations. 
Simultaneously to the disentanglement, the 
code optimizes for the orbital parameters of the system. The relative light ratios are assumed to be constant throughout the orbit and are 
fixed to those derived from the spectral analysis (Sect.\,\ref{subsec:specan}).
To determine the orbital elements, we used an initial orbit from S08. 
Since S08 had a much better phase coverage of the orbit, we adopt here their derived 
orbital parameters, with the exception of the secondary's amplitude $K_2$, which is determined in our study.

\section{Results}
\label{sec:results}

\begin{figure*}
\centering
\begin{subfigure}{\columnwidth}
  \centering
  \includegraphics[width=0.95\linewidth]{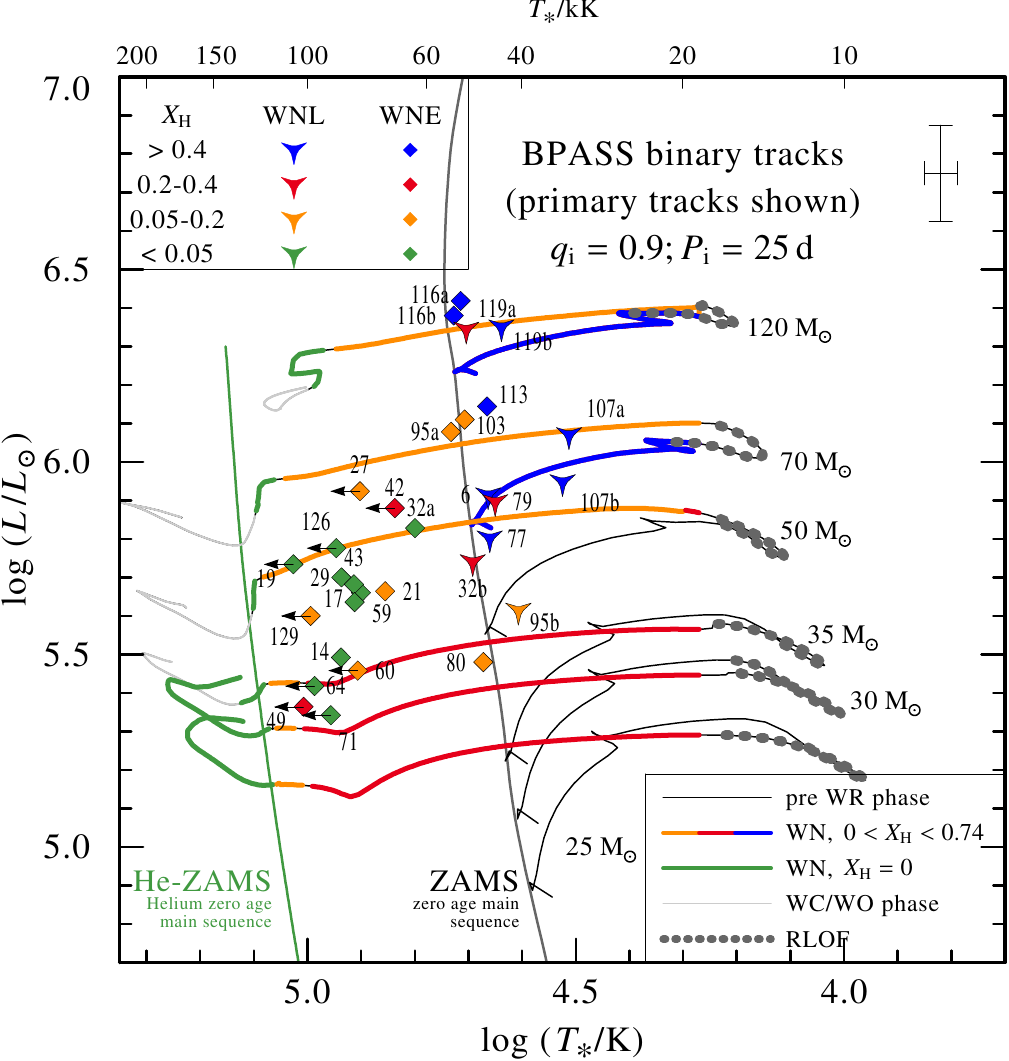}
%   \caption{s}
  \label{fig:sub1}
\end{subfigure}%
\begin{subfigure}{\columnwidth}
  \centering
  \includegraphics[width=0.95\linewidth]{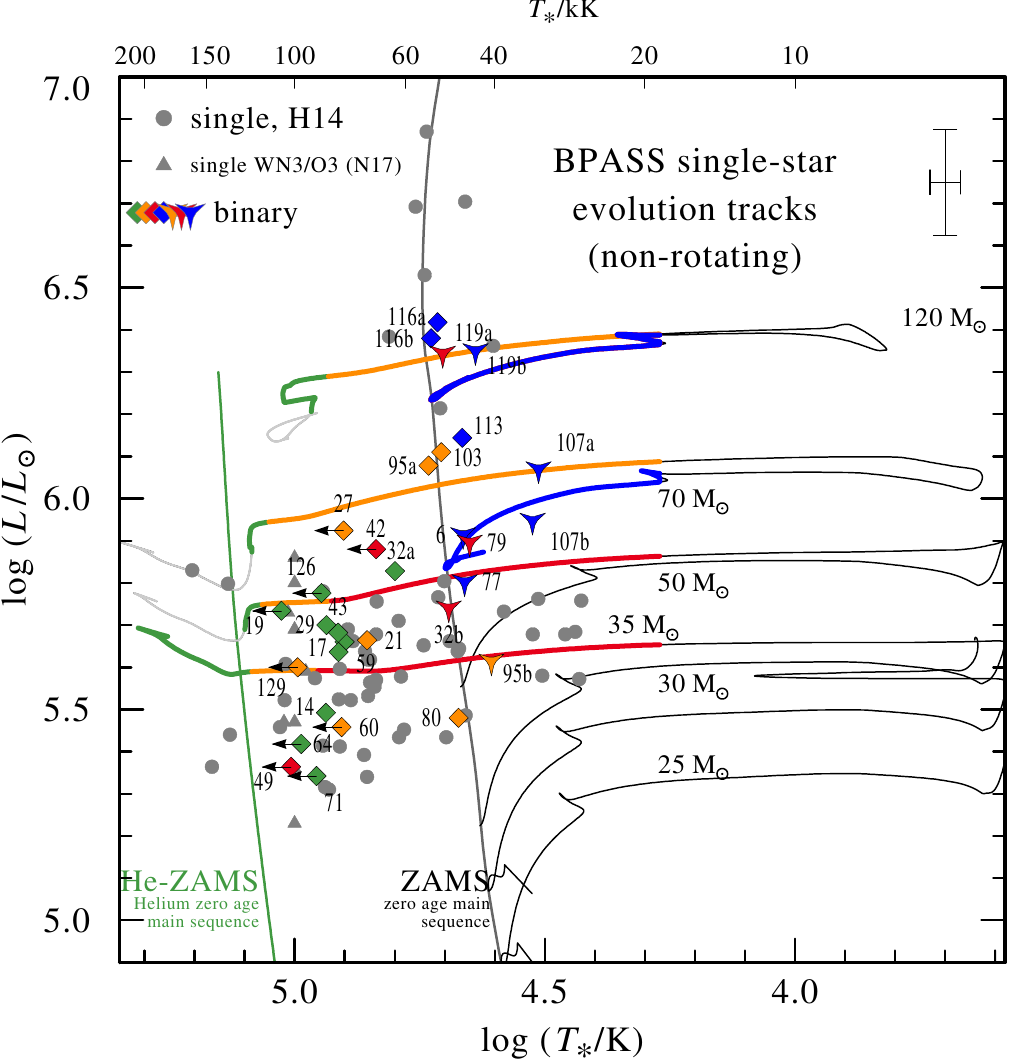}
%   \caption{A subfigure}
  \label{fig:sub2}
\end{subfigure}
\caption{{\it Left panel:} HRD positions of the LMC WR components. Labels correspond to the BAT99 catalog.
Plotted are a selection of binary evolution tracks \citep{Eldridge2008, Eldridge2016} calculated with the BPASS code for $Z = 0.008$. 
The tracks show the evolution of the primary star for several initial masses $M_{1,{\rm i}}$ and initial mass-ratio and period 
of $q_{\rm i}=0.9$ and $P_{\rm i}= 25\,$d, respectively.
The colours and symbols code the hydrogen abundance and WR type, as described in the legend.  
The WR phase is defined for $T_* > 20\,$kK. {\it Right panel:} as left panel, but showing BPASS tracks calculated 
for non-rotating single stars, and including the putatively 
single LMC WN stars (gray symbols, adopted from H14), and the WN/O3 stars (gray triangles, adopted from \citealt{Neugent2017}).
Little arrows imply lower bounds on $T_*$ (degeneracy domain / insufficient S/N).
}
\label{fig:HRD}
\end{figure*}

\subsection{Spectral analysis}
\label{subsec:specanres}

The derived stellar parameters for the WR binaries analysed here are given in Table\,\ref{tab:specan} and include the spectral 
type, effective temperature $T_*$ and $T_{2/3}$ (effective temperature at $\tau_\text{Ross} = 2/3$), surface gravity $\log g_*$, 
luminosity $\log L$, radius ($R_*$ and $R_{2/3}$), transformed radius $\log R_\text{t}$, terminal velocity 
$v_\infty$, mass-loss rate $\dot{M}$, Smith visual absolute magnitude $M_\text{v, Sm}$, fractional light ratio in the visual 
$f_\text{V}$, surface hydrogen and nitrogen mass fractions $X_\text{H}$ and $X_\text{N}$, projected rotation 
velocity $v \sin i$, spectroscopic mass $M_\text{spec}$ (see below), and reddening $E_\text{B-V}$.

For OB-type stars, $M_\text{spec}$ is calculated via $M_\text{spec} = G^{-1}\,g_*\,R_*^2$. 
For WR stars, $M_\text{spec}$ is calculated 
via mass-luminosity relations calculated for homogeneous stars by \citet{Graefener2011}. 
If $X_\text{H} \geq 0.4$, we specify the mass for a homogeneous star with the same 
$X_\text{H}$ and $\log L$, since the star is presumably young and on the main sequence. 
Otherwise, we give the mass for a pure He-star with the same $\log L$, 
since it is expected that the H-layer would be negligible in mass.

In cases where the wind parameters for the OB-type stars could not be derived, we adopted mass-loss 
rates from \citet{Vink2001}, and terminal velocities which scale as $v_\infty = 2.6\,v_{\rm esc}$ \citep{Lamers1995}. When only 
upper/lower limits could be derived, the final models were calculated using these limits.

In the left panel of Fig.\,\ref{fig:HRD}, we plot the HRD positions of the 31 WR components 
analysed in our study. BAT99\,72 is omitted due to its uncertain nature, and 
BAT99\,92 is omitted since it is found to be a WC star (see Sect.\,\ref{sec:comments}). 
In the right panel, we also include the positions of the putatively single
WR stars in the LMC, as derived by H14 and \citet{Neugent2017}. 
In Fig.\,\ref{fig:HRD}, we also show evolution tracks calculated with the BPASS code. The left panel shows 
evolution tracks for primaries of various masses in binaries with an initial mass-ratio of $q_{\rm i} = 0.9$ and 
$P_{\rm i} = 25\,$d. The right panel shows evolution tracks for the same initial masses, but for single stars. 

\begin{figure}[h]
\centering
  \includegraphics[width=0.5\textwidth]{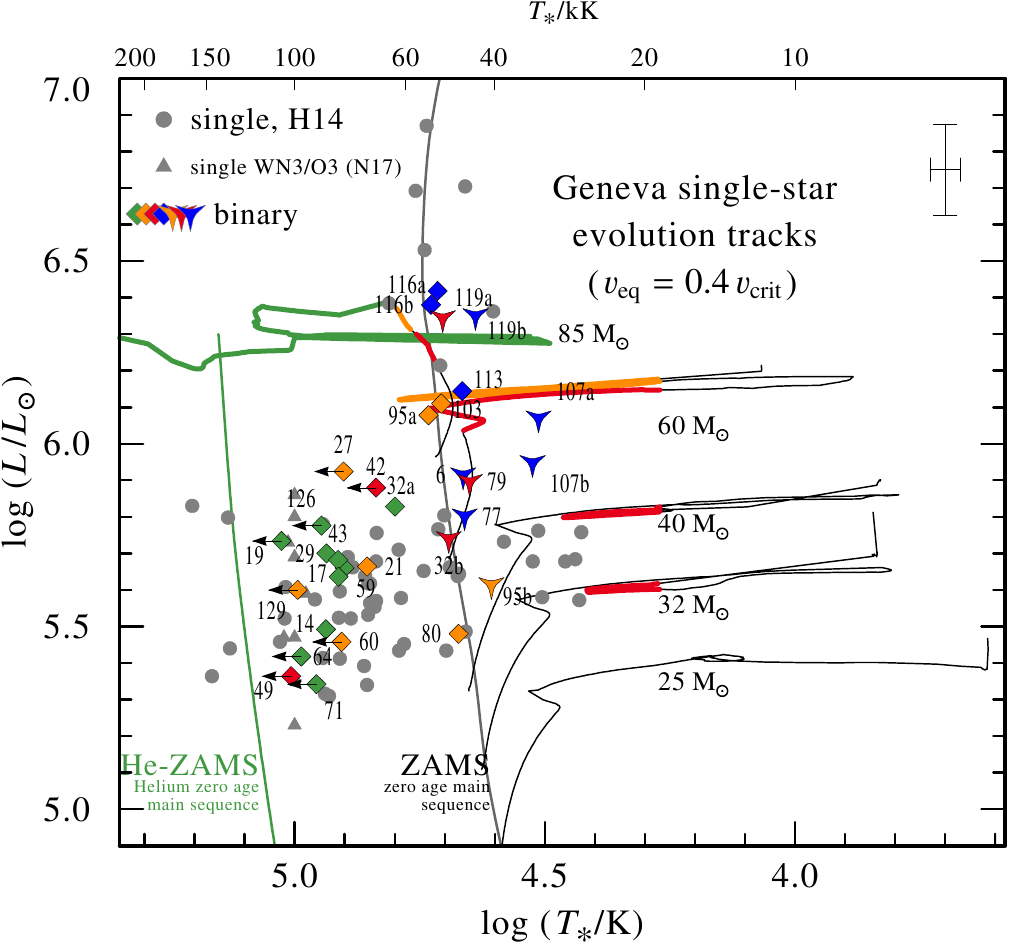}
  \caption{ Same as right panel of Fig.\,\ref{fig:HRD}, but this time showing tracks calculated with the Geneva code for single stars rotating (initially) at 40\% the critical 
  rotation velocity at $Z = 0.006$. Kindly provided by C.\ Georgy (priv.\ comm.)
  }
\label{fig:hrdGeneva}
\end{figure}

In Fig.\,\ref{fig:hrdGeneva}, we also show 
evolution tracks calculated with the Geneva code for single stars that rotate at 
40\% their critical rotation (Eggenberger et al.\ in prep.), kindly provided by C.\ Georgy (priv.\ comm.). The tracks were calculated at $Z=0.006$ (close to the BPASS value of 0.008) 
and first presented in \citet{Georgy2015}. They behave very differently 
from the BPASS tracks, also compared to previous generations of the Geneva tracks published by \citet{Meynet2005}. While a detailed comparison 
between the tracks is beyond the scope of this paper, the main difference likely originates in different mass-loss prescriptions between the codes. The Geneva tracks 
fail to reproduce even the most luminous cWR stars in the sample. We discuss the comparison between our results and the tracks in more detail in Sect.\,\ref{sec:disc}.

\renewcommand{\arraystretch}{1.2}
\begin{table*}[!htb]
\scriptsize 
\setlength\tabcolsep{2.5pt}
\caption{Derived parameters for LMC WN binaries with composite spectra}
\label{tab:specan}
\begin{center}
  \\ 
\hline
\noalign{}\hline\noalign{}
\end{tabular}
\tablefoot{
\tablefoottext{a}{Rotation is inferred indirectly from the round emission line profiles (see Appendix\,\ref{sec:comments}).}\tablefoottext{b}{Parameters adopted from \citet{Tehrani2019}; our results are comparable, but with the reddening law used here, we obtain $\log L_{1,2} = 6.31,6.20\,[L_\odot]$ (see Appendix\,\ref{sec:comments}).}
}
\end{center} 
\end{table*}

\subsubsection{Spectral classification}
\label{subsubsec:specclass}

For the spectral classification of OB-type stars, we used quantitative schemes by Sana et al. (in prep.), which are 
extensions of schemes published by 
\citet{Mathys1988, Mathys1989}, \citet{Walborn1990}, and \citet{Walborn2002}
for OB-type stars. For WR stars, we adopt previous classifications by \citet{Neugent2018} unless noted otherwise. 
The classifications are generally similar to those given by FMG03 and S08, with the exception that several WN4b stars in the latter 
studies became WN3 in \citet{Neugent2018}.
For ``slash'' WR stars, we used morphological 
classification schemes by \citet{Crowther2011} and \citet{Massey2009}.
If an empirical spectral disentanglement was possible, we classified the disentangled spectra. 
Otherwise, we classified the individual \emph{model} spectra, 
which should supply a good representation to the observed spectrum of the star. In many cases, however, 
better data quality and a better phase coverage would be necessary to confirm the spectral types.

\subsubsection{Errors from the spectral analysis}
\label{subsubsec:errors}

Due to the high computational 
cost of full non-LTE model calculations, a rigorous $\chi^2$-fitting that covers all parameters is not feasible. 
However, given that the study relied on grids of models, one can estimate realistic statistical 
errors on the parameters. Regardless, the true error is dominated by systematic errors, which 
originate, among other things, in the uncertain 
wind velocity field \citep{Graefener2005, Sander2017}, clumping \citep{Feldmeier1995, Oskinova2007, Sundqvist2013}, 
atmosphere inflation \citep{Graefener2012b, Sanyal2015, Grassitelli2018, Ro2019}  and binary effects 
such as WWCs, mutual irradiation , and tidal deformations \citep{Moffat1988, Shenar2017}.
It is our belief that the large number of analysed systems compensates for these uncertainties.

For WR stars, 
the typical statistical errors on $T_*$ correspond to half a grid spacing, or $\sigma^\text{WR}_{T_*}{=}0.05\,$dex. 
Larger errors are possible for the hottest stars in our sample, or stars with very thick winds, which are found 
in a so-called degeneracy domain \citep[see discussion in][]{Todt2015}. In this domain, 
$T_*$ and $R_\text{t}$ cannot be derived independently.
Errors of 0.05\,dex in $R_\text{t}$ are typical for single stars, but are somewhat larger for WR binaries due to degeneracy 
with the relative light ratios. Depending on how well the latter could be constrained, errors on $R_\text{t}$ are conservatively estimated to be 
$\sigma^\text{WR}_{R_\text{t}}{=}$0.1\,dex, except for stars in the degeneracy domain, where $R_\text{t}$ values could be 
arbitrarily smaller. Terminal velocities of WR stars are derived to a typical accuracy of 
$\sigma^\text{WR}_{v_\infty}=$100-200\,\kms. Hydrogen mass fractions are determined with an accuracy of 
$\sigma^\text{WR}_{X_\text{H}}{\approx}0.1$.

For OB-type stars, the temperature can typically be derived to an accuracy of $2-3\,$kK, which is roughly twice our grid's spacing. The main reason 
for this relatively large error is contamination with WR features, which are often hard to disentangle, as well as degeneracies with $\log g$.
$\log g_*$ could be poorly constrained, because the Balmer absorption lines of the OB-type 
components are often filled with 
emission stemming from the WR star. Nevertheless, this parameter could be constrained to a 
certain degree since larger 
$\log g_*$ values result in larger equivalent widths for the Balmer lines. 
A typical uncertainty on $\log g_*$ amounts to $0.3\,$dex.  

Transformed radii are not a helpful quantity for OB-type stars, for which the mass-loss rate is derived directly. 
The mass-loss rates and terminal velocities could only be constrained for OB-type stars in some cases, 
depending on the data (see Appendix\,\ref{sec:comments} for a detailed account.). The errors on $\dot{M}$ for OB-type stars, when a value 
is given, are typically of the order of $0.3\,$dex. Errors on their terminal velocities, when such were derived, 
are about 200\,\kms. Finally, the light ratios could typically be derived at a $\approx$20\% level. 

The total luminosity and reddening could be well constrained from the data, especially when flux-calibrated UV spectra are available.
Despite the $T_*-R_\text{t}$ degeneracy mentioned above, the errors on $\log L$ are typically modest. This is because ``degenerate'' models 
(that is, models with different $T_*-R_\text{t}$ values but virtually identical spectra) produce almost identical SEDs, and therefore 
require almost identical luminosities. Models with larger $T_*$ would therefore have correspondingly smaller radii $R_*$ to preserve 
$\log L$. The errors on the luminosities are primarily dominated by errors on the light ratios. Together with the error 
from the SED fitting, this amounts to 0.1--0.15 in $\log L$, depending on how well constrained the 
light ratios are. The reddening $E_\text{B-V}$ can be derived to an accuracy of 
$0.02$\,mag in cases where UV spectra are present, and to $0.05\,$mag otherwise. Similarly, errors on the absolute visual magnitudes $M_\text{V}$ are 
affected by errors on the light ratios, and typically correspond to $0.15\,$mag. 

The remaining errors follow from error propagation. For $R_*$, this amounts typically to $0.5\,R_\odot$ for 
$R_* < 10\,R_\odot$ and 
$1\,R_\odot$ otherwise. Since the mass-loss rate of WR stars scales with $R_\text{t}^{-3/2}$, errors on WR mass-loss rates are of the order 
of 0.15-0.2\,dex. While $\dot{M}$ also depends on $R_*$ and $v_\infty$ and is thus subject to further errors, a change in $R_*$ tends to result 
in a corresponding change in $R_\text{t}$ in a way that conserves the value of $\dot{M}$. We therefore only account for errors on $R_\text{t}$ 
here. 

Errors on $M_\text{spec}$ for OB-type stars arise from errors on $R_*$ and $\log g_*$, the latter being especially large. This easily corresponds 
to a factor two uncertainty in the mass. As for WR stars, the errors can be estimated from the mass-luminosity relations used
(see Sect.\,\ref{subsec:specanres}) by considering the errors on $\log L$ and $X_\text{H}$.
Rotation velocities are determined to an accuracy of ${\approx}30-50$\,\kms, depending on the resolution of our 
data. For co-added spectra, typically only upper limits could be derived (see Sect.\,\ref{sec:obs}).

\subsubsection{Disentangled spectra}
\label{subsubsec:disen}

\begin{figure}[!htb]
\centering
  \includegraphics[width=0.5\textwidth]{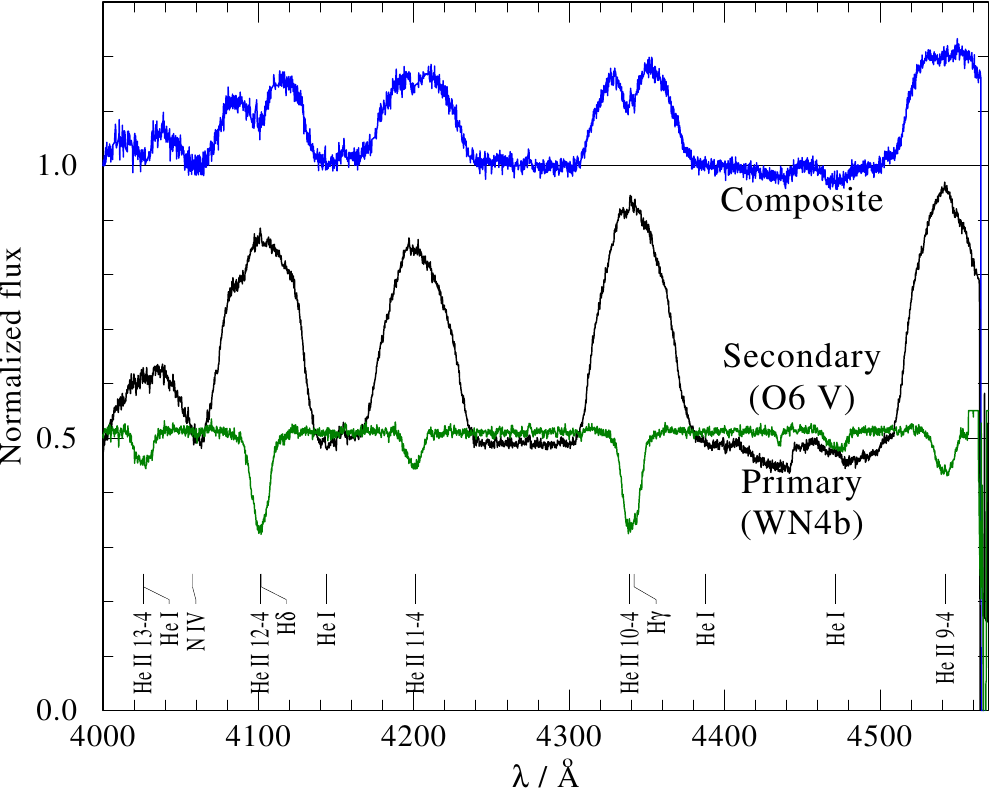}
  \caption{Disentanglement of BAT99\,19 using the shift-and-add technique. Shown are one of the observed composite FLAMES
  spectra (blue line) and the disentangled spectra for the WR primary (black line) and O-type secondary (green line). 
   The levels of the components' spectra reflect their fractional contribution to the total continuum flux.}
\label{fig:dis019}
\end{figure}

\begin{figure}[!htb]
\centering
  \includegraphics[width=0.5\textwidth]{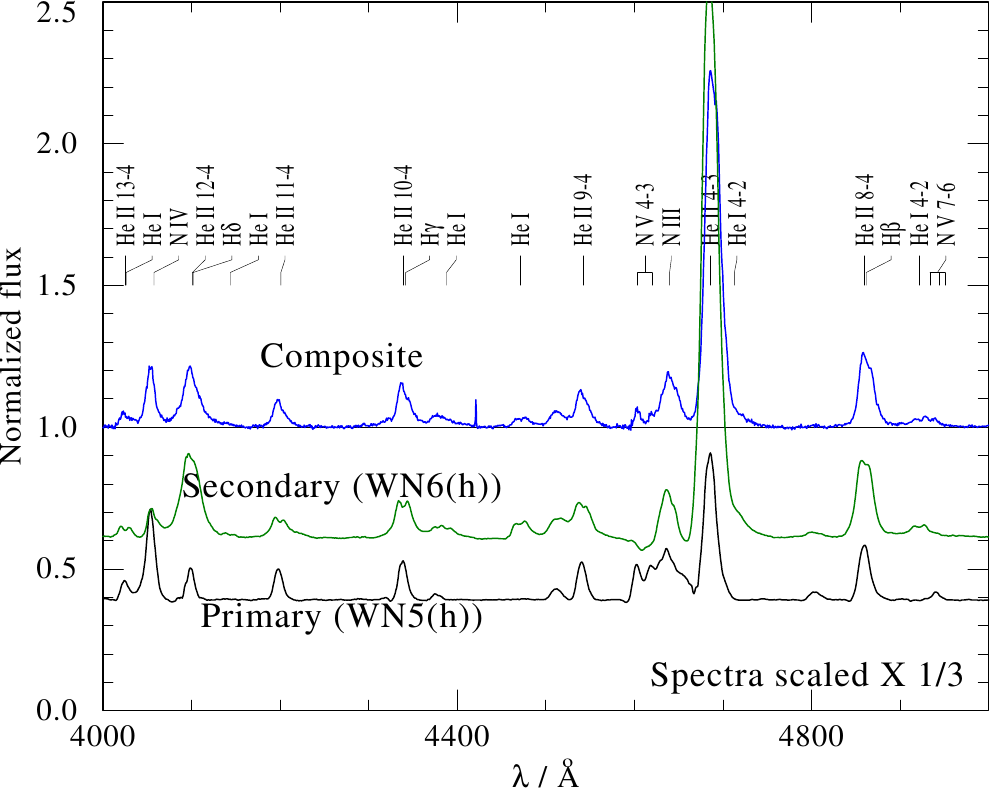}
  \caption{Same as Fig.\,\ref{fig:dis019}, but for BAT99\,32 and using the code \texttt{Spectangular} (see text for details).}
\label{fig:dis032}
\end{figure}

\begin{figure}[!htb]
\centering
  \includegraphics[width=0.5\textwidth]{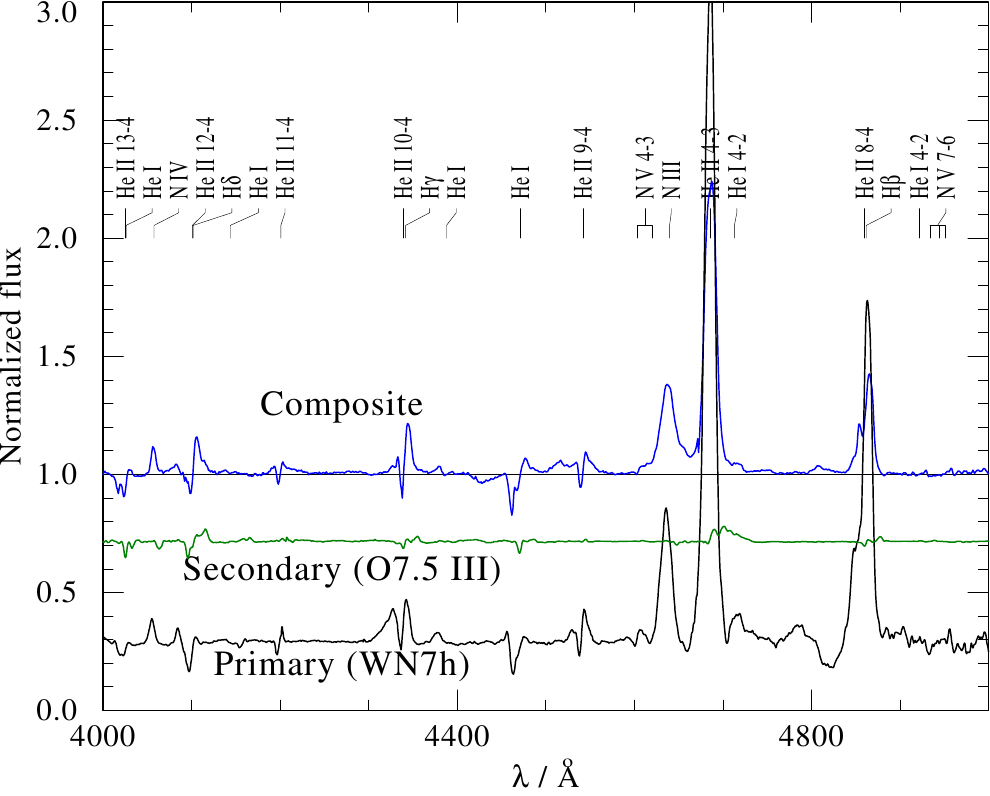}
  \caption{Same as Fig.\,\ref{fig:dis019}, but for BAT99\,77 and using the code \texttt{Spectangular} (see text for details).}
\label{fig:dis077}
\end{figure}

\begin{figure}[!htb]
\centering
  \includegraphics[width=0.5\textwidth]{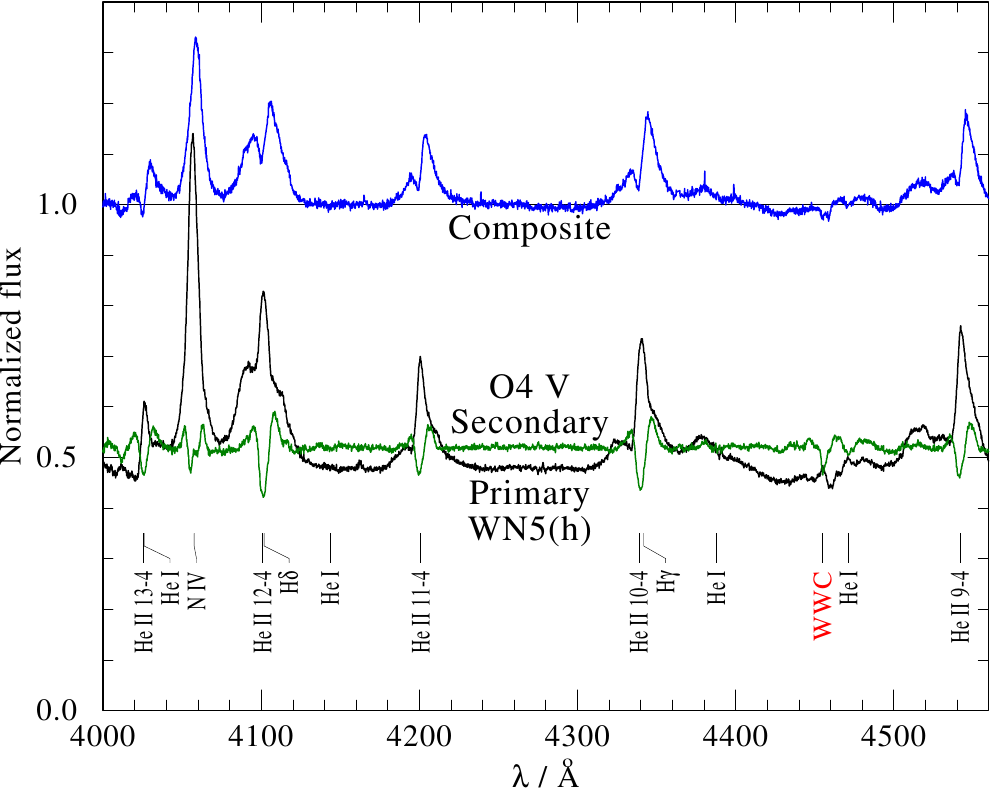}
  \caption{Same as Fig.\,\ref{fig:dis019}, but for BAT99\,103. }
\label{fig:dis103}
\end{figure}

\begin{figure}[!htb]
\centering
  \includegraphics[width=0.5\textwidth]{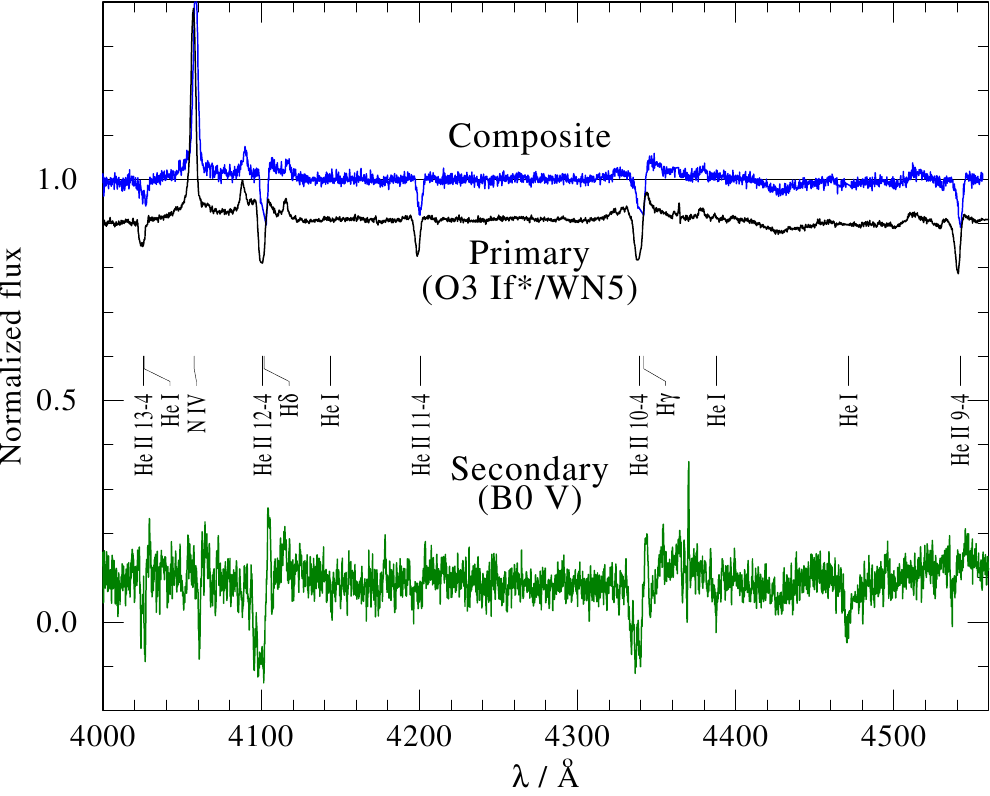}
  \caption{Same as Fig.\,\ref{fig:dis019}, but for BAT99\,113. The secondary's disentangled spectrum is binned at $1\AA$ to increase the 
  S/N.}
\label{fig:dis113}
\end{figure}

The results of our disentanglement of BAT99\,19, 32, 77, 103, and 113 are shown in Figs.\,\ref{fig:dis019}-\ref{fig:dis113}. In all cases, 
the observed composite spectra are corrected for the systematic velocities (see Table\,\ref{tab:orbitalpar}).
Note that the disentangled spectra are intrinsic to the components, i.e., they are corrected for line dilution. 

The disentangled spectra of BAT99\,19 (WN4b + O6~V, $P=18\,$d) imply a very large $v \sin i$ value for the secondary in excess
of 550\,\kms (Fig.\,\ref{fig:dis019}). This is by far the largest rotational velocity observed in our sample (Sect.\,\ref{subsec:rotvel}). 
Together with its period, this may imply that BAT99\,19 recently experienced a mass-transfer event. 
Interestingly, the spectral lines of the WR star are also peculiarly round and broad (Fig.\,\ref{fig:BAT19_round}). Such profiles
were attempted to be reproduced by assuming rotation of the WR star in \citet{Shenar2014}, requiring large co-rotation radii. 
It is not certain whether these line profiles are indeed related to rotation or not, but their shape is rare among WNE stars. 
Given that BAT99\,19 is an eclipsing system with such unique properties, we encourage its future study.

The disentangled spectra of the binary BAT99\,32 (WN5 + WN6(h) (+ abs), $P=1.9\,$d) - the shortest-period WR binary in our sample - appear to suggest that both components exhibit WR-like spectra (Fig.\,\ref{fig:dis032}). 
However, the disentangled spectrum of the secondary (green spectrum) is also suggestive of an additional absorption component, 
i.e., the system may be a triple. If BAT99\,32 is indeed a WR+WR binary, it would be an extremely important system to study.
Given its short period, it could be a promising candidate for a black-hole merger progenitor experiencing CHE. 
Alternatively, it could be a rare, short-period ms-WR + ms-WR system such as the 3.7\,d period Galactic WR binary \object{WR20a} \citep{Rauw2004}. 
Either way, these findings warrant additional studies of BAT99\,32.
Whether this result is real or a spurious effect from the limited number of spectra should be verified 
in future studies.

With only four phase-dependent spectra available, 
the results for BAT99\,77 (WN7h + O7~III, $P=3.0\,$d) are peculiar (Fig.\,\ref{fig:dis077}): The spectrum of the primary WR star seems reasonable
(albeit likely contaminated by the secondary's spectrum), 
but the spectrum of the secondary is almost featureless. Some features, such as the He\,{\sc ii}\,$\lambda 4686$ emission, are 
clearly biased due to contamination with the WR star. While some weak He\,{\sc i} lines are present, He\,{\sc ii} lines are almost 
completely absent, and the Balmer lines are extremely weak.  In contrast, He\,{\sc ii} absorption lines are clearly seen 
in the co-added spectrum at hand.  It is therefore likely that our results for this system, especially 
for the secondary, do not represent the component spectra well. Clearly, better data will be needed to properly disentangle 
this system.

The disentangled spectra of BAT99\,103  (WN5h + O3.5~V, $P=2.8\,$d) seem plausible, but significant deviations are seen in the residual spectra. These 
are most likely caused primarily by WWC, which is not accounted for in the disentanglement procedure. However, these features 
are not expected to interfere with the classification and analysis of the object.

The disentangling of BAT99\,113 (O3~If*/WN5 + B0~V) was especially challenging due to the faintness of the secondary, 
which contributes only 
10\% to the total light in the visual. Because of this, the S/N of the secondary's spectrum is relatively low. Nevertheless, 
clear signatures of He\,{\sc i} absorption can be seen in its spectrum. The Balmer lines are likely contaminated 
by the WR star: more data will be necessary to improve the disentanglement.

\subsection{Orbital analysis}
\label{subsec:orbanres}

\renewcommand{\arraystretch}{1}
\begin{table*}[!htb]
\setlength\tabcolsep{1pt}
\caption{Orbital parameters of LMC WN binaries with constrained orbits}
\label{tab:orbitalpar}
\begin{center}
{\renewcommand{\arraystretch}{1.3}
\resizebox{\textwidth}{!}{
\begin{tabular}{llcccccccccccccccc}
\hline\hline
BAT             &          Spectral type\tablefootmark{a}                       &        $P$                       &      $T_0$              &          $V_0$         & $e$              &        $\omega$     &     $K_{\rm WR}$        &   $K_2$ & $q(\frac{M_2}{M_{\rm WR}})$        & $M_{\rm WR} \sin^3 i$ & $M_2 \sin^3 i$     &   $i$ & & $M_2$  &   $M_{\rm WR}$   & $a_{\rm WR}$              & $a_2$             \\ 
                 &                                                                      &  [d]                                              &   [MJD]              &  [\kms]        &                  & [$^\circ$]   & [\kms]       & [\kms]   & $[M_\odot]$     &  $[M_\odot]$         & $[M_\odot]$        & [$^\circ$]   & & $[M_\odot]$    &  $[M_\odot]$                        & $[R_\odot]$        &  $[R_\odot]$      \\ 
\hline  
\multicolumn{18}{c}{SB2 solutions with constrained inclinations} \\ 
\hline
019\tablefootmark{b} &WN4${+}$O6\,V                                              &17.998(1)   &     51914.4${\pm0.2}$       &   258${\pm}7$     &   0 (fixed)  &   $194{\pm}5$             & 206${\pm}4$   & 115${\pm}5$   & 1.79${\pm}0.05$   & 22.1${\pm}2.6$   & 39.6${\pm}4.4$   & 86$^{+4}_{-3}$  &  $\rightarrow$  & 40$^{+5}_{-5}$  &  22$^{+3}_{-3}$  &  74$^{+4}_{-4}$  &  41$^{+3}_{-3}$  \\ 
119\tablefootmark{c} &\begin{tabular}{l}WN6h${+}$\\O3.5\,If*/WN7\end{tabular} &158.76(2)   &    56022.6${\pm0.2}$        &   270${\pm}5$     &   $0.788{\pm}0.007$    &  $61{\pm}7$       & 96${\pm}3$   & 95${\pm}4$   & 1.01${\pm}0.05$   & 13.3${\pm}1.1$   & 13.4${\pm}1.1$   & 39$^{+6}_{-6}$  &  $\rightarrow$  & 54$^{+36}_{-19}$  &  53$^{+36}_{-19}$  &  295$^{+57}_{-41}$  &  292$^{+60}_{-43}$  \\ 
129\tablefootmark{d} &WN3(h)${+}$O5\,V                                            &2.7689(2)   &    51945.927${\pm0.005}$    &   265${\pm}5$     &   0  (fixed)                   &   n/a             & 316${\pm}5$   & 193${\pm}6$   & 1.64${\pm}0.03$   & 14.3${\pm}1.5$   & 23.5${\pm}2.4$   & 73$^{+17}_{-14}$  &  $\rightarrow$  & 27$^{+14}_{-6}$  &  16$^{+8}_{-4}$  &  18$^{+3}_{-2}$  &  11$^{+2}_{-1}$  \\ 
\hline 
\multicolumn{18}{c}{SB2 solutions without constrained inclinations ($M_2$ adopted from BONNSAI evolution models)} \\ 
\hline 
049\tablefootmark{e} &WN3${+}$O8\,V                                           &31.69(3)   &     51943.0${\pm0.9}$       &   234${\pm}17$    &   $0.35{\pm}0.11$       &   $21{\pm}16$     & 104${\pm}13$   & 52${\pm}11$   & 2.0${\pm}0.25$   & 3.4${\pm}1.5$   & 6.8${\pm}2.8$   & 42$^{+10}_{-10}$  &  $\leftarrow$  & 23$^{+4}_{-3}$  &  11$^{+22}_{-8}$  &  92$^{+43}_{-27}$  &  46$^{+26}_{-16}$  \\ 
077\tablefootmark{b,f} &WN7${+}$O7.5\,III                                          &3.003(3)   &     52631.9${\pm0.1}$       &   333${\pm}8$    &   $0.32{\pm}0.02$       &   $7{\pm}4$      & 292${\pm}30$   & 176${\pm}30$   & 1.66${\pm}0.2$   & 10.2${\pm}3.4$   & 16.9${\pm}5.1$   & 51$^{+15}_{-11}$  &  $\leftarrow$  & 36$^{+9}_{-7}$  &  22$^{+30}_{-13}$  &  21$^{+7}_{-5}$  &  13$^{+5}_{-4}$  \\ 
103\tablefootmark{b} &WN5${+}$O4\,V                                            &2.7586(4)   &     53007.9${\pm0.1}$       &   220${\pm}10$    &   0  (fixed)                   &   n/a            & 266${\pm}8$   & 137${\pm}24$   & 1.94${\pm}0.18$   & 6.4${\pm}2.0$   & 12.3${\pm}3.3$   & 39$^{+7}_{-6}$  &  $\leftarrow$  & 51$^{+8}_{-9}$  &  26$^{+28}_{-15}$  &  23$^{+5}_{-4}$  &  12$^{+5}_{-4}$  \\ 
107\tablefootmark{b} &O6.5\,Iafc${+}$O6.5\,Iaf                                    &153.89(6)   &      56041.5${\pm0.4}$      &    262${\pm}1$    &   $0.49{\pm}0.01$      &    $130{\pm}2$   & 95${\pm}1$   & 118${\pm}2$   & 0.81${\pm}0.02$   & 56.5${\pm}2.1$   & 45.5${\pm}1.6$   & 74$^{+8}_{-13}$  &  $\leftarrow$  & 51$^{+15}_{-6}$  &  63$^{+25}_{-7}$  &  262$^{+31}_{-11}$  &  326$^{+40}_{-16}$  \\ 
113\tablefootmark{b} &O2\,If*/WN5${+}$B0\,V                                        &4.6965(2)   &      52993.5${\pm0.4}$      &    269${\pm}8$    &   0  (fixed)                  &    n/a           & 100${\pm}2$   & 310${\pm}11$   & 0.32${\pm}0.04$   & 25.3${\pm}2.8$   & 8.2${\pm}0.9$   & 52$^{+6}_{-5}$  &  $\leftarrow$  & 17$^{+2}_{-2}$  &  53$^{+20}_{-15}$  &  12$^{+2}_{-1}$  &  37$^{+5}_{-5}$  \\ 
116\tablefootmark{i} &WN5h${+}$WN5h                                                &154.55(5)   &       57671.2${\pm0.9}$       &  287${\pm5}$                &   $0.68{\pm}0.02$                   &   $20.9{\pm}3.8$             & 130${\pm}7$   & 141${\pm}6$   & 0.92${\pm}0.07$   & 65.3${\pm}7.3$   & 60.2${\pm}7.0$   & 51$^{+7}_{-5}$  &  $\leftarrow$  & 127$^{+17}_{-17}$  &  139$^{+21}_{-18}$  &  374$^{+56}_{-50}$  &  406$^{+56}_{-50}$  \\ 
\hline 
\multicolumn{18}{c}{WR + WR SB2 solutions without constrained inclinations ($i$ fixed to $57^\circ$)} \\ 
\hline 
032\tablefootmark{b,e} &\begin{tabular}{l}WN5(h)${+}$WN6(h):\\{+abs}\end{tabular}  &1.90756(1)   &    53011.6${\pm0.1}$        &  288${\pm}6$      &  $0.06{\pm}0.02$        &   $250{\pm}22$     & 120${\pm}3$   & 123${\pm}23$   & 0.98${\pm}0.19$   & 1.4${\pm}0.5$   & 1.4${\pm}0.4$   & 57 &   & 2 & 2 & 5 & 6 \\ 
095\tablefootmark{b} &WN5${+}$WN7                                             &2.111(2)   &    52999.8${\pm0.1}$        &  274${\pm}9$      &  0  (fixed)                   &   n/a             & 356${\pm}14$   & 162${\pm}20$   & 2.2${\pm}0.13$   & 9.5${\pm}2.3$   & 20.9${\pm}4.3$   & 57 &   & 35 & 16 & 18 & 8 \\ 
\hline 
\multicolumn{18}{c}{SB1 solutions with constrained inclinations ($M_2$ adopted from BONNSAI evolution models if possible) } \\ 
\hline 
% 006\tablefootmark{g} &\begin{tabular}{l}O3\,If*/WN7${+}$O7\,V\\ (+ (OB + OB)?)\end{tabular}&2.001185(5)   &      46505.84${\pm0.01}$    &    278${\pm}6$    &   0 (fixed)   &    n/a            & 320${\pm}7$   &  -  &  -  &  -  &  -  & 59$^{+31}_{-40}$  &    & 29$^{+6}_{-5}$  &  18$^{+29}_{-17}$  &  15$^{+26}_{-2}$  &  9$^{+71}_{-9}$  \\ 
006\tablefootmark{g} &\begin{tabular}{l}O3\,If*/WN7${+}$OB\\ +(O7~V: + ?)\end{tabular}&2.001185(5)   &      46505.84${\pm0.01}$    &    278${\pm}6$    &   0 (fixed)   &    n/a            & 320${\pm}7$   &  -  &  -  &  -  &  -  & -  &    &  -  &  -  &  -  &  -  \\ 
043\tablefootmark{e} &WN3${+}$O9\,V                                              &2.816(2)   &     51932.6${\pm0.3}$       &   287${\pm}9$     &   $0.07{\pm}0.05$       &   $211{\pm}43$    & 244${\pm}14$   &  -  &  -  &  -  &  -  & 68$^{+22}_{-19}$  &    & 17$^{+2}_{-2}$  &  14$^{+12}_{-11}$  &  15$^{+4}_{-2}$  &  12$^{+20}_{-10}$  \\ 
064\tablefootmark{e} &WN3${+}$O9\,V                                              &37.59(6)   &     51920.5${\pm2.6}$       &   235${\pm}21$    &   $0.16{\pm}0.09$       &   $29{\pm}26$     & 57${\pm}5$   &  -  &  -  &  -  &  -  & 80$^{+10}_{-8}$  &    & 17$^{+2}_{-2}$  &  65$^{+33}_{-24}$  &  43$^{+4}_{-4}$  &  163$^{+145}_{-79}$  \\ 
071\tablefootmark{e,j} &WN3${+}$O6.5\,V                                               &5.2081(5)   &       52314.5${\pm0.2}$     &      329${\pm}17$    &    $0.09{\pm}0.08$    &   $265{\pm}41$       & 227${\pm}40$   &  -  &  -  &  -  &  -  & 72$^{+18}_{-14}$  &    & 28$^{+4}_{-3}$  &  27$^{+39}_{-22}$  &  24$^{+8}_{-5}$  &  24$^{+61}_{-20}$  \\ 
\hline 
\multicolumn{18}{c}{SB1 solutions without constrained inclinations (masses unconstrained) } \\ 
\hline 
012\tablefootmark{f} &O3~If*/WN6&3.2358(5)   &  52269.8${\pm0.1}$  & 431${\pm8}$  &  $0.35{\pm} 0.06$  & $-29{\pm}11$  & (74${\pm}5$)\tablefootmark{l}   &  -  &  -  &  -  &  -  &  - &    &  - &  - &  - &  - \\ 
029\tablefootmark{e} &WN3${+}$B1.5\,V                                              &2.2016(3)   &     51552.0${\pm0.3}$       &   351${\pm}9$     &   $0.16{\pm}0.13$       &   $-7{\pm}46$     & 59${\pm}8$   &  -  &  -  &  -  &  -  &  - &    &  - &  - &  - &  - \\ 
059\tablefootmark{e} &WN3${+}$O6\,III                                            &4.7129(7)   &     51954.6${\pm0.2}$       &   367${\pm}28$    &   $0.32{\pm}0.08$       &   $329{\pm}19$    & 30${\pm}3$   &  -  &  -  &  -  &  -  &  - &    &  - &  - &  - &  - \\ 
092\tablefootmark{f} &\begin{tabular}{l}WN3${+}$O6\,V\\ + B1\,Ia\end{tabular} &4.3125(6)   &      53000.0${\pm0.1}$      &    332${\pm}7$    &   0  (fixed)                  &    n/a           & 204${\pm}5$   &  -  &  -  &  -  &  -  &  - &    &  - &  - &  - &  - \\ 
099\tablefootmark{f} &O2.5\,If*/WN6                                                 &92.6(3)   &    53045.4${\pm1.3}$        &  337${\pm}16$     &   0  (fixed)                   &   n/a            & 91${\pm}19$   &  -  &  -  &  -  &  -  &  - &    &  - &  - &  - &  - \\ 
112\tablefootmark{h} &WN5h                                                        &8.2(1)   &         -                   &    389${\pm}21$   &   0  (fixed)                 &  n/a              & 42${\pm}20$   &  -  &  -  &  -  &  -  &  - &    &  - &  - &  - &  - \\ 
126\tablefootmark{d,k} &WN3${+}$O7\,V{+OB}                                            &25.5(4)   &    51114${\pm2}$            &   317${\pm}38$    &  $0.38{\pm}0.06$        &   $343{\pm}15$    & 27${\pm}2$   &  -  &  -  &  -  &  -  &  - &    &  - &  - &  - &  - \\ 
\hline                                                      
\end{tabular}}}                                                               
\tablefoot{                                                                   
\tablefoottext{a}{For spectral type references, see Table\,\ref{tab:catalog}} 
\tablefoottext{b}{This study}                                                 
\tablefoottext{c}{\citet{Shenar2017}}                                         
\tablefoottext{d}{\citet{Foellmi2006}}                                        
\tablefoottext{e}{FMG03}                                                    
\tablefoottext{f}{S08}                                                      
\tablefoottext{g}{\citet{Niemela2001}, \citet{Koenigsberger2003}}             
\tablefoottext{h}{\citet{Schnurr2009}}                                        
\tablefoottext{i}{Adopted from \citet{Tehrani2019} }\tablefoottext{j}{Alternative solution found for $P = 2.3264\,$d - given solution based on photometric period}                            
\tablefoottext{k}{Photometric period is 1.55296${\pm}1{\cdot}10^{-5}$\,d}     
\tablefoottext{l}{Derived by S08, but cannot be confirmed here: RVs contant within 2$\sigma\approx 10\,$\kms.}   
}                                                                               
\end{center}                                                                  
\end{table*}

\begin{figure}[h]
\centering
  \includegraphics[width=0.5\textwidth]{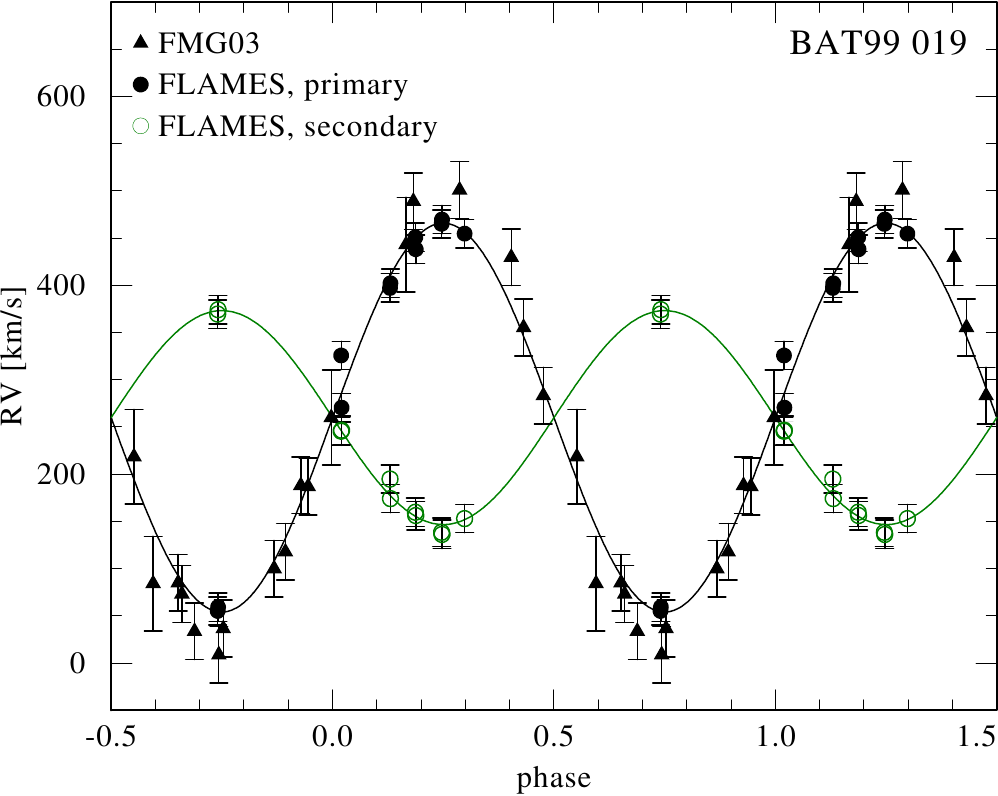}
  \caption{Orbital solution for BAT99\,019}
\label{fig:orbit_BAT019}
\end{figure} 

\begin{figure}[h]
\centering
  \includegraphics[width=0.5\textwidth]{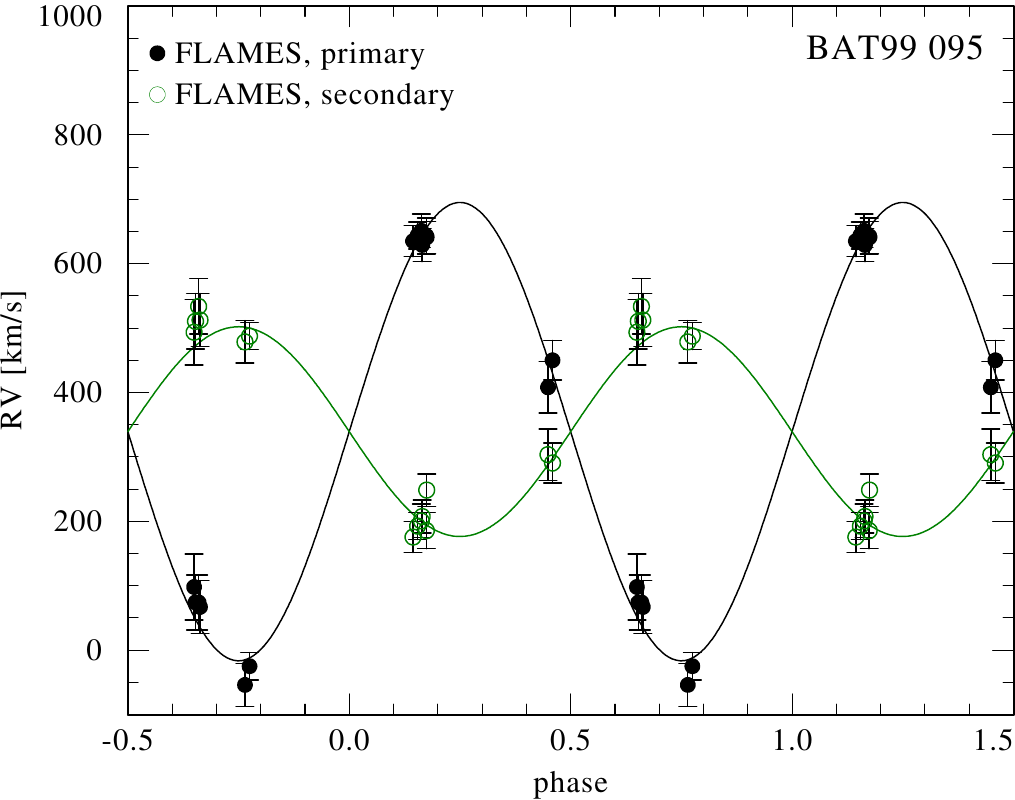}
  \caption{Orbital solution for BAT99\,95}
\label{fig:orbit_BAT095}
\end{figure}

\begin{figure}[h]
\centering
  \includegraphics[width=0.5\textwidth]{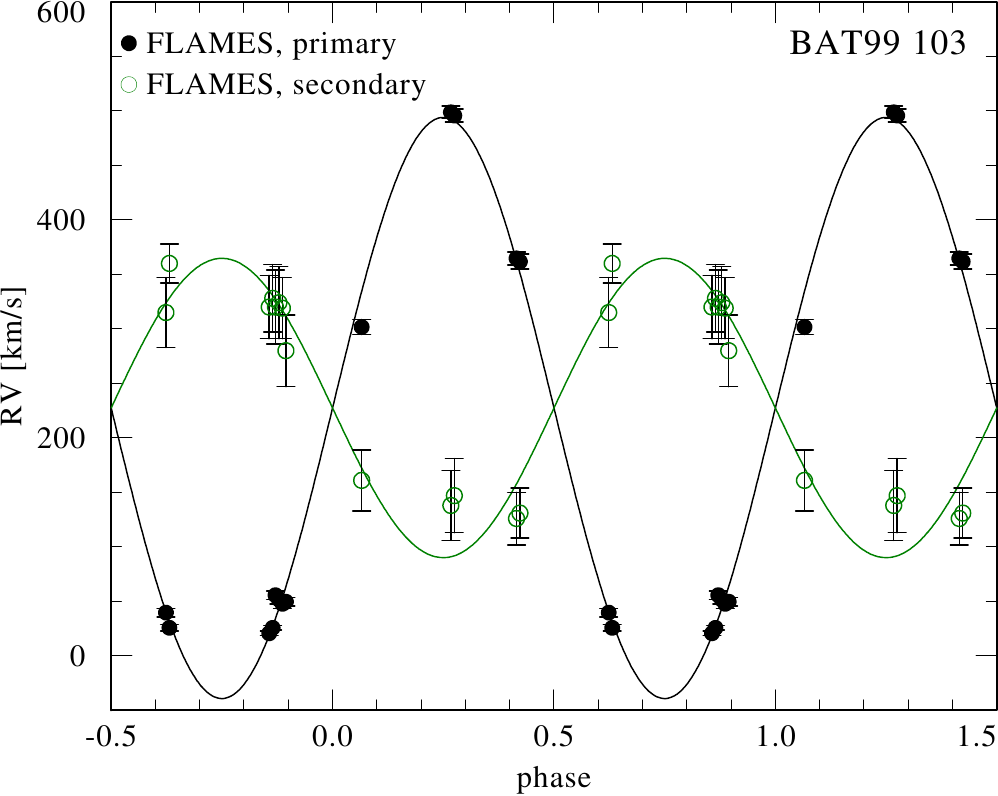}
  \caption{Orbital solution for BAT99\,103}
\label{fig:orbit_BAT103}
\end{figure} 

Many constraints on the orbital configurations of our targets are given by 
by FMG03 and S08. The vast majority of these solutions are SB1, i.e., they do not include the 
secondary's RV amplitude. Worse still, the orbital inclination $i$ is in most cases unknown.
In this study, we were able to derive SB2 solutions 
for seven systems: BAT99\,19, 32, 77, 95, 103, 107, and 113, where the orbital parameters of BAT99\,32 and 77 come from disentanglement
(see Sect.\,\ref{subsec:disen}), and those of BAT99\,19, 95, 103, 107 and 113 from the orbital analysis.  The RVs measured for BAT99\,12, 31, and 102 are constant 
within 3$\sigma$, while the nature of BAT99\,92 is uncertain (see Appendix\,\ref{sec:comments}).

\begin{figure}[h]
\centering
  \includegraphics[width=0.5\textwidth]{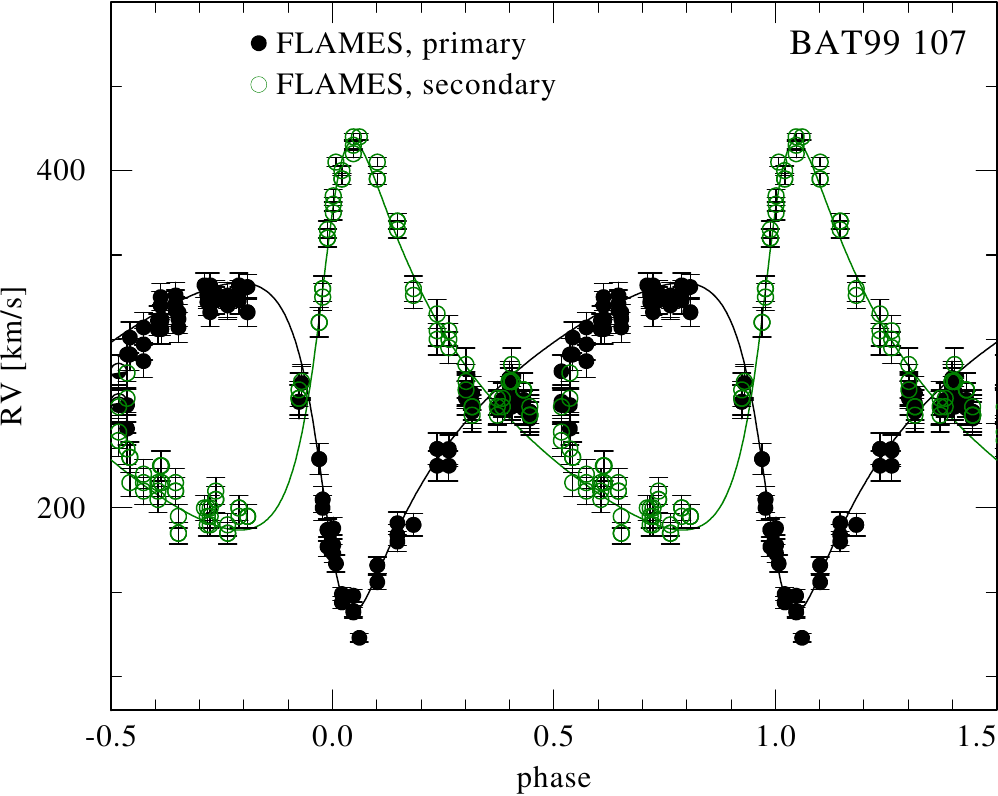}
  \caption{Orbital solution for BAT99\,107}
\label{fig:orbit_BAT107}
\end{figure}

\begin{figure}[h]
\centering
  \includegraphics[width=0.5\textwidth]{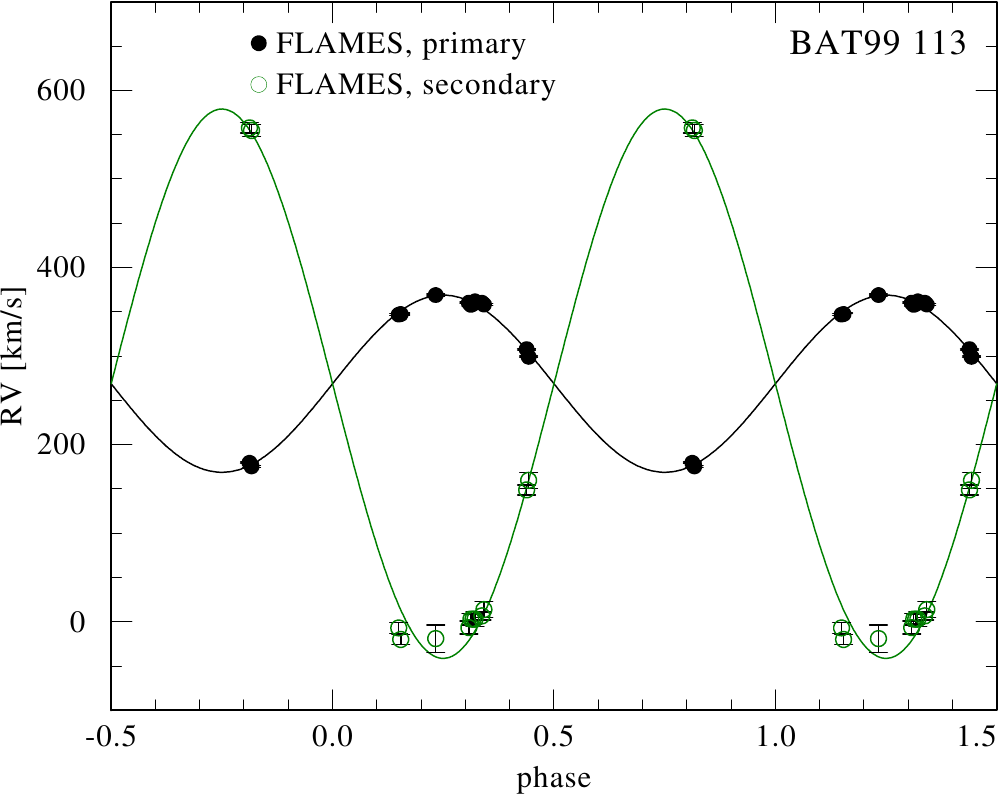}
  \caption{Orbital solution for BAT99\,113}
\label{fig:orbit_BAT113}
\end{figure}

The orbital parameters derived here and in previous studies are given in Table\,\ref{tab:orbitalpar}. The orbital solutions derived 
here for BAT99\,19, 95, 103, 107, and 113 are shown in Figs.\,\ref{fig:orbit_BAT019}-\ref{fig:orbit_BAT113}.
To constrain the minimum masses $M_{\rm WR} \sin^3 i$ and $M_2 \sin^3 i$ from the orbit, knowledge of the period $P$, eccentricity $e$, 
and the RV amplitudes $K_{\rm WR}$ and $K_2$ is needed. To constrain $M_{\rm WR}$ and $M_2$ and the semi-major axes $a_\text{WR}$ 
and $a_2$, the inclination $i$ is needed. Unfortunately, the full 
set of these parameters can only be measured in rare cases, e.g., SB2 eclipsing binaries. 
For the majority of our sample, only $P, e$, and $K_{\rm WR}$ could be constrained. Thus, 
Table\,\ref{tab:orbitalpar} is divided into five groups of objects, depending on the amount of information 
available on the system. 

The first group of objects (BAT99\,19, 119, and 129) in Table\,\ref{tab:orbitalpar} are SB2 binaries 
(i.e., $K_{\rm WR}$ and $K_2$ known) 
with constrained inclinations. The inclination of BAT99\,119 is constrained from polarimetry \citep{Shenar2017}. 
For the eclipsing systems BAT99\,19 and 129, a lower bound on $i$, $i_\text{min}$, 
is derived from the critical angle necessary to obtain eclipses
via an iterative solution of $\tan i > (R_1 + R_2)\,a(i)^{-1}$, using the stellar radii given in Tab.\,\ref{tab:specan}.  
% A full modeling of all available light curves of the systems is beyond the scope of this paper and will be a subject of a future paper that 
% focuses on eclipsing WR binaries (Shenar et al.\ in prep.). 
We then 
calculate $\langle \sin^3 i \rangle$ for $i_\text{min} < i < \pi/2$, from which $i$ is derived. The masses
are then derived from $\langle M_{\rm j} \rangle = M_{\rm j} \sin^3 i / \langle \sin^3 i \rangle$.
This is the only group of objects for which both $M_{\rm WR}$ and $M_2$ can be 
derived virtually without assumptions. 

The second group of objects (BAT99\,49, 77, 103, 107, 113, 116) also consists of SB2 binaries, 
but while $M \sin^3 i$ is constrained for both components, 
the inclinations are not. In this case, either $M_2$ or $i$ need to be adopted to retrieve 
$M_{\rm WR}$, which is one of the main purposes of this study.  We chose here to fix $M_2$ to the 
evolutionary mass predicted for the secondary with the derived physical parameters, given in Table\,\ref{tab:specan}. 
For this purpose, we used the BONNSAI\footnote{The BONNSAI web-service is available at www.astro.uni-bonn.de/stars/bonnsai} 
Bayesian statistics tool \citep{Schneider2014}. Using the  input of stellar parameters ($T_*, \log L, \log g, v \sin i$) and 
their corresponding errors from Table\,\ref{tab:specan}, the tool interpolates between evolutionary tracks 
calculated at LMC metallicity by \citet{Brott2011} and \citet{Koehler2015}
for stars with initial masses up to $500\,M_\odot$ and over a wide range of initial rotation velocities. Based on this set of evolution tracks, 
the BONNSAI tool predicts the most likely current mass of the secondary $M_2$ with corresponding errors.
With the orbital para\-meters and $M_2$ fixed, the inclination $i$ and the mass of the primary $M_{\rm WR}$ 
can be derived. We caution, however, that the results depend on the evolutionary models and the Bayesian algorithm. 
Given the potential systematics, we adopt an error on $M_2$ that is twice as large as given by the BONNSAI tool.

The third group consist of WR+WR SB2 binaries: BAT99\,32 and 95. Like the objects in the second group, 
they do not have constrained inclinations, but because both components appear to be WR stars, 
adopting their mass based on evolutionary models is uncertain.
We therefore adopt the inclination in this case, and fix $i$ to its mean statistical value so that 
$\sin^3 i = \left < \sin^3 i \right > = 3 \pi / 16$, or $i = 57^\circ$. 
The unconstrained values of $M_{\rm WR}$ and $M_2$ follow. BAT99\,116 is not included in this group (but in the previous one) 
because the components are ``ms-WR'' stars, for which evolutionary masses should be more relaible.

The fourth group of objects comprises binaries (BAT99\,6, 43, 64, 71) with constrained inclinations but unconstrained $K_2$, 
which in all cases come from reported eclipses in the systems.
In this case, we fix $M_2$ using the BONNSAI tool, exactly as done for the second group.  
The inclination is constrained just as for the first group. 

\begin{figure}
\centering
  \includegraphics[width=0.5\textwidth]{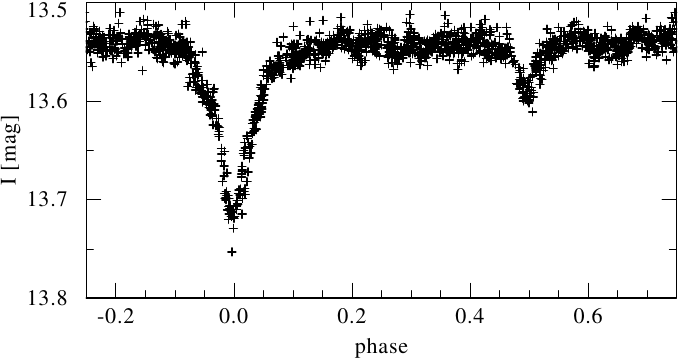}
  \caption{Light curve of BAT99\,19 folded with the parameters given in Table\,\ref{tab:orbitalpar}}
\label{fig:lc_BAT019}
\end{figure} 

\begin{figure}
\centering
  \includegraphics[width=0.5\textwidth]{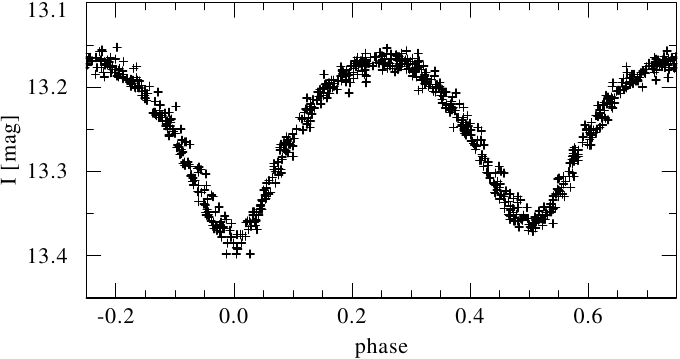}
  \caption{Light curve of BAT99\,126 folded with $P = 1.55296$\,d and $T_0 = 2100.55$ }
\label{fig:lc_BAT126}
\end{figure} 

The final group of objects, to which the majority of our sample belongs, 
contains systems that have neither $K_2$ nor $i$ constrained. In this case, we refrain from giving 
$M_1$ and $M_2$ since the errors are too large for yielding helpful information.

While the majority of masses derived here are plausible, the masses derived for the components of BAT99\,32 and BAT99\,64 are
unlikely to be correct. BAT99\,32 may have an exceptionally low inclination. For example,  $i{\approx}20^\circ$ 
would increase the masses  
from $2\,M_\odot$ to about $20\,M_\odot$. As for BAT99\,64, it is possible that the mass adopted 
for the secondary based on calibration 
with the BONNSAI tool is wrong, perhaps due to a previous mass-transfer event. 
We further note that the orbital solution of BAT99\,64 is based 
on relatively low-resolution data by FMG03, and may therefore require improvement. 
A more detailed discussion regarding the individual targets can be found in Appendix\,\ref{sec:comments}.

\subsection{Photometric variability}
\label{subsec:photo}

A few of our targets are eclipsing binaries (see Table\,\ref{tab:catalog}). MACHO lightcurves for eclipsing binaries have been 
presented by FMG03 for BAT99\,19, BAT99\,62, and BAT99\,129, as well as for the marginally eclipsing binaries BAT99\,43 and BAT99\,71, while 
a light curve for BAT99\,6 is presented by \citet{Niemela2001}. Here, we present two recently published 
OGLE light curves of BAT99\,19 and BAT99\,126 \citep{Graczyk2011}.

In Fig.\,\ref{fig:lc_BAT019}, the light curve of BAT99\,19 is folded with the orbital periods given in Table\,\ref{tab:orbitalpar}. As shown 
by FMG03, the system shows clear double eclipses. The unique nature of the WR star in this system, the very rapid rotation 
of the companion ($v_\text{eq} = 550-600\,$\kms), and the fact that it is an eclipsing binary, sets this target as a 
promising future candidate for focused studies of binary interaction.

In Fig.\,\ref{fig:lc_BAT126}, we show the OGLE light curve of BAT99\,126. \citet{Graczyk2011} published a period of 
$P = 1.55\,$d, which is used to plot the phase-folded light curve in Fig.\,\ref{fig:lc_BAT126}. 
Interestingly, 
this period is much shorter than the spectroscopic period of $P = 25.5$\,d reported by FMG03. This either suggests an error 
on the RV derivation of the components, or the presence of additional companions. A follow-up study of BAT99\,126 using newly acquired 
UVES spectra will soon be executed (Shenar et al.\ in prep.). 

\section{Discussion}
\label{sec:disc}

\subsection{Evolutionary status}
\label{subsec:evostat}

Below, we discuss several aspects involving the evolutionary status 
of the WR stars in the LMC.

\subsubsection{Single vs.\ binary}
\label{subsubsec:singbin}

The HRD positions of putatively single WN stars and the binary WN components, seen in Fig.\,\ref{fig:HRD}, seem to populate 
a similar regime on the HRD, with a few notable differences. While a few massive WN components in binaries 
 populate the $5.8 < \log L/L_\odot < 6.2$ region in the HRD (e.g., BAT99\,103, 113, 116), no single stars do. 
 Moreover, apparently-single WN stars reach luminosities of  up to $\log L/L_\odot = 6.9$ 
 (notably BAT99\,108 alias R\,136a1), while WN components in binaries 
 reach only $\log L/L_\odot = 6.4$. If not due to low-number statistics, 
 this may imply that the most massive stars are in fact the products of binary 
 mergers \citep{Crowther2010, DeMink2014}. 
 
All WN components in binaries have $T_* \gtrsim 40\,$kK, while effective temperatures of single WN stars go 
down to $T_* \approx 25\,$kK.  Since cooler WR stars typically occupy a larger volume, it is possible
that they are less likely to maintain their large radii in close binaries due to presence of the secondary star. Such late-type 
WN stars are therefore more likely to be stripped by a companion and appear hotter.

 Perhaps the most surprising result is that the bulk of apparently-single and binary WN stars spans the same 
 luminosity regime in the HRD, both reaching a minimum 
 luminosity of $\log L/L_\odot \approx 5.2-5.3$.  This means that there is no clear empirical evidence 
 suggesting that the binary channel enables lower-mass stars to enter the WR phase. 
 Naively, this also seems to suggest that there is no dividing mechanism operating in 
 the formation of the low-luminosity WR stars of both the  
 apparently single and the binary WN stars. We discuss this question in more 
 detail in  Sect.\,\ref{subsubsec:strippedstars}.

\subsubsection{Initial masses, ages, and evolutionary path}
\label{subsubsec:BinEvAn}

Evolved massive stars may reach radii $> 1000\,R_*$ during the red-supergiant phase, and as a consequence, (eccentric) massive binaries with 
periods of up to 10\,000\,d may interact during their lives. 
Given the relatively short periods of our targets ($P < 200\,$d), past binary 
interaction seems inevitable. 
The only way the components in our systems may have avoided mass-transfer 
is if the primary retained a small radius throughout its evolution. 
Generally, the more chemically-homogeneous a massive star is, the 
smaller it is going to be throughout its evolution.  The extreme case is described by chemically homogeneous 
evolution \citep[CHE,][]{Maeder1987b, Heger2000I, DeMink2009, Koenigsberger2014, Szecsi2015, Song2016}, in which the star never expands 
beyond its main-sequence radius. If the primary experiences CHE, mass-transfer is always avoided.
Usually, CHE is explained by invoking 
large initial rotation, which can efficiently mix the star. However, CHE can also be thought of as a proxy 
for increased homogeneity of massive stars, for which evidence is currently accumulating \citep[e.g.][]{Ramachandran2019, Higgins2019}. It should be further 
noted that, while evolution tracks always pass through the red-supergiant phase for all progenitor masses, no red-supergiants with progenitor masses $\gtrsim25\,M_\odot$ have ever 
been observed \citep[Humphreys-Davidson-limit:][]{Humphreys1979, Davies2018}. Hence, it is important to consider the possibility that the components of the systems have not interacted in the past.

To investigate the evolutionary paths of our targets, we therefore distinguish between three alternatives: 1. the primary expanded enough for 
mass-transfer to have occurred, 2. the primary experienced CHE and the secondary did not, and 3. both components experienced CHE. 
To consider these three alternatives, we follow a similar procedure as described in \citet{Shenar2016}.
To perform a systematic comparison 
between these three scenarios, we use a pre-calculated grid of evolution models calculated with the 
BPASS code \citep{Eldridge2008, Eldridge2016} for $Z=0.008$ (typical LMC metallicity). 
We stress that the efficiency and nature of mass-transfer is ``hard-coded'' in the 
BPASS models \citep[see, e.g.,][]{Eldridge2016}. There are many uncertainties involving the details of mass-transfer whose exploration 
is beyond the context of the current work. Here, we rather try to investigate whether or not the components interacted in the past, but 
encourage future studies to construct detailed models for the individual systems.

\begin{figure*}
\centering
\begin{subfigure}{0.67\columnwidth}
  \centering
  \includegraphics[width=\linewidth]{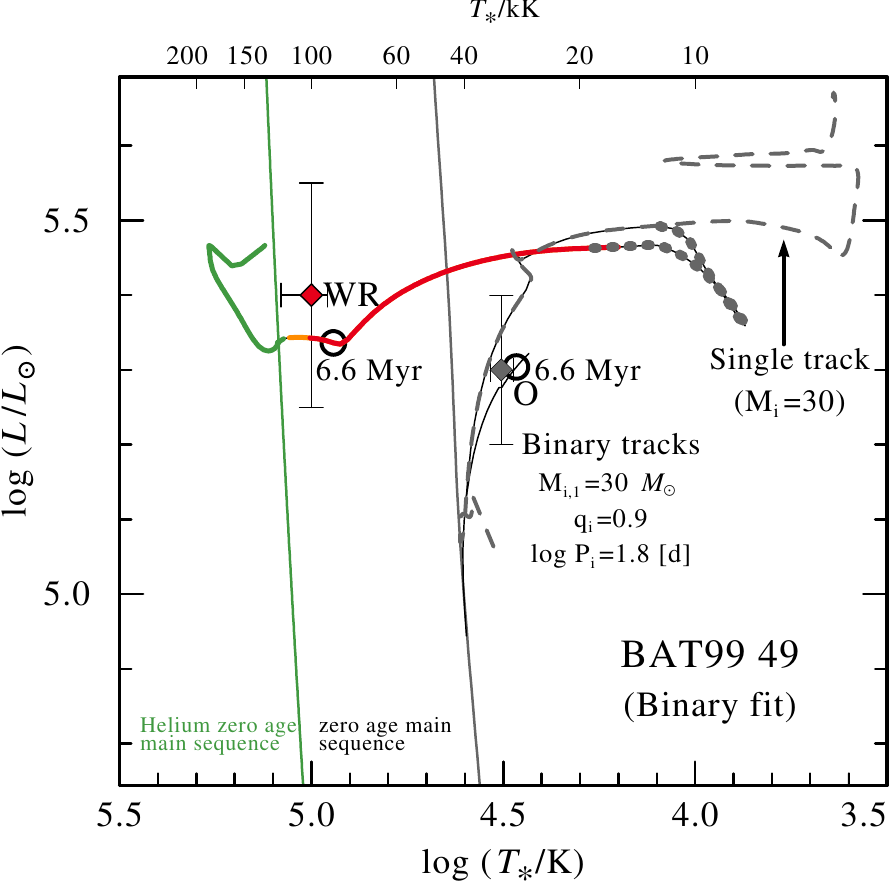}
%   \caption{s}
  \label{fig:BAT49hrdsub1}
\end{subfigure}%
\begin{subfigure}{.67\columnwidth}
  \centering
  \includegraphics[width=\linewidth]{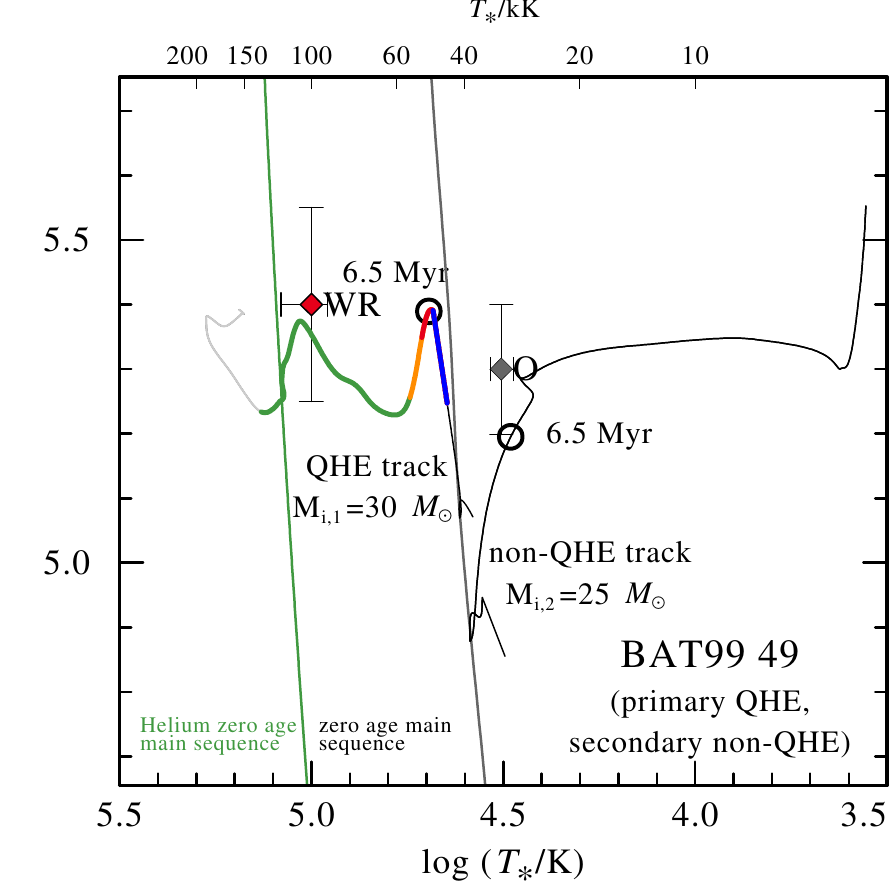}
%   \caption{A subfigure}
  \label{fig:BAT49hrdsub2}
\end{subfigure}
\begin{subfigure}{.67\columnwidth}
  \centering
  \includegraphics[width=\linewidth]{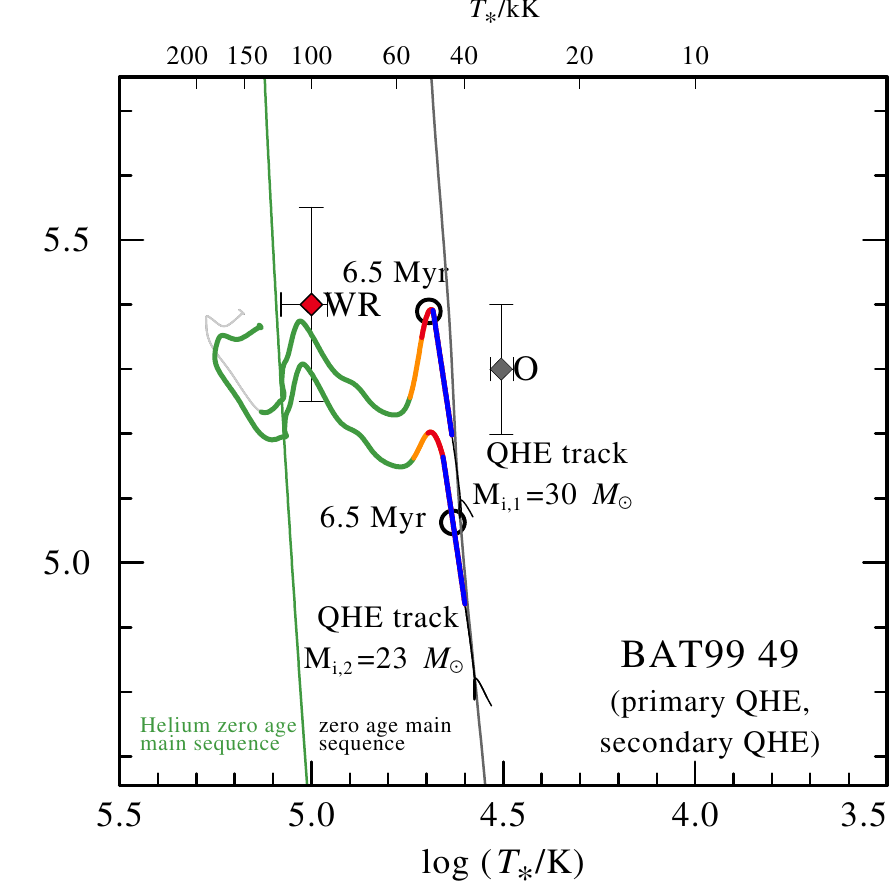}
%   \caption{A subfigure}
  \label{fig:BAT49hrdsub3}
\end{subfigure}
\caption{Best-fitting BPASS tracks and ages for BAT99\,49 for the case of 1.binary mass-transfer 2. the primary evolving homogeneously and the secondary non-homogeneously (middle panel), and 
3. both components evolving homogeneously (right panel). Colours and styles are as in Fig.\,\ref{fig:HRD}. The circles correspond to the derived ages.
Evidently, only the binary channel offers a consistent 
fit for BAT99\,49.}
\label{fig:BAT49hrd}
\end{figure*}

To explore the scenario in which the binary components did not evolve 
homogeneously and interacted via mass-transfer in the past,
we utilize a grid of BPASS binary tracks calculated for non-homogeneous binaries.  
Each track is defined by a set of three parameters: the initial mass of the primary $M_{\rm i, 1}$, the initial 
period $P_\text{i}$, and the initial mass ratio $q_\text{i} = M_\text{i,2} / M_\text{i,1}$. The tracks  
were calculated at intervals of $0.2$ on $0.2 \le \log P\,[\text{d}] \le 4$,  $0.2$ on $0.1 \le q_\text{i} \le 0.9$, 
and at unequal intervals of $5-30\,M_\odot$ on $10 < M_\text{i,1} < 150\,M_\odot$.
We then find the best-fitting binary track and age $t$ for each system by minimizing

\begin{equation}
 \chi^2\left(P_\text{i}, q_\text{i}, M_\text{i,\,1}, t\right) =\sum_{n=1}^{8} 
 \left(\frac{\text{O}_n - \text{E}_n\left(P_\text{i}, q_\text{i}, M_\text{i,1}, t\right)}{\sigma_n}\right)^2,
\label{eq:summin}
\end{equation}
where $\text{O}_n \in \left\{ \log T_\text{WR}, \log L_\text{WR}, \log T_2, \log L_2, M_\text{orb,WR}, M_\text{orb,\,2}\right.$
$\left. \log P, X_\text{H, WR} \right\}$ are the measured values for the
considered observables, and
$\text{E}_n\left(P_\text{i}, q_\text{i}, M_\text{i,1}, t\right)$ are the corresponding 
predictions of the evolutionary track defined by $P_\text{i}$, 
$q_\text{i}$, and $M_\text{i,1}$ at time $t$. $\sigma_n$ account both for measurement errors (Sect.\,\ref{subsubsec:errors}) 
and the grid spacing \citep[see details in][]{Shenar2016}.  We explore this scenario 
only for systems with constrained periods. Through this minimization procedure, we derive the initial 
masses (primary and secondary) and initial period for each system, as well as the age of the system. The best-fitting 
binary-evolution tracks for BAT99\,49 are shown in the leftmost panel of Fig.\,\ref{fig:BAT49hrd}.

Next, we consider the case in which the primary underwent CHE and the secondary did not.
For this purpose, 
we consider a grid of chemically-homogeneous BPASS tracks calculated for $Z=0.008$,
which run over the initial mass of the star at a spacing of $1\,M_\odot$ for $M_\text{i} \le 30\,M_\odot$, 
and $5-10\,M_\odot$ otherwise. We find the best-fitting initial mass and age for a homogeneously-evolving 
primary by minimizing

\begin{equation}
 \chi^2\left(M_\text{i}, t\right) =\sum_{n=1}^{4} 
 \left(\frac{\text{O}_n - \text{E}_n\left(M_\text{i}, t\right)}{\sigma_n}\right)^2,
\label{eq:summinsin}
\end{equation}
where $\text{O}_n \in \left\{ \log T_{*,\text{WR}}, \log L_\text{WR}, M_\text{orb,WR}, X_\text{H,WR} \right\}$. $\sigma_n$ have 
the same meaning as in Eq.\,(\ref{eq:summin}). We then repeat this procedure for the secondary for a grid of 
non-homogeneous single-star BPASS tracks. Assuming that the two components are coeval, we fix $t$ to the age 
derived for the WR star. A corresponding best-fitting track (at the age of the WR star) is then associated with the secondary. 
The corresponding best-fitting BPASS tracks for BAT99\,49 are shown in the middle panel of Fig.\,\ref{fig:BAT49hrd}. Evidently, 
this scenario cannot account for the properties of the system.

To explore the final alternative, in which both components undergo CHE, we perform the same minimization procedure as 
above, but this time scanning a grid of homogeneous models for the secondary. 
The corresponding best-fitting BPASS tracks for BAT99\,49 are shown in the rightmost panel of Fig.\,\ref{fig:BAT49hrd}. This 
scenario is also ruled out for BAT99\,49. Hence, relying on the BPASS tracks and our results, 
only past mass-transfer can explain the properties of BAT99\,49. Moreover, according to our solution, the WR primary 
is of the type b-WR, since it could only form via binary mass-transfer. The rough BPASS solution therefore suggests that 
BAT99\,49 started off as a $30\,M_\odot + 27\,M_\odot$ binary with an initial period of $P_\text{i} {\approx}60\,$d. 
6.2\,Myr after its formation, 
the system experienced a highly non-conservative case B mass-transfer via RLOF, during which about $10\,M_\odot$ were removed 
from the primary, which then entered the WR phase. 
Only $\approx 1\,M_\odot$ were accreted by the secondary. During this phase, the period of the orbit tightened 
from $P\approx 60\,$d to somewhat less than $\approx 30\,$d. The period since slowly increased due to wind mass-loss 
to the current observed value (32\,d),  6.6\,Myr after the formation of the system.

Initial masses, periods and ages for the binary scenario are given in Table\,\ref{tab:evtable} in cases where a solution could 
be found. Solutions were rejected if at least one of the observables did not fit with the track within 2$\sigma$ (``no solution''). 
If a solution exists for the CHE scenario, we also give the corresponding initial masses and age. In this case, we always 
choose the best-fitting CHE scenario (i.e., only the primary evolved via CHE, or both components evolved via CHE). In some 
cases, the CHE solution requires an initial mass that is above the grid's upper limit of $100\,M_\odot$. In this 
case, we state that a solution may exist for initial masses larger than $100\,M_\odot$. 
Additionally, for each WR star, we specify in Table\,\ref{tab:specan} which evolutionary channel 
(ms-,w-,bw-,or b-WR) is most consistent with the properties derived here according to the BPASS tracks.
We warn that this classification strongly depends on the evolution tracks being used (see Sect.\ref{subsubsec:strippedstars}).
An account for this classification for each target, as well as figures in the form of Fig.\,\ref{fig:BAT49hrd}, are given 
in Appendix\,\ref{sec:comments}.

About 1/3 of the WN stars in our sample are classified as \mbox{ms-WR} stars, based on their HRD positions, masses, 
and hydrogen mass-fractions. The rest are identified as cWR stars. 
For all cWR binaries in the sample but BAT99\,32, 95, we can rule out CHE for the secondary. 
Unlike the SMC, however, 
we find that it is in most cases impossible to tell - considering the errors of our analysis - whether the primary 
evolved homogeneously or not.
Because of the uncertain evolution channel of the primary, it is hard to accurately 
assess the incidence of b-WR or wb-WR stars among the cWR binaries 
that have interacted with a companion, which we estimate to be $45{\pm}30\%$. This very large error is a direct result of the uncertain evolution channel 
of the WR progenitor (CHE v.s. standard).
However, based on the BPASS 
tracks, we can estimate that only $12{\pm}7\%$ of the cWR stars in binaries formed purely due to binary interaction, i.e., 
$12\%$  are potentially b-WR stars.

\renewcommand{\arraystretch}{1.2}
\begin{table}[!htb]
\small
\caption{Derived initial masses, initial periods, and ages}
\label{tab:evtable}
\begin{center}
\begin{tabular}{l | c c c c | c c c}
\hline \hline 
BAT99    &                 \multicolumn{4}{c|}{Binary evolution}         &     \multicolumn{3}{c}{QHE}                         \\ 
         & $M_\text{WR,i} $ & $M_\text{2,i} $ & $\log P_\text{i}\,$ & Age      & $M_\text{WR,i}$ & $M_\text{2,i}$ & Age    \\ 
         &  $M_\odot$ & $M_\odot$ &  [d] &  [Myr]       & $ [M_\odot]$ & $M_\odot$ & Myr    \\ 
\hline
006 & 70 & 34 & 0.6 & 2.7 & 70 &  50   & 0.8\\ 
019 & 80 & 23 & 1.4 & 3.6 &  > 100 & - & - \\ 
029 & 35 & 10 & 0.8 & 5.5 & 100 &  10   & 3.5\\ 
032 & \multicolumn{4}{c|}{no solution} &  > 100 & - & - \\ 
043 & 40 & 19 & 0.6 & 5.0 & 100 &  20   & 3.5\\ 
049 & 30 & 26 & 1.8 & 6.6 &  \multicolumn{3}{c}{no solution} \\ 
059 & 50 & 34 & 1.2 & 4.3 & 100 &  40   & 4.2\\ 
064 & 40 & 19 & 1.8 & 5.3 & 50 &  20   & 6.1\\ 
071 & 40 & 27 & 1.0 & 5.3 & 35 &  22   & 8.1\\ 
077 & 50 & 44 & 0.6 & 3.0 & 60 &  50   & 1.0\\ 
092 & \multicolumn{4}{c|}{no solution} &  > 100 & - & - \\ 
095 & 70 & 34 & 0.6 & 3.6 &  > 100 & - & - \\ 
103 & \multicolumn{4}{c|}{no solution} & 100 &  50   & 2.5\\ 
107 & 70 & 62 & 2.2 & 3.0 &  > 100 & - & - \\ 
113 & 70 & 20 & 0.6 & 2.3 &  > 100 & - & - \\ 
116 & 150 & 134 & 2.2 & 0.5 &  > 100 & - & - \\ 
119 & \multicolumn{4}{c|}{no solution} &  > 100 & - & - \\ 
126 & 60 & 29 & 1.6 & 3.9 &  > 100 & - & - \\ 
129 & 40 & 27 & 1.0 & 5.1 & 100 &  30   & 4.2\\ 
\hline
\end{tabular}
\end{center}
\tablefoot{
{\it Binary evolution} values are derived from BPASS tracks to find the best-fitting evolution models, which correspond
to initial masses $M_{
m i}$ for both components, initial periods $P_{
m i}$, and ages. {\it QHE} values for
$M_{\rm i, 1}$ and the age are derived from single-star BPASS tracks that assume full mixing, thereby avoiding binary interaction.
The mass of the secondary is derived using the BONNSAI tool. If none is given, no solution using BONNSAI was found.
}
\end{table}

\subsubsection{The incidence of binary-stripped WR stars (b-WR)}
\label{subsubsec:strippedstars}

In Sect.\,\ref{sec:introduction}, we defined b-WR stars as WR stars that could only form via binary mass-transfer, 
and argued that they are expected to be common, especially at low metallicity.
Only a few promising candidates for b-WR stars exist. One prominent example is the so-called quasi WR (qWR) star 
\object{HD\,45166} \citep{Groh2008}, for which peculiar wind parameters and an exceptionally low inclination ($0.7^\circ$) were 
derived. 
Several low-mass (${\approx}1\,M_\odot$) 
O-type subdwarfs (SdO), which are believed to originate from binary 
mass-transfer, have been discovered near B-type stars (the putative mass-accretors), but their masses are too low to support 
a strong stellar wind and a corresponding WR-star appearance \citep[e.g.,][]{Wang2018}. 
So-called extreme helium stars \citep[e.g.][]{Jeffery2010} also do not fall into the mass/luminosity range that 
b-WR stars are expected to populate. 
While other peculiar WR stars have 
been suggested to originate from binary interaction \citep{Schootemeijer2018a, Neugent2017, Smith2018},
there is an apparent disagreement between the predicted abundance of b-WR stars and their observed number. Whether 
this disagreement points toward strong observational biases or flaws in population syntheses is still not clear.

We can roughly estimate the expected incidence of b-WR stars among a population of cWR stars. 
Marking with $M_\text{i,w-WR}$ the minimum initial
mass a star needs to enter the WR phase as a single star, with $M_\text{i,WR}$ the minimum initial mass 
for which a stripped star would appear as a WR star, with $f_\text{strip}$ the fraction of stars that would be stripped of their 
H-envelope by a companion, and $T_\text{cWR}$ the lifetime of the WR phase, we can estimate the expected incidence 
of purely binary-stripped (b-WR) stars among all cWR stars (single and binary) in a given population for a constant star-formation 
rate as follows:

\begin{equation}
\resizebox{0.5\textwidth}{!} 
{$
% a + b - \frac{a}{b}
\frac{N_\text{b-WR}}{N_\text{cWR}}  = 
\frac{\,\int\limits_{M_\text{i,WR}}^{M_\text{i,w-WR}} f_\text{strip}\,m^{-2.35}\,T_\text{WR}(m)\,{\rm d}m}
 {\int\limits_{M_\text{i,WR}}^{M_\text{i,w-WR}}f_\text{strip}\,m^{-2.35}\,T_\text{WR}(m)\,{\rm d}m + 
 \int\limits_{M_\text{i,w-WR}}^\infty m^{-2.35}\,T_\text{WR}(m)\,{\rm d}m} 
%  \int_{M_\text{i,w-WR}}^\infty} m^{-2.35}\,T_\text{cWR}(m)\,{\rm d}m} \approx 0.7, 
$}
\label{eq:bsWR}
\end{equation}
where the \citet{Salpeter1955} IMF was assumed.

We now need to estimate the values of $M_\text{i,w-WR}$ and $M_\text{i,WR}$.
Extrapolating mass-loss recipes published by \citet{Hainich2015} and constructing 
corresponding PoWR models implies that the WR phenomenon 
ceases below $\log L \approx 4.8\,[L_\odot]$ at LMC metallicity, corresponding to $M_\text{i,WR}\approx 15\,M_\odot$. 
In contrast, PoWR models calculated following 
recent recipes published by \citet{Vink2017} 
for optically-thin winds of stripped stars 
imply that the WR phenomenon stops below $\approx 5.5\,[L_\odot]$, 
corresponding to $M_{i,WR} \approx 30\,M_\odot$. Because these 
results are quite discrepant, we choose to estimate the minimum luminosity of WR stars empirically.
The HRD positions seen in this work, along with those published by H14 and \citet{Neugent2017}, 
imply that the WR phenomenon stops below $\log L \approx 5.2\,[L_\odot]$ in the LMC, corresponding roughly to $M_\text{i,WR}\approx 20-25\,M_\odot$. 
Based on the BPASS evolution tracks, stars with $M_\text{i} \gtrsim 35\,M_\odot$ can become WR stars as single stars, i.e., 
$M_\text{i,w-WR} \approx35\,M_\odot$. In stark contrast, the Geneva tracks (Fig.\,\ref{fig:hrdGeneva}) imply  
$M_\text{i,WR}\approx 70\,M_\odot$.
To obtain $T_\text{cWR}(m)$, we assume that it is equal to the lifetime of the core He-burning phase, which 
should be a good approximation for the order-of-magnitude estimate performed here.
$T_\text{cWR}(m)$ is estimated from Equation 79 in \citet{Hurley2002}.
Finally, we adopt $f_\text{strip} =0.33$, as  estimated by \citet{Sana2012}. We note that it is possible that $f_\text{strip}$ 
becomes smaller for larger masses due to, for example, their increased homogeneity. 
Plugging all of these in Eq.\,(\ref{eq:bsWR}), 
we obtain an expected incidence of 30\% b-WR stars among all cWR stars in the LMC for $M_\text{i, w-WR} = 35\,M_\odot$  (BPASS value),
and 70\% for $M_\text{i, w-WR} = 70\,M_\odot$ (Geneva value).

We now turn to estimating the \emph{observed} incidence of b-WR stars in the LMC. First, we need to consider 
the fact that the WC stars are not included in this work. However, since WC stars comprise about $18\%$ of 
the whole WR content in the LMC, we assume for our estimate that WN are representative of the whole population. 
Let us assume that the apparently-single WR stars truly formed as single stars. In this case, 
assuming the validity of the BPASS tracks ($M_\text{i,w-WR} = 35\,M_\odot$), 
the observed incidence of b-WR/cWR stars is merely $4\pm2\%$ in the LMC, a factor ten lower than expected ($\approx 30\%$).
If we take the Geneva value of $M_\text{i,w-WR} = 70\,M_\odot$ instead, virtually all cWR stars in our sample would be classified 
as b-WR, and we would obtain a total incidence of b-WR/cWR of $\approx 12\%$, which is about six times smaller than predicted 
($\approx 70\%$). Both of these reveal a strong discrepancy between observation and theory.
The immediate conclusion is that either 
the expected incidence of b-WR stars was strongly overestimated, or that the number of observed b-WR 
stars is strongly underestimated. We suggest that one of the following should hold:

\begin{enumerate}
 \item \emph{pre-WR mass-loss is widely underestimated in evolution codes.} Taking the small observed incidence of  b-WR stars at face 
 value, one way for Eq.\,\ref{eq:bsWR} to yield  similarly small fractions for the predicted incidence is by plugging in a lower value for $M_\text{i,w-WR}$. Values in the vicinity of $M_\text{i,w-WR} = 25\,M_\odot$ would make the denominator approach zero. 
 For example, the Brussels evolution code 
 predicts that single stars may already reach the WR phase intrinsically already for $M_\text{i,w-WR} \approx 25\,M_\odot$ in the LMC
\citep{Vanbeveren1998, Vanbeveren1998b}. Thus, 
enhancing pre-WR mass-loss rates can mitigate the apparent 
contrast between observation and theory. Porosity, which is often neglected in determinations of the mass-loss rate, can be one 
cause for a possible underestimation of $\dot{M}$ \citep[e.g.][]{Oskinova2007}. Additionally, underestimated mass-loss during the red-supergiant phase \citep{vanLoon2005}, 
or the lack of treatment of eruptive mass-loss during an LBV-like phase \citep{Owocki2017}, may be important to consider.
However, this does not  seem to agree with the multitude of studies suggesting 
a lowering of mass-loss rates throughout the evolution of massive stars \citep[e.g.][]{Puls2008b, Mauron2011, Vink2017}.

\item \emph{Mixing of stars with $M_\text{i} \gtrsim 20\,M_\odot$ is much more efficient than assumed in evolution codes.}
Mixing in massive stars is poorly constrained, especially for $M_{\rm i} \gtrsim20\,M_\odot$ 
\citep[e.g.][]{Higgins2019, Schootemeijer2019}.
Mixing increases the size of the stellar core and reduces the size of the envelope that needs to be stripped in order 
for the star to enter the cWR phase. Therefore, it effectively reduces the value of $M_\text{i,w-WR}$, increases the 
efficiency of single-star evolution in forming cWR stars, and reduces the importance of binary interaction in forming 
WR stars. While observed
distributions of rotational velocities  \citep[e.g.][]{Ramirez2013, Sabin-Sanjulian2017, Ramirez2017} 
render rotationally-induced mixing unlikely for the population as a whole, alternative mixing 
processes \citep[e.g., gravity waves,][]{Aerts2018, Bowman2019} that are usually not included in evolution models 
may play an important role in the evolution of massive stars. 

\item \emph{The majority of LMC WR stars - both apparently-single and binary - are products of binary interactions.} 
If mass-loss/mixing cannot be further increased to explain the existence of apparently-single 
low-luminosity WR stars, a solution 
involving binary mass-transfer seems hard to avoid.
\citet{Vanbeveren1998b} predicted that $\approx 80\%$ 
of the apparently-single WR stars were affected by past binary interaction.
However, it requires of us to explain the apparent lack 
of companions for putatively single WR stars in the LMC.
Several binary channels that produce apparently-single 
WR stars exist. One may involve envelope stripping by low-mass 
companions \citep[e.g.][]{Paczynski1976, Podsiadlowski1992, Schootemeijer2018a}. 
Instead of a bright mass-gainer, these stars may host 
low-mass stars that have stripped the envelopes of the WR progenitors during CEE. Another possibility 
involves three-body interactions in triple systems, ejecting the stripped WR star from the original binary and producing 
a truly single WR star with binary history
\citep[e.g.][]{Hut1983, Toonen2016}.
A third alternative is that the WR stars were the original secondaries in binaries. 
After the primary exploded as a SN (or directly collapsed to 
form a BH), the secondary would eventually fill its Roche lobe and a second mass-transfer phase 
would initiate. Such a system may eventually 
appear as a low luminosity WR star. However, systems that survive both mass-transfer 
phases are expected to be very rare \citep[e.g.][]{Vandenheuvel2017}.
Moreover, for these systems to appear as single stars requires
both an inhibition  mechanism of the anticipated X-ray emission as well as orbital RV shifts that fall below the detection 
threshold. Finally, it is possible that single low-luminosity WR stars may form through the merging of two stars, during which 
the merger product loses much of its outer H-rich envelope. \end{enumerate}

\subsection{Mass-luminosity relation}
\label{subsec:masslum}

\begin{figure}
\centering
  \includegraphics[width=0.5\textwidth]{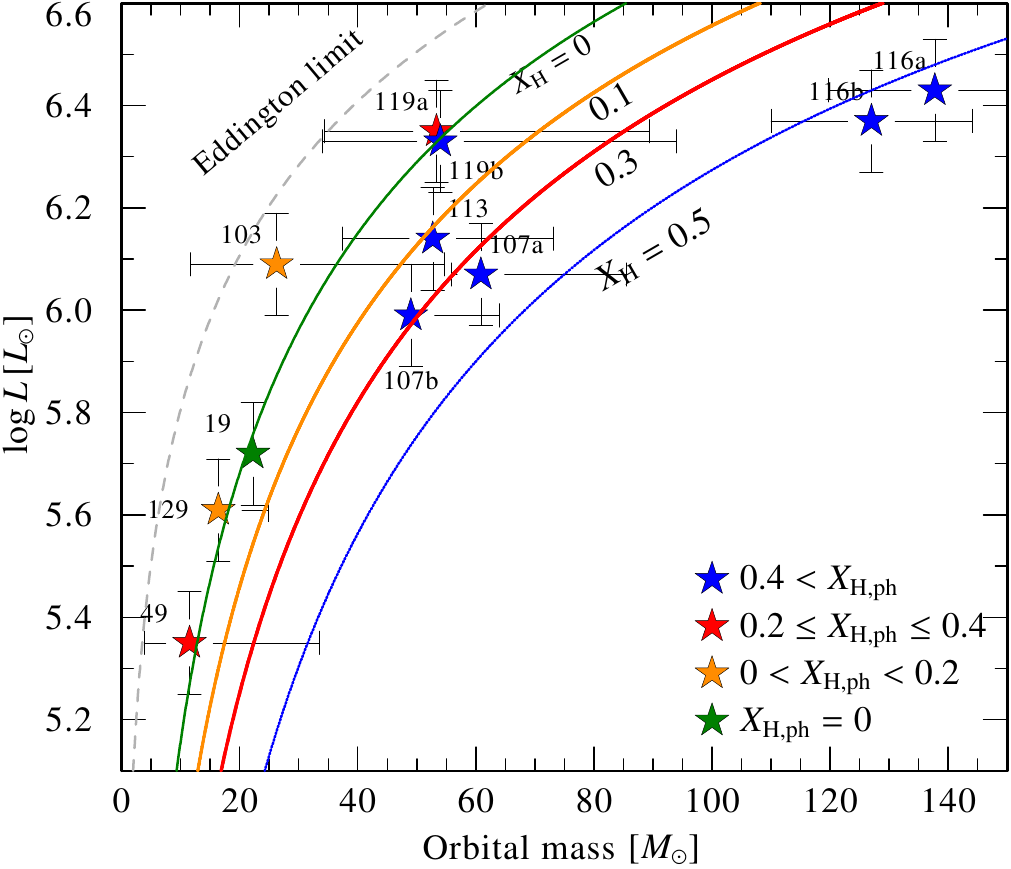}
  \caption{The positions of the WR components on a $M - \log L$ diagram (symbols) compared to 
  MLRs calculated for homogeneous stars \citep{Graefener2011}, depicted by solid curves. 
  The colours correspond to the hydrogen content (see legend). The Eddington limit 
  calculated for a fully ionized helium atmosphere is also plotted (gray dashed line).}
\label{fig:MLR}
\end{figure}

In Fig.\,\ref{fig:MLR}, we show the positions of the WR components in a mass-luminosity diagram. The luminosities are taken 
from Table\,\ref{tab:specan}, while the masses are taken from Table\,\ref{tab:orbitalpar}. Only measurements with constrained errors 
are shown. Also plotted are mass-luminosity relations calculated for homogeneous stars with different hydrogen mass fractions 
by \citet{Graefener2011}, as well as the Eddington limit for an atmosphere composed of helium. 

It is evident that the observations roughly follow the trends of the mass-luminosity relations. Considering 
the uncertainties, more data (spectroscopic, 
photometric, polarimetric) are necessary to reduce the errors on the orbital masses.
However, statistically, it appears that WN stars containing some hydrogen lie 
above their respective mass-luminosity relation. That is, the stars are overluminous compared to a homogeneous star with 
the same amount of hydrogen. This suggests that the majority of these stars are not 
homogeneous and likely core He-burning.
A similar result was obtained for the SMC sample \citep{Shenar2016}. 

\subsection{Rotational velocities of the OB-type companions}
\label{subsec:rotvel}

\begin{figure}
\centering
  \includegraphics[width=0.5\textwidth]{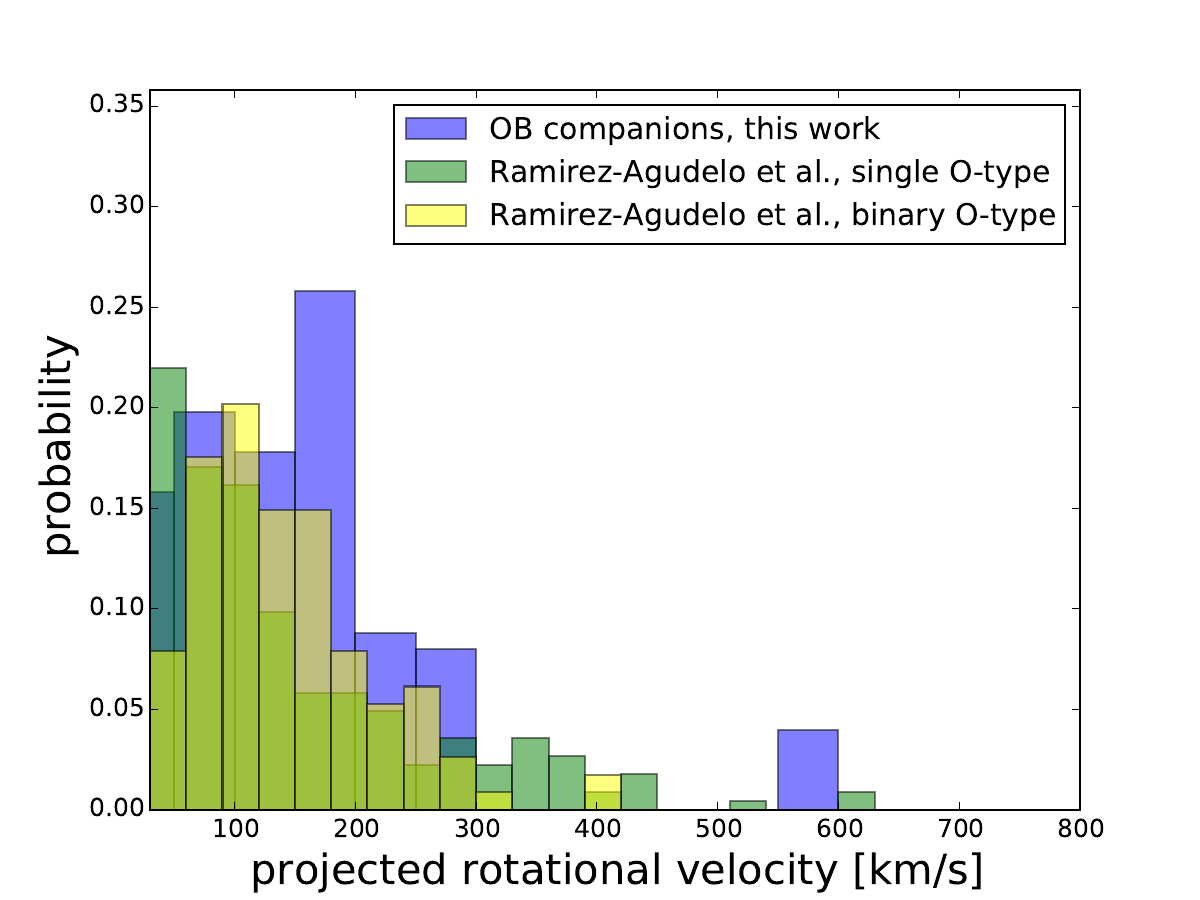}
  \caption{Normalized distribution of $v \sin i$ values for the 26 OB-type companions of WN stars measured in this work
  (blue), compared with $v \sin i$ distributions for apparently single and binary OB-type stars in the Tarantula 
  region, adopted from \citet{Ramirez2013, Ramirez2015}.}
\label{fig:vsinis}
\end{figure}

When mass-transfer in binaries occurs, companions not only accrete mass, but also angular momentum. 
Mass accretors are expected to reach near-critical rotation \citep{Packet1981}.
It is therefore expected that, if the OB-companions of the WR binaries in our sample accreted mass in the past, they would 
exhibit large rotational velocities.

In Fig.\,\ref{fig:vsinis}, we plot a normalized histogram of the projected rotational velocities measured for 26 OB-type companions 
in our study in bins of 50\,\kms. In cases where only upper limits $v \sin i_{\rm up}$ 
could be derived, a flat contribution for all bins with $v < v \sin i_{\rm up}$ is assumed. Evidently, the distribution 
peaks around 150\,\kms, and quickly drops beyond $200\,$\kms. 
Similar results were recently obtained by \citet{Shara2017} and \citet{Vanbeveren2018} for several Galactic WR+O binaries. 
A single outlier - BAT99\,19 - reaches the very large value 
of $550\,$\kms. While its edge-on geometry ($i\approx90^\circ$) favours a large $v \sin i$ value, inclination 
effects are expected to lead to an overall shift of ${\approx} 50\,$\kms~for the general distribution, leaving 
BAT99\,19 a clear outlier in terms of its rotational velocity.

Fig.\,\ref{fig:vsinis} also includes $v \sin i$ measurements for 216 apparently-single O-type stars 
and 114 O-type primaries of spectroscopic binaries in the Tarantula region, adopted from 
\citet{Ramirez2013} and \citet{Ramirez2015}, respectively. It is apparent that our sample, which is much smaller 
than the Tarantula samples, exhibits more rapid rotation on average than the Tarantula single-star distribution, 
and is comparable to the Tarantula binary distribution. 
This is compatible with the fact that many companions of WR stars in our sample 
accreted mass and hence angular momentum from their companions. However, it is also apparent that most  
companions in our sample are far from being critical rotators, with $v_\text{crit}$ being of the order of $600-700\,$\kms.

\begin{figure}
\centering
  \includegraphics[width=0.5\textwidth]{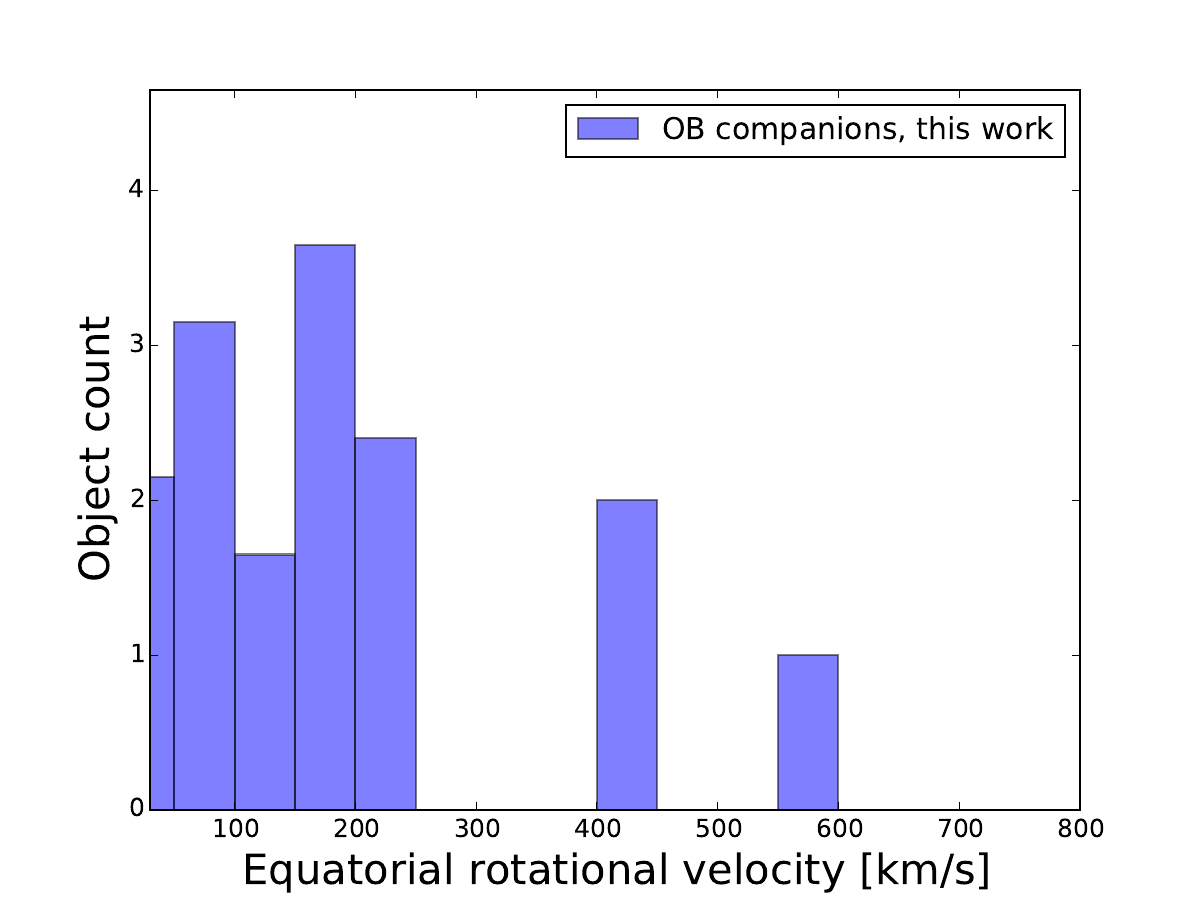}
  \caption{Histogram of $v_\text{eq}$ for the 16 OB-type companions of WN stars with constrained $v \sin i$ 
  (Table\,\ref{tab:specan}) and $i$ (Table\,\ref{tab:orbitalpar}), assuming an alignment of the rotational and orbital axes. Upper limits contribute equally 
  to all lower bins, hence the fractional counts.
  }
\label{fig:veqs}
\end{figure}

In Fig.\,\ref{fig:veqs}, we show a histogram of the estimated equatorial rotation velocities $v_\text{eq}$ for 
16 OB-companions in our sample for which $v \sin i$ and $i$ are both constrained, assuming that 
the orbital inclination is identical to the rotational inclination. Values for $v \sin i$ are 
taken from Table\,\ref{tab:specan} and for $i$ from Table\,\ref{tab:orbitalpar}. Accounting for projection 
effects, three WR binaries with rapidly rotating companions are revealed: BAT99\,19, 49, and 103.

As for the remainder of the sample, if the OB-companions accreted mass, 
their rotations seem to have been slowed down over the course of ${\approx}0.5\,$Myr 
(typical half-lifetime of a WR star). For example, BAT99\,129 appears to be a post mass-transfer 
system in which the companion rotates with a moderate speed of $v_\text{eq}{\approx}200\,$\kms. 
It has been speculated \citep[e.g.,][]{Vanbeveren2018} that the braking of the rotation occurs due to strong magnetic fields arising during the accretion process, 
which slows down the rotation of the mass accretors \citep{Meynet2011}. 
The fact that the companion of BAT99\,19 is still such a rapid rotator may suggest that the mass-transfer only occurred 
recently. 

\subsection{A prescription for WN mass-loss rates}
\label{subsec:masslossprisc}

\begin{figure}
\centering
  \includegraphics[width=0.5\textwidth]{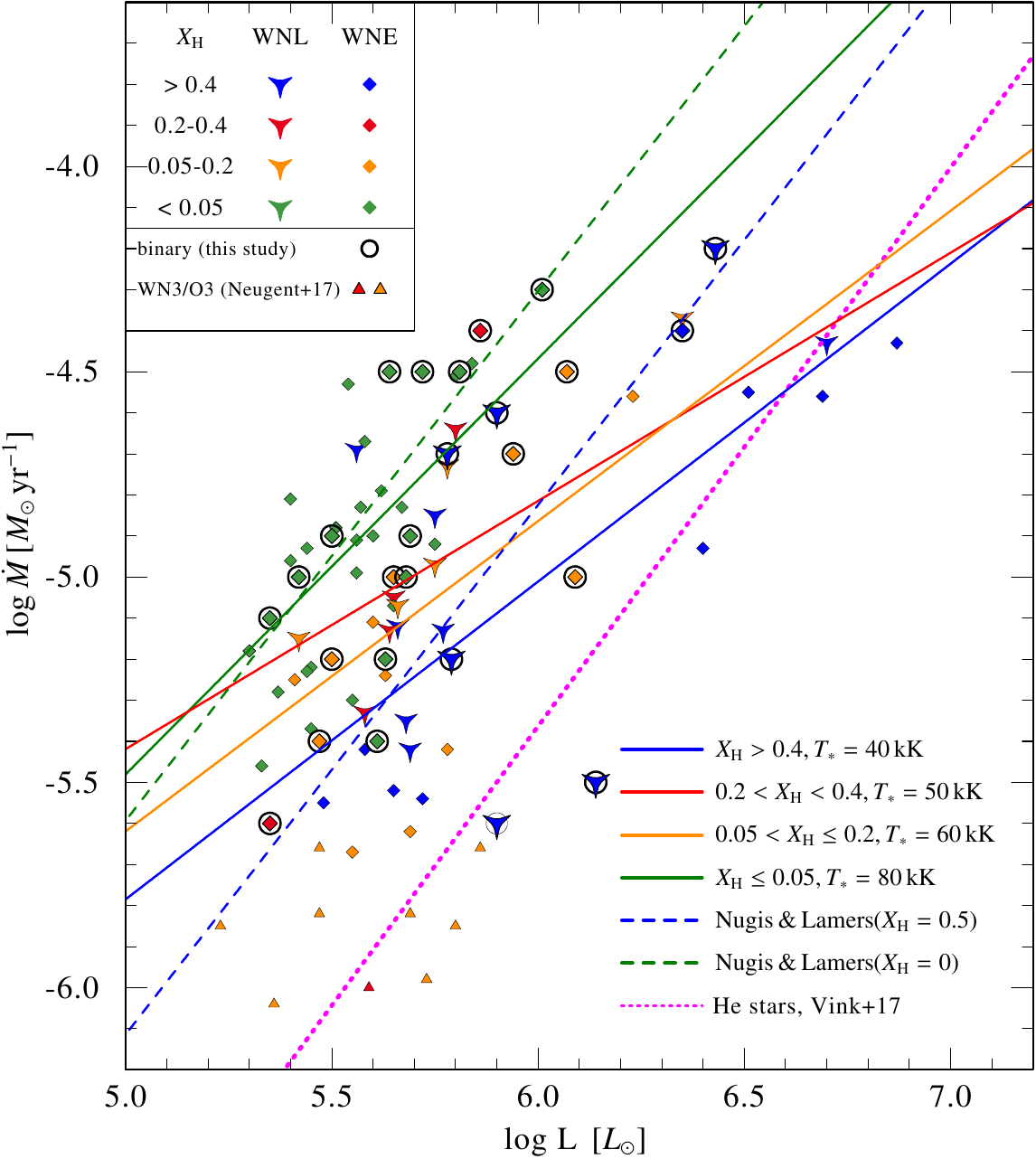}
  \caption{The positions of the single (adopted from H14) and binary (this work) 
  LMC WN sample on the $\log L - \log \dot{M}$ plane. 
  The meaning of the colours and symbols are as in Fig.\,\ref{fig:HRD}, with the binaries marked with a circle 
  (see legend). Also shown are the WN3/O3 stars analysed by \citet{Neugent2017}. Plotted are the projection 
  of Eq.\,\ref{eq:MdotZ} onto the LMC metallicity and different hydrogen content, where there temperature is 
  fixed to typical values per subtype.
  We also plot the commonly-used relations published by \citet{Nugis2000} for the LMC and for H-rich/free stars 
  and the theoretical relation published by \citet{Vink2017}.
  }
\label{fig:MdotL}
\end{figure}

Combining the results for the single WN stars in the SMC with those obtained 
for single WN stars in the the Milky Way \citep{Hamann2006}, the Andromeda galaxy
M31 \citep{Sander2014}, and the LMC (H14), \citet{Hainich2015} derived an 
empirical prescription for the mass-loss rates of WN stars as a function of $L$, $T_*$, $X_\text{He}$, 
and $Z$. Here, we repeat this exercise 
by including the binary WN stars in the SMC \citep{Shenar2016,Shenar2018} and LMC (this work). Moreover, the parameters of the 
single Galactic WN stars were recently revised based on the new Gaia distances \citep{Hamann2019}, and are therefore updated in our analysis.
We assume $Z= 0.018, 0.014, 0.006, 0.0012$ for M31, the MW, the LMC, and the SMC, respectively
\citep[see thorough discussion in][]{Hainich2015}.
As \citet{Hainich2015}, we utilize a $\chi^2$ fitting 
procedure between the observed parameters and the following linear 
relation with the five unknowns $C_\text{i}, i\in{1,2,3,4,5}$:

\begin{equation}
 \log{\dot{M}} = C_1 + C_2 \log L + C_3 \log T_* + C_4 \log X_\text{He} + C_5 \log Z.
\label{eq:MdotZ} 
\end{equation}

\renewcommand{\arraystretch}{1.5}
\begin{table}
{\scriptsize
\setlength\tabcolsep{2.5pt}
\caption{Coefficients for the $\log \dot{M}$ prescription, Eq.\,\ref{eq:MdotZ}}
\label{tab:MdotLpars}
\begin{center}
{
\begin{tabular}{lccccccc}
\hline\hline
Subtype           & N & $C_1$      &          $C_2$          &        $C_3$                &      $C_4$            &         $C_5$   & $\sigma$  \\ 
\hline
All & 183 & -6.22 & 0.74 $\pm$ 0.07 & -0.21 $\pm$ 0.16 & 1.42 $\pm$ 0.20 & 0.83 $\pm$ 0.09 & 0.28 \\ 
$X_{\rm H} \ge 0.4$ & 42 & -6.50 & 0.79 $\pm$ 0.10 & -0.37 $\pm$ 0.39 &  - & 0.68 $\pm$ 0.19 & 0.25 \\ 
$0.2 < X_{\rm H} < 0.4$ & 32 & -4.02 & 0.60 $\pm$ 0.17 & -0.74 $\pm$ 0.40 &  - & 0.43 $\pm$ 0.20 & 0.30 \\ 
$0.05 < X_{\rm H} \le 0.2$ & 43 & -3.84 & 0.76 $\pm$ 0.16 & -0.78 $\pm$ 0.39 &  - & 0.81 $\pm$ 0.24 & 0.30 \\ 
$X_{\rm H} \le 0.05$ & 66 & -8.13 & 1.01 $\pm$ 0.19 & -0.06 $\pm$ 0.28 &  - & 0.95 $\pm$ 0.19 & 0.26 \\ 
\hline
\end{tabular}}
\end{center}}
\end{table}

The resulting coefficient and their formal standard deviations are listed in Table\,\ref{tab:MdotLpars}, along with the 
total sample size.
In Fig.\,\ref{fig:MdotL}, we show the relation projected to the LMC metallicity and different ranges of hydrogen 
mass-fractions. The temperatures, which are of secondary importance for the mass-loss rates, are fixed to typical 
values per subtype for this illustration (see legend). 
We also plot the commonly used prescriptions by \citet{Nugis2000} for hydrogen rich/free WR stars, 
projected onto the LMC metallicity. Finally, we plot the relation recently published by \citet{Vink2017} for stripped He-stars at $Z=Z_{\rm LMC}$, 
which applies for optically-thin winds.

The (almost linear) dependence of $\dot{M}$ on $Z$ ($\dot{M} \propto Z^{0.8}$) is in very good agreement with 
\citet{Vink2005}, and slightly shallower than the value reported by \citet{Hainich2015}. The reminder of the parameters 
broadly agree with those published by \citet{Hainich2015}.

The relation derived has a standard deviation of about 0.3\,dex (factor 2). Including systematics, this 
scatter $\sigma$ is compatible with typical analysis errors and therefore probably reflects them.
In Fig.\,\ref{fig:MdotL}, we also include the so-called WN3/O3 stars, analysed by \citet{Neugent2017}. It is immediately 
clear from Fig.\,\ref{fig:MdotL} that these stars are outliers. Relative to their reported luminosities, they have very 
weak mass-loss rates, almost comparable to those of early O-type stars. 
It is likely that the winds of these stars obey a different relation due to the fact that their winds 
are optically thin. Indeed, the magnitudes of their mass-loss rates agree much better with prescriptions by \citet{Vink2017}. Notably, however, the slope 
of the relation as a function of $\log L$ seems to be shallower empirically compared to the predicted relation.

\section{Summary}
\label{sec:sum}

This study focused on the population of WN binaries in the LMC. 
Through spectroscopy of these objects, our aim was 
to provide an unprecedented test for our understanding of the evolution and  formation of 
WR stars at low metallicity, focusing on the role of binary nature and chemically-homogeneous evolution. 
We conclude the following:

\begin{itemize}
 \item Of the 44 binary candidates investigated (out of ${\approx}130$ WN stars in the LMC), 
 the spectra of 28 are recognized as composite, with the status of BAT99\,72 uncertain. 
 Five systems are potential SB1s (BAT99\,12, 99, 105, 112, and 114), but 
 their binary status is not certain. The binary nature 
 of the remaining 11 binary candidates cannot be confirmed. 
 \item About 1/3 of the our sample are on, or close to, the main sequence (ms-WR), with the remainder 
 being He-burning WR stars (cWR). 
 \item Notable systems include: 1.\ the 18\,d period eclipsing WN4+O6\,V system BAT99\,19, with the most rapidly 
 rotating secondary (550\,\kms), 2.\ BAT99\,32 and 95, which may host two WR stars in short orbit and are therefore potential evolved GW progenitors, 
 3.\ BAT99\,126 (WR+O), whose light curve suggests a contact configuration in a 1.5\,d orbit together with a spectroscopic 
 period of 25\,d, implying a triple configuration, and 4.\ BAT99\,12 and 99, which 
 appear to be X-ray bright SB1 WR binaries, and may therefore be candidates for WR + compact object systems.
 \item We can reject chemically-homogeneous evolution for almost all OB-type companions of the WN stars, 
 but the status of the primaries is less certain.  
 \item Based on our results and the BPASS grid of evolution models, 
 $45{\pm}30\%$ of the cWR stars in binaries transferred mass to their companion 
 (i.e., $45{\pm}30\%$ are b-WR or wb-WR stars). The very large uncertainty follows from the uncertain expansion of the WR progenitor (CHE or quasi-CHE v.s.\ standard evolution)
 However, only  $12{\pm}7\%$ of the cWR components in binaries can be explained through binary-interaction alone (12\% are b-WR). Assuming that the apparently-single 
 WR stars truly formed as single stars, this amounts to an observed
 fraction of  $4{\pm}2\%$ b-WR stars of the whole WN content in the LMC, 
 compared to the theoretical prediction of $30-40\%$. A similar contradiction between 
 the observation and theory is obtained when comparing to the Geneva evolution models (see Sect.\,\ref{subsubsec:strippedstars}).
 \item Projected rotational velocities of the OB-type secondaries are found to be larger than observed for 
 single O-type stars (typically $150-200\,$\kms), but usually far from break-up (with BAT99\,19, 49, and 
 103 as exceptions to the rule). This potentially suggests the presence of a braking mechanism 
 of rotation, e.g., through stellar winds or magnetic fields, 
 \item The upper-end of the HRD is populated by apparently-single WR stars reaching estimated current 
 masses of ${\approx} 300\,M_\odot$, while orbital and evolutionary masses derived for WR stars in binaries reach 
 ${\approx} 130\,M_\odot$, possibly implying that the most massive stars observed in the LMC are mergers.
 \item Otherwise, the single WN stars and binary WN components span a similar regime on the HRD. Both the apparently-single 
 and binary WN stars  down to a minimum luminosity of ${\approx} 5.2-5.3\,L_\odot$ that is not reached by standard single-star tracks. WN stars in binaries 
 are not observed at lower luminosities than single WN stars on average. We conclude that either pre-WR mass-loss (mainly 
 during the red supergiant phase) and/or mixing is strongly underestimated in evolution codes, 
 or that the majority of cWR stars in the LMC - both apparently single as well as binary - 
 are products of binary interaction.  
\end{itemize}

% Despite tremendous efforts, we still cannot claim to understand with certainty the past (and future) 
% evolution of WR stars. 
The few detailed studies performed on specific WR systems suggest that, in the long run, 
we will have to push for high data quality to further reduce measurement uncertainties and possible biases. Moreover, 
deeper multiplicity surveys will be required to determine the bias-corrected binary fraction of WR stars, and to determine 
whether the apparently-single WR stars are truly single.

If mass-loss is the primary agent that leads to the formation of WR stars in the LMC, then mass-loss 
rates and/or mixing prior to the WR phase in the mass-range $20-60\,M_\odot$ in evolution codes such as the Geneva code, 
MESA, and even BPASS, are strongly underestimated. This, however, 
counteracts the trend of the recent decades, where mass-loss rates have been reported to 
be systemically lower than originally thought. 
 If an increase of mass-loss and/or mixing is not supported by future studies, the consequence would be that the bulk of cWR stars in the LMC would be located below 
the threshold for forming WR stars through self-stripping (cf.\ Fig.\,\ref{fig:HRD}, right panel, and Fig.\,\ref{fig:hrdGeneva}). 
It would then be difficult to imagine a solution to this problem that does not involve some sort of binary interaction. That is, one would be forced to conclude 
that the majority of cWR stars (apparently-single and binary) were formed through binary interaction.
With evidence continuing to grow that 
binary interaction dominates the evolution of massive stars, this is an exciting prospect. 
However, extraordinary claims require extraordinary 
evidence, and no such evidence could be established here.

\begin{acknowledgements}
We thank the anonymous referee for their help in improving our manuscript.
T.S. acknowledges support from the German Verbunsforschung (DLR) grant 50 OR 1612 and from 
the European Research Council (ERC) under the European Union's DLV-772225-MULTIPLES Horizon 2020 research and innovation programme.
% T.S 
% would like to thank A.\ Gilkis,  A.\ Menon, and E.\ Laplace for their helpful insights on this study.
A.A.C.S. is supported by the Deutsche Forschungsgemeinschaft (DFG) under grant HA 1455/26 and would like to 
thank STFC for funding under grant number ST/R000565/1.
V.R. is grateful for financial support from Deutscher Akademischer Austauschdienst (DAAD), as a part of Graduate School Scholarship Program. 
A.M. is grateful for financial aid from NSERC (Canada) and FQRNT (Quebec).
LMO acknowledges support by the DLR grant 50 OR 1508. 
Some of the data presented in this paper were obtained from the Mikulski Archive for Space Telescopes (MAST). 
STScI is operated by the Association of Universities for Research in Astronomy, Inc., under NASA contract NAS5-26555
This research made use of the VizieR catalogue access tool, CDS, Strasbourg, France.
The original description of the VizieR service was published in A\&AS 143, 23. 
% Support for MAST for non-HST data is provided by the NASA Office of Space Science 
% via grant NNX09AF08G and by other grants and contracts. 
\end{acknowledgements}

\bibliography{literature}

\Online
\begin{appendix}
% % 
\section{Comments on individual targets}
\label{sec:comments}
% 
% 
% {\bf [Daniel Comment: State values when giving differences]}
% 
In the few paragraphs below, we give a short overview on each system, and 
discuss specific issues related to their analysis. 

\emph{\bf BAT99\,6} 
was originally classified as O6-7 + WN5-6 by \citet{Walborn1977}. This $\approx 2.0\,$d period 
binary system (Sk $-67^\circ18$) 
was reclassified as O3\,f*+O by \citet{Niemela2001}. The latter authors suggest that the system is composed 
of four stars due to the presence of a second period of 19\,d in the RVs of the He\,{\sc i} absorption lines.
In contrast, \cite{Koenigsberger2003} showed that the flux level of the available IUE spectra is not consistent with more than 
two luminous massive stars in the system. Moreover, 
the light curve published by \citet{Niemela2001} is suggestive of a contact configuration, making BAT99\,6 a potential  
candidate for a GW progenitor. This is in line with the relatively high projected rotational 
velocity measured for the primary ($250\,$\kms). 

\begin{figure*}
\centering
\begin{subfigure}{0.67\columnwidth}
  \centering
  \includegraphics[width=\linewidth]{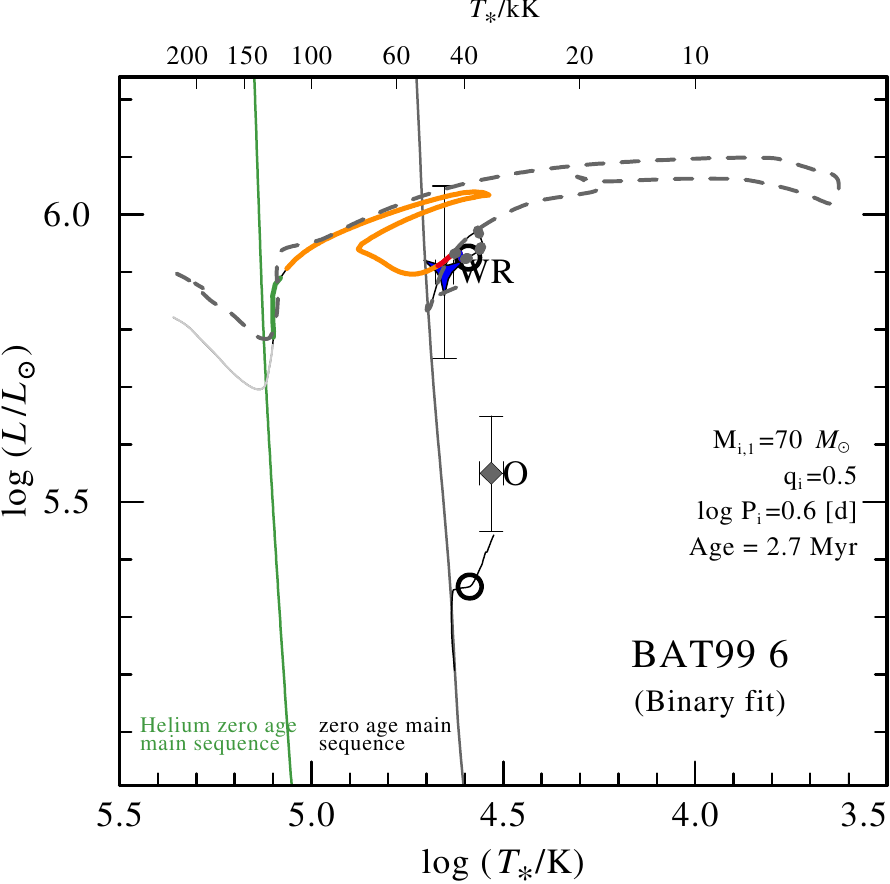}
%   \caption{s}
  \label{fig:BAT6hrdsub1}
\end{subfigure}%
\begin{subfigure}{.67\columnwidth}
  \centering
  \includegraphics[width=\linewidth]{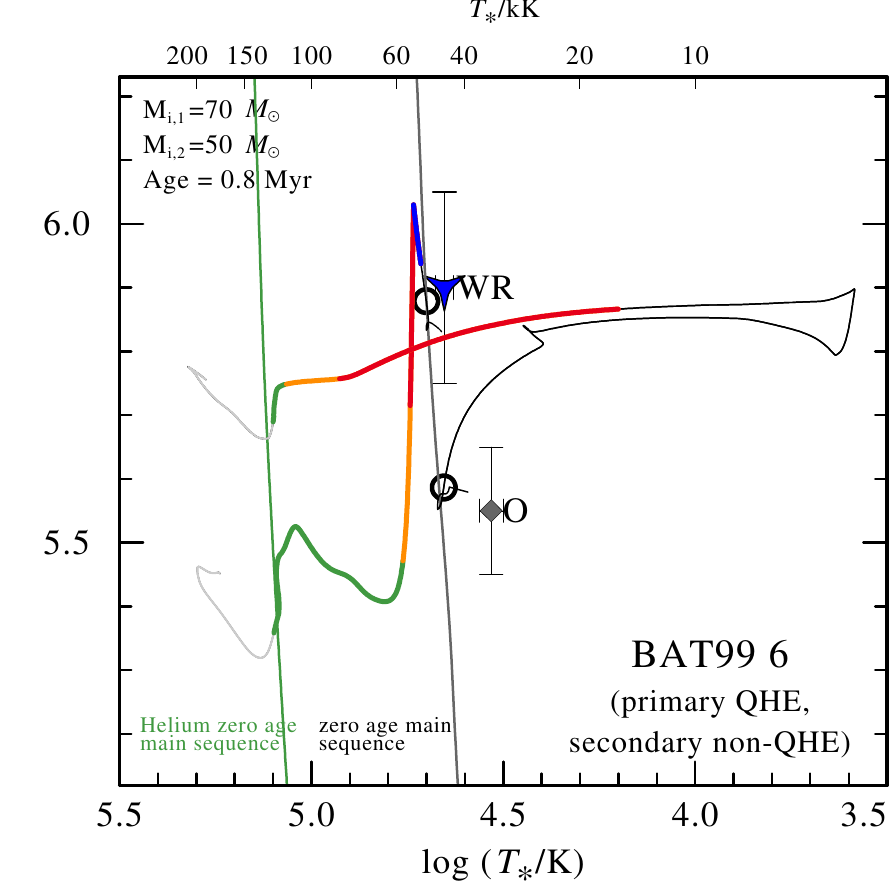}
%   \caption{A subfigure}
  \label{fig:BAT6hrdsub2}
\end{subfigure}
\begin{subfigure}{.67\columnwidth}
  \centering
  \includegraphics[width=\linewidth]{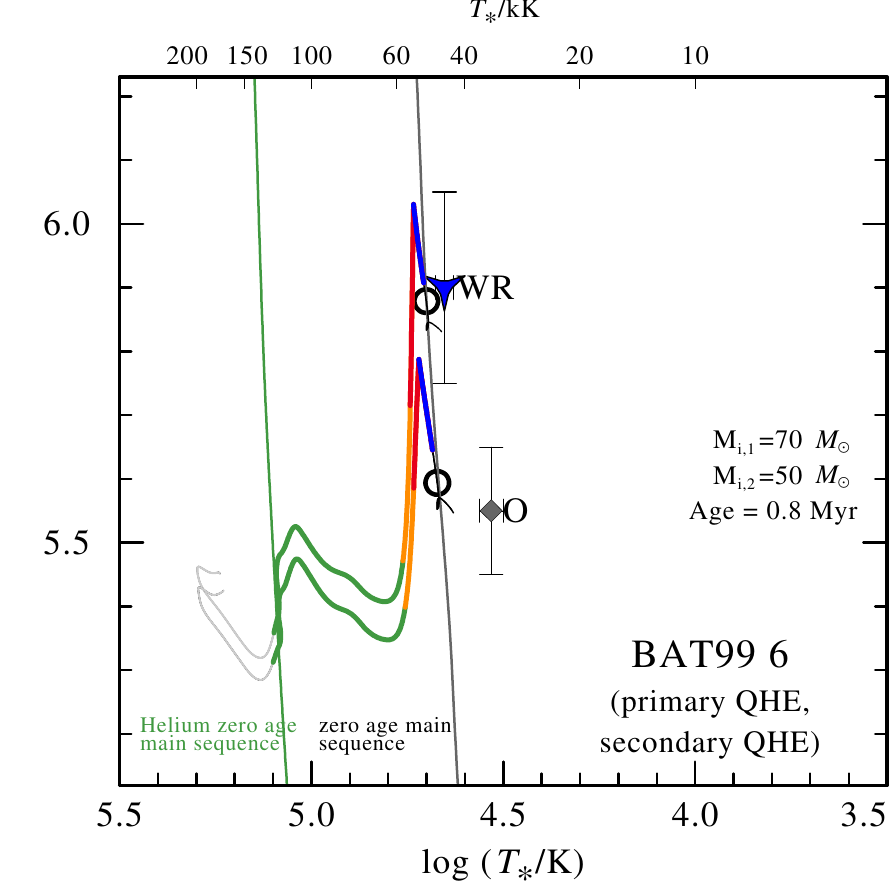}
%   \caption{A subfigure}
  \label{fig:BAT6hrdsub3}
\end{subfigure}
\caption{As Fig.\,\ref{fig:BAT49hrd}, but for BAT99\,6.}
\label{fig:BAT6hrd}
\end{figure*}

While the available FEROS spectra are indeed indicative of more than two sources contributing to the spectra, 
they do not enable us to unambiguously conclude this, let alone to derive their parameters. 
Therefore, we analyse the system as a binary. The results for the primary (WR) component should be 
reliable, while the secondary's should be taken with caution, since it might be representative of further sources in the spectrum.
For example, to avoid a saturated P-Cygni C\,{\sc iv} profile in the UV, the mass-loss rate of the secondary was fixed to a 
value that is much smaller than expected for its spectral type. This is possibly due to further sources present in the system.

Since the primary exhibits a P Cygni H$\beta$ profile, 
and given the slight dominance of the N\,{\sc iv} over the N\,{\sc iii} emission, 
we reclassify the primary as O3~If*/WN7, following morphological scheme by \citet{Crowther2011}. 
A classification of the model spectrum for the secondary implies the spectral class O7~V.
Motivated by \citet{Niemela2001}, we classify the system therefore as (O3~If*/WN7 + OB) + (O7~V + ?).

The source was observed and detected by the {\em XMM-Newton} X-ray 
observatory.  According to the {\em XMM-Newton} Serendipitous Source 
Catalog (3XMM DR8 Version), the observed flux in 0.2-12\,keV band is 
$F_{\rm X}\approx 7\times 10^{-15}$\,erg\,cm$^{-2}$\,s$^{-1}$. At the LMC 
distance, the X-ray luminosity corrected for the ISM redding is $L_{\rm 
X}\approx 3\times 10^{33}$\,erg\,s$^{-1}$.
We suggest that BAT99\,6 is a colliding
wind binary, where the copious X-rays are produced by the collision of the winds of the two components.
The X-ray luminosity of BAT99\,6 is rather high
compared to other colliding wind binaries \citep[e.g.][]{Oskinova2005},
but still not as high as expected in the case of accreting compact objects. 

Our results suggest that the primary in the system is on (or close to) the main sequence, 
i.e., it is a ms-WR star. The HRD position of the secondary does not support its CHE
(see rightmost panel of Fig.\,\ref{fig:BAT6hrd}). However, we cannot 
rule out CHE for the primary. We caution that our results may 
be biased by the presence of further components in the system, which are not accounted for here.
Due to the complexity of this system (potentially high-order multiplicity, 
contact configuration), it is very likely that the BPASS models are too simplistic. We encourage further 
dedicated studies of this important system.

\begin{figure}
\centering
  \includegraphics[width=0.5\textwidth]{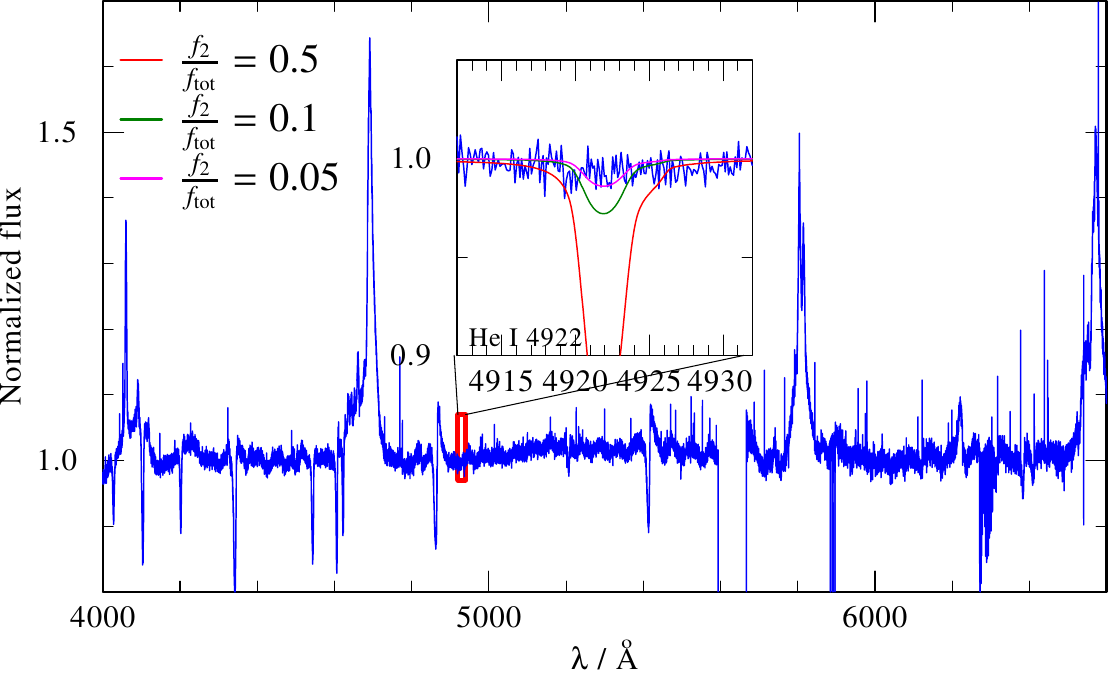}
  \caption{A UVES observation of BAT99\,12 (blue line). A zoom-in on a prominent and isolated He\,{\sc i} also includes
  models for $T = 20\,$kK dwarfs ($\log g = 4.0\,$[cgs]) that are 
  diluted to 50\%, 10\%, and 5\% (red, green, and pink lines, respectively). The models 
  assume a typical rotation of $100\,$\kms. No features corresponding to a late-type massive secondary can be spotted.}
\label{fig:BAT12_UVES}
\end{figure}

\emph{\bf BAT99\,12}
was classified as O2~If*/WN5 \citep{Crowther2011}. This star was reported by 
S08 to exhibit a period of $\approx 3.2\,$d, with $K_1 = 80\,$\kms~and $e=0.34$, and was thus considered a 
confirmed binary by H14. However, upon 
careful examination of the high quality UVES (see Fig.\,\ref{fig:BAT12_UVES}) and HST spectra at hand, no spectral features are found which 
can be associated with a companion. 
% Moreover, with the three CTIO spectra at hand, we cannot confirm
% a significant RV variability within 3$\sigma$ in lines other than He\,{\sc ii}\,$\lambda 4686$, which was used by S08 to derive 
% the orbital period. 
The observed spectra are very well reproduced by a single component
\citep{Doran2013, Hainich2014}. 

The RVs derive in this study from the CTIO spectra are consistent within 3$\sigma$ with constant RV, and we cannot confirm 
the marginal period derived by S08. If 
this object truly is a binary with a period of $P=3.2$\,d, it must be seen at a very low inclination or exhibit 
a mass ratio far from unity.
If there is a companion in this system, it either exhibits a very similar spectrum, or contributes weakly to the 
total light. 
The mass function of the system ($f = 0.17\,M_\odot$) implies that, for a primary 
mass $\gtrsim 40\,M_\odot$, as is expected for its spectral type, the secondary would need to be quite massive  
($M_2 \gtrsim 6\,M_\odot$). However, we can rule out the presence of a late-type massive star contributing more 
than ${\approx}5\%$ to the total light, as illustrated in Fig.\,\ref{fig:BAT12_UVES}. This results in implausible 
low luminosities for a secondary.

We conclude that this object is likely a single star, 
and that the RV scatter observed for it in the He\,{\sc ii}\,$\lambda 4686$ may be a result of 
periodic activity intrinsic to the star, for example, via corotating interaction regions 
\citep[CIRs][]{Cranmer1996, Kaper1999, St-Louis2009}. We cannot reject the possibility that the object is a WR binary hosting 
a compact object, presumably a BH. 

\emph{\bf BAT99\,14}
was classified by FMG03 as WN4o(+OB) because of the presence of absorption lines in its 
spectrum. Despite the binary appearance, these authors could not measure significant RV shifts, and associated the
OB-type component in the spectrum with a visual companion situated roughly 6'' from the WR star. However, photometry obtained 
for BAT99\,14 suggests it is contaminated by the presence of a cool and luminous K-type supergiant, 
(see Fig\,\ref{fig:BAT014}). This is clearly the visual companion referred to by FMG03, which is seen as a very bright 
source in photos taken in optical surveys such as DSS2. 
Moreover, in images obtained through the GALEX UV survey \citep{Lee2011},  
the dominant visual companion is no longer seen in the UV, as is expected for a K-type star. Instead, the presence of another 
source becomes evident, only vaguely seen in visual images. This could potentially be the OB-type companion that contaminates 
the spectrum, which would suggest BAT99\,14 is a not a close binary system. Future spectroscopy campaigns involving BAT99\,14
should attempt to resolve this system.

We fit the photometry of the system using three components: a WR model, an O-type model, and a K-type model. 
For the K star, we used a synthetic MARCS spectrum calculated for 
MK-type stars \citep{Gustafsson2008}, retrieved from the POLLUX archive \citep{Palacios2010}. The best fit to the 
SED is obtained for $T_3 = 4200\,$K, and, at the distance of the LMC, $\log L_3 = 5.4\,[L_\odot]$ and an absolute 
visual magnitude of $M_{V,3} = -7.9\,$mag. Based on calibrations by \citet{Allen1973}\footnote{http://xoomer.virgilio.it/hrtrace/Allen.htm}, 
this magnitude is potentially too bright for a K-type star. 
We therefore suggest that the tertiary K-star is in fact a line-of-sight 
contamination with a galactic K star. 

The spectra at hand are not contaminated by the bright K-type star, as we carefully checked. 
Without a spectrum for this star, we cannot classify it, but calibrations of $T_*$ and $\log L$  
by \citet{Allen1973} suggest it is a K4~I star.
In the spectra, clear He\,{\sc i,ii} absorption features can be seen, which enable an approximate derivation of the secondary's temperature 
($T_2 = 33\,$kK) and light ratio. The gravity cannot be accurately determined from the spectra, but 
$\log g = 4.0\pm0.3$\,[cgs] provides a satisfactory fit and is consistent with the derived 
luminosity of the secondary. Based on the UV data, some constraints for the secondary's wind parameters were derived. 
Classification of the synthetic spectrum of the secondary gives a spectral type of O9~V.

The WR primary's HRD position is not reached by the single-star BPASS tracks. Since no period could be derived for this 
system by FMG03, we cannot derive a binary evolution model from the BPASS grid. However, given its HRD position, we 
classify the primary as a b-WR star. More observations will be necessary to conclude whether the components of BAT99\,14 
are close enough to have interacted in the past.

\emph{\bf BAT99\,17}
was never before reported as a binary candidate, and the RVs derived by FMG03 were reported to be consistent with a single star.
However, the spectrum of this object shows clear He\,{\sc i} absorption lines which cannot originate from the hot WR primary. 
We believe that this is strong evidence for the presence of a cooler companion star in the spectrum.
With the spectra at hand, only an estimation of the secondary's parameters are possible. No He\,{\sc ii} lines can be seen, 
suggesting that the secondary is a late-type massive star with $T_* \le 30\,$kK. Lower temperatures give rise to N\,{\sc ii} 
and other low-ionization lines that are not observed. The relative faintness of the secondary suggests that it is a main sequence
star.  More data will be necessary 
to determine whether this system forms a spectroscopic binary, and whether it has undergone interaction in the past. Based 
on our results, we classify it as w/wb-WR.

\emph{\bf BAT99\,19}
is an eclipsing binary system with a period of $P = 18.0\,$d, originally classified as WN4b+OB? by \citet{Smith1996}, later revised to WN4b+O5: by FMG03. 
Very recently, \citet{Zasche2016} performed a lightcurve analysis of the system to derive several physical and orbital parameters. However, their 
derived temperature for the secondary ($T_2 \approx 26\,kK$) is not compatible at all with the spectral type of the secondary, and 
their implausible results, as they admit themselves, are very likely biased by the simplistic assumptions made in their work. However, their work 
confirms that the inclination of the system must be very close to $i = 90^\circ$.

In this study, we profited strongly from available FLAMES spectra, which offer phase coverage of the system. These spectra enabled 
us to derive an SB2 orbital solution for the system and disentangle the composite spectra in the spectral range $3960 - 4550\,$\AA. The results are shown 
in Sects.\,\ref{subsec:orban} and \ref{subsec:disen}. A classification of the disentangled spectrum 
implies a spectral type of O6~V for the secondary, which is consistent with its derived 
stellar parameters ($T_* = 40\,$kK, $\log g = 4.0\,$[cgs]).  
Its luminosity is found to be lower than expected for an O6~V type by about $0.2$\,dex \citep{Martins2005}.

\begin{figure*}
\centering
\begin{subfigure}{0.67\columnwidth}
  \centering
  \includegraphics[width=\linewidth]{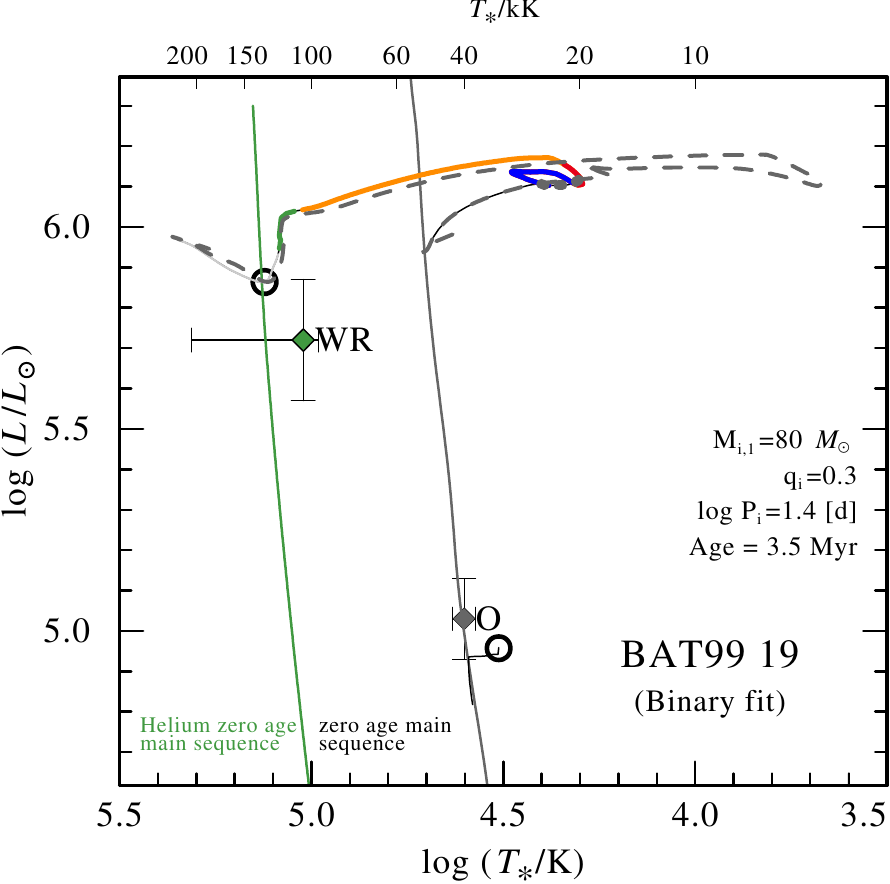}
%   \caption{s}
  \label{fig:BAT19hrdsub1}
\end{subfigure}%
\begin{subfigure}{.67\columnwidth}
  \centering
  \includegraphics[width=\linewidth]{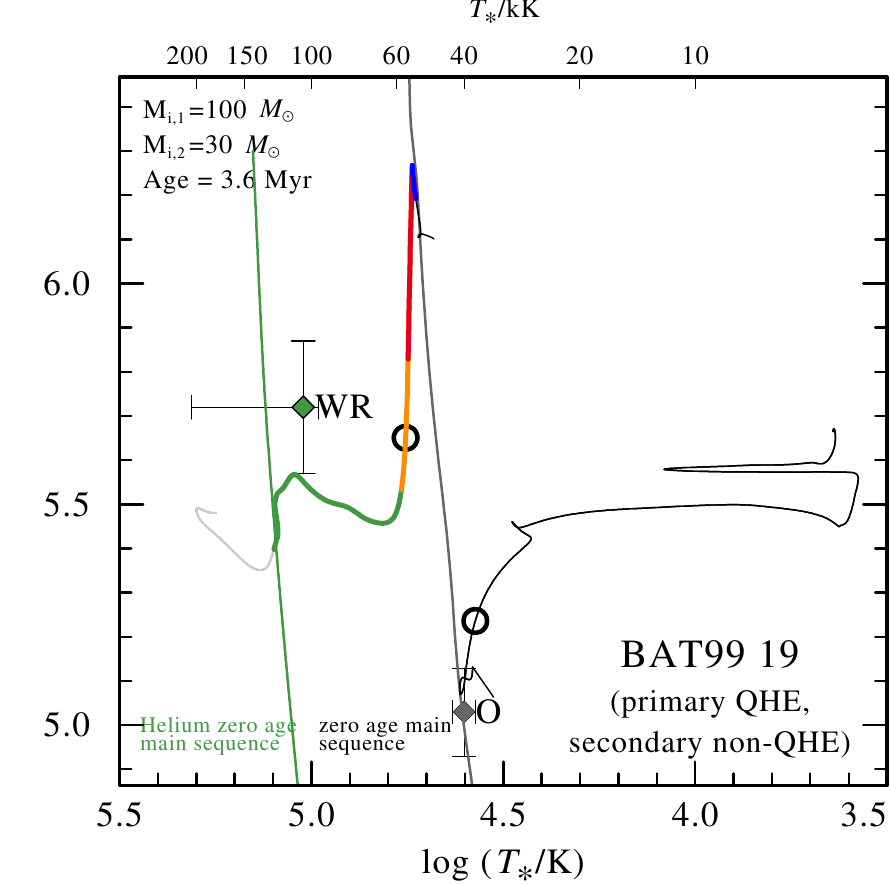}
%   \caption{A subfigure}
  \label{fig:BAT19hrdsub2}
\end{subfigure}
\begin{subfigure}{.67\columnwidth}
  \centering
  \includegraphics[width=\linewidth]{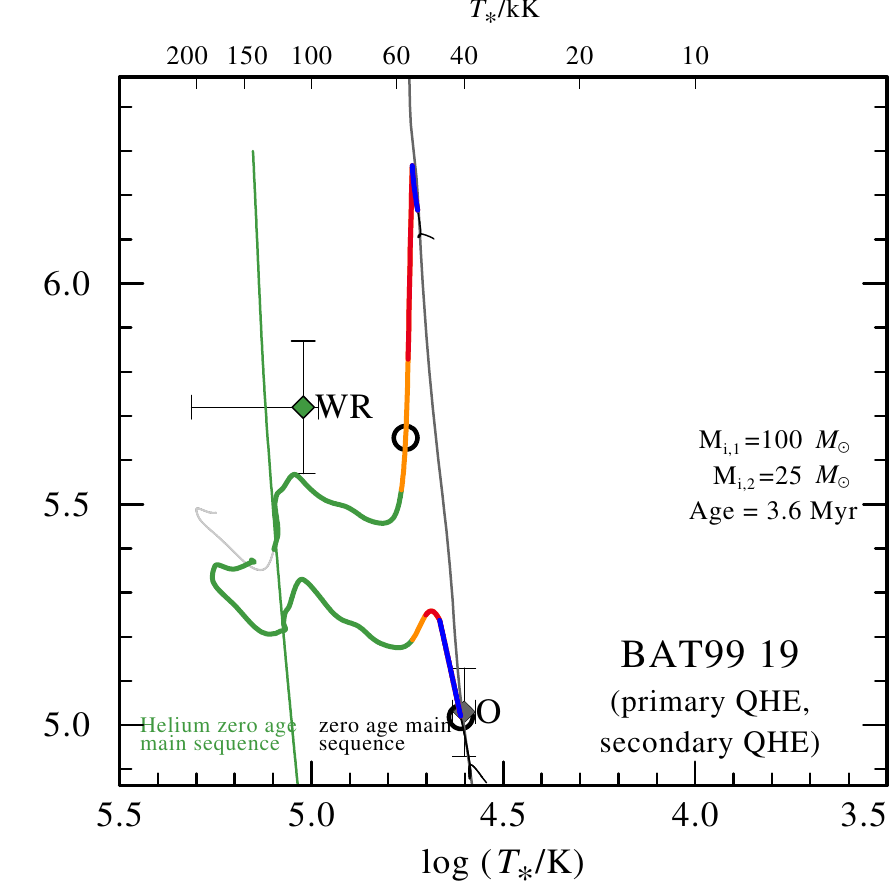}
%   \caption{A subfigure}
  \label{fig:BAT19hrdsub3}
\end{subfigure}
\caption{As Fig.\,\ref{fig:BAT49hrd}, but for BAT99\,19.}
\label{fig:BAT19hrd}
\end{figure*}

A particularly interesting fact in this system is the very rapid rotation measured for 
the secondary: $v \sin i \approx 550\,$\,\kms, much larger than the orbitally-synchronized rotation 
of ${\approx}20\,$\kms. This could be a very strong indication 
that mass transfer had occurred in this system in the recent past \citep[e.g.][]{Shara2017}.

Peculiarly, the emission lines of the primary WR star are unusually round, which has led previous studies to suggest that such WR stars may be rapid 
rotators themselves \citep{Hamann2006, Shenar2014}. Retaining the round profiles in the models is only possible when adopting substantial 
surface rotation velocities comparable to that of the secondary ($\approx 600\,$\kms), and more importantly, large co-rotation radii that reach up to 
$r \approx 5\,R_*$. Either way, the spectrum implies a terminal velocity of at least 2300\,\kms. 
The possibility of strong magnetic fields supporting such co-rotation was thoroughly discussed by \citet{Shenar2014}. 
These fields were not yet measured in WR stars \citep{Chevrotiere2013, Chevrotiere2014, Chene2019}.
The question as to the potential rotation of the WR component remains open until more data are available. At any rate, the assumption 
 of rapid rotation of the WR star does not alter 
our main results regarding this system, which should be subject to more studies in the future.

\begin{figure}
\centering
  \includegraphics[width=0.5\textwidth]{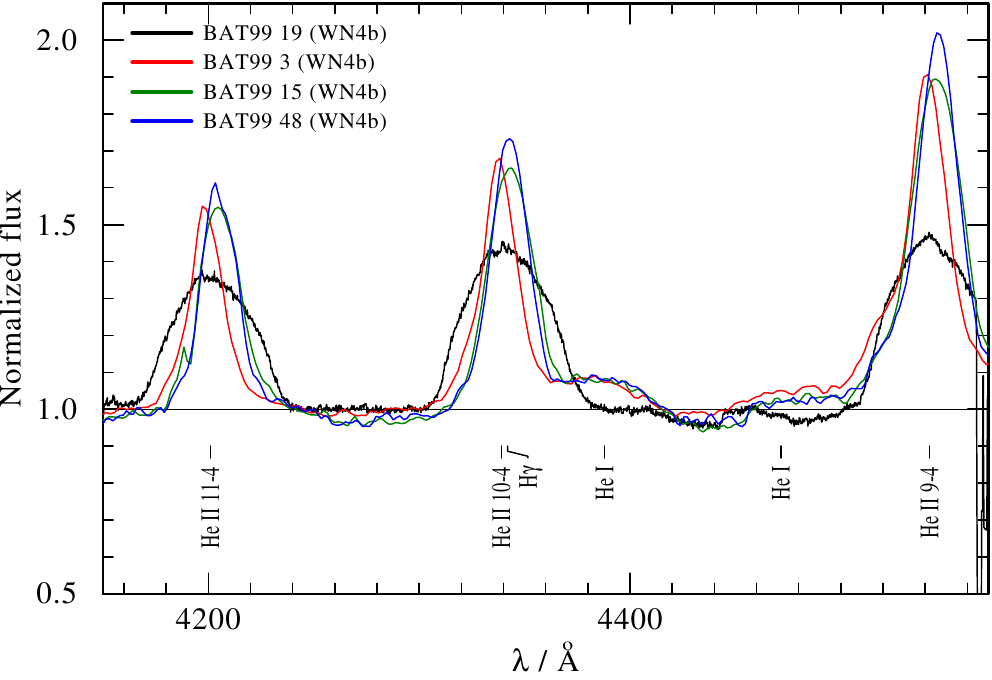}
  \caption{A comparison between the disentangled spectrum of BAT99\,19 and observed spectra of 
  BAT99\,3, 15, and 48 (see legend). Despite their identical spectral type (WN4b), BAT99\,19 shows peculiarly broad and round lines.}
\label{fig:BAT19_round}
\end{figure}

The best evolutionary fit to all parameters, including the orbital masses, is obtained 
when assuming non-homogeneous evolution and hence past mass-transfer (Fig.\,\ref{fig:BAT19hrd}). However, as for BAT99\,6, we cannot rule out CHE
(Fig.\,\ref{fig:BAT19hrd}).
The rapid rotation of the secondary strongly suggests that it accreted mass in the past, but it may also 
support its homogeneous evolution. The primary's HRD position suggests that 
it was massive enough to
have been formed as a single star, and we therefore classify it as w/wb-WR.

\emph{\bf \object{BAT99\,21}:}
was classified as WN4+OB by \citet{Breysacher1981}, and later reclassified as WN4o+OB by FMG03. While FMG03 detect 
marginal RV variations, they cannot infer a period. Since BAT99\,21 has a known visual companion located 
at an angular separation of 2'', the latter authors conclude that BAT99\,21 is likely not a short period 
binary but rather a visual one. However, the visual OB companion dominates in the optical and therefore 
biased previous derivations of the WR component's parameters.

FMG03 suggested that the primary's atmosphere is hydrogen free. This is confirmed by our modelling efforts.
The secondary's absorption features are clearly apparent in the co-added optical spectrum. 
Because H14 neglected the dominant companion in their analysis, 
the WR primary was found to be significantly 
more luminous (by $\approx 0.6\,$dex) than reported here. A classification of the model spectrum suggests 
the secondary is an O9\,III star.

Similarly to BAT99\,14, although the presence of two components in its spectrum is clear, no period could be 
derived by FMG03, with only marginal $\sigma_\text{RV}$ reported. Therefore, it is not known whether BAT99\,21
is an interacting binary. Since the primary's HRD position is covered by the $M_\text{i} = 35\,M_\odot$ single-star BPASS 
tracks, we classify it 
as w/wb-WR.

\emph{\bf BAT99\,27}
was reported by FMG03 to be a visual binary, with the WR primary strongly diluted by its B supergiant companion. FMG03 classified
this system as WN5b(+B1 Ia) and argued that the two stars likely do not form a close binary. \citet{Neugent2018} reclassified the system to 
WN4 + B~I.
While some evidence 
for sinusoidal RV variations was found in the system, no strict periodicity could be inferred. 
The relative contribution of the B component was estimated from the relative strengths of the N\,{\sc v}\,$\lambda \lambda 1239, 1243$ resonance 
doublet and He\,{\sc i} lines in the optical, as well as from an estimation of the amount of dilution compared 
to single WR stars of similar physical parameters. 
In this work, 
we find that the system's composite spectrum can be well reproduced assuming a WR and a B-supergiant component. 
Our tests also show that it could be easy for an O-type spectrum to ``hide'' in the low-resolution 
optical spectrum at hand. Although the WR companion was suggested to have peculiarly round emission lines which may imply wind co-rotation 
\citep[H14][]{Shenar2014}, we do not find a notable discrepancy between our standard non-rotating models 
and the observations.
Accounting for the B supergiant reduces the  luminosity of the WR star to $\log L = 5.8\,[L_\odot]$,  more than 1\,dex compared to H14. 

If the WR primary has interacted with a companion in the past, the companion is very unlikely to be the B1~Ia star observed 
in the spectrum. Given the lack of additional confirmed components, 
the HRD position of the WR primary, and the lack of evidence for additional stripping (other than wind-stripping), 
we classify the primary as w-WR.

\emph{\bf BAT99\,29} is a confirmed WN4b+OB binary with a 2.2\,d period (FMG03),  later reclassified to WN3+OB by \citet{Neugent2018}.
As already argued by H14,  the secondary 
is much fainter than the primary in the optical. The secondary is clearly seen in a few He\,{\sc i} lines, primarily $\lambda 4471$. 
Judging by the amount of dilution and the strengths of its He\,{\sc i} lines, 
it contributes roughly 20\% to the total light of the system. Our synthetic spectrum suggests a spectral 
type of B1~V for the secondary.

\begin{figure*}
\centering
\begin{subfigure}{0.67\columnwidth}
  \centering
  \includegraphics[width=\linewidth]{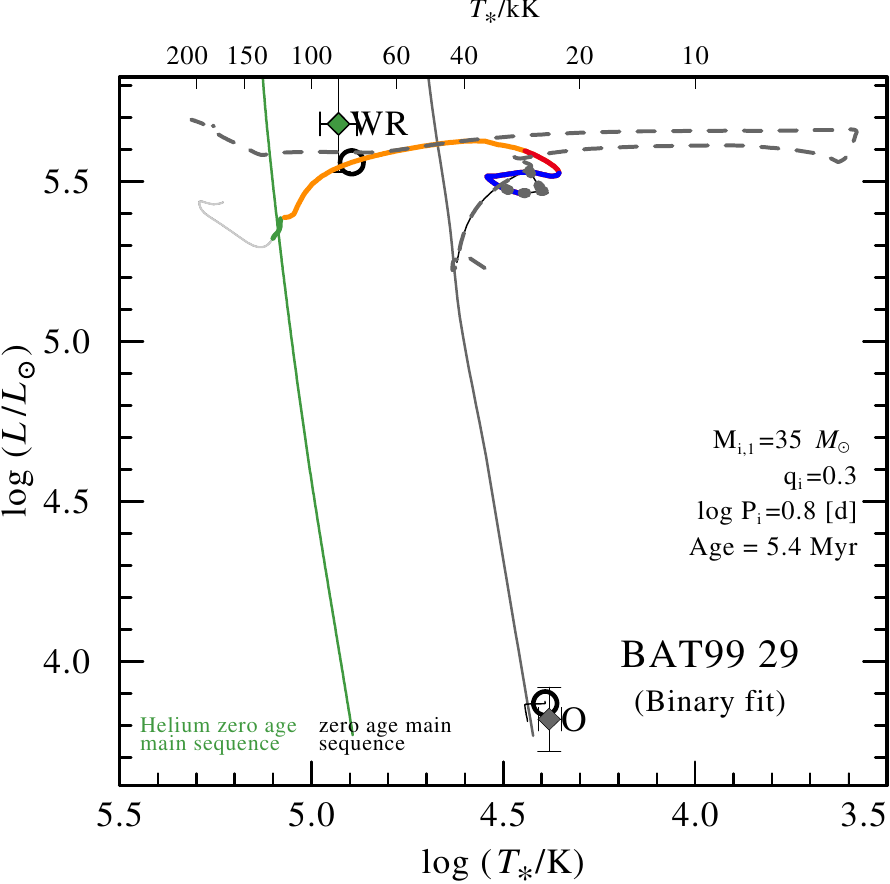}
%   \caption{s}
  \label{fig:BAT29hrdsub1}
\end{subfigure}%
\begin{subfigure}{.67\columnwidth}
  \centering
  \includegraphics[width=\linewidth]{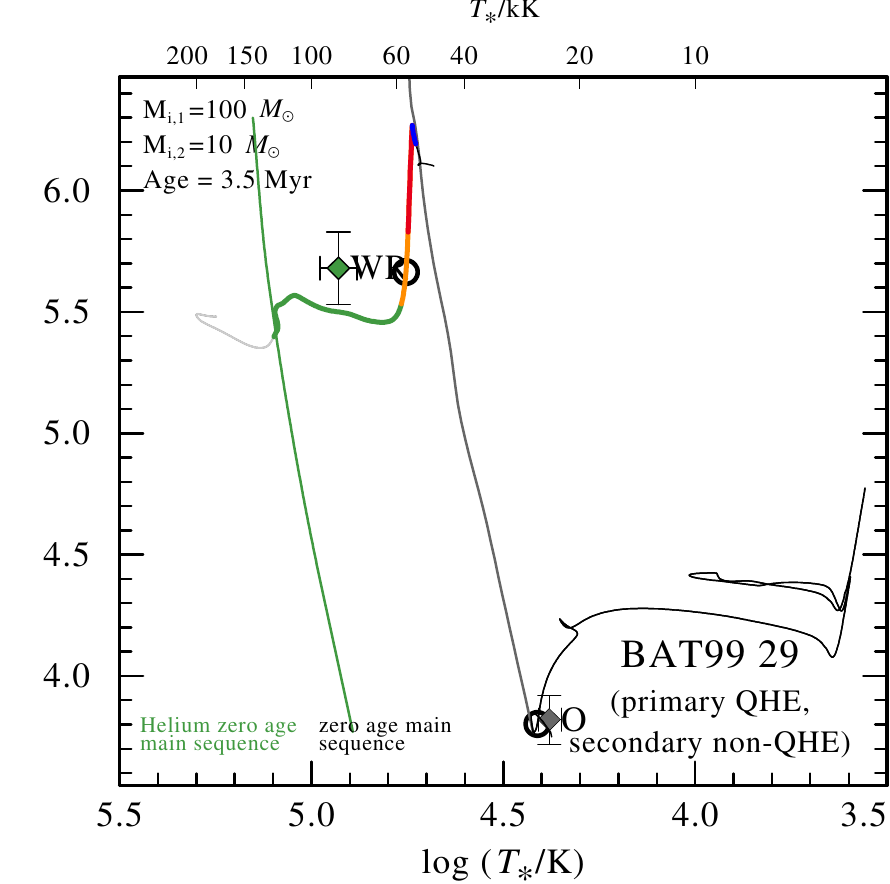}
%   \caption{A subfigure}
  \label{fig:BAT29hrdsub2}
\end{subfigure}
\begin{subfigure}{.67\columnwidth}
  \centering
  \includegraphics[width=\linewidth]{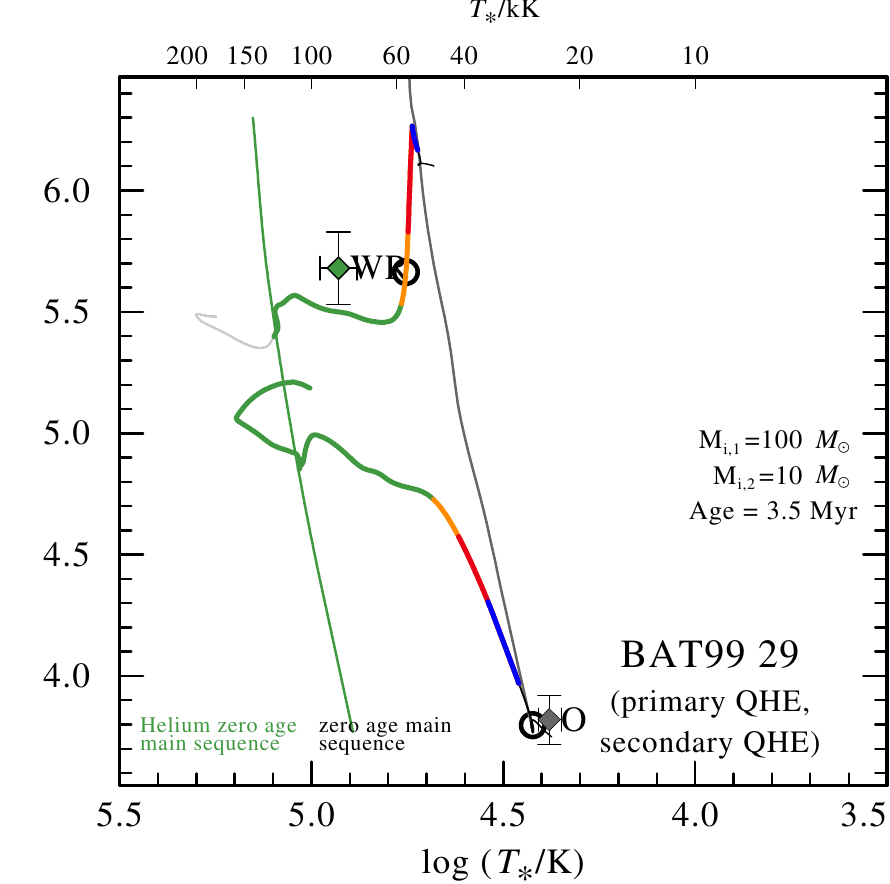}
%   \caption{A subfigure}
  \label{fig:BAT29hrdsub3}
\end{subfigure}
\caption{As Fig.\,\ref{fig:BAT49hrd}, but for BAT99\,29.}
\label{fig:BAT29hrd}
\end{figure*}

Adopting a typical mass for the secondary ($9\,M_\odot$) and an inclination of $57^\circ$ 
(Table\,\ref{tab:orbitalpar}) implies a peculiarly large mass for the primary ($90\,M_\odot$), which 
in turn suggests that the adopted mass for the secondary is wrong. Therefore, 
deriving an evolutionary channel for this system is somewhat speculative. Nevertheless, 
given the proximity of the companions, it is very likely that they interacted. The best-fitting
BPASS binary track is shown in Fig.\,\ref{fig:BAT29hrd}. Based on its HRD position, we classify the WR primary as wb-WR.

\emph{\bf \object{BAT99\,31}} was classified as WN3b by \citet{Smith1996}, was later reclassified WN4b 
by FMG03, and later reclassified yet again to WN3 \citep{Neugent2018}. The phase coverage of this system in the latter study was poor and, except for one data point, the RVs 
are consistent with BAT99\,31 being a single WR star. This is supported by the FLAMES spectra 
at hand, which show no sign for RV shifts. Moreover, no traces of a second companion can be seen in the spectrum 
of the star. Some diffuse X-ray emission in the system was reported by FMG03, but the origin of this emission 
does not seem to be related to a binary companion, which cannot be confirmed in this study. We therefore 
omit it from our analysis.

\emph{\bf \object{BAT99\,32}} was classified as WN6(h) in the original BAT99 catalog, later confirmed by S08. S08 
were able to confirm a period of $P = 1.9$\,d, making it the shortest-period WN binary in the LMC. However, S08 were 
not able to infer the spectral type of the secondary from their spectra.

The CTIO spectra at hand allowed us to disentangle the optical spectrum of BAT99\,32 (see Fig.\,\ref{fig:dis032}). However, 
the small number of spectra does not allow us to perform the disentanglement unambiguously. Moreover, our results suggest
that the system comprises three components: Two emission-line stars and one absorption-line star. With only five spectra
at hand, the disentangled spectra should be taken with caution. Here, we analyse the object as a WR + WR binary.

Since FMG03 have a much better phase coverage of the system, we adopt their orbital parameters. However, the 
secondary semi-amplitude $K_2$ could not be derived by FMG03, and is therefore adopted from the disentanglement procedure. 
This results in very low minimum masses of $2.4\,M_\odot$ for both components. This could only be compensated 
for by an inclination of  $i{\approx 20}^\circ$ or smaller. However, better data would be necessary to better constrain 
the orbital parameters of the system.

From the disentangled spectra, we can derive approximate temperatures for the primary and secondary, the latter 
being some 15kK hotter than the former. The light ratio in this system was very difficult to derive, because unique features which belong 
to the primary (e.g.,  N\,{\sc iv}\,$\lambda 4060$) 
or to the secondary (e.g.,  N\,{\sc iii}\,$\lambda \lambda 4634, 4641$) are sensitive to the mass-loss rates. 
Moreover, the possible presence of a ``step'' 
in the saturated resonance P-Cygni line C\,{\sc iv} in the UV is ambiguous due to neighbouring iron lines. Therefore, 
the light ratios were determined primarily by calibrating the combined models with lines that typically show 
a global EW, such as the N\,{\sc iv}\,$\lambda 4060$ line. Clear emission excess in lines such as 
H$\delta$ and He\,{\sc i}\,$\lambda 5875$ stemming from WWC can be seen \citep[c.f.,][]{Shenar2017}.
Based on our models, we reclassify the system to WN5(h) + WN6(h):(+abs). 

\begin{figure*}
\centering
\begin{subfigure}{0.67\columnwidth}
  \centering
  \includegraphics[width=\linewidth]{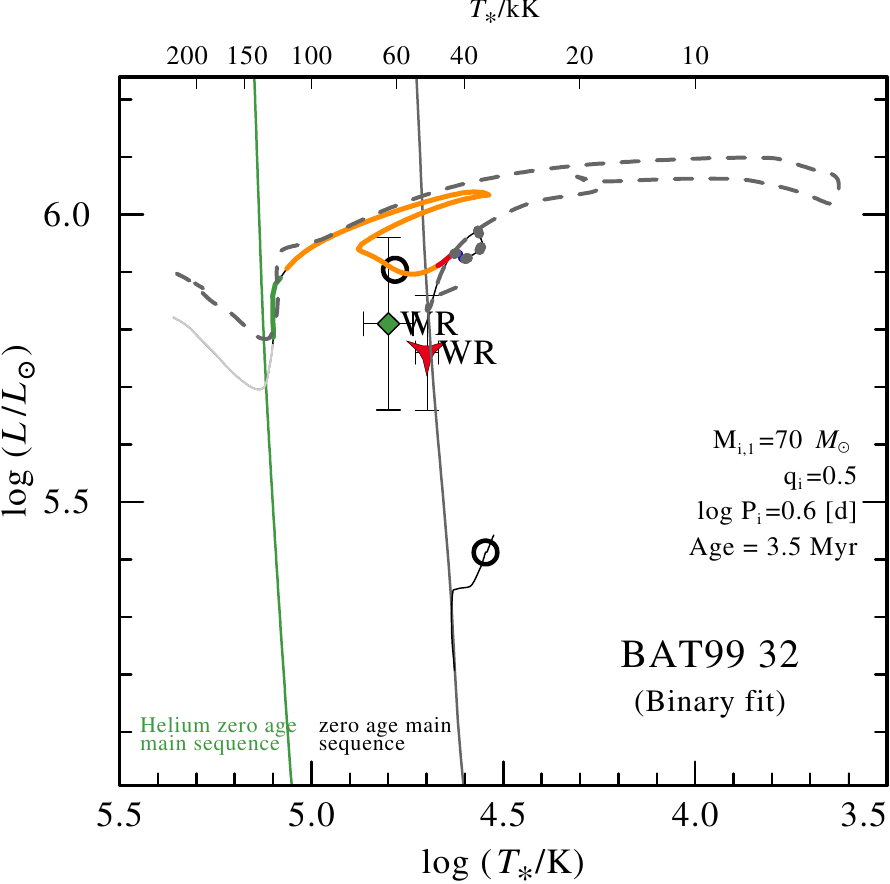}
%   \caption{s}
  \label{fig:BAT32hrdsub1}
\end{subfigure}%
\begin{subfigure}{.67\columnwidth}
  \centering
  \includegraphics[width=\linewidth]{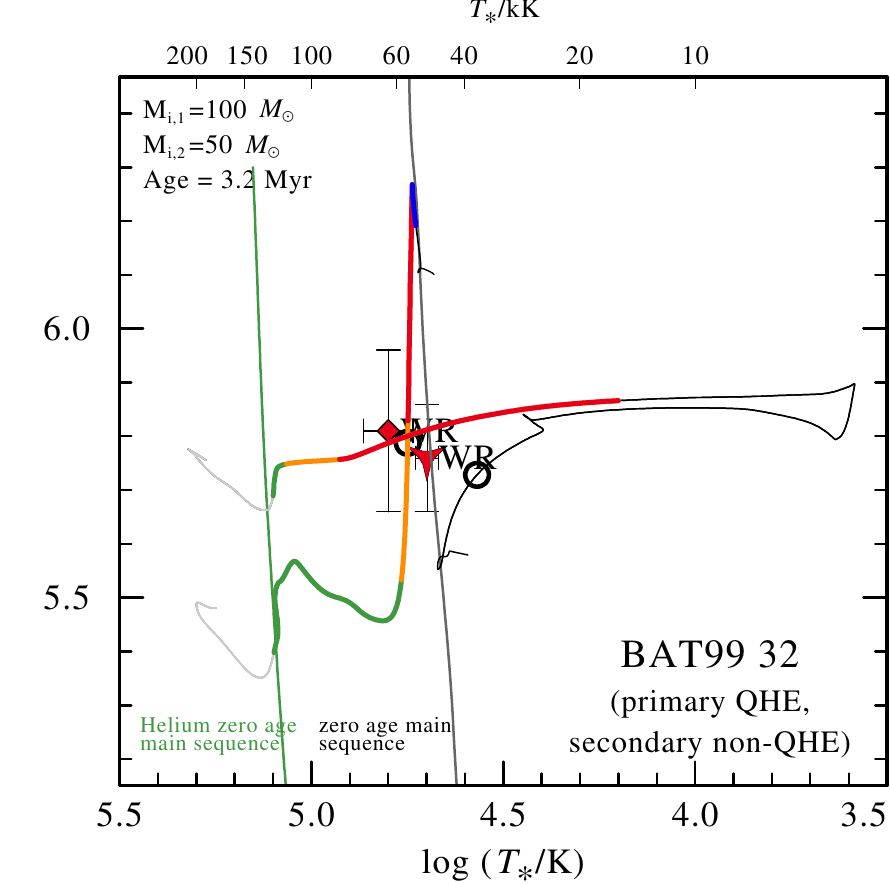}
%   \caption{A subfigure}
  \label{fig:BAT32hrdsub2}
\end{subfigure}
\begin{subfigure}{.67\columnwidth}
  \centering
  \includegraphics[width=\linewidth]{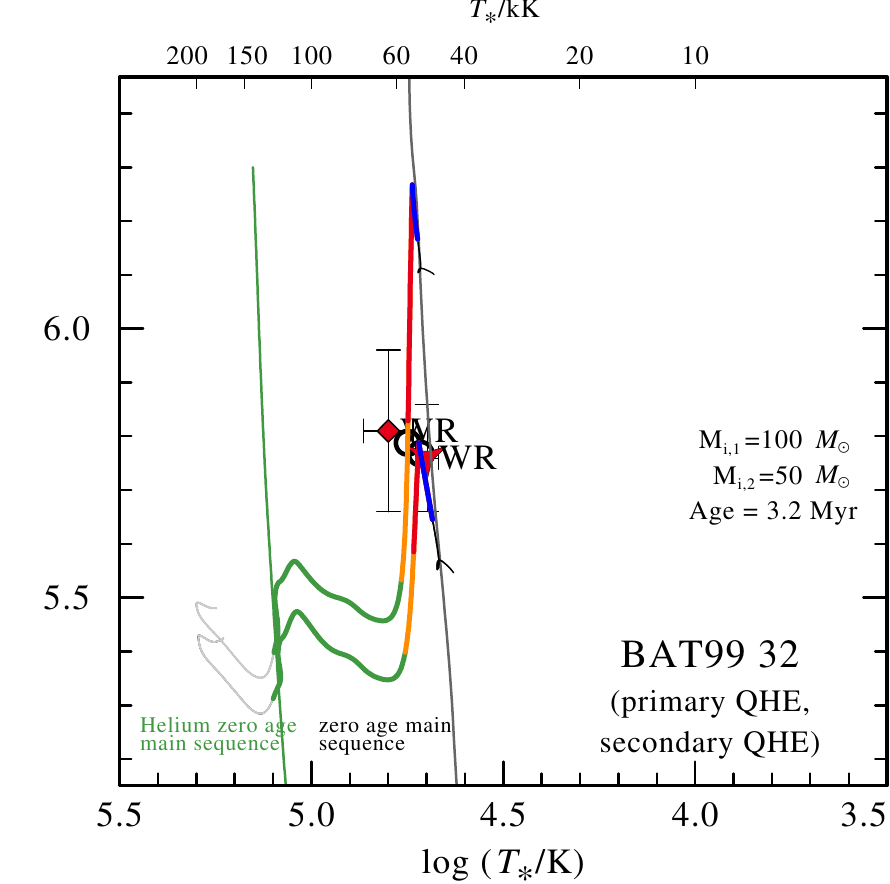}
%   \caption{A subfigure}
  \label{fig:BAT32hrdsub3}
\end{subfigure}
\caption{As Fig.\,\ref{fig:BAT49hrd}, but for BAT99\,32.}
\label{fig:BAT32hrd}
\end{figure*}

Given that BAT99\,32 is not only the shortest period WR binary in the LMC, but also potentially contains two WR stars as companions, 
we suggest that this system should be carefully observed and studied in the future.  The system may be 
a promising candidate for an evolved GW progenitor undergoing CHE.
Only a scenario in which both components evolve via CHE can provide a consistent fit to the 
derived parameters (Fig.\,\ref{fig:BAT32hrd}). However, considering the contradictory results 
found in this study (e.g., implausibly low orbital masses), we prefer to not overspeculate 
until the necessary data will become available, and we therefore do not give an evolutionary solution to this system. 

\emph{\bf \object{BAT99\,36}} was classified as WN4b/WCE by \citet{Smith1996}, confirmed by FMG03. Recently, it was reclassified again 
to WN4b/WCE+OB by \citet{Neugent2018}. Already \citet{Crowther1995} argued that faint traces of a secondary companion may be seen on top of the 
Balmer lines in the optical spectrum of the object. However, the lack of observed RV variation make BAT99\,36 
consistent with it being a single star (FMG03). While very faint traces for a companion 
are potentially seen in the spectrum, the quality of the data do not enable us to analyse the system unambigiously. We therefore omit it from our analysis.

\emph{\bf \object{BAT99\,40}} was originally classified as WN4o+O in the BAT99 catalogue because of seemingly 
strong absorption features in the spectrum. 
FMG03 later reported a rather high X-ray luminosity 
of $L_\text{X} \approx 5\cdot10^{33}\,$\lum, which could suggest the presence of colliding winds. However, 
the RV scatter reported for the object is consistent with it being a single star. Moreover, FMG03
reclassified the object to WN4(h)a, suggesting that the absorption features are blue-shifted (P-Cygni like) and thus 
intrinsic to the WR star. This claim can be confirmed by our study. The object is therefore omitted from our sample.

\emph{\bf \object{BAT99\,42}} is a known visual binary classified as  WN5b(h) + B3~I \citep[e.g.,][]{Smith1996}. 
A quick inspection of the system's spectrum reveals very weak emission lines, which implies a strong 
dilution of the WR star.
\citet{Seggewiss1991} reported a photometric variability 
of the system with a period of $P \approx 30\,$d, but no spectroscopic counterpart at this period 
is reported by FMG03. 
A high X-ray luminosity of $L_\text{X} \approx 5\cdot10^{34}\,$\lum was reported for the system by FMG03, which could 
potentially arise from the presence of colliding winds. 
Based on more recent HST images, H14 showed that the slit FMG03 used to acquire their spectra, which are used
in this study, included at least three dominant sources. However, the spectra at hand do not allow 
for a disentanglement of these three sources. Instead, we treat the system as a WR+B-supergiant binary for the spectral 
analysis. Since the WR-star is accompanied by more than one close OB companion on the 
sky and it is not obvious which one of these participates in a close orbit with the WR star, 
the parameters derived for the companion - especially its luminosity - should be taken with caution.

The relative contribution of the WR star was derived based on the strength of the Balmer lines as well as diagnostic 
He\,{\sc i} lines in the spectrum. The N\,{\sc v} resonance line in the UV, stemming solely from the WR star, also helped 
to constrain the relative light ratio of the components. The lack of apparent He\,{\sc ii} lines and the 
SED of the system implies that the secondary is cool.
The temperature derived here for the WR component is comparable to that derived by H14. In contrast, 
when accounting for the secondary companion, the luminosity of the WR star drops by about 2\,dex compared to the
very high luminosity of $\log L \approx 8\,[L_\odot]$ reported by H14.  While still high, the luminosity derived here 
is in line with other hydrogen rich stars 
of similar spectral types. Because H14 neglected the dilution caused by the B supergiant, the transformed radius 
is almost an order of magnitude smaller in this study. The discrepancy in the mass-loss 
rate compared to H14 is less extreme due to the counteracting correlations $\dot{M} \propto L^{1/3}$ and 
$\dot{M} \propto R_\text{t}^{-3/2}$. 

The WR component in the system has a unique round-shaped emission profile that is reminiscent of rotation profiles. Therefore, 
H14 adopted $v \sin i \approx 2000$\,\kms~to reproduce it. Since such a high value implies very large co-rotation radii 
\citep[see][]{Shenar2014}, we do not assume this rotation here, hence the larger terminal velocity derived in this study. 
However, we note that the 
models cannot fully reproduce the round shape of the He\,{\sc ii}\,$\lambda 4686$ line using the standard $\beta$-law.

More data would be needed to derive a consistent evolutionary model for the system. Given the high luminosity 
of the primary and the current lack of evidence pointing towards a close, interacting companion, 
we classify the WR primary as w-WR here.

\emph{\bf \object{BAT99\,43}}  was classified as WN4o+OB by FMG03, who detect a very short period of $P = 2.8\,$d for this 
binary, and later to WN3+OB by \citet{Neugent2018}. We can confirm the presence of an O-type star in the spectrum, which is most clearly apparent in faint 
He\,{\sc i} lines. The temperature and light ratios were roughly estimated from the overall strength of these lines 
compared to that of He\,{\sc ii} lines. As reported by H14, the companion does not dominate in the optical, although it does 
contribute a non-negligible amount of flux, with an estimated ratio of $F_\text{V, O} / F_\text{V, WR} = 0.7$. 
As a result, the luminosity 
derived here for the WR star is $\approx 0.15$\,dex lower, while the effective temperature  is $\approx10\,$kK 
higher than derived by H14.

We cannot conclusively derive an evolutionary channel for the system. Since the binary fit
fails to reach a hydrogen-free phase at the observed position, and 
given the short period of the system, it is possible that 
both components experience CHE. Since the primary is massive 
enough to become a WR star as a single star, we classify it as w-WR.

\begin{figure*}
\centering
\begin{subfigure}{0.67\columnwidth}
  \centering
  \includegraphics[width=\linewidth]{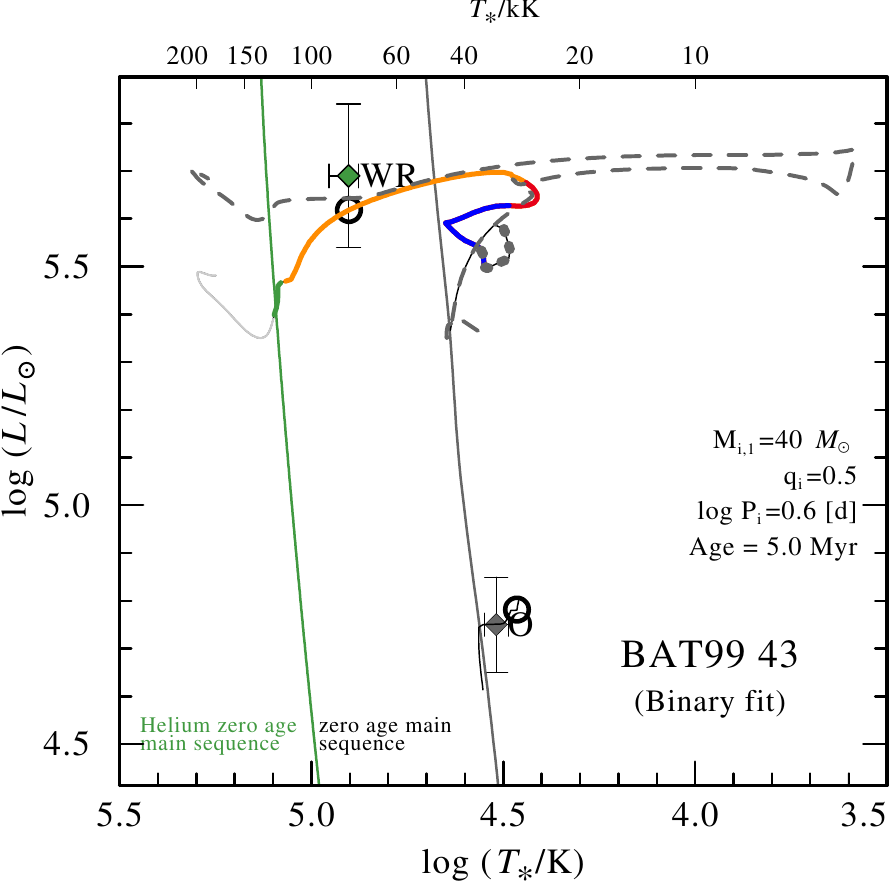}
%   \caption{s}
  \label{fig:BAT43hrdsub1}
\end{subfigure}%s
\begin{subfigure}{.67\columnwidth}
  \centering
  \includegraphics[width=\linewidth]{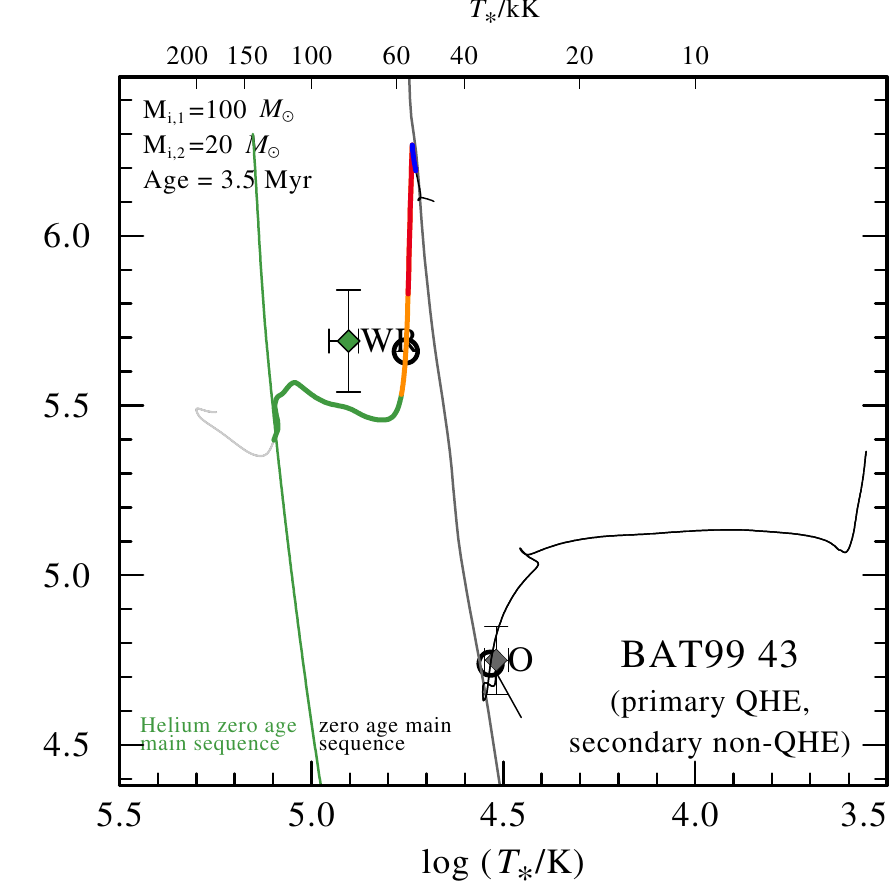}
%   \caption{A subfigure}
  \label{fig:BAT43hrdsub2}
\end{subfigure}
\begin{subfigure}{.67\columnwidth}
  \centering
  \includegraphics[width=\linewidth]{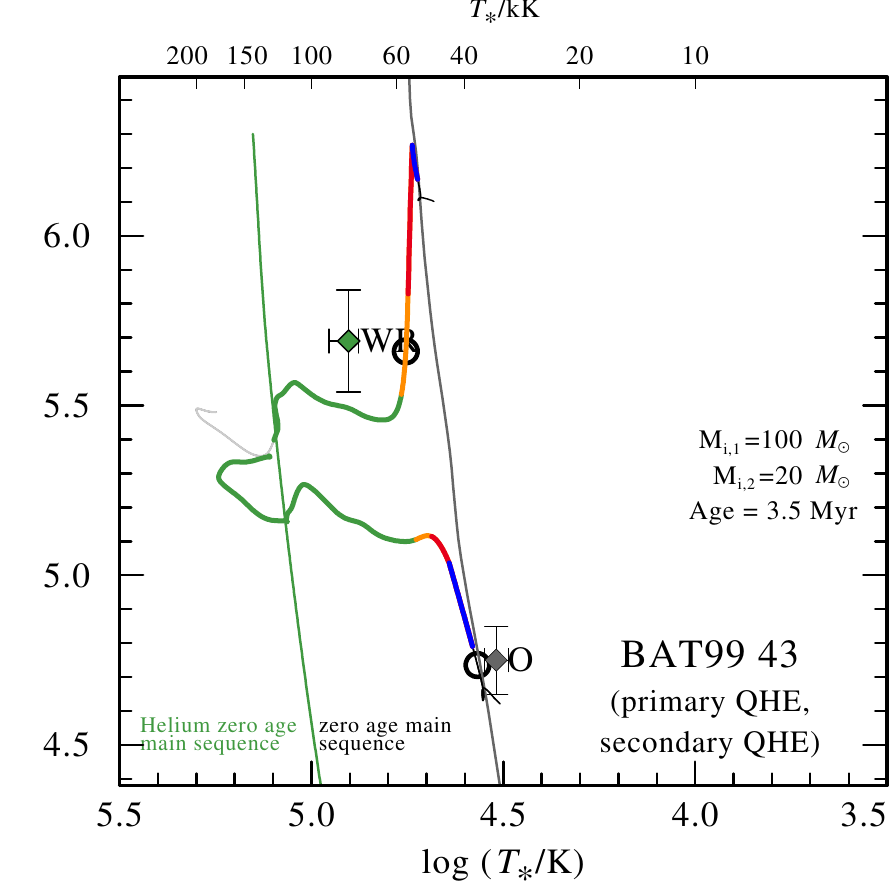}
%   \caption{A subfigure}
  \label{fig:BAT43hrdsub3}
\end{subfigure}
\caption{As Fig.\,\ref{fig:BAT49hrd}, but for BAT99\,43.}
\label{fig:BAT43hrd}
\end{figure*}

\emph{\bf \object{BAT99\,47}} was originally classified as WN3 in the Brey81 catalog,  
later updated to WN3b by FMG03. A substantial X-ray luminosity ($\log L_\text{X} \approx 33.6\,$[\lum]) 
was reported by \citet{Guerrero2008b}, which led H14 to treat BAT99\,47 as a binary candidate. However, 
the object's RVs are fully consistent with it being single, and no traces for a companion can be seen 
in the available spectra. We thus omit BAT99\,47 from our analysis sample.

\emph{\bf \object{BAT99\,49}} was first confirmed as a binary by \citet{Niemela1991}, who reported a period 
of $P = 33.9$\,d for the system. It was later classified by FMG03 as WN4:b + O8 V. 
Clear He\,{\sc i} absorption features are apparent in the spectrum. With 
only N\,{\sc v} emission features apparent in the optical spectrum, a high temperature ($\approx 100\,$kK) 
is implied for the WR star, suggesting that the companion is responsible for the He\,{\sc i} absorption features.
Several He\,{\sc ii} absorption features can also be unambiguously attributed to the secondary, constraining 
its temperature. The light ratio was determined from the overall strength of the 
He\,{\sc i}, {\sc ii}, and Balmer lines. 

The UV observations do not reveal a significant P-Cygni signature of the C\,{\sc iv} resonance line. To suppress this line 
in the O-star model, a rather low mass-loss rate, as well as standard X-ray emission, were needed. This results in a too-strong absorption 
in the Balmer lines, especially in H$\alpha$. We suspect that these lines are contaminated by nebular emission, hence the apparent discrepancy.
A classification of the model spectrum results results in the spectral type O8~V.

As illustrated in Fig.\,\ref{fig:BAT49hrd}), 
the BPASS tracks strongly suggest that the system experienced mass-transfer in the past. The CHE tracks (both for the primary 
alone as well as for both components) completely fail to reproduce the system's properties (middle and right panels of 
Fig.\,\ref{fig:BAT49hrd}). According to our solution, the WR primary could only form through binary interaction, and 
we classify it as b-WR.

% 
% % 
\emph{\bf \object{BAT99\,59}} was classified as a binary (WN4 + OB) in the original Brey81 catalog, based 
on clear absorption features apparent on top of the combined Balmer+Pickering emission lines.
The companion was presumed to be a B-type star by \citet{Smith1996}. FMG03 reclassified the system 
to WN4b + O8 and reported marginal evidence for a periodic RV signal with a period of 
$P \approx 4.7$\,d, with the RV amplitude of the WR star comparable to their detection limit ($\approx 30\,$\kms). Finally, the system was classified 
to WN3+OB by \citet{Neugent2018}.

The light ratio can be estimated from the relative strengths of the Balmer absorption lines, as well as 
temperature-insensitive lines such as the C\,{\sc iii} line complex at $\lambda \approx 1170\,$\AA. The temperature of the 
WR star is well constrained by the presence of the N\,{\sc iv}\,$\lambda 4060$ and N\,{\sc v}\,$\lambda \lambda 4604,
4620$ doublet.  The temperature of the secondary is constrained from the presence of He\,{\sc i} absorption lines, and
He\,{\sc ii} absorption features overlaid on the He\,{\sc ii} emission. The low resolution of the spectra only allow 
for a rough estimate of the surface gravity and rotation velocity of the secondary. Based on its 
synthetic spectrum, we classify the secondary as O6.5~III.

\begin{figure*}
\centering
\begin{subfigure}{0.67\columnwidth}
  \centering
  \includegraphics[width=\linewidth]{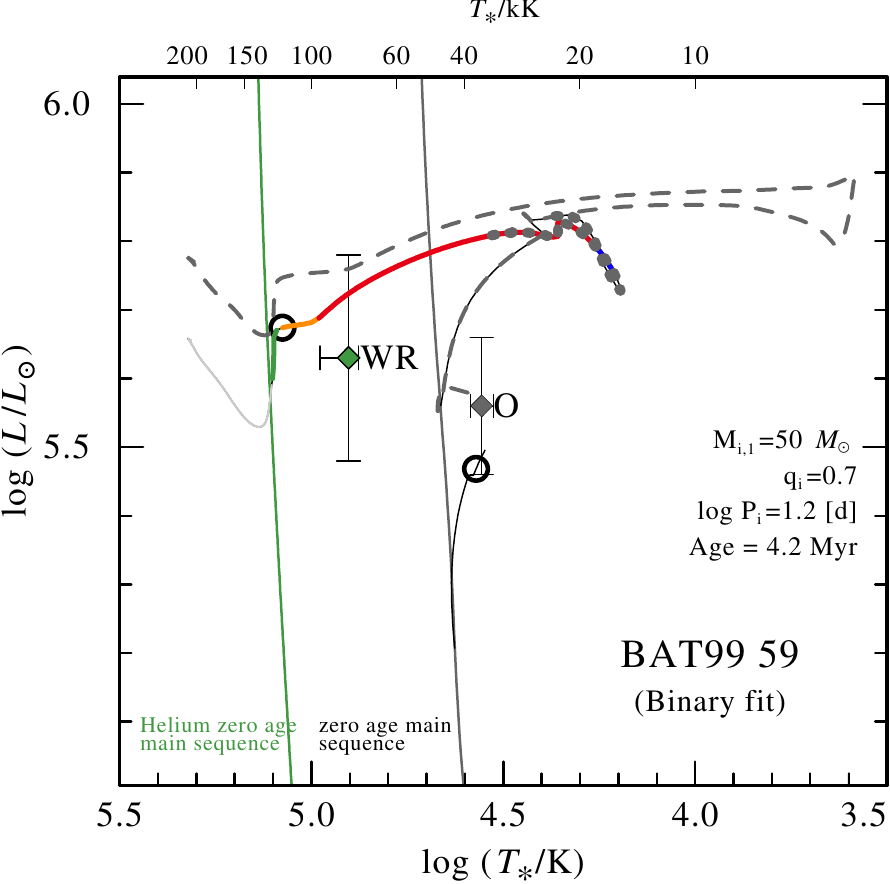}
%   \caption{s}
  \label{fig:BAT59hrdsub1}
\end{subfigure}%
\begin{subfigure}{.67\columnwidth}
  \centering
  \includegraphics[width=\linewidth]{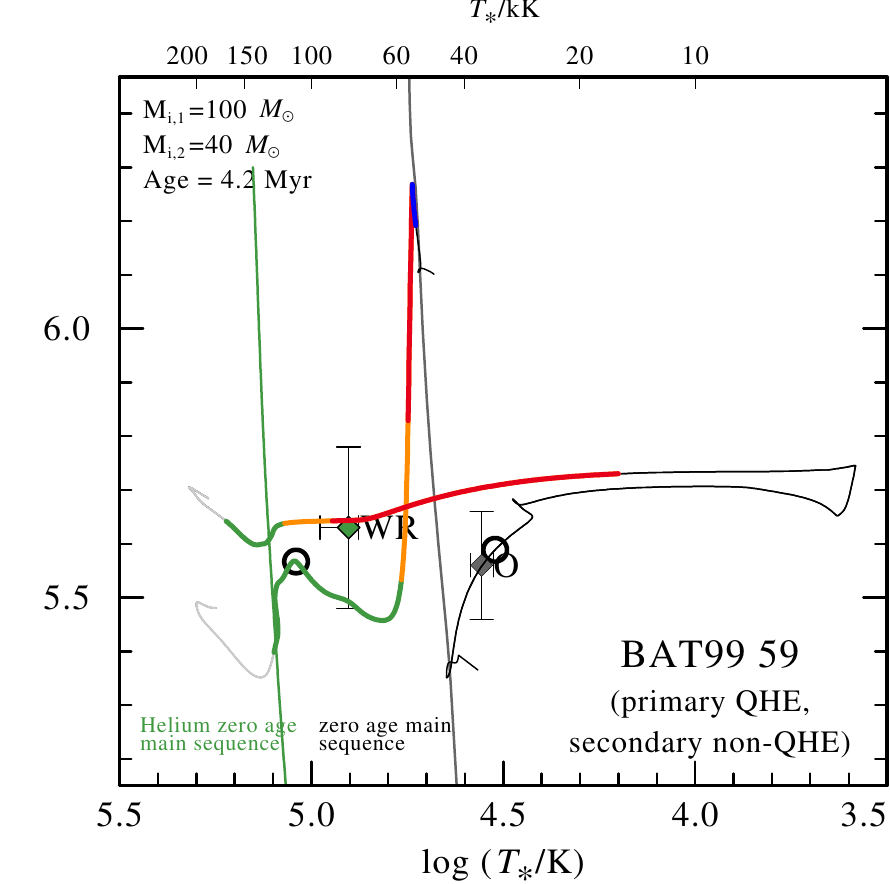}
%   \caption{A subfigure}
  \label{fig:BAT59hrdsub2}
\end{subfigure}
\begin{subfigure}{.67\columnwidth}
  \centering
  \includegraphics[width=\linewidth]{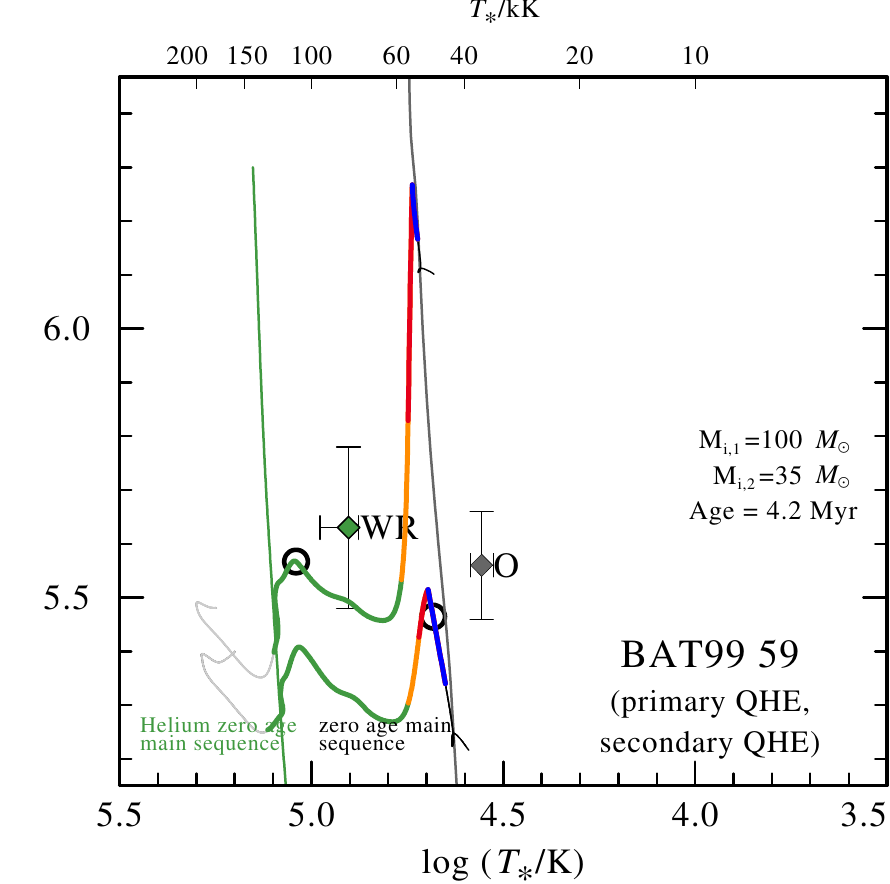}
%   \caption{A subfigure}
  \label{fig:BAT59hrdsub3}
\end{subfigure}
\caption{ As Fig.\,\ref{fig:BAT49hrd}, but for BAT99\,59.}
\label{fig:BAT59hrd}
\end{figure*}

Evolution-wise, we cannot statistically discern between past mass-transfer  and 
the primary undergoing CHE (Fig.\,\ref{fig:BAT59hrd}), although we can rule out that the secondary evolved via CHE, 
making the CHE channel less likely overall. Given our solution, we classify the WR star as wb-WR.

\emph{\bf \object{BAT99\,60}}  was classified by \citet{Breysacher1981} as WN3 + OB based on absorption features in the spectrum, 
but this classification was rejected by FMG03, who reclassified it as WN4(h)a, claiming that the absorption originated in 
the WR star. Furthermore, the RVs derived by FMG03 are consistent with a single star. Therefore, this object was not considered 
a binary candidate by H14. 
However, upon inspection of the spectrum, it is obvious that absorption lines belonging 
to He\,{\sc i} cannot originate in the hot WR star. This is supported by the recent assignment of the spectral type WN3+OB by \citet{Neugent2018}.
The presence of a second, cooler star is thus evident from the spectrum, and 
we therefore include this target in our sample. With only one epoch, the analysis can only be done in a rough manner from the ratio of the He lines and their overall 
strength. The low RV scatter implies either that the inclination of the system is low, 
or that it is not a spectroscopic binary. Without 
knowledge of the configuration of the system, we cannot derive an evolutionary scenario for the system. However, 
the luminosity of the WR star puts it in the regime of b-WR stars, which motivates our tentative formation channel 
classification.

\emph{\bf \object{BAT99\,64}} was originally classified WN4+OB? in the Brey81 catalogue due to the presence of 
He\,{\sc i} absorption lines in the spectrum of a seemingly early-type WR star. The star was later reclassified 
by FMG03 to WN4o+O9, who detected periodic RV variations in the system with a period of $P = 37.6\,$d, and later revised again to WN3+O by \citet{Neugent2018}. Furthermore, 
the system portrays a single eclipse, presumably occurring when the O companion occults the WR companion, with the photometric 
and spectroscopic periods in agreement (see Figure 10 in FMG03). Assuming the WR star is fully occulted by its companion during 
eclipse, the magnitude difference of $\Delta V \approx 0.4\,$mag implies a flux ratio in the optical of roughly $F_\text{V,O} / F_\text{V, WR} \approx 2$. 
This ratio agrees very well with the ratio derived spectroscopically. This stands in contrast to the claims of H14, who argued that the 
companion does not contribute substantially to the optical spectrum.

The temperature of the WR star can be constrained well by the presence of strong 
N\,{\sc v} and weak N\,{\sc iv} emissions. With only faint traces of He\,{\sc ii} absorption, the O companion is confirmed to be a 
late O-type star. As a result of accounting for binarity, the luminosity of the WR star is found 
to be $\approx 0.7\,$dex smaller in this study compared to the single-star analysis performed by H14. A classification 
of the synthetic spectrum of the secondary implies it is an O9~V star.

\begin{figure*}
\centering
\begin{subfigure}{0.67\columnwidth}
  \centering
  \includegraphics[width=\linewidth]{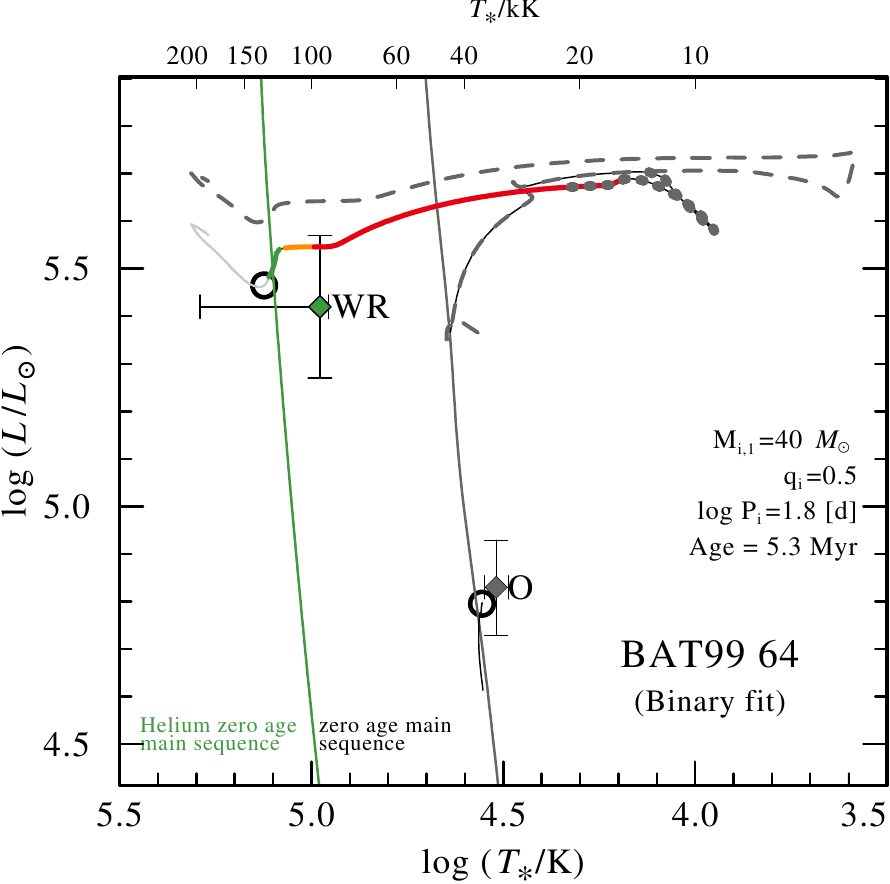}
%   \caption{s}
  \label{fig:BAT64hrdsub1}
\end{subfigure}%
\begin{subfigure}{.67\columnwidth}
  \centering
  \includegraphics[width=\linewidth]{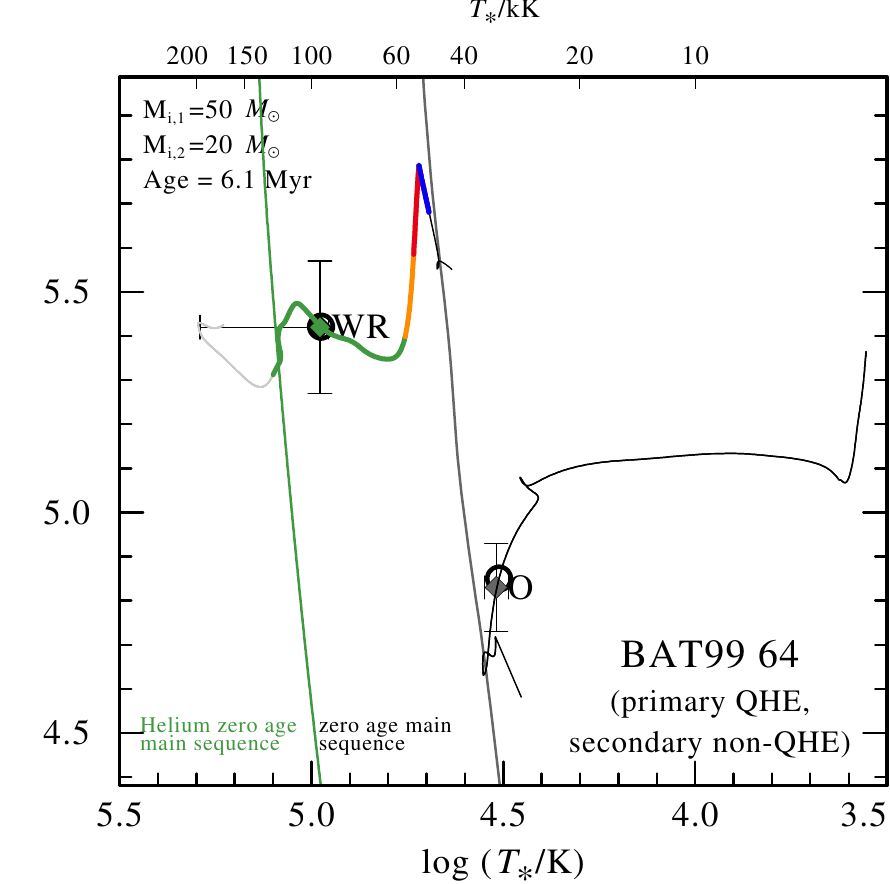}
%   \caption{A subfigure}
  \label{fig:BAT64hrdsub2}
\end{subfigure}
\begin{subfigure}{.67\columnwidth}
  \centering
  \includegraphics[width=\linewidth]{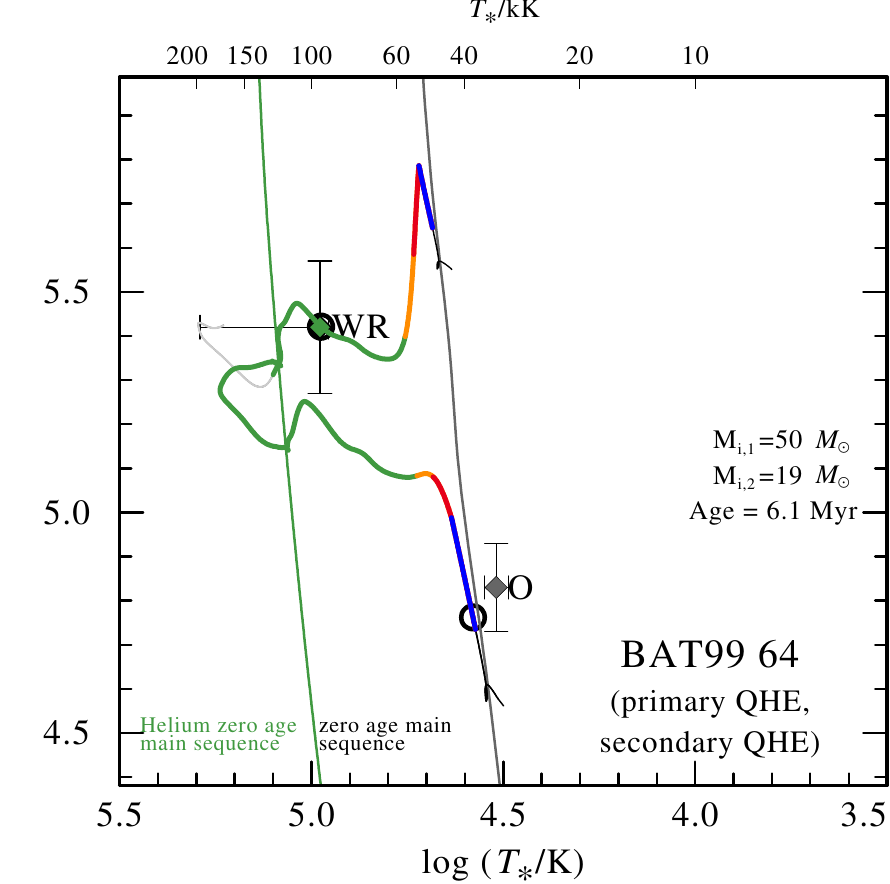}
%   \caption{A subfigure}
  \label{fig:BAT64hrdsub3}
\end{subfigure}
\caption{As Fig.\,\ref{fig:BAT49hrd}, but for BAT99\,64.}
\label{fig:BAT64hrd}
\end{figure*}

We cannot be confident of the evolutionary path of 
BAT99\,64 (Fig.\,\ref{fig:BAT64hrd}). Both the binary channel as well as the CHE channel for the primary yield statistically 
acceptable results (within 2$\sigma$). However, the primary's HRD position implies that, 
in the likely case that it did not undergo CHE, it entered the WR phase via binary interaction, 
and we therefore classify it as w/b-WR.

\emph{\bf \object{BAT99\,67}} was  classified as WN5o + OB by \citet{Smith1996} owing to 
absorption features superposed with the WR emission in Balmer lines. However, FMG03 reclassified 
the object to WN5ha, arguing that the absorption is strongly blue-shifted and thus forms intrinsically in the wind of the WR star.
Moreover, despite having reported an X-ray luminosity of $\log L_\text{X} = 33.3\,$[\lum], which potentially implies the presence of WWC, 
FMG03 did not detect significant periodic RV variability, concluding that BAT99\,67 is likely not a close binary. By inspecting the 
spectra at hand, we could find no clear spectroscopic traces for a secondary companion. 
We therefore omit BAT99\,67 from our analysis.

\emph{\bf \object{BAT99\,71}} was detected by FMG03 to be a binary on the basis of the system's large RV variations and was 
classified as WN4 + O8, later revised to WN3+abs by \citet{Neugent2018}. While a spectroscopic period of $P = 2.3\,$d was originally found, a very faint eclipse is visible 
in their photometric dataset when folding the data with a $P = 5.2\,$d period. 
This period was shown by FMG03 to be consistent with the RV variability as well, and 
therefore likely represents the true period of the system (see discussion in FMG03).

In our study, we derive a higher temperature for the secondary 
than expected for an O8 star, owing to the absence of several weak He\,{\sc i} absorption lines. A classification of the model 
spectrum implies an O6.5~V star.
The light ratio can only be roughly constrained based on the overall strength of the companion's features. The temperature derived 
here for the WR companion is significantly higher than derived by H14, which is likely a consequence of accounting for the binary nature 
of the system in this study. Likewise, with the O6.5~V component contributing $\approx80\%$ in the optical, the luminosity derived here for the 
WR star is $0.6\,$dex smaller than derived by H14. We confirm that the WR primary is hydrogen free. 

\begin{figure*}
\centering
\begin{subfigure}{0.67\columnwidth}
  \centering
  \includegraphics[width=\linewidth]{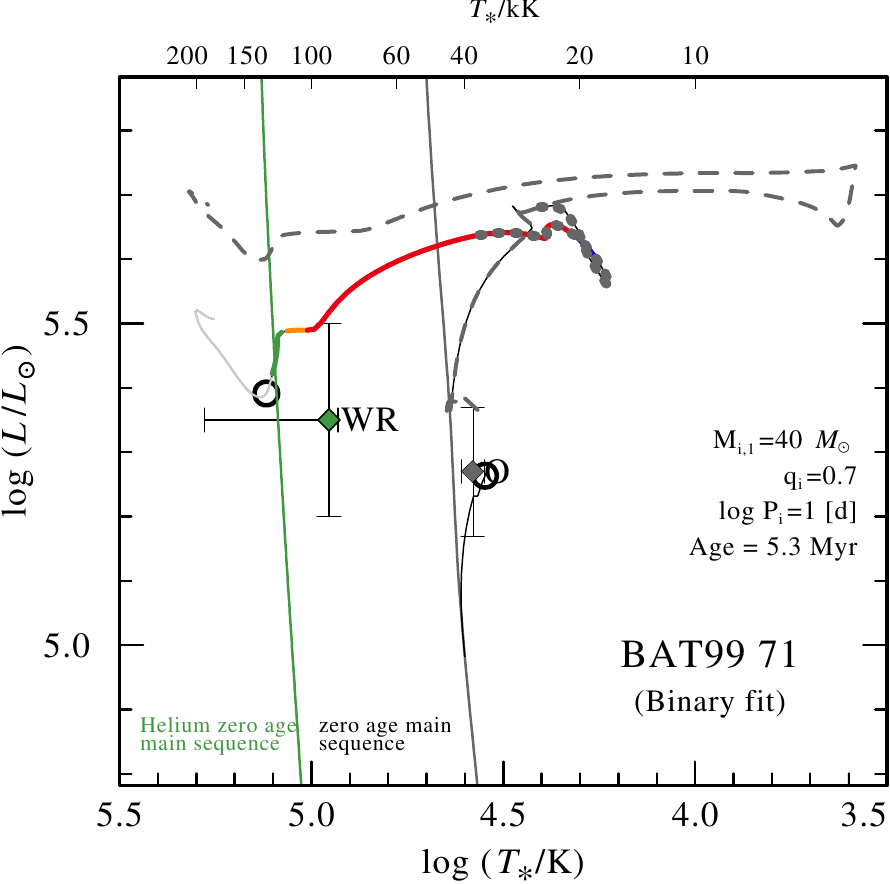}
%   \caption{s}
  \label{fig:BAT71hrdsub1}
\end{subfigure}%
\begin{subfigure}{.67\columnwidth}
  \centering
  \includegraphics[width=\linewidth]{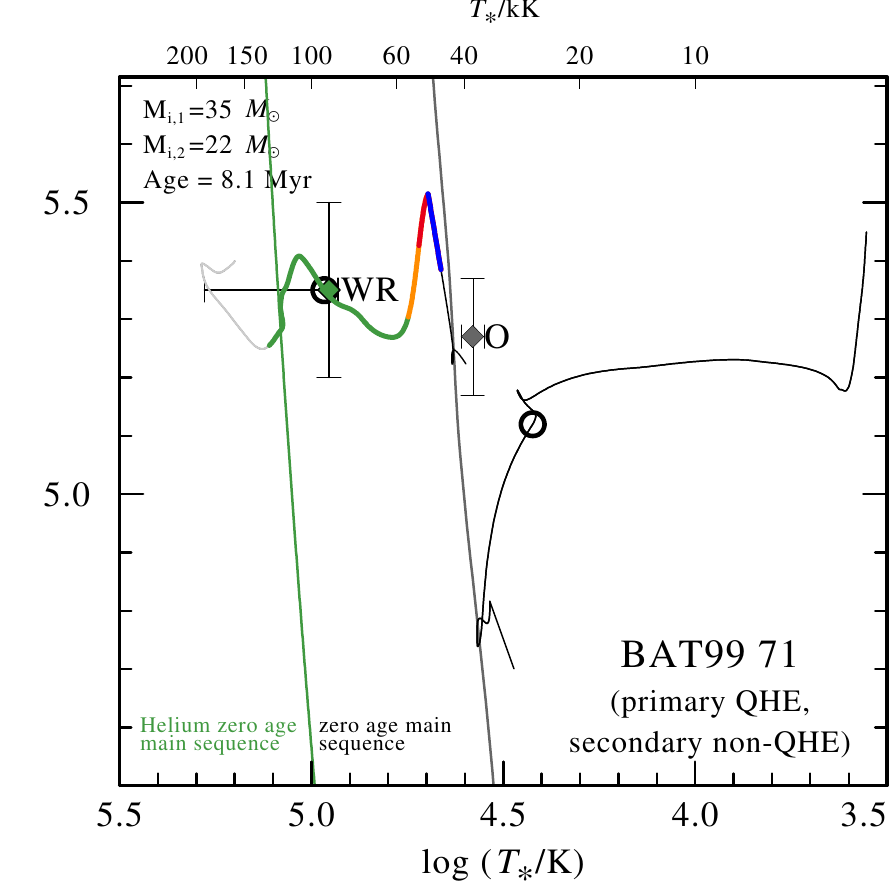}
%   \caption{A subfigure}
  \label{fig:BAT71hrdsub2}
\end{subfigure}
\begin{subfigure}{.67\columnwidth}
  \centering
  \includegraphics[width=\linewidth]{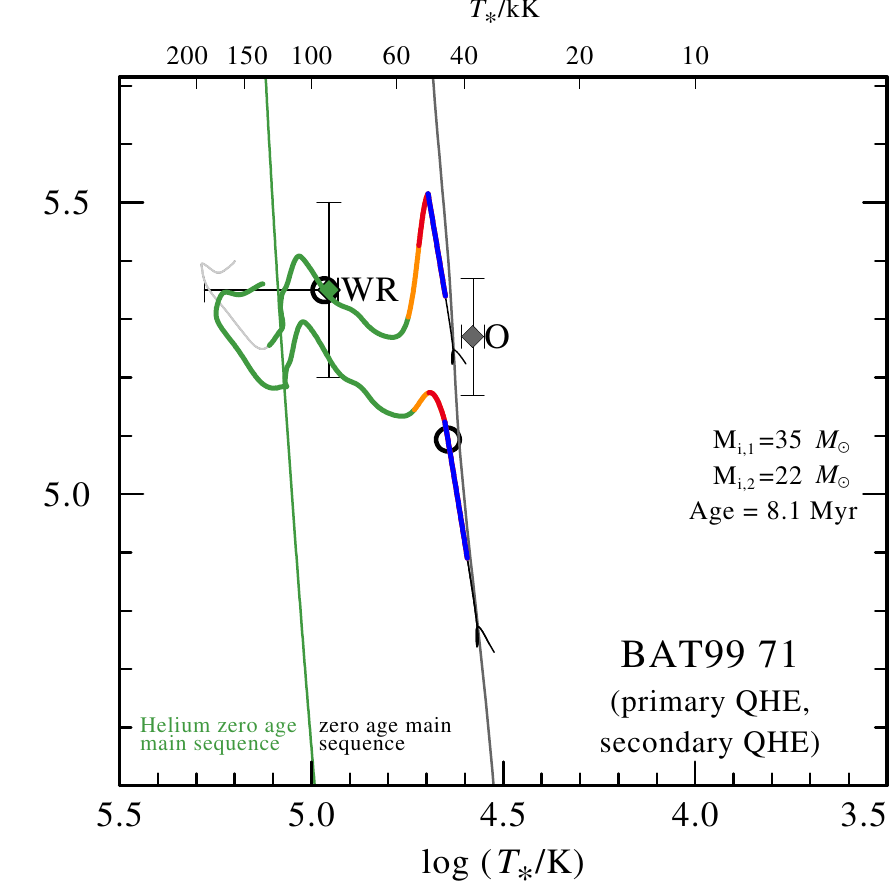}
%   \caption{A subfigure}
  \label{fig:BAT71hrdsub3}
\end{subfigure}
\caption{As Fig.\,\ref{fig:BAT49hrd}, but for BAT99\,71.}
\label{fig:BAT71hrd}
\end{figure*}

As for BAT99\,64, both the binary channel as well as the CHE channel for the primary 
yield statistically acceptable results (Fig.\,\ref{fig:BAT71hrd}). From our results, we conclude
that the primary is either a w-WR star or a wb-WR star.

\emph{\bf \object{BAT99\,72}} was suggested to be a medium-period ($P \approx 100-1000\,$d) binary by FMG03, who designated 
it as a WN4h+O3: binary on the basis of absorption features they detected in their dataset. The optical spectra 
at hand are indeed suggestive of absorption features, but it is hard to tell whether they belong to the WR star itself 
or to a companion. No He\,{\sc i} features can be seen, which immediately implies that the putative companion 
must be of an early-type. It is not obvious that the WR component suffers from line dilution: some putatively single 
WN4h stars (e.g., \object{BAT99\,25}) show comparable line strengths, while others (e.g., \object{BAT99\,40}) may 
suggest some dilution. 

With our data, we cannot discern between the quality of fit when including a model spectrum for an O3~V star.
However, when accounting for binarity, the resulting 
luminosities of both components end up being very low, ${\approx}5.1\,$dex for both. This is especially puzzling 
for an O3~V star, which is expected to be significantly more luminous. Moreover, this makes the WR component 
in BAT99\,72 the least-luminous WR star in our sample. While this could make it a promising candidate for a 
WR star stripped in a binary, we warn that these results should be taken with caution and verified in future work.

\emph{\bf \object{BAT99\,77}} was classified as WN7ha by S08, who detected a clear periodic RV variability with 
$P = 3.0\,$d. The system is found in a very crowded region containing several massive stars, causing a strong contamination 
of several IUE datasets, as well as photometric data. Unlike H14, in this study we chose archival IUE spectra and photometry that correspond to the 
lowest flux level measured. This yields much more realistic parameters for the WR star. 

\begin{figure*}
\centering
\begin{subfigure}{0.67\columnwidth}
  \centering
  \includegraphics[width=\linewidth]{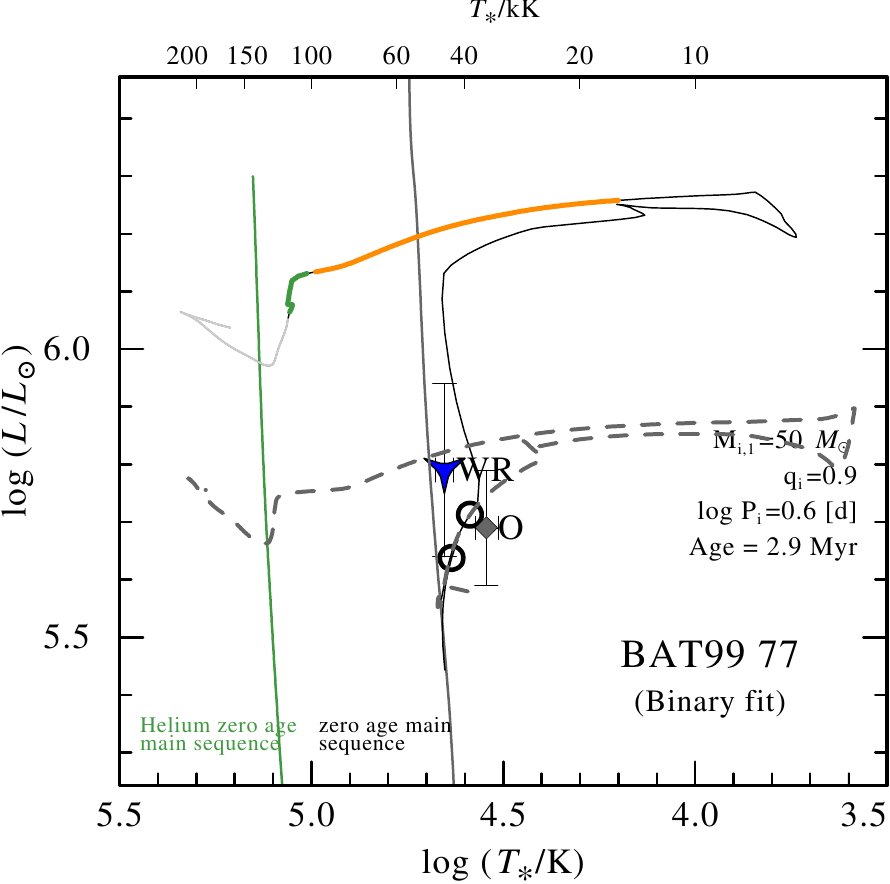}
%   \caption{s}
  \label{fig:BAT77hrdsub1}
\end{subfigure}%
\begin{subfigure}{.67\columnwidth}
  \centering
  \includegraphics[width=\linewidth]{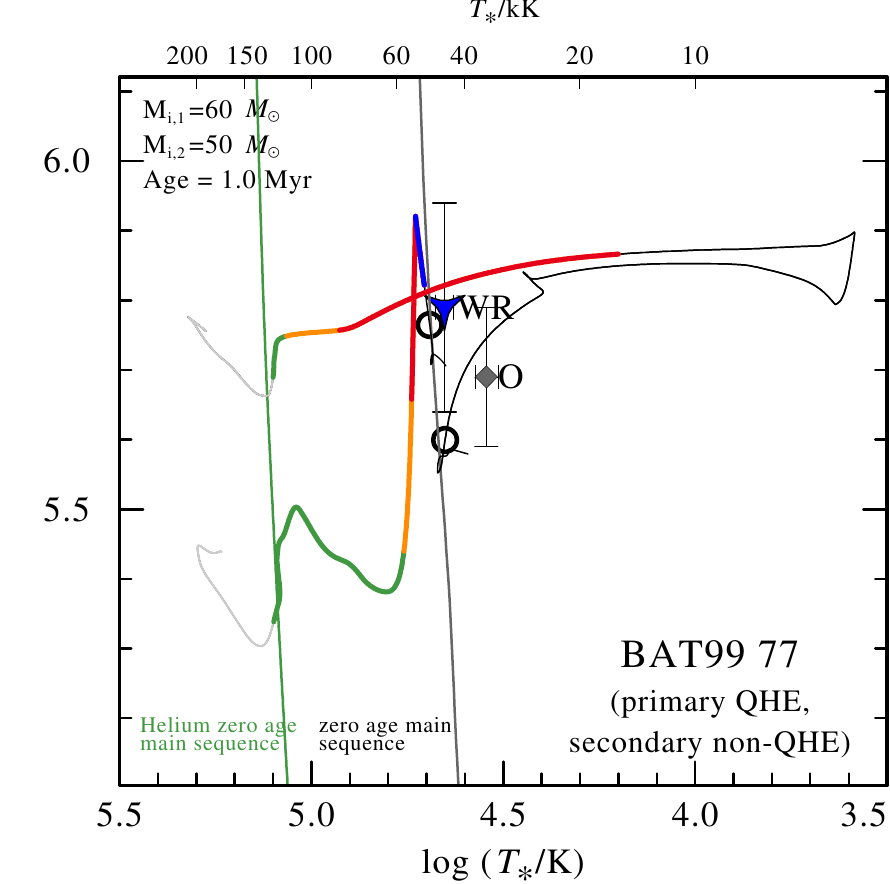}
%   \caption{A subfigure}
  \label{fig:BAT77hrdsub2}
\end{subfigure}
\begin{subfigure}{.67\columnwidth}
  \centering
  \includegraphics[width=\linewidth]{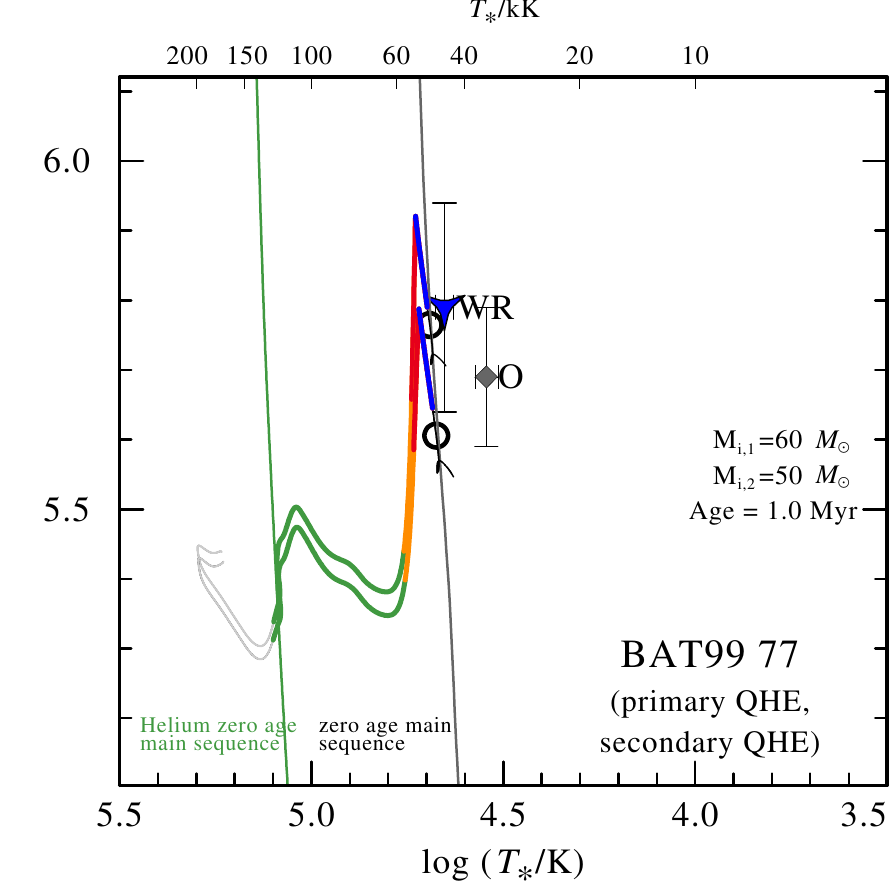}
%   \caption{A subfigure}
  \label{fig:BAT77hrdsub3}
\end{subfigure}
\caption{As Fig.\,\ref{fig:BAT49hrd}, but for BAT99\,77.}
\label{fig:BAT77hrd}
\end{figure*}

The WR star's temperature can be well constrained based on the strong N\,{\sc iii} 
emission lines and the weak N\,{\sc iv}\,$\lambda 4060$ emission.
Because of the WR component's relatively low temperature, absorption features belonging solely to the secondary star 
are very hard to identify unambiguously, leaving its parameters virtually unconstrained. Compared to other WN7h stars, the He\,{\sc ii} 
and Balmer emission lines are about five times weaker in the spectrum of BAT99\,77. Therefore, we assume the companion contributes 
$80\%$ to the optical spectrum. A satisfactory fit is obtained for a temperature of $T = 35\,$kK for the secondary, but this should be seen 
as a very rough estimate. The implied luminosity of the secondary is very high ($5.7\,$dex), suggesting that it is an O7~III star. 
The gravity was fixed based on calibrations by \citet{Martins2005}. Finally, the hydrogen content 
in the WR star is found to be very high, $X_\text{H} = 0.7$, 
as reported by H14. Most importantly, the WR star's luminosity is reduced in this study by more 
than 1\,dex compared to that reported by H14. 
However, more data will be necessary to unambiguously analyse this system.

None of the evolutionary scenarios we explore here provide a good fit to the system's parameters, 
as is evident from Fig.\,\ref{fig:BAT77hrd}. In fact, all binary tracks at the relevant parameter regime of 
short periods and large masses end up merging, hence the rise in luminosity for the best-fitting 
binary track in the leftmost panel of Fig.\,\ref{fig:BAT77hrd}. The O companion appears to be evolved, but none 
of the tracks manage to reproduce its evolved state simultaneously with the non-evolved appearance of the WR star. 
This may suggest that more components are present in the system. Given the high hydrogen content of the WR primary 
($X_\text{H} \approx 0.7$), we classify it as a ms-WR star, although the luminosity is somewhat lower than 
for other ms-WR stars in our sample ($\log L = 5.8\,[L_\odot]$)

\emph{\bf \object{BAT99\,78}} is situated in a very tight region of massive stars situated in the cluster \object{HDE 269828}, classified as 
WN6 by FMG03 and as WN4 by \citet{Neugent2018}.
The spectra obtained by FMG03 were, according to them, contaminated by several sources. However, 
\citet{Walborn1999} extracted an uncontaminated HST spectrum of the WR component. The spectrum does not reveal any immediate evidence 
for binarity. Furthermore, FMG03 were not able to detect a periodic RV variability in BAT99\,78. 
The N\,{\sc iv} features, especially N\,{\sc iv}\,$\lambda 4060$, are potentially suggestive of a double peak profile, which could 
imply that a second WR-like star is hidden in the spectrum. However, more data will be necessary to confirm/reject this hypothesis.
We therefore omit this object from our analysis.

\emph{\bf \object{BAT99\,79}} was classified as WN7h+OB in the original BAT99 catalogue owing to clear absorption 
features in the spectrum. However, no significant RV variability was reported by S08. \citet{Crowther1997} estimated 
that significant contribution stems from the secondary star, although this statement was not confirmed by H14. 

Our analysis, however, implies that the companion contributes $\approx 60\%$ to the spectrum in the optical, causing clear 
absorption features in the Balmer lines and several strong He\,{\sc i} lines. The primary WR star is in itself a rather 
cool star, exhibiting He\,{\sc i} lines with blue-shifted absorption. The presence of N\,{\sc iv} and N\,{\sc iii} emission 
lines enabled us to assess its temperature. The light ratio is constrained from the overall strength of the O-type features.
The temperature of the secondary can be constrained from the presence of strong He\,{\sc i} absorption features and very 
weak He\,{\sc ii} absorption, however, its $\log g$ can only be roughly constrained. Based on the strength 
of absorption in the Balmer lines, low values $\log g \lesssim 3.4\,$[cgs] provide a better fit to the data. 
Based on a classification of the secondary's synthetic spectrum, we tentatively classify it as O9~I.
The HRD position of the WR primary, along with its very large hydrogen content (70\%), 
is consistent with it being a ms-WR star.

\emph{\bf \object{BAT99\,80}} was classified as O4~If/WN6 in 
the BAT99 catalog, suggesting that the clear He\,{\sc i} 
absorption lines in its spectrum are intrinsic to the WR star. In contrast, 
S08, who reclassified the target to WN5h:a, suggested that the spectrum of the WR star is diluted by a secondary star, 
causing the apparent absorption. However, S08 were not able to detect 
significant RV variability. On the other hand, a substantial X-ray luminosity 
of $\log L_\text{X} \approx 33.9$\,[\lum] was reported by \citet{Guerrero2008}, potentially related 
to the presence of WWC in the system. 

Inspection of the optical spectrum leads us to believe that the object is indeed a binary. The WR star 
is hot enough to show a strong N\,{\sc iv}\,$\lambda 4060$ emission and a faint N\,{\sc v}\,$\lambda \lambda 4604,
4620$ emission. At this temperature ($\approx 50\,$kK), it does not seem possible to reproduce the multitude 
of He\,{\sc i} absorptions. The companion is therefore cool. 
The N\,{\sc iii} feature may be partially attributed 
to the O companion. Although the solution found here provides a better fit to the spectrum, we cannot fully exclude 
the fact that BAT99\,80 may be a single star. Our solution should be regarded as an alternative solution to the one 
presented by H14, who do not assume a secondary star, and therefore derive a significantly higher luminosity. A classification 
of the secondary's synthetic spectrum implies  an O9.5~V spectral class. The primary's HRD position, compared to the 
BPASS tracks, lead us to classify it as w/wb-WR.

\emph{\bf \object{BAT99\,82}} was classified as WN3b in the BAT99 catalog. This spectral type was later 
confirmed by FMG03, who did not detect a periodic RV variation for this object. A rather high 
X-ray luminosity of $\log L_\text{X} \approx 33.2$\,[\lum] was reported by \citet{Guerrero2008}, which motivated 
H14 to mark this object as a binary candidate. However, no feature in the spectrum implies the presence of an 
additional companion. We therefore omit this object from our analysis.

\emph{\bf \object{BAT99\,92}} was originally classified as WN3 + B1 I in the original \citet{Breysacher1981} catalog, later revised 
to WN6 + B1 I in the BAT99 catalog. S08 reclassified the primary yet again to WN3. The latter authors 
reported a period of $P = 4.3\,$d for the system, consistent with the period given in the BAT99 catalog. 
Since the absorption features of the B1 I component seem to maintain a constant RV, it was suggested 
that the system contains an additional companion orbiting the WR star at a short orbit, motivating 
S08 to classify the system as WN3:b (+ O) + B1 Ia. The very high brightness of the system is consistent 
with this conjecture.
Moreover, as noted in the BAT99 catalog, 
the object portrays a very strong emission of the C\,{\sc iv} resonance doublet in the UV, which 
suggests that it might be a rare transition-type WN/C star \citep{Conti1989}. Alternatively, it could imply that 
a WC star is also present in the system. Lastly, the optical recombination emission lines are extremely broad 
and round, a fact which motivated H14 to assume a significant rotation of the WR star.

\begin{figure}[h]
\centering
  \includegraphics[width=0.5\textwidth]{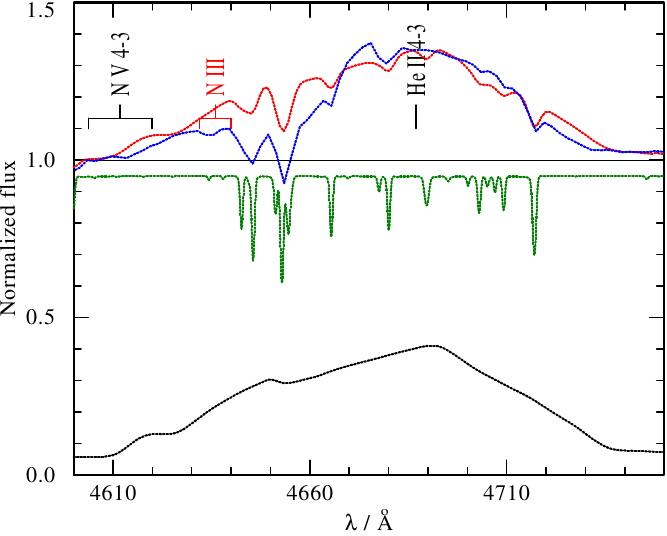}
  \caption{A comparison between the observed CTIO spectrum of BAT99\,92 (blue) and the 
  synthetic composite spectrum (red), comprising the WC model (black) and B-type model (green). 
  The wide WR features and the absorption originating in the B-type companion 
  give rise to a feature that resembles the  N\,{\sc iii} doublet, which is likely erroneous.}
\label{fig:BAT92_WC}
\end{figure}

We strongly believe that BAT99\,92 is, in fact, a WC binary. The reason for the previous assignment of the 
spectral class WN is the presence of a feature that resembles the N\,{\sc iii}\,$\lambda \lambda 4634, 4641$ emission 
line, which is characteristic for cooler WN stars. However, as we illustrate in Fig.\,\ref{fig:BAT92_WC}, this 
feature belongs to the He\,{\sc ii}\,$\lambda 4686$ complex, and only appears to be a separate line due to the absorption 
of the B-type companion. This, together with the strong C\,{\sc iv}\,$\lambda 5812$ optical line 
and C\,{\sc iv} resonance line in the UV, gives strong support that the primary is a WC star, which we classify 
as WC4.

The limited resolution of the data only allow for an approximate solution to the system.
The primary's temperature is estimated primarily from the lack of C\,{\sc iii} spectral features, setting a low bound 
of roughly 90\,kK at this parameter regime. The temperature of the second component is estimated to $23\,$kK from the plethora of 
metal lines that are observed, albeit in low resolution. 
The light-ratio of the components is estimated from the overall strength of the secondary's 
spectral features. The strong H$\alpha$ emission, which is seen in the spectrum taken by 
\citep{Torres1988} implies that the B1~Ia secondary has a very high mass-loss rate of $\log \dot{M} = -5.1$\,\smy. 
The luminosities of both components are found to be very large - $\log L \approx 6.0\,[L_\odot]$. This could indicate 
that further components are present in the spectra. However, higher-resolution data will be necessary to establish this.

Since BAT99\,92 appears not to host a WN component, we do not investigate its evolutionary path here, but will include 
it in future studies of the WC content of the LMC.

\emph{\bf \object{BAT99\,93}} was classified as O3 If/WN6 in the BAT99 catalog, later revised 
to O3 If* by \citet{Evans2011}. Although S08 could not detect a periodic RV variability, 
it was considered a binary candidate by H14 due to the X-ray emission detect from the object 
by \citet{Guerrero2008}. The optical spectrum at hand, however, shows no features that can be 
attributed to a secondary star. We therefore omit this object from our analysis.

\emph{\bf \object{BAT99\,95}} (\object{VFTS 402}) was first classified as a  spectroscopic binary by S08, who inferred a 
period of $P = 2.1\,$d for the system, and classified it as WN7h. The companion in the system was believed to be an 
OB-type star \citep{Evans2011, Bestenlehner2014}, but no concrete evidence for this was found. In fact, based on archival FLAMES spectra, 
we find evidence that 
this system is composed of \emph{two} WR stars: a cooler, more massive WN star, which we will refer to as the secondary, and a
hotter, less massive WN star, which we will refer to as the primary. 
We choose this nomenclature because evolutionary-wise, the hotter 
star was likely the more massive component originally. Another reason for this choice is that the 
RV curve derived by S08 corresponds to the motion of the hotter component. 

\begin{figure}[h]
\centering
  \includegraphics[width=0.5\textwidth]{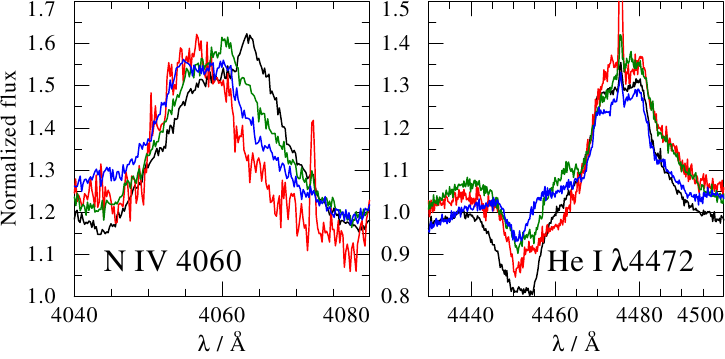}
  \caption{Four FLAMES observations of BAT99\,95 at phases $\phi = 0.18, 0.57, 0.65, 0.88$. The motion of the primary, hotter WR star 
  can be seen in the N\,{\sc iv}\,$\lambda 4060$ line, following the motion in other He\,{\sc ii} lines. He\,{\sc i} lines such as 
  He\,{\sc i}\,$\lambda 4471$ do not follow this motion, and even depict a slight anti-phase behavior (see trough of black and red emission wing 
  in the right panel).}
\label{fig:BAT95_FLAMES}
\end{figure} 

To understand why the companion is most likely a WR star, we refer the reader to Fig.\,\ref{fig:BAT95_FLAMES}, which shows
the N\,{\sc iv}\,$\lambda 4060$ line together with the 
He\,{\sc i}\,$\lambda 4471$ line in four archival FLAMES observations. 
The N\,{\sc iv} line traces the same velocities as the He\,{\sc ii} lines, and therefore traces the 
RV curve derived by S08. The profile variations of the N\,{\sc iv}\,$\lambda 4060$ 
line are suggestive of two emission components that move in anti-phase.
The He\,{\sc i} lines 
show a less-pronounced anti-phased behavior, and generally do not follow the same behavior of the He\,{\sc ii} lines. 
The origin of the He\,{\sc i} is, we believe, primarily the secondary star. A non-negligible contribution from a WWC region
is likely in this line.
It may also responsible for the strong N\,{\sc iii} emission seen 
in the optical spectrum, though without phase coverage beyond $5500\,\AA$, we can only speculate that this is the case.
Another argument against an OB-type companion is that, if assumed to be present, a typical OB-type companion would strongly 
dilute the emission lines 
of the WR star. To reproduce the emission lines in their observed strengths, one would need to assume extremely large mass-loss 
rates in the excess of 
$\log \dot{M} = -4$\,\smy, which is hard to motivate physically. Lastly, the combination of faint N\,{\sc v}, strong N\,{\sc iv}, and 
very strong N\,{\sc iii} in the optical spectrum is very hard to reproduce using just one WR model.

We derived the RVs of both companions based on a 2D cross-correlation with the N\,{\sc iv} line using a PoWR model 
as a template for both stars. However, we caution that the fit quality was not satisfactory for all phases and significant 
systematic errors may be present.
The motion of the cooler secondary should be best monitored with the  N\,{\sc iii}\,$\lambda \lambda 4634, 4641$ doublet.
Unfortunately, only one phase of FLAMES observations is available that covers this doublet, so a careful derivation of the RV curves of both components 
is currently not possible. We encourage deeper observations of BAT99\,95 in future studies.

The stellar parameters derived from our analysis are generally very uncertain, and the fit quality is poor. The temperatures could be fairly well constrained based on the 
nitrogen and helium balance, but the light ratios, and therefore the mass-loss rates, are strongly degenerate. 
A classification of the model spectra suggests the spectral classes WN5(h) and WN7(h) for the primary and secondary, respectively 
\citep{Smith1996}.
The light ratio was primarily based on the 
observed strength of the N\,{\sc iv} line compared to its strength in single WR stars of a similar spectral type (WN5h). 
However, we caution that this may be wrong 
due to a potential contribution of the primary to this line.  

It is possible that our results are affected by the presence of a strong WWC signature, especially at lines belonging to 
low ionization stages (e.g., He\,{\sc i}). More data will be necessary to obtain an adequate disentanglement of this system.

\begin{figure*}
\centering
\begin{subfigure}{0.67\columnwidth}
  \centering
  \includegraphics[width=\linewidth]{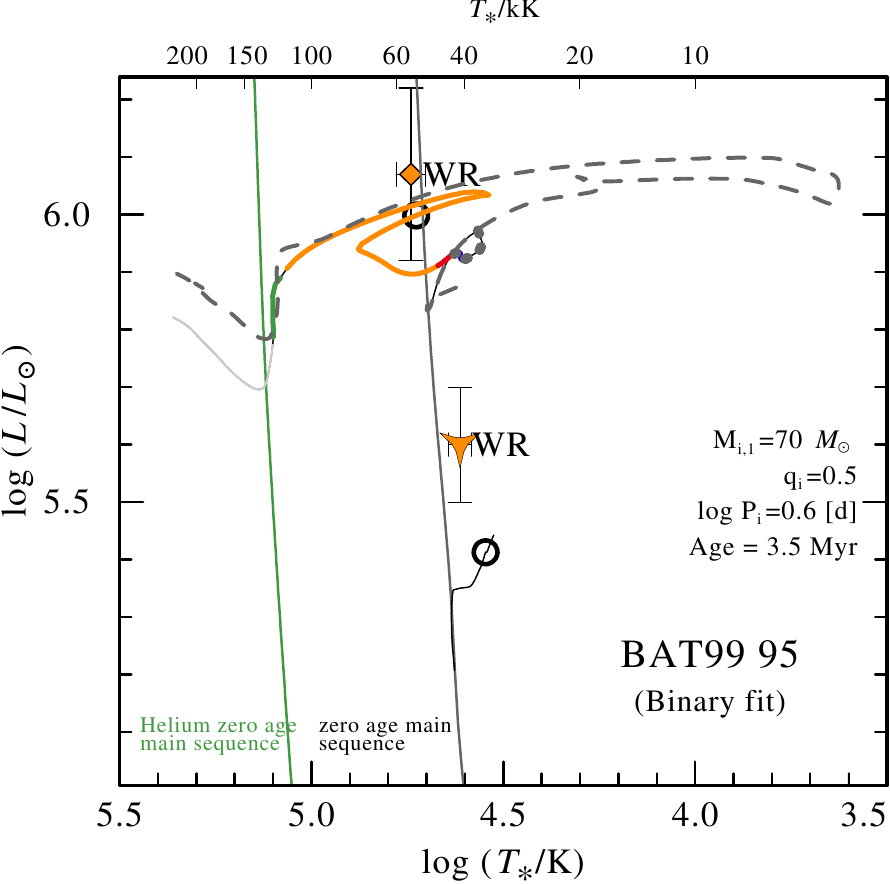}
%   \caption{s}
  \label{fig:BAT95hrdsub1}
\end{subfigure}%
\begin{subfigure}{.67\columnwidth}
  \centering
  \includegraphics[width=\linewidth]{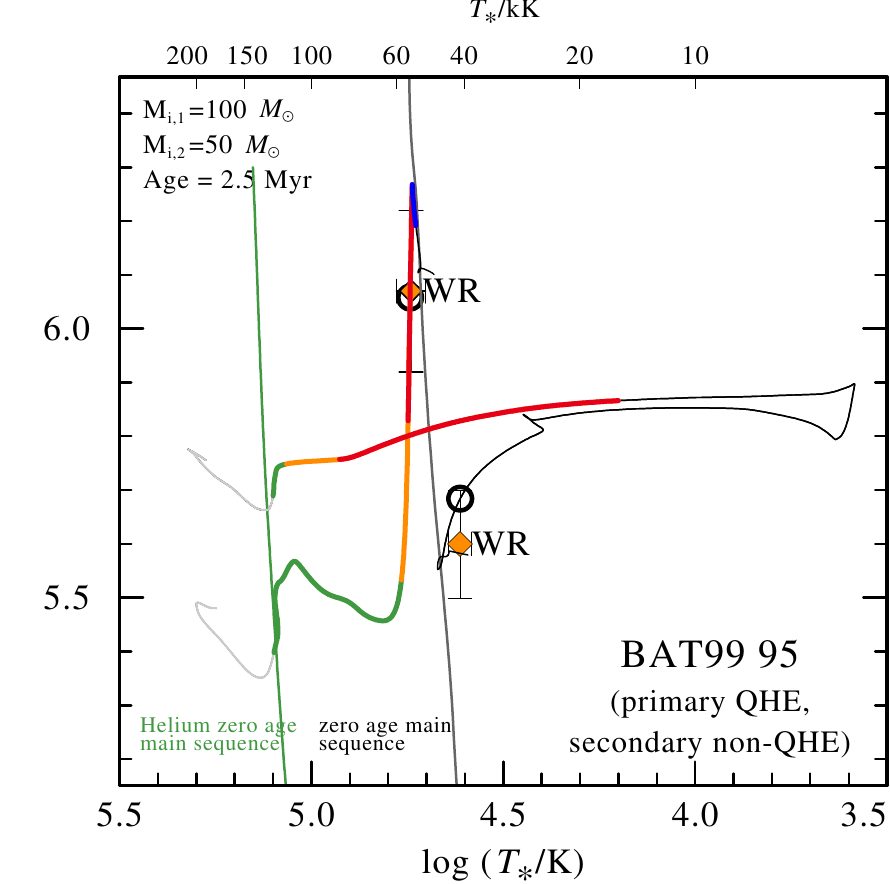}
%   \caption{A subfigure}
  \label{fig:BAT95hrdsub2}
\end{subfigure}
\begin{subfigure}{.67\columnwidth}
  \centering
  \includegraphics[width=\linewidth]{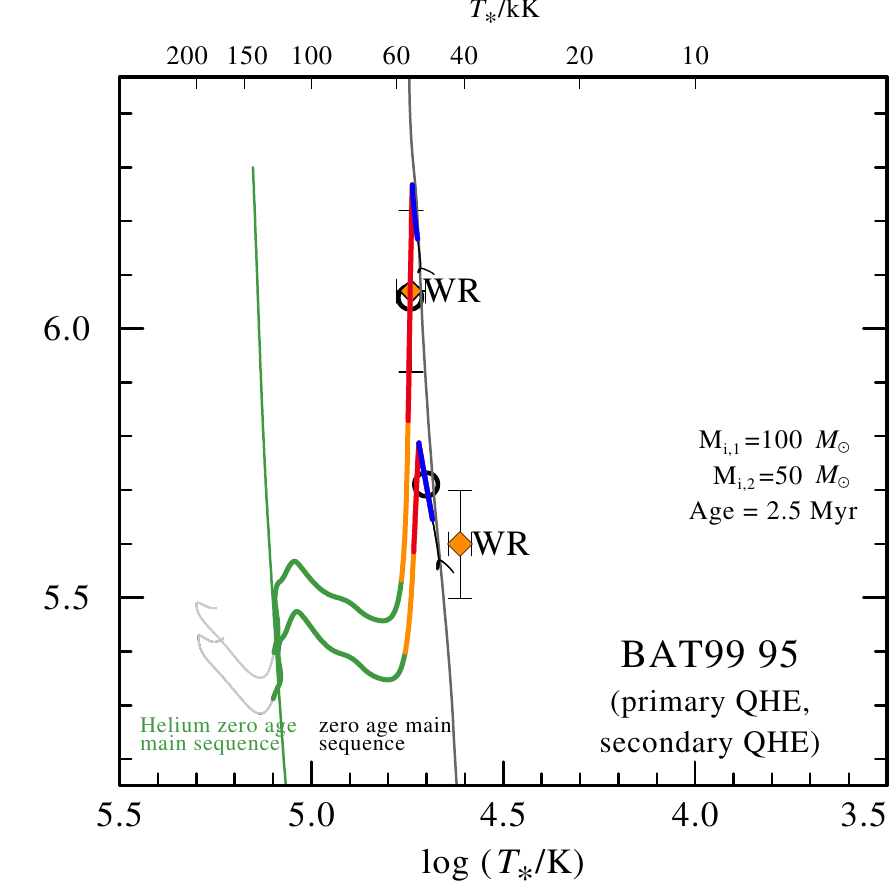}
%   \caption{A subfigure}
  \label{fig:BAT95hrdsub3}
\end{subfigure}
\caption{As Fig.\,\ref{fig:BAT49hrd}, but for BAT99\,95.}
\label{fig:BAT95hrd}
\end{figure*}

Given the large hydrogen mass fractions of the components, we classify them as 
ms-WR stars. However, it should be noted that our analysis suffers from large uncertainties 
due to the uncertain nature of the components and the lack of spectral coverage.

\emph{\bf \object{BAT99\,99}} This star was classified as O2.5~If*/WN6 in the original BAT99 
catalog. A relatively high X-ray luminosity of $L_\text{X} \approx 10^{34}$\,\lum~was derived for this object by \citet{Guerrero2008}, potentially 
implying the presence of either WWCs or an accreting object. S08 derived a rather long period of 93\,d for the system. From the available spectra, 
we cannot find any evidence for a companion, either in the optical or in the UV. A low-mass star is virtually excluded from the mass function 
(see orbital parameters by S08). Considering the X-ray luminosity, 
the most likely companion is either a very hot OB-type star with a similar absorption spectrum as the primary (e.g., O3~V star), or 
an accreting BH. However, more data are necessary to test this 
hypothesis.

\emph{\bf \object{BAT99\,100}} (\object{VFTS 1001}) is situated in a very crowded region in the Tarantula nebula and was classified 
as WN6h by S08. The object was detected in X-rays, although the source is rather faint \citep{Guerrero2008}. Furthermore, no RV 
variations were found by S08. In the available spectra, we cannot detect features that can be attributed to a secondary star. We therefore 
omit this star from our sample.

\emph{\bf \object{BAT99\,102}}  (\object{VFTS 507}) is another star found in a crowded region in the Tarantula nebula, situated 
about 1'' away from the stars BAT99\,101 and BAT99\,103. The target was classified as WN6 by S08. The latter authors cannot recover 
the 2.76\,d binary period reported for the system by \citet{Moffat1989}. In fact, they recover this period for BAT99\,103, implying that the two sources 
were confused by \citet{Moffat1989}. 

We retrieved archival FLAMES spectra for the object. The spectra cover a database of about a year, but no RV variability can be detected. We therefore
conclude that BAT99\,102 is likely a single star. 

\emph{\bf \object{BAT99\,103}} (\object{VFTS 509}, \object{RMC 140b}) is located in the immediate vicinity of BAT99\,101 and 102. It was classified as 
WN6 by S08 and later as WN5(h)+O by \citet{Evans2011}. S08 inferred a 2.76\,d period for this object from RV variations (see also notes for BAT99\,102), and 
a mild detection of X-rays was reported by \citet{Guerrero2008b}. \citet{Bestenlehner2014} performed a spectral analysis of the WR component, but did not account 
for the secondary. 

Several phase-dependent FLAMES spectra were available for our study and 
enabled the disentanglement and orbital analysis 
of the system (cf.\ Figs.\,\ref{fig:dis103} and \ref{fig:orbit_BAT103}). Unfortunately, the spectra do 
not provide a good coverage of the orbit in RV space. The disentangled spectra 
are plausible, but some artifacts appear to contaminate the disentangled spectrum of the secondary. These are most likely 
due to nebular line contamination, as well as WWC signatures. Nevertheless, the disentangled spectrum enables 
the classification of the secondary and implies the spectral type O3.5\,V. This agrees with the classification 
of our model spectrum.

\begin{figure*}
\centering
\begin{subfigure}{0.67\columnwidth}
  \centering
  \includegraphics[width=\linewidth]{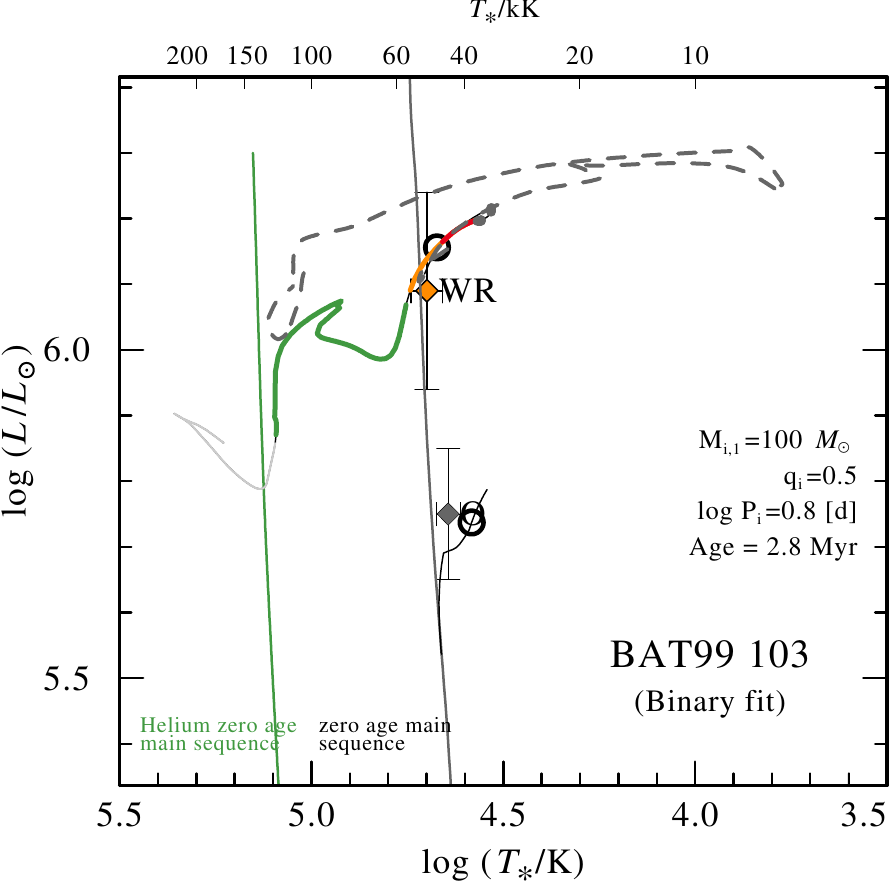}
%   \caption{s}
  \label{fig:BAT103hrdsub1}
\end{subfigure}%
\begin{subfigure}{.67\columnwidth}
  \centering
  \includegraphics[width=\linewidth]{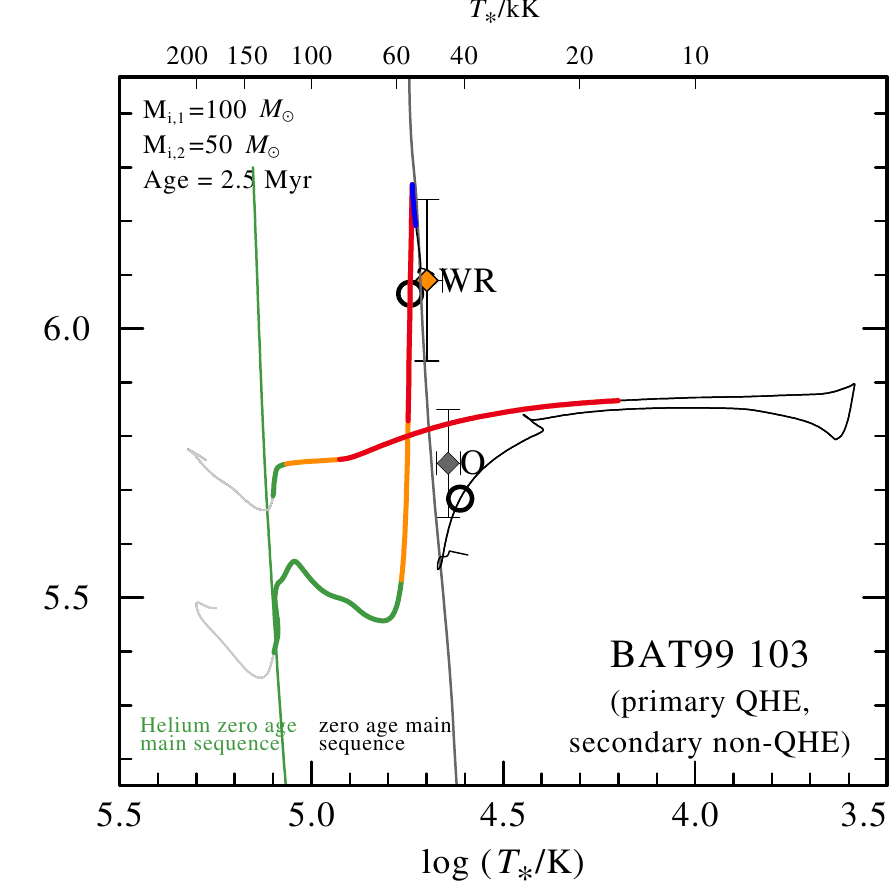}
%   \caption{A subfigure}
  \label{fig:BAT103hrdsub2}
\end{subfigure}
\begin{subfigure}{.67\columnwidth}
  \centering
  \includegraphics[width=\linewidth]{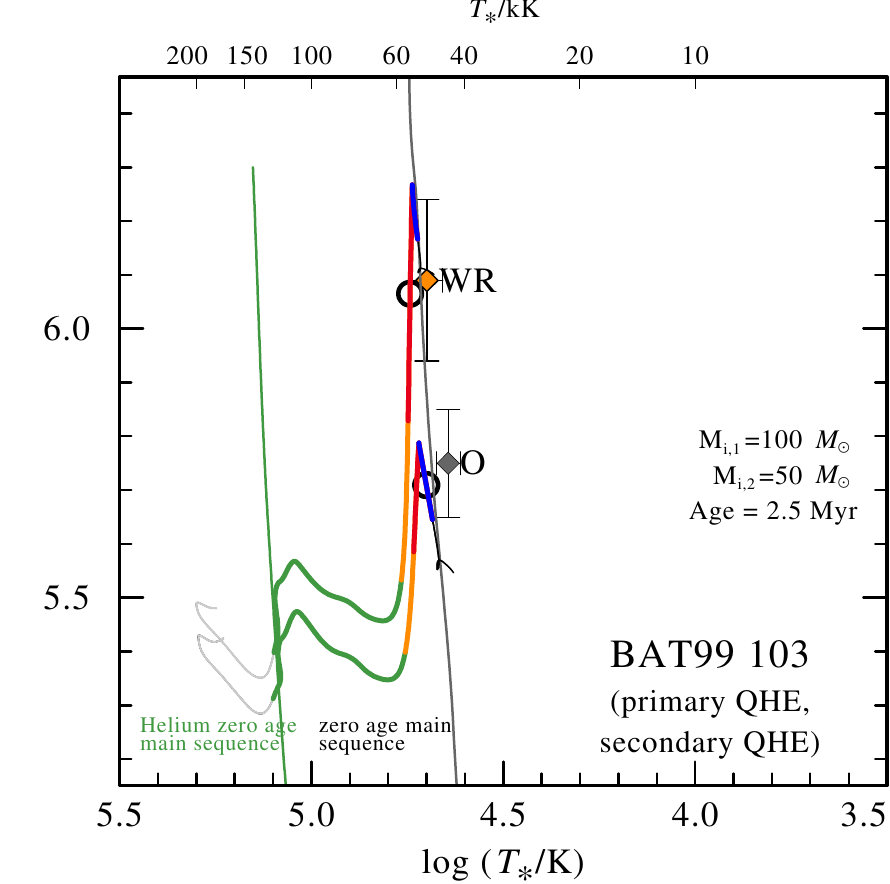}
%   \caption{A subfigure}
  \label{fig:BAT103hrdsub3}
\end{subfigure}
\caption{As Fig.\,\ref{fig:BAT49hrd}, but for BAT99\,103.}
\label{fig:BAT103hrd}
\end{figure*}

The effective temperature of the WR component could be well constrained from 
the presence of all three nitrogen ionization stages N\,{\sc iii-v}. The light ratio 
was determined from the overall strength of the O features. 
Since all the features of the O companion are entangled with the primary, we caution that the value 
of the light ratio is subject to a large error, as is the mass-loss rate of the WR component. 
The wind parameters of the primary follow from the strength and shape of the emission lines.

We cannot find a binary-evolution track that reproduces the system's parameters (Fig.\,\ref{fig:BAT103hrd}).
The best-fitting binary track reaches a current period of $P \approx 10\,$d, which is four times longer than observed. 
Tracks with shorter initial periods end up merging and fail to reproduce the system's observables. 
Based on BPASS, only a channel in which the primary avoided interaction provides 
a consistent fit to the observations.
The estimated equatorial 
rotation velocity of the secondary ($v\sin i = 200\,$\kms\,and $v_\text{eq} \approx $450\,\kms) is consistent with a past mass-accretion phase, 
but at the short 2.7\,d period may also suggest a homogeneous evolution history.
Given the uncertainties involved in the evolution of such short-period binaries, we classify the primary as w-WR.

\emph{\bf \object{BAT99\,105}} was classified as a transition type O3~If*/WN6 in the original BAT99 catalog, later revised to WN7 by 
S08,  and finally to O2~If* by \citet{Neugent2018}. S08 could not infer a periodic RV variability despite the mild RV scatter ($\sigma_\text{RV} \approx 35\,$\kms). H14 considered 
this object a binary candidate based on its rather high X-ray luminosity for the object of $\log L_\text{X} = 33.4\,$[\lum] \citep{Guerrero2008}.
After a careful inspection of the available spectra (UVES and HST), we could not detect any clear signs for a companion in the spectrum. 
More data will be needed to reject the presence of a massive companion in the spectrum, however. We therefore omit this object from our sample.

\emph{\bf \object{BAT99\,107}} (\object{VFTS 527}, \object{RMC 139})  was classified by S08 as WNL/Of, i.e., a transition-type star. \citet{Moffat1989} 
reported a 52.7\,d period for this object, but this could not be confirmed by S08. Recently, \citet{Taylor2011} analysed high quality FLAMES spectra and derived 
a spectral and orbital solution for the system. They conclude that the system comprises two massive O-type stars (O6.5~Iafc + O6~Iaf) with a period 
of $P = 153.9\,$d. The long period explains why S08 could not infer it, with their survey being sensitive to periods up to 100\,d.

\begin{figure}[h]
\centering
  \includegraphics[width=0.5\textwidth]{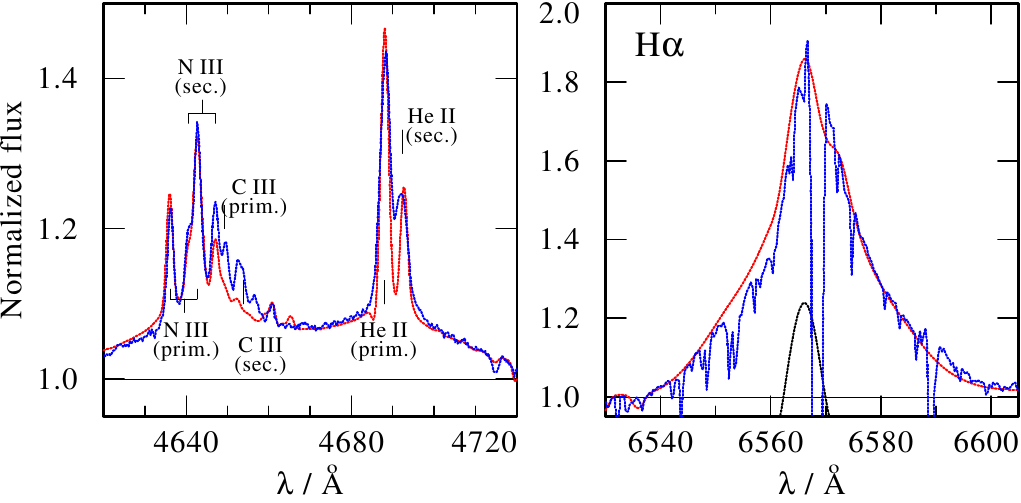}
  \caption{A zoom-in of the main emission features observed in the optical spectra of BAT99~107, showing an averaged X-SHOOTER observation (blue line) at phase 
  $\phi = 0.05$ (maximum Doppler separation), compared to our composite binary PoWR fit (red line). The narrow emission lines in the left panel probe the inner 
  wind velocity field, while H$\alpha$ shown in the right panel probes the outer velocity field.}
\label{fig:BAT107_FLAMES}
\end{figure} 

The analysis of this system is fairly straightforward, since the motion of the components is easily 
seen in the spectra. The parameters derived here agree well with those reported by \citet{Taylor2011}. The latter 
authors do not state their derived mass-loss rates. The N\,{\sc iii} and He\,{\sc ii} emission lines are narrow and originate in both stars - this can be easily 
seen from their Doppler motion (see Fig.\,\ref{fig:BAT107_FLAMES}, left panel). To reproduce this narrow emission profile, 
$\beta$ values of 2 or higher were necessary  for the power index of the velocity law. Here, we adopt $\beta = 2$ for both stars. These lines 
form very close to the stellar photosphere and therefore contain little information regarding the actual wind velocity. Fortunately, the H\,$\alpha$ line is much broader 
and probes the outer velocity field (Fig.\,\ref{fig:BAT107_FLAMES}, right panel). Without UV observations, however, the terminal velocity is subject to a large error.
The mass-loss rates derived here for both stars are of the order of $\log \dot{M} \approx -5\,$\smy, with the primary's mass-loss slightly larger. Importantly, no hydrogen 
depletion (or helium enrichment) can be deduced from the observations, but clear signs for a strong 
nitrogen enrichment are present (by a factor of about 50), as well as strong oxygen depletion.
Since this objects was originally considered to be a WR star, it was kept in our sample. 
If the components were a WR star, it would clearly belong to the ms-WR class - and we therefore 
classify them accordingly.

\emph{\bf \object{BAT99\,111}} (\object{RMC 136b}) was classified as WN9ha in the original BAT99 catalog, later 
updated to O4~If+ by \citet{Massey1998} and finally to O4If/WN8 by \citet{Crowther2016}. H14 considered this system a binary 
candidate based on supposed X-ray detection by \citet{Townsley2006}. However, the source identified as BAT99\,111 by \citet{Townsley2006} is 
separated by 0.3-0.5'' from BAT99\,111, depending on whether the coordinates from the BAT99 catalogue or HST images are assumed. In the dense region of R\,136, this separation 
is highly significant. \citet{Guerrero2008} confirm that BAT99\,111 is not associated with an X-ray point source. Interestingly, a comparison between two UV HST 
spectra taken on the 01-02-1996 (PI:Heap) and the 07-4-2012 (PI:Crowther) are suggestive of a RV shift of  $\approx 40\,$\kms, which could imply Doppler motion. More observations
will be needed to confirm this, however. For now, we omit this star from our sample.

\emph{\bf \object{BAT99\,112}} (\object{RMC 136c}) is another object situated close to the core of the \object{R136} cluster, classified as WN5h by 
\citet{Crowther1998}  and recently to WN4.5h by \citet{Neugent2018}. \citet{Schnurr2009} reported a marginal detection of periodic RV variations with $P = 8.2\,$d with an amplitude of $\approx 40\,$\kms. Moreover, 
the object portrays a very high X-ray luminosity of $\log L_\text{X} \approx 34.8\,$[\lum] \citep{Townsley2006, Guerrero2008}, suggesting the possible 
presence of WWC \citep{Schnurr2009}. Despite these indications toward binarity, we cannot confirm the presence of a secondary star from the single HST spectrum at hand. More optical 
observations are needed to uncover possible features of the secondary. We therefore omit this star from our sample.

\emph{\bf \object{BAT99\,113}} (\object{VFTS 542}, \object{MK 30}) is a transition type star classified by \citet{Crowther2011} 
as O2~If*/WN5. S08 inferred
periodic RV variations with $P = 4.7\,$d. H14 and later \citet{Bestenlehner2014} 
provided a single-star analysis of the object, and suggested that the secondary's 
contribution to the spectrum is likely low. 

We carefully inspected the FLAMES spectra at hand, 
mostly covering the binary's conjunction phases, i.e., highest Doppler shifts. 
We identified very weak spectral signatures moving in anti-phase to the WR primary in the lines 
He\,{\sc i}\,$\lambda 4388$ and $\lambda 4471$. Using these lines, we were able to measure the RVs of the secondary and disentangle 
the spectrum. The orbital solution is shown in Fig.\,\ref{fig:orbit_BAT113}, and 
the disentangled spectra are shown in Fig.\,\ref{fig:dis113}. The secondary's spectrum shows clear signatures 
of He\,{\sc i} lines, and potentially very weak He\,{\sc ii} lines. We tentatively classify it as B0~V, but better data will be necessary 
to validate this.
From calibration of the disentangled spectrum with the intrinsic strength of He\,{\sc i} lines for B0~V stars, we can derive a light 
ratio of $f_{V,2}/f_{\rm tot} = 0.1$. The parameters found for the secondary agree with its tentative spectral type.

\begin{figure*}
\centering
\begin{subfigure}{0.67\columnwidth}
  \centering
  \includegraphics[width=\linewidth]{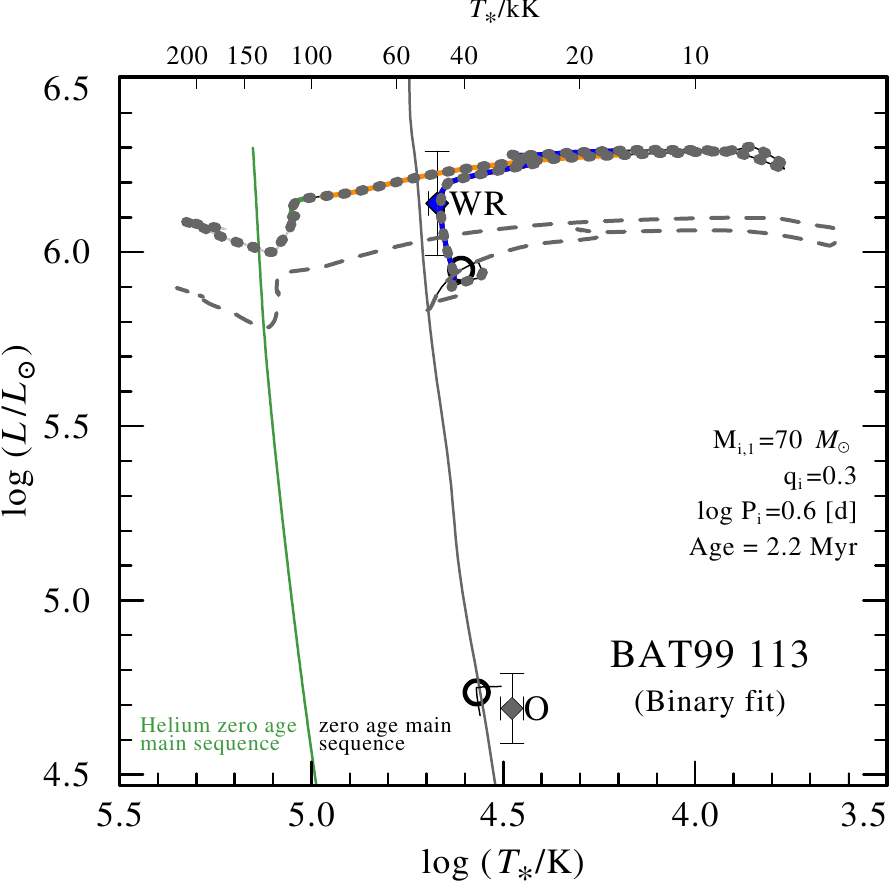}
%   \caption{s}
  \label{fig:BAT113hrdsub1}
\end{subfigure}%
\begin{subfigure}{.67\columnwidth}
  \centering
  \includegraphics[width=\linewidth]{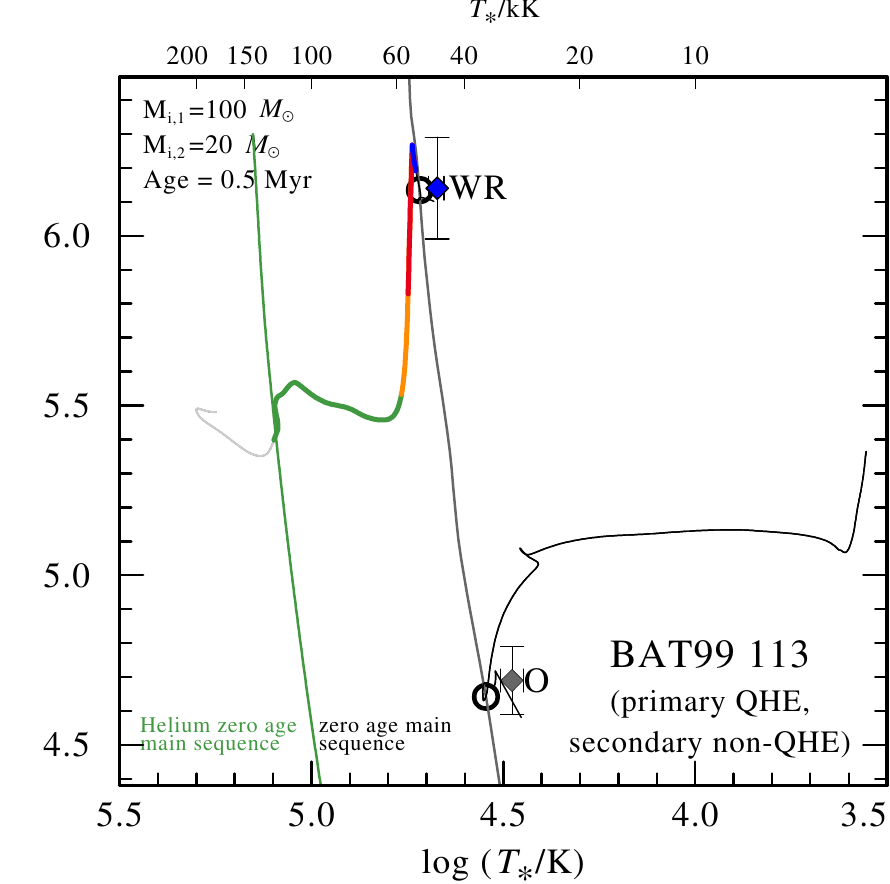}
%   \caption{A subfigure}
  \label{fig:BAT113hrdsub2}
\end{subfigure}
\begin{subfigure}{.67\columnwidth}
  \centering
  \includegraphics[width=\linewidth]{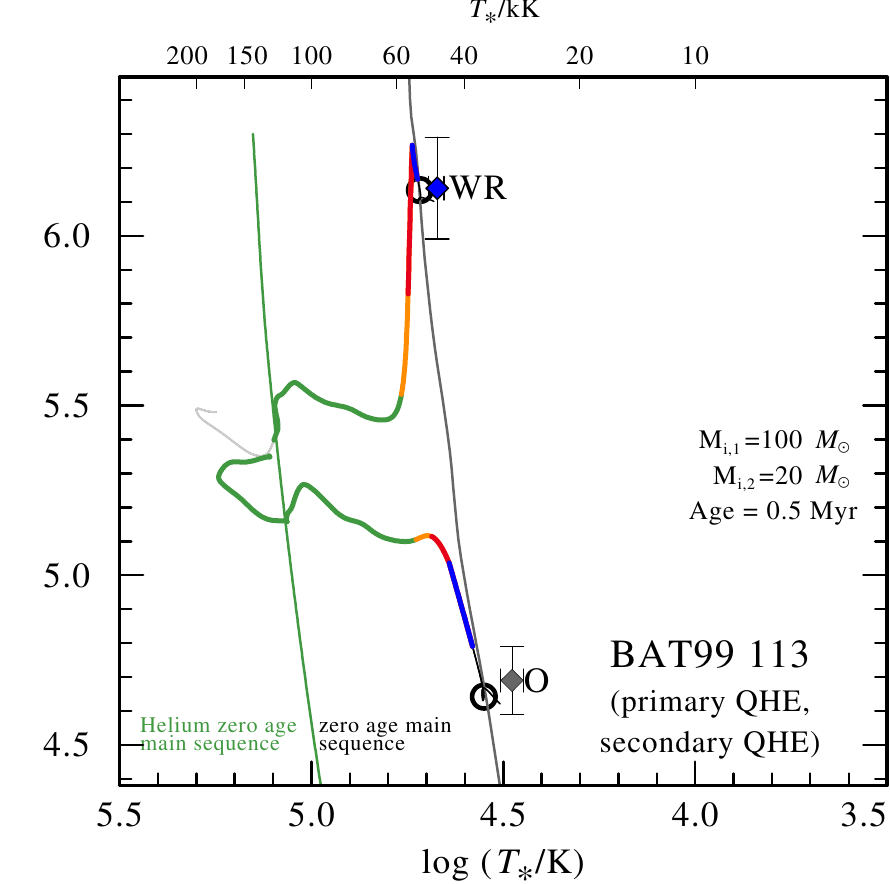}
%   \caption{A subfigure}
  \label{fig:BAT113hrdsub3}
\end{subfigure}
\caption{Fig.\,\ref{fig:BAT49hrd}, but for BAT99\,113.}
\label{fig:BAT113hrd}
\end{figure*}

Fig.\,\ref{fig:BAT113hrd} illustrates the best-fitting BPASS tracks of BAT99\,113 
for the three scenarios considered here. All relevant binary tracks end up merging shortly 
after the currently observed evolutionary phase. In contrast, a merger is avoided in the CHE 
scenario. It is not possible to determine which of these two fit the data better. Given 
its HRD position, however, we classify the WR primary as w/wb-WR.

\emph{\bf \object{BAT99\,114}} (\object{VFTS 545}, \object{MK 35}) is another transition type star with an identical spectral type to BAT99\,113 \citep{Crowther2011, Evans2011}.
The object is a binary candidate on the basis of its X-ray luminosity of $\log L_\text{X} = 33.4\,$[\lum] \citep{Guerrero2008}. 
However, while S08 could find an RV variability which exceeds their cutoff of 20\,\kms ($\sigma_\text{RV} = 23\,$\kms), they could not infer a period. 
H14 and \citet{Bestenlehner2014} provided a single-star fit to the object. Based on the single HST spectrum we possess for this object, we cannot 
find any clear indications for the presence of a binary companion in the system. We therefore omit this object from our sample.

\emph{\bf \object{BAT99\,116}} was classified as WN5h:a by S08, who detected 
non-periodic RV variability ($\sigma_\text{RV} = 33\,$\kms) for the object. It is one 
of the strongest X-ray sources among the WR stars in the LMC, with $\log L_\text{X} = 35.3$\,[\lum], and was hence considered a binary candidate by H14. 
\citet{Pollock2018} performed an analysis of a {\it Chandra} X-ray light curve and derived a period of $P{=}155.1$\,d for the system. The X-ray light curve 
is very suggestive of a WWC binary, implying that both components exhibit significant stellar winds.

Recently, \citet{Tehrani2019} published an orbital and spectroscopic analysis of BAT99\,116, finding 
it to be potentially the most massive binary weighed. 
Since these authors had much better spectra and a better phase coverage, we decide to adopt their parameters here. 
In Fig.\,\ref{fig:BAT116}, we show the PoWR models calculated with those parameters, showing that the agreement is good. 
Their orbital parameters were adopted in Table\,\ref{tab:orbitalpar}, as well as the light ratio of the two components. 
However, with the reddening law used here, we are led to lower luminosities of 
$\log L_1 = 6.31\,[L_\odot]$ and $\log L_2 = 6.20\,[L_\odot]$, which has some consequence 
on the mass. 
\citet{Tehrani2019} used a reddening law specifically derived for the cluster 30 Dorados by 
\citet{Maiz2014}. In this study, we chose to work with a homogeneous reddening law for the whole LMC, which may introduce some systematic differences 
between our study and that of \citet{Tehrani2019}. It is likely that the values derived by \citet{Tehrani2019} are more accurate and are therefore 
adopted here, but it is important to keep in mind that a systematic difference is possible, and hence lower luminosities (and masses). 
Since BAT99\,116 is reportedly the most massive binary ever weighed \citep{Tehrani2019}, it would be very important to attempt to derive 
its inclination independently, for example, through polarimetric studies.

The two WR components clearly belong to the ms-WR class. The future evolution of the system 
is thoroughly discussed by \citet{Tehrani2019}.

\emph{\bf \object{BAT99\,119}} (\object{RMC 145}, \object{VFTS 695})   is a WWC binary consisting of two massive ms-WR stars (WN6h + O3.5~If/WN7) 
that are members of an eccentric $158.7\,$d period binary. The system was thoroughly analysed by \citet{Shenar2017}, to which we refer for details.

\emph{\bf \object{BAT99\,126}} was classified as WN4b+OB by FMG03  and more recently to WN3+O7 by \citet{Neugent2018}. \citet{Testor1998} suggested the object is a binary based on its spectral appearance. FMG03 later 
detected a periodic RV variability with $P =25.5\,$d, although the RV scatter is comparable to their detection threshold. Moreover, the object portrays significant 
X-rays ($\log L_\text{X} \approx 33\,$[\lum], \citealt{Guerrero2008}), which is further suggestive of a binary nature.  Interestingly, the OGLE lightcurve of the system 
reveals that this object comprises a contact binary with a period of $P = 1.5\,$d. Whether or not the members of this contact system coincide with the WR binary is 
unclear. However, newly acquired UVES data should help uncover the true configuration of this important system in future work (Shenar et al.\ in prep.).

\begin{figure*}
\centering
\begin{subfigure}{0.67\columnwidth}
  \centering
  \includegraphics[width=\linewidth]{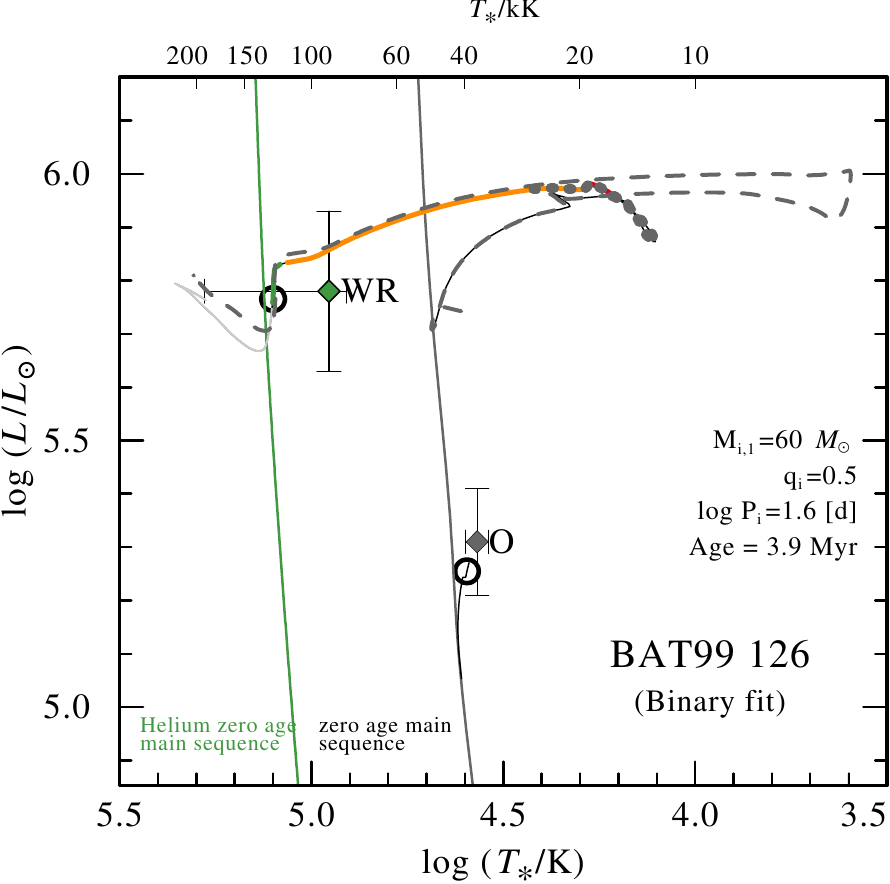}
%   \caption{s}
  \label{fig:BAT126hrdsub1}
\end{subfigure}%
\begin{subfigure}{.67\columnwidth}
  \centering
  \includegraphics[width=\linewidth]{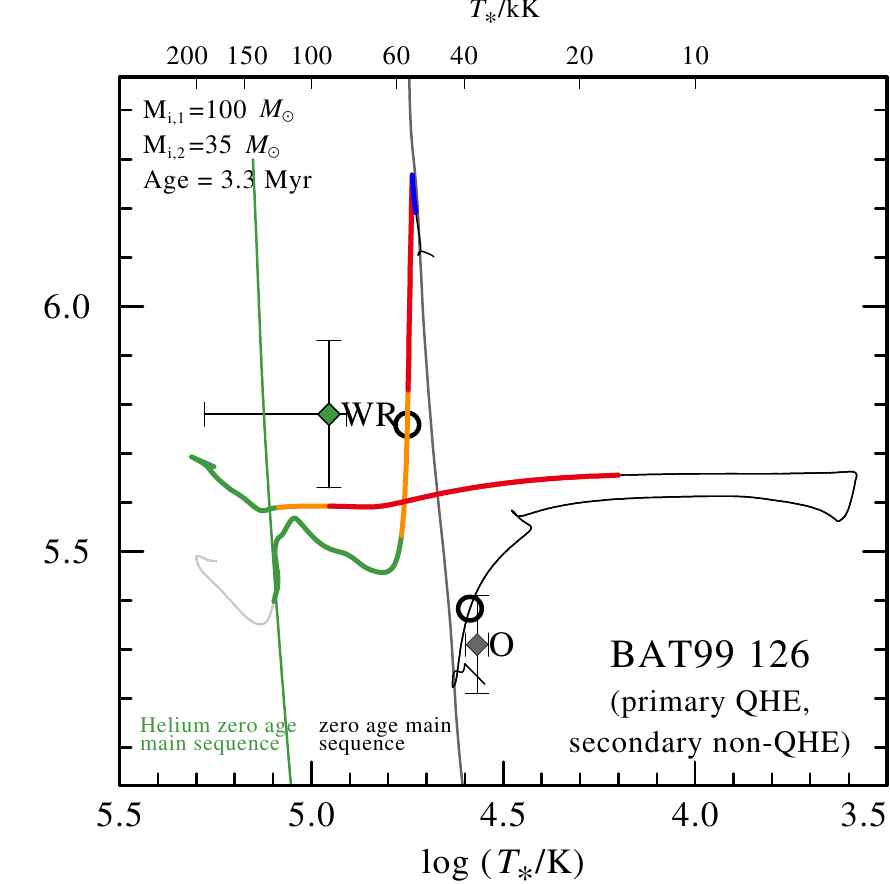}
%   \caption{A subfigure}
  \label{fig:BAT126hrdsub2}
\end{subfigure}
\begin{subfigure}{.67\columnwidth}
  \centering
  \includegraphics[width=\linewidth]{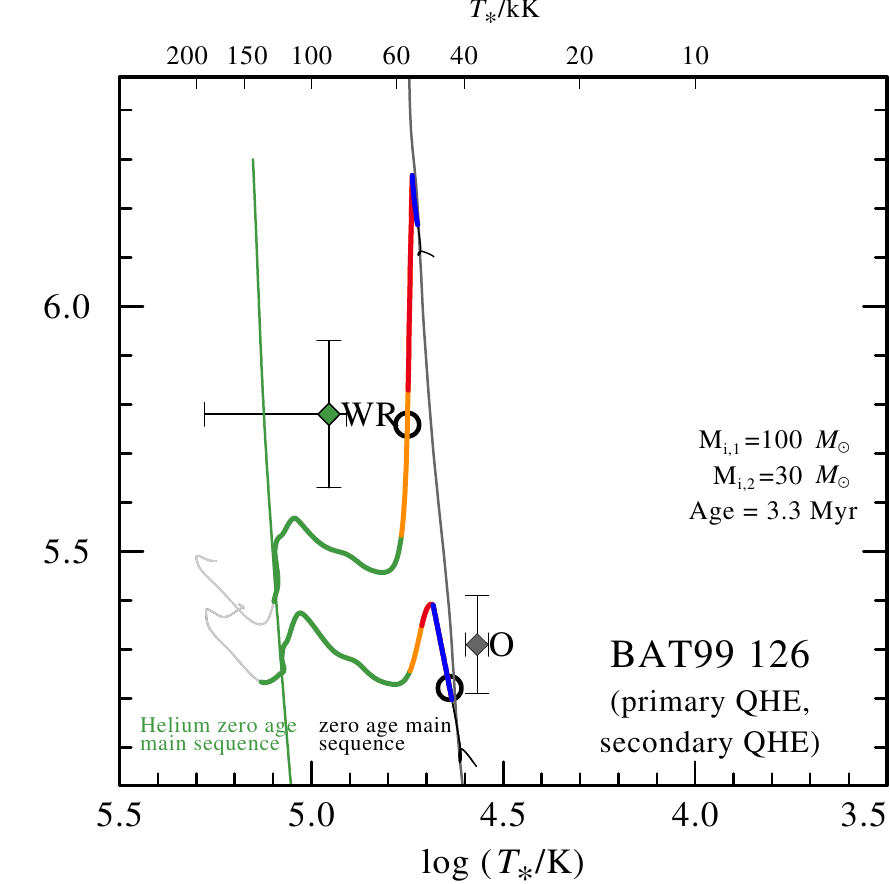}
%   \caption{A subfigure}
  \label{fig:BAT126hrdsub3}
\end{subfigure}
\caption{As Fig.\,\ref{fig:BAT49hrd}, but for BAT99\,126.}
\label{fig:BAT126hrd}
\end{figure*}

In the only spectrum at hand, we indeed find 
clear spectral signatures belonging to a secondary star in He\,{\sc i} lines, as well as He\,{\sc ii} lines, which are entangled with those of the WR primary. From the
ionization balance, we derive a temperature of $37\,$kK for the secondary. 
The model spectrum of the secondary gives the spectral type O6.5~V. However, since this system may be a triple, we caution that these results 
may be subject to systematic errors. Based on our comparisons with the BPASS tracks
(Fig.\,\ref{fig:BAT126hrd}), we classify the WR primary as w/wb-WR.

\emph{\bf \object{BAT99\,129}} is an eclipsing WR binary with a period of $2.8$\,d \citep{Wyrzykowski2003}. It 
was most recently classified as WN3(h)a + O5~V by \citet{Foellmi2006}, who derived an orbital solution for the system, and disentangled its spectrum. Thanks to these past efforts, 
the task of analysing the system was rather straight forward, and we find a good agreement with the light ratio reported by \citet{Foellmi2006}, and our derived 
stellar parameters agree well with their reported spectral types.

\begin{figure*}
\centering
\begin{subfigure}{0.67\columnwidth}
  \centering
  \includegraphics[width=\linewidth]{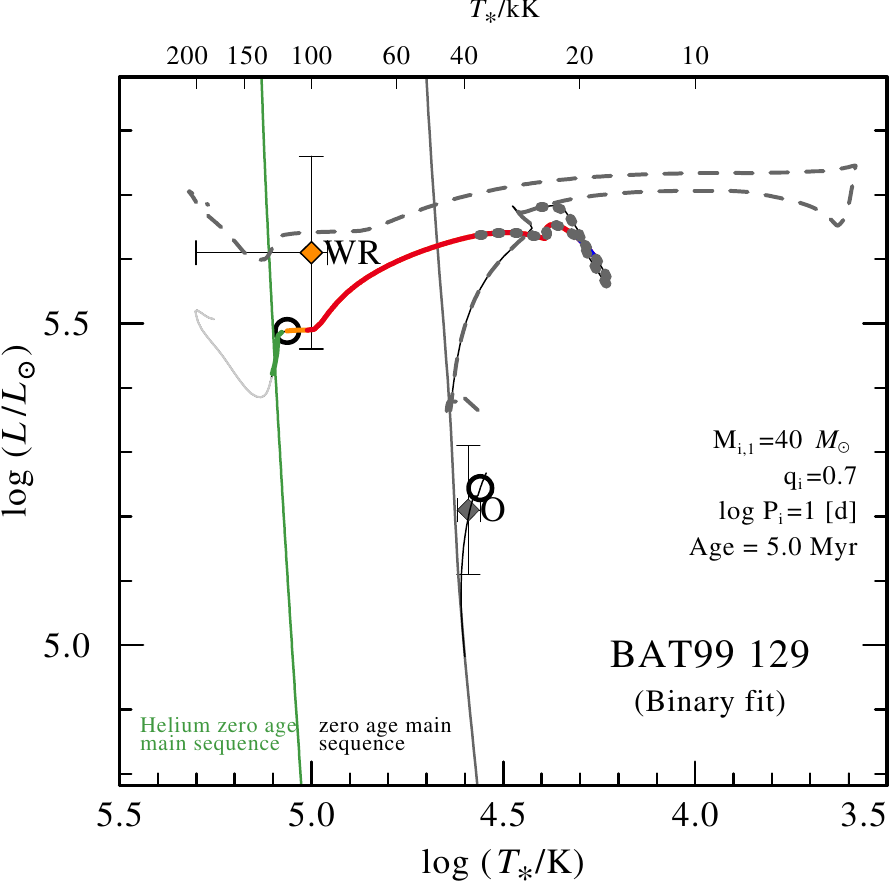}
%   \caption{s}
  \label{fig:BAT129hrdsub1}
\end{subfigure}%
\begin{subfigure}{.67\columnwidth}
  \centering
  \includegraphics[width=\linewidth]{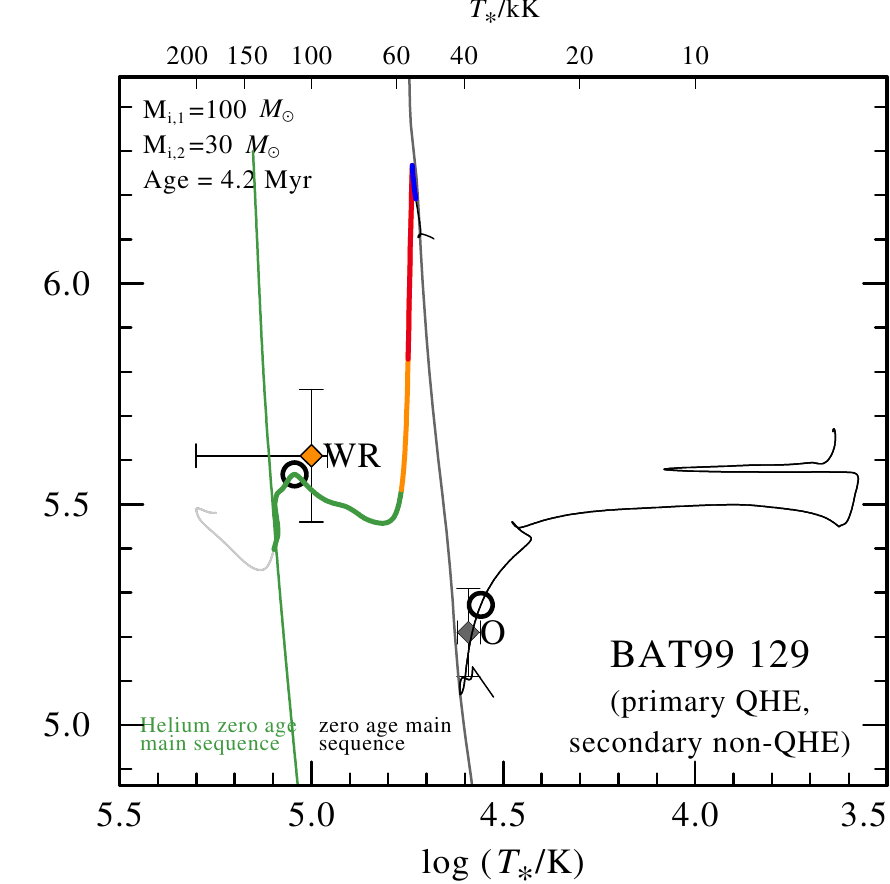}
%   \caption{A subfigure}
  \label{fig:BAT129hrdsub2}
\end{subfigure}
\begin{subfigure}{.67\columnwidth}
  \centering
  \includegraphics[width=\linewidth]{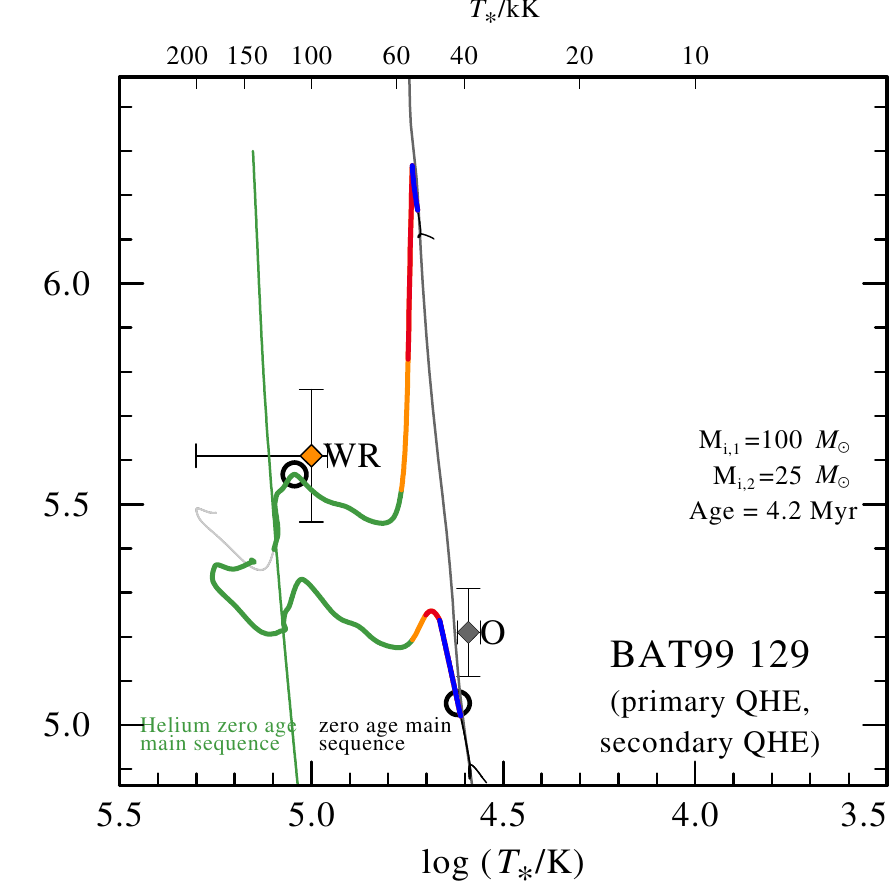}
%   \caption{A subfigure}
  \label{fig:BAT129hrdsub3}
\end{subfigure}
\caption{As Fig.\,\ref{fig:BAT49hrd}, but for BAT99\,129.}
\label{fig:BAT129hrd}
\end{figure*}

Only tracks that account for past mass-transfer can account for the properties of the system 
(Fig.\,\ref{fig:BAT129hrd}). Due to this, and since the WR primary is found to have had a large enough initial 
mass to enter the WR phase intrinsically, we classify it as wb-WR.

\section{Spectral fits}
\label{sec:specfits}

\include{specfits}

\section{Observation log}
\label{sec:obslog}

\setlength\tabcolsep{5pt}
\begin{table}
\caption{Observation log for Far UV}
\begin{center}
{\tiny
\begin{tabular}{c|ccccccc}
        \hline 
        \hline
        BAT99           &  Ins. & $R$ & S/N & MJD    & $\phi$ & RV1  & RV2 \\
        \hline

\multirow{1}{*}{6}       & FUSE & 20k & 50 & 52827.45 &  0.88    & (300)   & (300)     \\
\multirow{1}{*}{12}      & FUSE & 20k & 50 & 52478.86 &  0.61    & (450)   & -       \\
\multirow{1}{*}{14}      & -    & -   & -  & -        &  -       & -       & -       \\
\multirow{1}{*}{17}      & -    & -   & -  & -        &  -       & -       & -       \\
\multirow{1}{*}{19}      & FUSE & 20k & 50 & 53469.88 &  0.41    & (250)   & (250)       \\
\multirow{1}{*}{21}      & -    & -   & -  & -        &  -       & -       & -       \\
\multirow{1}{*}{27}      & FUSE & 20k & 50 & 52170.26 &  -       & (230)   & (230)      \\
\end{tabular}
}
\end{center}
\label{tab:ObsLogFUV}
\end{table}

\setlength\tabcolsep{5pt}
\begin{table}
\caption{Observation log for UV}
\begin{center}
{\tiny
\begin{tabular}{c|ccccccc}
        \hline 
        \hline
        BAT99            &  Ins. & $R$ & S/N & MJD    & $\phi$  & RV1  & RV2 \\
        \hline
\multirow{1}{*}{6}       &  IUE  & 15k &  10 & 43922.62 & 0.58    & (300)     & (300)     \\
\multirow{1}{*}{12}      & HST & 4k   & 100 & 52805.86 &  0.67      & (450)     & -     \\
\multirow{1}{*}{14}      & IUE & 1k & 50 & 48404.92  &  0.42      & (350     & (350)      \\
\multirow{1}{*}{17}      & IUE & 1k & 50 & 47286.91 &  -          & (300)   & (300)      \\
\multirow{1}{*}{19}      & IUE & 15k & 10 & 52827.45 &  0.88      & (270)   & (270)      \\
\multirow{1}{*}{21}      & -    & -   & -  & -        &  -       & -       & -          \\
\multirow{1}{*}{27}      & IUE  & 15k & 10 & 49607.00 &  -     & (230)      & (230)       \\
\end{tabular}
}
\end{center}
\label{tab:ObsLogUV}
\end{table}

\setlength\tabcolsep{2pt}
\begin{table}
\caption{Observation log for Optical}
\begin{center}
{\tiny
\begin{tabular}{c|ccccccc}
        \hline \hline
        BAT99                 &  Ins.\ & $R$ & S/N & MJD    & $\phi$   & RV1  & RV2 \\
        \hline

\multirow{1}{*}{6}       &  FEROS  & 48k &  100 & 53912.81 & 0.06  & (300)     & (300)     \\
                         &  FEROS  & 48k & 100  & 53913.90 & 0.61  & (300)     & (300)     \\                
\multirow{1}{*}{12}      & UVES & 40k  &  80    & 53338.28    &  0.21    & 440${\pm6}$\tablefootmark{a}  & -      \\
                         & CTIO & 2.4k & 110    & 53718.10    &  0.59    & 429${\pm5}$\tablefootmark{a}  & -      \\
                         & CTIO & 2.4k & 120    & 53720.08    &  0.20    & 438${\pm5}$\tablefootmark{a}  & -      \\                         
                         & CTIO & 2.4k & 170    & 53722.17    &  0.84    & 437${\pm5}$\tablefootmark{a}  & -      \\                  
\multirow{1}{*}{14}      & CTIO & 2.4k & 400    & co-added  &  -      & (250)     & -      \\
\end{tabular}
 }
\tablefoot{                                                                   
\tablefoottext{a}{1D cross-correlation in the region 4000-4600$\AA$} 
\tablefoottext{b}{2D cross-correlation in the region 4000-4600$\AA$} 
\tablefoottext{c}{1D cross-correlation of N\,{\sc iv}\,$\lambda 4060$} 
}      
\end{center}
\label{tab:ObsLogOpt1}
\end{table}

\setlength\tabcolsep{2pt}
\begin{table}
\caption{Observation log for Optical (continued)}
\begin{center}
{\tiny
\begin{tabular}{c|ccccccc}
        \hline \hline
        BAT99                 &  Ins.\ & $R$ & S/N & MJD    & $\phi$   & RV1  & RV2 \\
        \hline
\multirow{1}{*}{79}      & CTIO & 2.4k & 400    & co-added &  -      & (300)  & (300)      \\
\multirow{1}{*}{80}      & CTIO & 2.4k & 400    & co-added &  -      & (300)  & (300)      \\
\multirow{1}{*}{82}      & CTIO & 2.4k & 400    & co-added &  -      & - & -            \\
\multirow{1}{*}{86}      & CTIO & 2.4k & 400    & co-added &  -      & - & -            \\
\multirow{1}{*}{92}      & CTIO & 2.4k & 205    & 53718.26071  &  0.91  & 270$\pm11$\tablefootmark{d}    & 250$\pm3$\tablefootmark{a}  \\
                         & CTIO & 2.4k & 150    & 53719.27191  &  0.14  & 229$\pm17$\tablefootmark{d}    & 240$\pm3$\tablefootmark{a} \\                         
                         & CTIO & 2.4k & 195    & 53722.21941  &  0.83  & 294$\pm12$\tablefootmark{d}    & 247$\pm3$\tablefootmark{a}   \\
\end{tabular}
 }
\tablefoot{                                                                   
\tablefoottext{d}{1D cross-correlation with He\,{\sc ii}\,$\lambda 4686$} 
\tablefoottext{e}{2D cross-correlation with N\,{\sc iv}\,$\lambda 4060$} 
}      
\end{center}
\label{tab:ObsLogOpt2}
\end{table}

\setlength\tabcolsep{2pt}
\begin{table}
\caption{Observation log for Optical (continued)}
\begin{center}
{\tiny
\begin{tabular}{c|ccccccc}
        \hline \hline
        BAT99                 &  Ins.\ & $R$ & S/N & MJD    & $\phi$   & RV1  & RV2 \\
        \hline
\multirow{1}{*}{107}    &FLAMES& 7k   & 350  & 54748.265 &   0.59 &     306 $\pm10$\tablefootmark{b}  & 211  $\pm11$\tablefootmark{b}      \\
                         &FLAMES& 7k   & 310  & 54748.287  &  0.59 &     306 $\pm9$\tablefootmark{b}  & 215  $\pm10$\tablefootmark{b}       \\                         
                         &FLAMES& 7k   & 310  & 54748.311  &  0.59 &     311 $\pm9$\tablefootmark{b}  & 203  $\pm8$\tablefootmark{b}      \\                         
                         &FLAMES& 7k   & 60  & 54749.212  &  0.60 &      320 $\pm10$\tablefootmark{b}  & 225  $\pm10$\tablefootmark{b}                 \\
                         &FLAMES& 7k   & 290  & 54749.233  &  0.60 &     325 $\pm10$\tablefootmark{b}  & 225  $\pm11$\tablefootmark{b}                 \\
                         &FLAMES& 7k   & 280  & 54837.122  &  0.17 &     190 $\pm5$\tablefootmark{b}  &  330  $\pm6$\tablefootmark{b}       \\
                         &FLAMES& 7k   & 230  & 54837.143  &  0.17 &     190 $\pm6$\tablefootmark{b}  &  326  $\pm4$\tablefootmark{b}             \\
\end{tabular}
 }
\end{center}
\label{tab:ObsLogOpt3}
\end{table}
             
\setlength\tabcolsep{2pt}
\begin{table}
\caption{Observation log for Optical (continued)}
\begin{center}
{\tiny
\begin{tabular}{c|ccccccc}
        \hline \hline
        BAT99                 &  Ins.\ & $R$ & S/N & MJD    & $\phi$   & RV1  & RV2 \\
        \hline       
                         &FLAMES& 7k   & 230  & 56316.229 &  0.78 &     327  $\pm4$\tablefootmark{b}  &  203  $\pm5$\tablefootmark{b}      \\
                         &FLAMES& 7k   & 200  & 56347.013 &  0.98 &     177  $\pm5$\tablefootmark{b}  &  361  $\pm5$\tablefootmark{b}      \\
                         &FLAMES& 7k   & 230  & 56347.025 &  0.98 &     187  $\pm6$\tablefootmark{b}  &  365  $\pm5$\tablefootmark{b}      \\
                         &FLAMES& 7k   &  60  & 56347.037 &  0.98 &     187  $\pm5$\tablefootmark{b}  &  360  $\pm5$\tablefootmark{b}      \\
                         &FLAMES& 7k   & 190  & 56349.021 &  0.00 &     188  $\pm11$\tablefootmark{b} &  385  $\pm9$\tablefootmark{b}     \\
                         &FLAMES& 7k   & 190  & 56349.033 &  0.00 &     173  $\pm10$\tablefootmark{b} &  375  $\pm10$\tablefootmark{b}     \\
                         &FLAMES& 7k   & 240  & 56349.045 &  0.00 &     178  $\pm12$\tablefootmark{b} &  380  $\pm11$\tablefootmark{b}     \\
\end{tabular}
 }
\end{center}
\label{tab:ObsLogOpt2}
\end{table}
                         
\setlength\tabcolsep{2pt}
\begin{table}
\caption{Observation log for Optical (continued)}
\begin{center}
{\tiny
\begin{tabular}{c|ccccccc}
        \hline \hline
        BAT99                 &  Ins.\ & $R$ & S/N & MJD    & $\phi$   & RV1  & RV2 \\
        \hline     
                         &FLAMES& 7k   & 170  & 56719.021 &  0.40 &     269  $\pm10$\tablefootmark{b} &  278  $\pm10$\tablefootmark{b}     \\        
                         &FLAMES& 7k   & 100  & 56719.033 &  0.40 &     260  $\pm11$\tablefootmark{b} &  287  $\pm9$\tablefootmark{b}     \\
                         &FLAMES& 7k   & 190  & 56719.044 &  0.40 &     269  $\pm9$\tablefootmark{b} &  275  $\pm9$\tablefootmark{b}     \\
                         &FLAMES& 7k   & 150  & 56723.172 &  0.43 &     264  $\pm12$\tablefootmark{b} &  270  $\pm10$\tablefootmark{b}     \\
                         &FLAMES& 7k   & 180  & 56723.184 &  0.43 &     259  $\pm10$\tablefootmark{b} &  279  $\pm11$\tablefootmark{b}     \\
                         &FLAMES& 7k   & 170  & 56723.196 &  0.43 &     264  $\pm9$\tablefootmark{b} &  278  $\pm10$\tablefootmark{b}     \\
                        &X-SHOOTER& 7k   & 200  & 55588.19 &  0.05 &     123  $\pm4$\tablefootmark{b} &  420  $\pm4$\tablefootmark{b}        \\
\end{tabular}
 }
\tablefoot{                                                                   
\tablefoottext{f}{1D cross-correlation with He\,{\sc i}\,$\lambda 4472$} 
}       
\end{center}
\label{tab:ObsLogOpt4}
\end{table}

  % -----------------------------------  Table 1
\begin{table}
\caption{Summary of CTIO observations for BAT99-12 taken in the framework of this study. 
Parameters are: Number of observations $N$ per night, exposure time per single exposure, and SNR of nightly average.}
\begin{tabular}{llll}
\hline  \noalign{\smallskip}
Night &  Obs. & $t_{\text{exp}}$ & SNR\\
 Dec 2005 & N x& (sec) & 4260 $\pm$ 10\,\AA\\
 \noalign{\smallskip}\hline  \noalign{\smallskip}
 13. & 3 x& 1800 & 360\\
 14. & 0 x&  & \\
 15. & 3 x& 1500 & 330\\
 16. & 0 x&  & \\
 17. & 3 x& 1500 & 340\\
 \noalign{\smallskip} \hline
\end{tabular}
\label{tab12}
\end{table}

  % -----------------------------------  Table 2
\begin{table}
\caption{As Table \ref{tab12} but for BAT99-32.}
\begin{tabular}{llll}
\hline  \noalign{\smallskip}
Night &  Obs. & $t_{\text{exp}}$ & SNR\\
 Dec 2005 & N x& (sec) & 4260 $\pm$ 10\,\AA\\
 \noalign{\smallskip}\hline  \noalign{\smallskip}
 13. & 3 x& 750 & 450\\
 14. & 3 x& 750 & 460\\
 15. & 3 x& 750 & 320\\
 16. & 3 x& 750 & 380\\
 17. & 3 x& 750 & 520\\
 \noalign{\smallskip} \hline
\end{tabular}
\label{tab32}
\end{table}

  % -----------------------------------  Table 3
\begin{table}
\caption{As Table \ref{tab12} but for BAT99-77.}
\begin{tabular}{llll}
\hline  \noalign{\smallskip}
Night &  Obs. & $t_{\text{exp}}$ & SNR \\
 Dec 2005 & N x& (sec) & 4270 $\pm$ 10\,\AA\\
 \noalign{\smallskip}\hline  \noalign{\smallskip}
 13. & 3 x& 900, (2x)1500 & 380\\
 14. & 3 x& 900, (2x)1500 & 430\\
 15. & 2 x& 900, 1500 & 230\\
 16. & 0 x&  & \\
 17. & 3 x& 900, (2x)1500 & 400 \\
 \noalign{\smallskip} \hline
\end{tabular}
\label{tab77}
\end{table}

  % -----------------------------------  Table 4
\begin{table}
\caption{As Table \ref{tab12} but for BAT99-92.}
\begin{tabular}{llll}
\hline  \noalign{\smallskip}
Night &  Obs. & $t_{\text{exp}}$ & SNR \\
 Dec 2005 & N x& (sec) & 4050 $\pm$ 10\,\AA\\
 \noalign{\smallskip}\hline  \noalign{\smallskip}
 13. & 2 x& 450, 900 & 330\\
 14. & 2 x& 450, 900 & 330\\
 15. & 0 x&  & \\
 16. & 0 x&  & \\
 17. & 2 x& 700 & 270 \\
 \noalign{\smallskip} \hline
\end{tabular}
\label{tab92}
\end{table}

  % -----------------------------------  Table 5
\begin{table}
\caption{As Table \ref{tab12} but for BAT99-95.}
\begin{tabular}{llll}
\hline  \noalign{\smallskip}
Night &  Obs. & $t_{\text{exp}}$ & SNR \\
 Dec 2005 & N x& (sec) & 4270 $\pm$ 10\,\AA\\
 \noalign{\smallskip}\hline  \noalign{\smallskip}
 13. & 3 x& 1200, (2x)1000 & 360\\
 14. & 3 x& 1000 & 340\\
 15. & 0 x&  & \\
 16. & 0 x&  & \\
 17. & 3 x& 1000 & 350 \\
 \noalign{\smallskip} \hline
\end{tabular}
\label{tab95}
\end{table}

  % -----------------------------------  Table 6
\begin{table}
\caption{As Table \ref{tab12} but for BAT99-103.}
\begin{tabular}{llll}
\hline  \noalign{\smallskip}
Night &  Obs. & $t_{\text{exp}}$ & SNR \\
 Dec 2005 & N x& (sec) & 4270 $\pm$ 10\,\AA\\
 \noalign{\smallskip}\hline  \noalign{\smallskip}
 13. & 3 x& 1200 & 460\\
 14. & 3 x& 1200 & 320\\
 15. & 0 x&  & \\
 16. & 0 x&  & \\
 17. & 3 x& 1100 & 200 \\
 \noalign{\smallskip} \hline
\end{tabular}
\label{tab103}
\end{table}

  % -----------------------------------  Table 7
\begin{table}
\caption{As Table \ref{tab12} but for BAT99-113.}
\begin{tabular}{llll}
\hline  \noalign{\smallskip}
Night &  Obs. & $t_{\text{exp}}$ & SNR \\
 Dec 2005 & N x& (sec) & 4270 $\pm$ 10\,\AA\\
 \noalign{\smallskip}\hline  \noalign{\smallskip}
 13. & 3 x& 1200, (2x)1500 & 200\\
 14. & 0 x&  & \\
 15. & 3 x& 1200, (2x)1500 & 360\\
 16. & 0 x&  & \\
 17. & 3 x& 1200, (2x)1500 & 330 \\
 \noalign{\smallskip} \hline
\end{tabular}
\label{tab113}
\end{table}

\end{appendix} 

\end{document}